\documentstyle[10pt,epsf,epsfig,dp_delphititle]{dp_delphi}
\makeindex
\def\DpPaperGroup{PH--EP}
\def\DpPaperRef{2007--029}
\def\DpDate{9 July 2007}
\def\DpAuthors{DELPHI Collaboration}
\def\DpSubmit{(Accepted by Eur. Phys. J. C)}
\def\DpTitle{{
 Higgs boson searches in CP-conserving and CP-violating 
 MSSM scenarios with the DELPHI detector
}}
\def\DpComment{}
\def\DpEMail{}

\newcommand{\hp} {\mbox{$ {\mathrm H}^+ \,$}}
\newcommand{\hm} {\mbox{$ {\mathrm H}^- $}}

\newcommand{\MA} {\mbox{$ m_{\mathrm A} $}}
\newcommand{\MH} {\mbox{$ m_{\mathrm H} \, $}}
\newcommand{\mh} {\mbox{$ m_{\mathrm h} $}}
\newcommand{\mA} {\mbox{$ m_{\mathrm A} $}}
\newcommand{\mHp} {\mbox{$ m_{{\mathrm H}^{\pm}}$}}
\newcommand{\mH} {\mbox{$ m_{{\mathrm H}}$}}
\newcommand{\mHu} {\mbox{$ m_{{\mathrm H_1}}$}}
\newcommand{\mHd} {\mbox{$ m_{{\mathrm H_2}}$}}

\newcommand{\mtop} {\mbox{$ m_{\mathrm {top}} $}}

\newcommand{\HH}{\mbox{$ {\mathrm H}^+{\mathrm H}^- $}}
\newcommand{\WW} {\mbox{${\mathrm W}^+{\mathrm W}^-$}}
\newcommand{\ZZ} {\mbox{${\mathrm Z}{\mathrm Z}$}}
\newcommand{\HZ} {\mbox{${\mathrm {HZ}}$}}
\newcommand{\hA} {\mbox{$ {\mathrm {hA}}$}}
\newcommand{\HA} {\mbox{$ {\mathrm {HA}}$}}
\newcommand{\hZ} {\mbox{$ {\mathrm {hZ}}$}}
\newcommand{\HiZ} {\mbox{$ {\mathrm {H_i Z}}$}}
\newcommand{\HuZ} {\mbox{$ {\mathrm {H_1 Z}}$}}
\newcommand{\HdZ} {\mbox{$ {\mathrm {H_2 Z}}$}}
\newcommand{\HuHd} {\mbox{$ {\mathrm {H_1 H_2}}$}}
\newcommand{\HuHt} {\mbox{$ {\mathrm {H_1 H_3}}$}}
\newcommand{\gaga}{\mbox{$\gamma \gamma$}}
\newcommand{\ee}{\mbox{${\mathrm e}^+{\mathrm e}^-$}}
\newcommand{\llb}{\mbox{${\mathrm l}^+{\mathrm l}^-$}}
\newcommand{\mm}{\mbox{$\mu^+ \mu^-$}}
\newcommand{\mumu} {\mbox{$ \mu^+ \mu^- $}}

\newcommand{\qqbar}{\mbox{${\mathrm q}\bar{\mathrm q}$}}
\newcommand{\bbbar}{\mbox{${\mathrm b}\bar{\mathrm b}$}}
\newcommand{\ccbar}{\mbox{${\mathrm c}\bar{\mathrm c}$}}
\newcommand{\nunubar}{$\nu \bar{\nu}\;$}

\newcommand{\toto}{\mbox{$\tau^+ \tau^-$}}
\newcommand{\tautau}{\mbox{$\tau^+ \tau^- $}}

\newcommand{\nnqq}{$\nu\bar{\nu}{\mathrm q \bar{q}}$ }

\newcommand{\tautauqq}{\mbox{$\tau^+ \tau^- $\qqbar }}
\newcommand{\tautaubb}{\mbox{$\tau^+ \tau^- $\bbbar }}
\newcommand{\hAtt}{${\mathrm{hA}}\rightarrow$\toto \qqbar}

\newcommand{\Abb}{A $\rightarrow $ \bbbar}
\newcommand{\Acc}{A $\rightarrow $ \ccbar}
\newcommand{\Aany}{A $\rightarrow$ any}
\newcommand{\Agg}{A $\rightarrow \gamma \gamma $}
\newcommand{\App}{A $\rightarrow $ 2 prongs}
\newcommand{\Ahad}{A $\rightarrow $ hadrons}
\newcommand{\Att}{A $\rightarrow $ \toto}
\newcommand{\Zqq} {${\mathrm{Z}} \rightarrow$ \qqbar}
\newcommand{\Zany} {Z $\rightarrow$ any}
\newcommand{\Zanyt} {Z $\rightarrow$ any but \toto}
\newcommand{\Zll} {${\mathrm{Z}} \rightarrow$ \llb}
\newcommand{\Ztt} {${\mathrm{Z}} \rightarrow$ \toto}
\newcommand{\Zem} {${\mathrm{Z}} \rightarrow$ \ee, \mm}
\newcommand{\Znn} {${\mathrm{Z}} \rightarrow$ \nunubar}
\newcommand{\hAA} {${\mathrm{h}} \rightarrow {\mathrm{AA}}$}
\newcommand{\Hcas} {${\mathrm{H_2}} \rightarrow {\mathrm{H_1 H_1}}$}
\newcommand{\cascg} {${\mathrm{AA}} \rightarrow \gamma \gamma $}
\newcommand{\cascv} {${\mathrm{AA}} \rightarrow {\mathrm V}^0 {\mathrm V}^0$}
\newcommand{\cascp} {${\mathrm{AA}} \rightarrow$ 4 prongs}
\newcommand{\casca} {${\mathrm{AA}} \rightarrow $ any}
\newcommand{\casch} {${\mathrm{AA}} \rightarrow $ hadrons}
\newcommand{\casct} {${\mathrm{AA}} \rightarrow $ \toto \toto}
\newcommand{\cascb} {${\mathrm{AA}} \rightarrow $ \bbbar \bbbar}
\newcommand{\cascc} {${\mathrm{AA}} \rightarrow$ \ccbar \ccbar}
\newcommand{\hbb}{h $\rightarrow $ \bbbar}
\newcommand{\hqq}{h $\rightarrow $ \qqbar}
\newcommand{\htt}{h $\rightarrow $ \toto}
\newcommand{\hvz}{h $\rightarrow {\mathrm V}^0$}
\newcommand{\hpp}{h $\rightarrow $ 2 prongs}
\newcommand{\hmono}{h $\rightarrow $ jet}
\newcommand{\hjj}{h $\rightarrow $ jet jet}
\newcommand{\hhad}{h $\rightarrow $ hadrons}
\newcommand{\rgr}{$\rightarrow$}

\newcommand{\Hdbb}{H$_2$~\rgr~\bbbar}

\newcommand{\MeVcc} {\mbox{${\mathrm{MeV}}/c^2 $}}
\newcommand{\GeV} {\mbox{${\mathrm{GeV}} $}}
\newcommand{\GeVc} {\mbox{${\mathrm{GeV}}/c $}}
\newcommand{\GeVcc} {\mbox{${\mathrm{GeV}}/c^2 $}}
\newcommand{\TeVcc} {\mbox{${\mathrm{TeV}}/c^2 $}}

\newcommand{\cscs }{\mbox{$\mathrm{c \bar{s} \bar{c} s} \,$}}
\newcommand{\tntn }{\mbox{$\tau^+ \nu_{\tau} \tau^- {\bar{\nu}}_{\tau} \,$}}
\newcommand{\cstn }{\mbox{$\mathrm{c s} \tau \nu_{\tau} \,$}}
\newcommand{\watn }{\mbox{$\mathrm{W^*A} \tau \nu_{\tau} \,$}} 
\newcommand{\wawa }{\mbox{$\mathrm{W^*AW^*A} \,$}}

\newcommand{\lumi}{\mbox{$\cal L$}}

\newcommand{\likear}{\mbox{$\cal Q$}}

\newcommand{\CLb}{\mbox{$\rm CL_{\rm b}$}}
\newcommand{\CLs}{\mbox{$\rm CL_{\rm s}$}}
\newcommand{\CLsb}{\mbox{$\rm CL_{\rm s + b}$}}
\newcommand{\rs}{\mbox{$\sqrt{s}$}}

\newcommand{\tbeta} {\mbox{$\tan \beta$}}
\newcommand{\sinab} {\mbox{$\sin (\alpha-\beta)$}}
\newcommand{\cosab} {\mbox{$\cos (\alpha-\beta)$}}
\newcommand{\alp} {\mbox{$\alpha$}}

\def\EPJ#1#2#3{{ Eur. Phys. J.} {\bf{C#1}} (#2) #3}
\def\CPC#1#2#3{{ Comp. Phys. Comm.} {\bf{#1}} (#2) #3}
\def\NPB#1#2#3{{ Nucl.~Phys.} {\bf{B#1}} (#2) #3}
\def\PLB#1#2#3{{ Phys.~Lett.} {\bf{B#1}} (#2) #3}
\def\PRD#1#2#3{{ Phys.~Rev.} {\bf{D#1}} (#2) #3}

\def\ZPC#1#2#3{{ Z.~Phys.} {\bf C#1} (#2) #3}

\def\NIMA#1#2#3{{ Nucl.~Instr.~and~Meth.} {\bf A#1} (#2) #3}

\begin{document}
\makeatletter
\newcount\@tempcntc
\def\@citex[#1]#2{\if@filesw\immediate\write\@auxout{\string\citation{#2}}\fi
  \@tempcnta\z@\@tempcntb\m@ne\def\@citea{}\@cite{\@for\@citeb:=#2\do
    {\@ifundefined
       {b@\@citeb}{\@citeo\@tempcntb\m@ne\@citea\def\@citea{,}{\bf ?}\@warning
       {Citation `\@citeb' on page \thepage \space undefined}}%
    {\setbox\z@\hbox{\global\@tempcntc0\csname b@\@citeb\endcsname\relax}%
     \ifnum\@tempcntc=\z@ \@citeo\@tempcntb\m@ne
       \@citea\def\@citea{,}\hbox{\csname b@\@citeb\endcsname}%
     \else
      \advance\@tempcntb\@ne
      \ifnum\@tempcntb=\@tempcntc
      \else\advance\@tempcntb\m@ne\@citeo
      \@tempcnta\@tempcntc\@tempcntb\@tempcntc\fi\fi}}\@citeo}{#1}}
\def\@citeo{\ifnum\@tempcnta>\@tempcntb\else\@citea\def\@citea{,}%
  \ifnum\@tempcnta=\@tempcntb\the\@tempcnta\else
   {\advance\@tempcnta\@ne\ifnum\@tempcnta=\@tempcntb \else \def\@citea{--}\fi
    \advance\@tempcnta\m@ne\the\@tempcnta\@citea\the\@tempcntb}\fi\fi}
 
\makeatother
\begin{titlepage}
\pagenumbering{roman}
\CERNpreprint{\DpPaperGroup}{\DpPaperRef} 
  \begin{flushright}
  \large DAPNIA--07--150\\
  \end{flushright}
\date{{\small\DpDate}} 
\title{\DpTitle} 
\address{\DpAuthors} 

\begin{shortabs} 
\noindent
%
\noindent

This paper presents the final interpretation of the results from
{\sc DELPHI} on the searches for Higgs bosons in the Minimal 
Supersymmetric extension of the Standard Model ({\sc MSSM}). 
A few representative scenarios are considered, that include
CP conservation and explicit CP violation in the Higgs sector. 
The experimental results encompass the searches for
neutral Higgs bosons at {\sc LEP1} and {\sc LEP2} in final states as
expected in the {\sc MSSM}, as well as {\sc LEP2} searches for charged
Higgs bosons and for neutral Higgs bosons decaying into
hadrons independent of the quark flavour. 
The data reveal no significant excess with respect to
background expectations. The results are translated into excluded
regions of the parameter space in the various scenarios.
In the CP-conserving case, these lead to
limits on the masses of the lightest scalar and
pseudoscalar Higgs bosons, h and A, and on $\tan \beta$. The
dependence of these limits on the top quark mass is discussed.
Allowing for CP violation reduces the experimental sensitivity to
Higgs bosons. It is shown that this effect depends strongly
on the values of the parameters responsible for CP 
violation in the Higgs sector.

\end{shortabs}

\vfill

\begin{center}
\DpSubmit \ \\ 
\DpComment \ \\
\DpEMail \ \\
\end{center}
\vfill
\clearpage

\headsep 10.0pt

\addtolength{\textheight}{10mm}
\addtolength{\footskip}{-5mm}
\begingroup
%
\newcommand{\DpName}[2]{\hbox{#1$^{\ref{#2}}$},\hfill}
\newcommand{\DpNameTwo}[3]{\hbox{#1$^{\ref{#2},\ref{#3}}$},\hfill}
\newcommand{\DpNameThree}[4]{\hbox{#1$^{\ref{#2},\ref{#3},\ref{#4}}$},\hfill}
\newskip\Bigfill \Bigfill = 0pt plus 1000fill
\newcommand{\DpNameLast}[2]{\hbox{#1$^{\ref{#2}}$}\hspace{\Bigfill}}
%
\footnotesize
\noindent
\DpName{J.Abdallah}{LPNHE}
\DpName{P.Abreu}{LIP}
\DpName{W.Adam}{VIENNA}
\DpName{P.Adzic}{DEMOKRITOS}
\DpName{T.Albrecht}{KARLSRUHE}
\DpName{R.Alemany-Fernandez}{CERN}
\DpName{T.Allmendinger}{KARLSRUHE}
\DpName{P.P.Allport}{LIVERPOOL}
\DpName{U.Amaldi}{MILANO2}
\DpName{N.Amapane}{TORINO}
% ...
\normalsize
\endgroup
\newpage

\titlefoot{Department of Physics and Astronomy, Iowa State
     University, Ames IA 50011-3160, USA
    \label{AMES}}
\titlefoot{IIHE, ULB-VUB,
     Pleinlaan 2, B-1050 Brussels, Belgium
    \label{BRUSSELS}}
\titlefoot{Physics Laboratory, University of Athens, Solonos Str.
     104, GR-10680 Athens, Greece
    \label{ATHENS}}
\titlefoot{Department of Physics, University of Bergen,
     All\'egaten 55, NO-5007 Bergen, Norway
    \label{BERGEN}}
\titlefoot{Dipartimento di Fisica, Universit\`a di Bologna and INFN,
     Via Irnerio 46, IT-40126 Bologna, Italy
    \label{BOLOGNA}}
\titlefoot{Centro Brasileiro de Pesquisas F\'{\i}sicas, rua Xavier Sigaud 150,
     BR-22290 Rio de Janeiro, Brazil
    \label{BRASIL-CBPF}}
\titlefoot{Inst. de F\'{\i}sica, Univ. Estadual do Rio de Janeiro,
     rua S\~{a}o Francisco Xavier 524, Rio de Janeiro, Brazil
    \label{BRASIL-IFUERJ}}
\titlefoot{Coll\`ege de France, Lab. de Physique Corpusculaire, IN2P3-CNRS,
     FR-75231 Paris Cedex 05, France
    \label{CDF}}
\titlefoot{CERN, CH-1211 Geneva 23, Switzerland
    \label{CERN}}
\titlefoot{Institut de Recherches Subatomiques, IN2P3 - CNRS/ULP - BP20,
     FR-67037 Strasbourg Cedex, France
    \label{CRN}}
\titlefoot{Now at DESY-Zeuthen, Platanenallee 6, D-15735 Zeuthen, Germany
    \label{DESY}}
\titlefoot{Institute of Nuclear Physics, N.C.S.R. Demokritos,
     P.O. Box 60228, GR-15310 Athens, Greece
    \label{DEMOKRITOS}}
\titlefoot{FZU, Inst. of Phys. of the C.A.S. High Energy Physics Division,
     Na Slovance 2, CZ-180 40, Praha 8, Czech Republic
    \label{FZU}}
\titlefoot{Dipartimento di Fisica, Universit\`a di Genova and INFN,
     Via Dodecaneso 33, IT-16146 Genova, Italy
    \label{GENOVA}}
\titlefoot{Institut des Sciences Nucl\'eaires, IN2P3-CNRS, Universit\'e
     de Grenoble 1, FR-38026 Grenoble Cedex, France
    \label{GRENOBLE}}
\titlefoot{Helsinki Institute of Physics and Department of Physical Sciences,
     P.O. Box 64, FIN-00014 University of Helsinki, 
     \indent~~Finland
    \label{HELSINKI}}
\titlefoot{Joint Institute for Nuclear Research, Dubna, Head Post
     Office, P.O. Box 79, RU-101 000 Moscow, Russian Federation
    \label{JINR}}
\titlefoot{Institut f\"ur Experimentelle Kernphysik,
     Universit\"at Karlsruhe, Postfach 6980, DE-76128 Karlsruhe,
     Germany
    \label{KARLSRUHE}}
\titlefoot{Institute of Nuclear Physics PAN,Ul. Radzikowskiego 152,
     PL-31142 Krakow, Poland
    \label{KRAKOW1}}
\titlefoot{Faculty of Physics and Nuclear Techniques, University of Mining
     and Metallurgy, PL-30055 Krakow, Poland
    \label{KRAKOW2}}
\titlefoot{Universit\'e de Paris-Sud, Lab. de l'Acc\'el\'erateur
     Lin\'eaire, IN2P3-CNRS, B\^{a}t. 200, FR-91405 Orsay Cedex, France
    \label{LAL}}
\titlefoot{School of Physics and Chemistry, University of Lancaster,
     Lancaster LA1 4YB, UK
    \label{LANCASTER}}
\titlefoot{LIP, IST, FCUL - Av. Elias Garcia, 14-$1^{o}$,
     PT-1000 Lisboa Codex, Portugal
    \label{LIP}}
\titlefoot{Department of Physics, University of Liverpool, P.O.
     Box 147, Liverpool L69 3BX, UK
    \label{LIVERPOOL}}
\titlefoot{Dept. of Physics and Astronomy, Kelvin Building,
     University of Glasgow, Glasgow G12 8QQ, UK
    \label{GLASGOW}}
\titlefoot{LPNHE, IN2P3-CNRS, Univ.~Paris VI et VII, Tour 33 (RdC),
     4 place Jussieu, FR-75252 Paris Cedex 05, France
    \label{LPNHE}}
\titlefoot{Department of Physics, University of Lund,
     S\"olvegatan 14, SE-223 63 Lund, Sweden
    \label{LUND}}
\titlefoot{Universit\'e Claude Bernard de Lyon, IPNL, IN2P3-CNRS,
     FR-69622 Villeurbanne Cedex, France
    \label{LYON}}
\titlefoot{Dipartimento di Fisica, Universit\`a di Milano and INFN-MILANO,
     Via Celoria 16, IT-20133 Milan, Italy
    \label{MILANO}}
\titlefoot{Dipartimento di Fisica, Univ. di Milano-Bicocca and
     INFN-MILANO, Piazza della Scienza 3, IT-20126 Milan, Italy
    \label{MILANO2}}
\titlefoot{IPNP of MFF, Charles Univ., Areal MFF,
     V Holesovickach 2, CZ-180 00, Praha 8, Czech Republic
    \label{NC}}
\titlefoot{NIKHEF, Postbus 41882, NL-1009 DB
     Amsterdam, The Netherlands
    \label{NIKHEF}}
\titlefoot{National Technical University, Physics Department,
     Zografou Campus, GR-15773 Athens, Greece
    \label{NTU-ATHENS}}
\titlefoot{Physics Department, University of Oslo, Blindern,
     NO-0316 Oslo, Norway
    \label{OSLO}}
\titlefoot{Dpto. Fisica, Univ. Oviedo, Avda. Calvo Sotelo
     s/n, ES-33007 Oviedo, Spain
    \label{OVIEDO}}
\titlefoot{Department of Physics, University of Oxford,
     Keble Road, Oxford OX1 3RH, UK
    \label{OXFORD}}
\titlefoot{Dipartimento di Fisica, Universit\`a di Padova and
     INFN, Via Marzolo 8, IT-35131 Padua, Italy
    \label{PADOVA}}
\titlefoot{Rutherford Appleton Laboratory, Chilton, Didcot
     OX11 OQX, UK
    \label{RAL}}
\titlefoot{Dipartimento di Fisica, Universit\`a di Roma II and
     INFN, Tor Vergata, IT-00173 Rome, Italy
    \label{ROMA2}}
\titlefoot{Dipartimento di Fisica, Universit\`a di Roma III and
     INFN, Via della Vasca Navale 84, IT-00146 Rome, Italy
    \label{ROMA3}}
\titlefoot{DAPNIA/Service de Physique des Particules,
     CEA-Saclay, FR-91191 Gif-sur-Yvette Cedex, France
    \label{SACLAY}}
\titlefoot{Instituto de Fisica de Cantabria (CSIC-UC), Avda.
     los Castros s/n, ES-39006 Santander, Spain
    \label{SANTANDER}}
\titlefoot{Inst. for High Energy Physics, Serpukov
     P.O. Box 35, Protvino, (Moscow Region), Russian Federation
    \label{SERPUKHOV}}
\titlefoot{J. Stefan Institute, Jamova 39, SI-1000 Ljubljana, Slovenia
    \label{SLOVENIJA1}}
\titlefoot{Laboratory for Astroparticle Physics,
     University of Nova Gorica, Kostanjeviska 16a, SI-5000 Nova Gorica, Slovenia
    \label{SLOVENIJA2}}
\titlefoot{Department of Physics, University of Ljubljana,
     SI-1000 Ljubljana, Slovenia
    \label{SLOVENIJA3}}
\titlefoot{Fysikum, Stockholm University,
     Box 6730, SE-113 85 Stockholm, Sweden
    \label{STOCKHOLM}}
\titlefoot{Dipartimento di Fisica Sperimentale, Universit\`a di
     Torino and INFN, Via P. Giuria 1, IT-10125 Turin, Italy
    \label{TORINO}}
\titlefoot{INFN,Sezione di Torino and Dipartimento di Fisica Teorica,
     Universit\`a di Torino, Via Giuria 1,
     IT-10125 Turin, Italy
    \label{TORINOTH}}
\titlefoot{Dipartimento di Fisica, Universit\`a di Trieste and
     INFN, Via A. Valerio 2, IT-34127 Trieste, Italy
    \label{TRIESTE}}
\titlefoot{Istituto di Fisica, Universit\`a di Udine and INFN,
     IT-33100 Udine, Italy
    \label{UDINE}}
\titlefoot{Univ. Federal do Rio de Janeiro, C.P. 68528
     Cidade Univ., Ilha do Fund\~ao
     BR-21945-970 Rio de Janeiro, Brazil
    \label{UFRJ}}
\titlefoot{Department of Radiation Sciences, University of
     Uppsala, P.O. Box 535, SE-751 21 Uppsala, Sweden
    \label{UPPSALA}}
\titlefoot{IFIC, Valencia-CSIC, and D.F.A.M.N., U. de Valencia,
     Avda. Dr. Moliner 50, ES-46100 Burjassot (Valencia), Spain
    \label{VALENCIA}}
\titlefoot{Institut f\"ur Hochenergiephysik, \"Osterr. Akad.
     d. Wissensch., Nikolsdorfergasse 18, AT-1050 Vienna, Austria
    \label{VIENNA}}
\titlefoot{Inst. Nuclear Studies and University of Warsaw, Ul.
     Hoza 69, PL-00681 Warsaw, Poland
    \label{WARSZAWA}}
\titlefoot{Now at University of Warwick, Coventry CV4 7AL, UK
    \label{WARWICK}}
\titlefoot{Fachbereich Physik, University of Wuppertal, Postfach
     100 127, DE-42097 Wuppertal, Germany \\
\noindent
{$^\dagger$~deceased}
    \label{WUPPERTAL}}
\addtolength{\textheight}{-10mm}
\addtolength{\footskip}{5mm}
\clearpage
\headsep 30.0pt
\end{titlepage}
%
\pagenumbering{arabic} 
\setcounter{footnote}{0} %
\large

%
\section{Introduction}

   This paper presents the final interpretation of the Higgs boson
search results from {\sc DELPHI} in the framework of representative
scenarios of the
Minimal Supersymmetric Standard Model ({\sc MSSM}).
With respect to the previous {\sc MSSM} interpretation published 
in Ref.~\cite{ref:pap00}, this analysis uses an enlarged set of 
experimental results, updated calculations of
{\sc MSSM} radiative corrections and covers more
scenarios, including models with {\sc CP} violation in the Higgs sector.

As compared with the Standard Model, the {\sc MSSM}
has an extended Higgs sector with two doublets of complex Higgs fields,
leading to five physical Higgs bosons, of which three are neutral. 

\begin{figure}[htbp]
\begin{center}
\begin{tabular}{cc}
\epsfig{figure=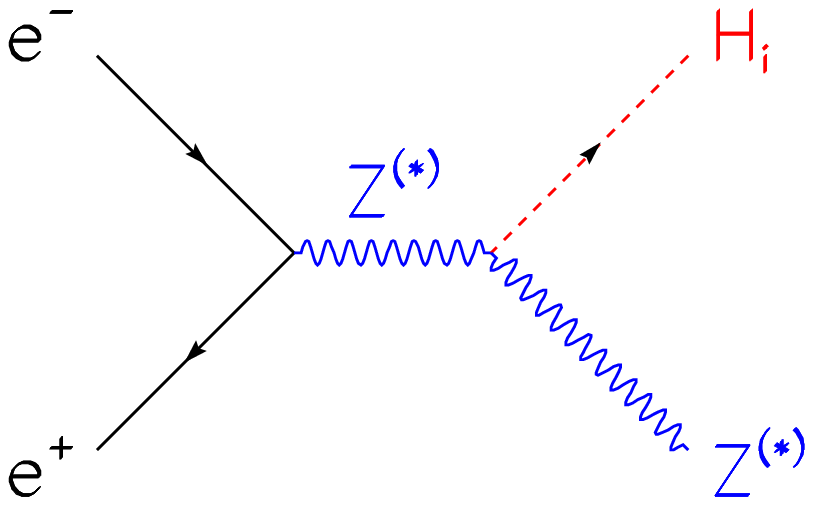,height=5cm} &
\epsfig{figure=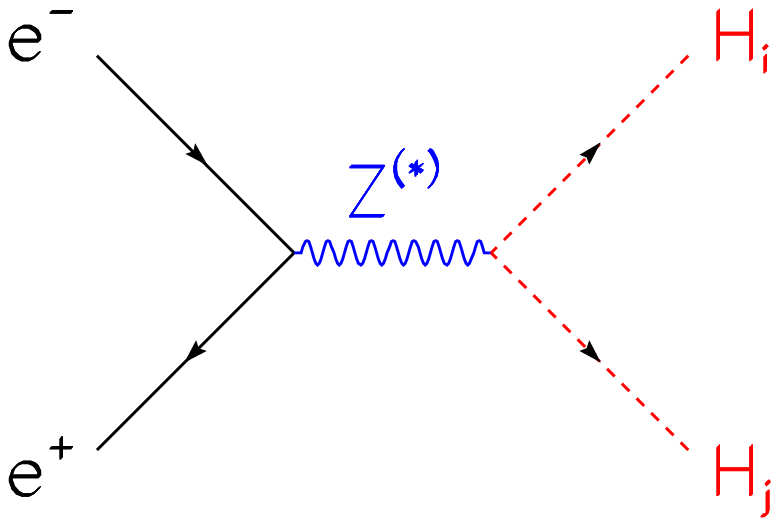,height=5cm} \\
\end{tabular}
\caption[]{
    Main production processes of {\sc MSSM} neutral Higgs bosons
    at {\sc LEP}. Left: associated production of a Z and a  
    Higgs boson, which must be one of the {\sc CP}-even scalars (h or H)
    if {\sc CP} is conserved or any Higgs boson (H$_1$, H$_2$, H$_3$) 
    in the contrary case. 
    At {\sc LEP1}, the intermediate Z is on-shell and the
    final Z is off-shell, while it is the reverse at {\sc LEP2}.
    Right: pair-production of neutral Higgs bosons. 
    If {\sc CP} is conserved, one of them must
    be  {\sc CP}-even (h or H) and the other one is the
    {\sc CP}-odd pseudo-scalar A. If {\sc CP} is not conserved, 
    the pair can be any couple of different scalars
    among H$_1$, H$_2$ and H$_3$.
    The intermediate Z is on-shell at {\sc LEP1}.}
\label{fig:diagrams}
\end{center}
\end{figure}

If {\sc CP} is conserved, two of the three 
neutral Higgs bosons are CP-even. They are denoted h, for the lighter one, 
and H. The third one is a  {\sc CP}-odd pseudo-scalar, denoted A. 
In \ee\ collisions, the dominant production mechanisms are 
the s-channel processes described in Fig.~\ref{fig:diagrams}, that
is the associated production of a Z and a  {\sc CP}-even Higgs boson
and the pair production of either {\sc CP}-even boson together
with the {\sc CP}-odd scalar. 
These processes are complemented by additional
t-channel diagrams in the final states where a {\sc CP}-even Higgs boson 
is produced with neutrinos or electrons, which proceed through \WW\ and 
\ZZ\ fusions, respectively. These diagrams and their interference with 
the \HiZ\ process have an impact on the production cross-section at
masses around the \HiZ\ kinematic threshold. 
At {\sc LEP2} energies, the only significant effect is from \WW\ fusion
which doubles the neutrino \HiZ\ cross-section at the kinematic 
threshold.
Finally, charged Higgs bosons, \hp\ and \hm, are 
produced in pairs through a diagram similar to that in  
Fig.~\ref{fig:diagrams}, right, via exchange of a Z boson or a photon.

\begin{figure}[htbp]
\begin{center}
\begin{tabular}{c}
\epsfig{figure=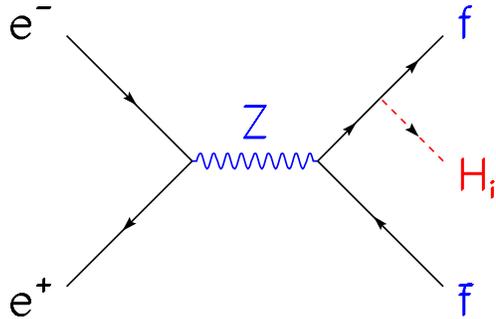,height=5cm} 
\end{tabular}
\caption[]{
    Additional production process of {\sc MSSM} neutral Higgs bosons
    at {\sc LEP}. The radiation of a Higgs boson off a Z boson
    decay fermion gives a detectable signal only at {\sc LEP1}.
    This signal is exploited in the case of {\sc CP} violation.}
\label{fig:yukawa}
\end{center}
\end{figure}

Although {\sc CP} is conserved at tree level in the {\sc MSSM}, radiative
corrections can introduce {\sc CP} violation through stop and sbottom loops,
leading to
changes in the neutral Higgs boson sector~\cite {ref:mssmpheno}.
If {\sc CP} is not conserved,
the three neutral Higgs bosons are no longer pure {\sc CP} eigenstates
but mixtures of {\sc CP}-even and {\sc CP}-odd components. They
are usually denoted H$_1$, H$_2$ and H$_3$, in increasing mass. 
The main production mechanisms are the same as in the {\sc CP} conserving 
case, except that, {\it a priori}, any scalar
can be produced in association with a Z boson or through \WW\ and 
\ZZ\ fusions, and any couple of different Higgs bosons can be pair-produced. 
The main phenomenological difference with respect to the 
{\sc CP}-conserving case
lies in the strength of the couplings of the Z boson to the Higgs scalars. 
In significant regions of the parameter space, {\sc CP} violation turns off
the otherwise dominant coupling between the Z boson and the lightest
Higgs boson. 
In that case, if none of the other processes of 
Fig.~\ref{fig:diagrams} are possible (due e.g. to kinematics), 
the dominant Higgs boson production mechanism at {\sc LEP} becomes the Yukawa 
process of Fig.~\ref{fig:yukawa}. Of the two phases of {\sc LEP}, only
{\sc LEP1} has a significant sensitivity to this process. 
In the Standard Model, the corresponding cross-sections are negligible, 
e.g. a fraction of a pb  for a few~\GeVcc\ Higgs boson. 
In the {\sc MSSM}, these can be 
enhanced by up to three orders of magnitude with respect to their 
Standard Model values, leading to detectable signals which become valuable
in the case of {\sc CP} violation.

  The decay properties of the Higgs bosons are moderately
affected by {\sc CP} violation, at least in the 
range of masses accessible at {\sc LEP}, that is up to masses around
100~\GeVcc~\cite {ref:mssmpheno}. In most of the {\sc MSSM}
parameter space of the scenarios studied hereafter, the 
three neutral Higgs bosons decay mainly into the pair of heaviest fermions 
kinematically permitted, even if {\sc CP} is not conserved.
Below the \mumu\ threshold,
a Higgs boson would decay into \gaga\ or \ee\ pairs with a significant
lifetime. Above the \mumu\ threshold, the lifetime is negligible and 
Higgs bosons decay at the primary vertex. 
Up to a mass of 3~\GeVcc\, the main decays are
into \mumu\ pairs and also into hadronic channels with a large proportion
of two-prong final states. Above 3~\GeVcc\, the dominant decays are 
successively into \ccbar, \tautau\ and finally \bbbar\ pairs for
Higgs boson masses above 12~\GeVcc.
Besides these decays into fermions, there 
are also regions of the parameter space where one neutral Higgs boson 
can undergo cascade decays to a pair of Higgs bosons, as for 
example \hAA\ if {\sc CP} is conserved or \Hcas\ in the contrary case.
In some cases, especially if {\sc CP} is not conserved,
this mode dominates over the decays into 
{\sc SM} particles. 
In the scenarios considered in this paper, charged Higgs bosons have a mass
above 60~\GeVcc\ and 
decay either into the pair of heaviest fermions allowed by kinematics, that is
into cs or $\tau\nu$ pairs, or into a W$^*$ and a light Higgs
boson, e.g. into a W$^*$A pair if {\sc CP} is conserved.
Finally, these scenarios do not allow neutral or charged Higgs boson decays 
into supersymmetric particles such as sfermions, charginos or invisible 
neutralinos. Note that searches for neutral Higgs bosons decaying into 
invisible products were performed at {\sc LEP}, as 
reported in Ref.~\cite{ref:invis}.

The different decay channels define the topologies that were
searched for to cover the {\sc MSSM} parameter region
kinematically accessible at {\sc LEP} energies. 
These topologies are described in Section~\ref{sec:channels}. 
Section~\ref{sec:tools} summarizes 
the definition and techniques related to 
confidence levels used in the statistical interpretation of the searches. 
The eight 
{\sc CP}-conserving {\sc MSSM} benchmark scenarios studied in this analysis are
presented  in Section~\ref{sec:benchmark} and the results obtained
in these scenarios when combining all searches are given in 
Section~\ref{sec:results}. Similarily, the {\sc CP}-violating scenarios
and the corresponding results are covered in Sections~\ref{sec:cpv}
and ~\ref{sec:resucpv}.
The top quark mass has a significant impact on the properties
of the Higgs bosons (e.g. mass spectrum of the neutral Higgs bosons, 
{\sc CP}-violating effects). Results are thus derived for several values 
of this mass, namely: \mtop~=~169.2, 174.3, 179.4 and 183.0~\GeVcc, which
were defined after the measurement of the top quark mass
at the Tevatron, run I~\cite{ref:mtop_old}.
Of the two values close to the present experimental measurement
of \mtop~=~170.9~$\pm$~1.1~$\pm$~1.5~\GeVcc~\cite{ref:mtop}, 
174.3~\GeVcc\ gives the most conservative results 
and thus was chosen as
a reference in most of the exclusion plots and to quote absolute mass 
and \tbeta\ limits. 
Readers interested in similar analyses at {\sc LEP}
are referred to Ref.~\cite{ref:opal,ref:hwg}.

\section{Search channels}\label{sec:channels}

\begin{table}[p]
{\normalsize
\vspace{-1.5cm}
\begin{center}
\begin{tabular}{clcccc}     \hline 
\rs & final state & range & \lumi\      & disc. & ref. \\
(GeV) &           & (\GeVcc)   & (pb$^{-1}$) & info. & \\ \hline
 \multicolumn{6}{c}{\hZ\ with direct decays} \\ \hline
91 &  \Zem
          & $< 0.21$   & 2.5 & no & \cite{ref:vlow} \\
91 &  (\hvz) (\Zany)               
          & $< 0.21$   & 2.5 & no & \cite{ref:vlow} \\
91 &  (\hpp) (\Zqq)
          & $0.21 - 2.$  & 0.5 & no & \cite{ref:low} \\
91 &  (\hmono) (\Zem)
          & $1. - 20.$ & 0.5 & no & \cite{ref:low} \\
91 &  (\hjj) (\Zll, \nunubar)
          & $>12.$ & 3.6 & no & \cite{ref:jjlow} \\
91 &  (\hjj) (\Zem, \nunubar) 
          & $>35.$ & 33.4 & no & \cite{ref:jjhigh} \\
\bf{161,172} & (\hbb)(\Zany), (\htt)(\Zqq)  
            & $>40.$ & 19.9 & 1d &\cite{ref:pap96} \\
\bf{183} &   (\hbb)(\Zany), (\htt)(\Zqq)  
            & $>55.$ & 52.0 & 1d & \cite{ref:pap97} \\
\bf{189} &   (\hbb)(\Zany), (\htt)(\Zqq)   
            & $>65.$ & 158.0 & 2d &\cite{ref:pap98} \\
\bf{192-208} & (\hbb)(\Zany)   
            & $>12.$ & 452.4 & 2d &\cite{ref:pap99,ref:pap00} \\
\bf{192-208} & (\htt)(\Zqq)   
            & $>50.$ & 452.4 & 2d &\cite{ref:pap99,ref:pap00} \\
189-208 & (\hhad)(\Zany\ but \tautau)
            & $>4.$  & 610.4 & mix &\cite{ref:papflbl} \\
\hline
\multicolumn{6}{c}{\hA\ with direct decays} \\\hline
91        & 4 prongs       
            & $>0.4$ & 5.3 & no & \cite{ref:4prongs} \\
91        & \toto\ hadrons       
            & $>8.$ & 0.5 & no & \cite{ref:ha89} \\
91        & \toto\ jet jet       
            & $>50$ & 3.6 & no & \cite{ref:jjlow} \\
91        & \bbbar \bbbar, \bbbar \ccbar 
            & $>30.$ & 33.4 & no & \cite{ref:ha92} \\
91        & \toto \bbbar 
            & $>16.$ & 79.4 & no & A.1 \\
91        & \bbbar \bbbar 
            & $>24.$ & 79.4 & no & \cite{ref:2hdm} \\
\bf{133}       &  \bbbar \bbbar
            & $>80.$ & 6.0 & no & \cite{ref:pap95} \\
\bf{161,172} &  \bbbar \bbbar, \toto \bbbar   
            & $>80.$ & 20.0 & 1d &\cite{ref:pap96} \\
\bf{183}       &  \bbbar \bbbar, \toto \bbbar   
            & $>100.$ & 54.0 & 1d & \cite{ref:pap97} \\
\bf{189}       &  \bbbar \bbbar, \toto \bbbar  
            & $>130.$ & 158.0 & 2d &\cite{ref:pap98} \\
\bf{192-208} &  \toto  \bbbar  
            & $>120.$ & 452.4 & 2d &\cite{ref:pap99,ref:pap00} \\
\bf{192-208} &  \bbbar \bbbar  
            & $>80.$ & 452.4 & 2d &\cite{ref:pap99,ref:pap00} \\
189-208 &  \toto \toto     
            & $>8.$ & 570.9 & 1d & \cite{ref:2hdm} \\
189-208 &  \bbbar \bbbar    
            & $>24.$ & 610.2 & no & \cite{ref:2hdm} \\
189-208 & hadrons
            & $>8.$  & 610.4 & mix &\cite{ref:papflbl} \\
\hline
\multicolumn{6}{c}{\hZ\ or \hA\ with \hAA\ cascade} \\\hline
91        & \Zqq                  
            & $<0.21$ & 16.2 & no & \cite{ref:dallas} \\
91        & (\cascv) (\Zanyt)                  
            & $<0.21$ & 9.7 & no & \cite{ref:dallas} \\
91        & (\cascg) (\Zany\ or \Agg)                  
            & $<0.21$ & 12.5 & no & \cite{ref:dallas} \\
91        & (\cascp) (\Zany\ or \App)                  
            & $>0.21$ & 12.9 & no & \cite{ref:dallas} \\
91        & (\casch) (\Znn\ or \Ahad)                  
            & $>0.21$ & 15.1 & no & \cite{ref:dallas} \\
91        & (\casct) (\Znn\ or \Att)                  
            & $>3.5$ & 15.1 & no & \cite{ref:dallas} \\
\bf{161,172} & (\casca) (\Zqq, \nunubar or \Aany)    
            & $>20.$ & 20.0 & 1d &\cite{ref:pap96} \\
\bf{183}      & (\cascb) (\Zqq)
            & $>12.$ & 54.0 & 1d & \cite{ref:pap97} \\
\bf{192-208} & (\cascb, \bbbar \ccbar, \ccbar \ccbar) (\Zqq)   
            & $>12.$ & 452.4 & 2d &\cite{ref:pap99,ref:pap00} \\
192-208 &  (\cascc ) (\Zqq)    
            & $>4.$ & 452.4 & 2d &  A.2 \\
189-208 & (\cascb) (\Zqq\ or \Abb)   
            & $>12.$ & 610.2 & no &\cite{ref:2hdm} \\
\hline
\multicolumn{6}{c}{ffh or ffA Yukawa production} \\\hline
91        &\bbbar (\htt), \bbbar (\Att) 
            & $4.-50.$ & 79.4 & no & \cite{ref:2hdm} \\
91        &\bbbar (\hbb), \bbbar (\Abb) 
            & $11.-50.$ & 79.4 & no & \cite{ref:2hdm} \\
91        &\toto (\htt), \toto (\Att) 
            & $4.-50.$ & 79.4 & no & \cite{ref:2hdm} \\

\hline
\multicolumn{6}{c}{\HH} \\\hline
189-208 & \cscs, \cstn, \watn, \wawa 
            & $>40.$ & 610.4 & 2d & \cite{ref:paphphm} \\
189-208 & \tntn
            & $>40.$ & 570.8 & 1d & \cite{ref:paphphm} \\
\hline
\end{tabular}
\caption[]{
    List of signals expected from {\sc MSSM} Higgs bosons that 
    were searched for in the {\sc DELPHI} data sample. Indicated for 
    each signal are the centre-of-mass energy, 
    final state, analysed mass range, integrated luminosity,
    level of discriminant information included in the confidence level
    estimates (none, one- or two-dimensional) and the reference 
    where details of the analysis are published. Here h and A 
    denote any neutral Higgs boson allowed to be produced
    in each of the indicated production processes.
    The mass range applies
    to \mh\ for \hZ\ production, to \mh+\mA\ for \hA\ production,
    to \mA\ for \hAA\ processes, to the Higgs boson mass for
    either Yukawa process and to \mHp\ for
    \HH\ production. When no upper bound is given, the 
    limit imposed by kinematics or vanishing branching fractions must be 
    understood. 
}
\label{tab:channels}
\end{center}
}
\end{table}

The different analyses performed to search for neutral and charged 
Higgs bosons in the whole {\sc LEP1} and {\sc LEP2}
{\sc DELPHI} data samples are summarized in 
Table~\ref{tab:channels} which lists the final states, mass ranges, 
integrated luminosities
and the references for more details about the selections 
and their performance. 
Two channels, the \tautaubb\ signal at {\sc LEP1}
and the (h~\rgr~\cascc)~(\Zqq) signal at A masses below the
\bbbar\ threshold, were analysed for this paper, using selections
already published. The efficiencies and the references for the
selections can be found in the Appendices 1 and 2 of this paper. 
In the Table, the notations h and A which label
the different analysis channels must be understood as generic notations
for any pair of neutral Higgs bosons that could be produced in each
of the production processes listed in the Table. As an example,
the hZ analyses, originally designed to search for the 
{\sc CP}-even h boson in {\sc CP}-conserving scenarios,
can be applied to search for the
second {\sc CP}-even Higgs boson, H, as well as for
any of the three Higgs scalars in {\sc CP}-violating scenarios.
It must be noted that the kinematic properties of the signal processes
are only slightly affected by {\sc CP}-violation, since, when {\sc CP} is
not conserved, the production processes still proceed through the
{\sc CP}-even and {\sc CP}-odd components of the neutral Higgs
bosons, as explained in Ref.~\cite{ref:hwg}. The same topological searches
can thus be applied whether {\sc CP} is conserved or not.

As compared with our previous publication~\cite{ref:pap00}, the following
changes were introduced in the experimental results used. 
The {\sc MSSM} interpretation in Ref.~\cite{ref:pap00} 
relied only on searches performed at {\sc LEP2} at masses above 12~\GeVcc\
in \mh\ in the \hZ\ process, with either direct or cascade decays, and 
above 40~\GeVcc\ in \mh, \mA\ in the \hA\ channels, with only direct
decays of the Higgs bosons. The corresponding channels 
have their \rs\ values in bold characters in Table~\ref{tab:channels}. 
Scans of
the {\sc MSSM} parameter space were thus restricted to \mA\ above 
12~\GeVcc\ and assumed the published {\sc LEP1} 
limits\footnote{\mh$>$44~(46)~\GeVcc\ 
when \mh\ is above (below) the ${\mathrm{AA}}$ threshold~\cite{ref:ha92}} 
to be valid. 
Including all {\sc LEP1} results, which have a sensitivity
starting from vanishing h and A masses, and the additional {\sc LEP2}
searches of Ref.~\cite{ref:2hdm}, whose sensitivity in the \hA\ mode 
complements that of the two other sets of results, allows 
scans of the {\sc MSSM} parameter space to be performed
with no restriction on masses.
Moreover, some of the analyses of Ref.~\cite{ref:2hdm} cover production
processes which are negligible if {\sc CP} is conserved but are 
enhanced by {\sc CP} violation, such as Yukawa processes or the
production of \toto\toto\ final states. 
Adding the searches for neutral Higgs bosons decaying into hadrons of 
any flavour~\cite{ref:papflbl} is expected to provide  
sensitivity in scenarios where the Higgs boson decays into \bbbar\  
would vanish. 
As their mass coverage starts at low mass, 
these analyses also increase the experimental sensitivity to
Higgs bosons below the \bbbar\ threshold, a region otherwise covered 
only by analyses of subsets of the {\sc LEP1} data. 
Finally, the charged Higgs boson searches~\cite{ref:paphphm}
help in a few {\sc CP}-conserving scenarios in the low \mA\ region
where the charged bosons are kinematically accessible at {\sc LEP2}.

  Moreover, our previous interpretation was dealing only with 
the production of the two lightest Higgs bosons, 
the h and A scalars in {\sc CP}-conserving scenarios.
In this analysis, the production
of the third boson, if kinematically accessible, is also
accounted for, which can lead to a significant gain in sensitivity 
in restricted areas of the parameter space.
In {\sc CP}-conserving scenarios, this leads to
including the \HZ\ and \HA\ signals besides the usual \hZ\ and
\hA\ processes, while in {\sc CP}-violating models, the 
H$_2$Z and H$_1$H$_3$ signals are taken into account in addition
to the dominant H$_1$Z and H$_1$H$_2$ channels (the two other 
processes, H$_3$Z and  H$_2$H$_3$ being out of reach). 

\section{Tools for the statistical analysis}\label{sec:tools}

When scanning over the parameter space of a model, confidence levels are 
computed at each point to test the compatibility of data with the
hypothesis of background only and with that of background plus 
signal as expected from the model. Throughout this section, 
the notations h, H and A must be understood as generic notations
for the three neutral Higgs bosons of any type of {\sc MSSM} scenario.

\subsection{Confidence level definitions and calculations}
The confidence levels are calculated using a modified frequentist
technique based on the extended maximum likelihood ratio~\cite{ref:alex}
which has also been adopted by the {\sc LEP} Higgs working group.
The basis of the calculation is the likelihood ratio test-statistic, \likear:
\[ \ln\likear = -S + \sum_i \ln \frac{s_i+b_i}{b_i} \]
where $S$ is the total signal expected and $s_i$ and $b_i$ are the
signal and background densities for event $i$.
These densities are constructed using either expected rates only
or also additional discriminant information, which can be one- or
two-dimensional. Table~\ref{tab:channels} presents the level
of discriminant information for each channel:
{\sc LEP1} results rely on rates only, while {\sc LEP2} results
mix channels without or with discriminant information.
As an example, in neutral Higgs boson channels with discriminant
information, the first variable is the reconstructed Higgs boson mass 
in the \hZ\ analyses and the sum of the reconstructed h and A masses 
in the \hA\ analyses, while the second variable, if any, is 
channel-dependent, as specified in the references listed in the Table. 
Charged Higgs analyses use discriminant information in a similar 
way~\cite{ref:paphphm}.
The searches for Higgs bosons decaying hadronically encompass analyses 
without or with 1d discriminant information together with analyses 
whose selections vary with the mass hypothesis~\cite{ref:papflbl}.

The observed value of \likear\ is compared 
with the expected Probability Density Functions (PDFs) for \likear, 
which are built using Monte Carlo sampling under the assumptions that 
background processes
only or that both signal and background are present. The confidence 
levels \CLb\ and \CLsb\ are their integrals from $-\infty$ to the 
observed value of \likear.
Systematic uncertainties in the rates of signal or background events are taken 
into account in the calculation of the PDFs for \likear\ by 
randomly varying the expected rates while generating the
distribution~\cite{ref:CousinsAndHighland},
which has the effect of broadening the expected \likear\ distribution and 
therefore making extreme events seem more probable.

\CLb\ is 
the probability of obtaining a result as background-like
or more so than the one observed if the background hypothesis is correct.
Similarly,
the confidence level for the hypothesis that both signal and background 
are present, \CLsb, is the probability, in this hypothesis,
to obtain more background-like results than those observed. 
The quantity \CLs\ is defined as the ratio of these
two probabilities, \CLsb/\CLb. It is not a true
confidence level, but a conservative pseudo-confidence level
for the signal hypothesis.
All exclusions discussed hereafter use \CLs\ and require it to be
5\% for an exclusion confidence of 95\%. As using \CLs\
instead of \CLsb\ is conservative, the rate of fake exclusions
is ensured to be below 5\% when \CLs\ is equal to 5\%.

\subsection{Estimation of expected signal and background densities}
\label{sec:pdf}

The expected signal and background densities, which are required 
to check the consistency of the data with the background and signal 
processes have two components: the overall normalization which sets
the expected rates and the Probability Density Functions (PDF) of the 
additional discriminant information, if any.

  The expected background and signal rates were
calculated from the number of simulated events 
passing the cuts. For the signal the efficiencies derived from 
simulations at given mass points had to be interpolated
to estimate efficiencies at Higgs boson masses which were
not simulated. In most cases this was done using one polynomial
or if necessary two polynomials, one
to describe the slow rise, and a second to handle the kinematic cut-off,
which can be much more abrupt.
For the cases where two signal masses must
be allowed,  a two-dimensional  parameterization was used.

 The shapes of the PDFs were derived using 
histograms which are taken from the simulated events.
In the case of two-dimensional PDFs these distributions were
smoothed using a two-dimensional kernel, which consists of  a 
Gaussian distribution with a small component of a longer 
tail~\cite{ref:smoothing}.
The global covariance of the distribution was used to determine the 
relative scale factors of the two axes. The width of the kernel
varied from point to point, such that the statistical error on 
the estimated background processes was constant at 20\%. 
Finally multiplicative correction factors (each a one-dimensional
distribution for one of the two dimensions of the PDF) were derived
such that when projected onto either axis the PDF has the same distribution 
as would have been observed if it had been projected onto the axis 
first and then smoothed.
This makes better use of the simulation statistics if there are features 
which are essentially one-dimensional, such as mass peaks.
The error parameter fixed to 20\% was an important choice. It was set by
dividing the background simulation into two subsamples, generating a 
PDF with one and using the other to test for over-training by calculating 
the \CLb\ obtained from simulation of background events. This should be 
0.5 if the results are not to be biased, and a value of 20\% for
the error gave the closest approximation to 0.5 in all channels. Examples
of smoothed two-dimensional PDFs are given in Fig.~\ref{fig:adding_H}.

  The signal simulations were made at fixed Higgs boson masses, but in order
to test a continuous range of masses, interpolation 
software~\cite{ref:alex-interp} was used to create signal PDFs at 
arbitrary masses. In the last year of operation, {\sc LEP} energy was
varied continuously while simulations were made at fixed beam energies.
The same interpolation software was used to create signal and background 
PDFs at the correct centre-of-mass energies~\cite{ref:pap00}.
The interpolation was done by linearly interpolating the cumulative 
distributions taking as a parameter the signal mass or the centre-of-mass
energy. The procedure 
has been tested over ranges up to 40~\GeVcc\ in mass while the actual 
shifts in the simulations were up to 0.3~\GeV\ in $\sqrt{s}$,
and 5~\GeVcc\ in mass for the \hZ\ signals overall, but less
than 0.5~\GeVcc\ for Higgs boson masses between 113.5 and 116.5~\GeVcc.
For the \hA\ channels, the actual shifts were 5~\GeVcc\ in either mass
for Higgs boson masses between 80 and 95~\GeVcc\ and up to 20~\GeVcc\
elsewhere. Comparisons of simulated and interpolated distributions 
for a given mass were made in all channels and showed good agreement.


\subsection{The case of non-independent channels}\label{sec:overlap}

  When combining the results in all channels to derive confidence
levels, only independent channels must be included, which requires some 
special treatment for a few non-independent cases. 

\begin{table}[htbp]
{\small
\begin{center}
\begin{tabular}{c|c|lcc}     \hline
\multicolumn{4}{c}{a) different signals - one analysis with mass 
hypothesis-independent selections} \\\hline
analysis & \rs\ (GeV) & signals added & ref. \\\hline
ffh four-b & 91 
           & (\bbbar h \rgr\ \bbbar \bbbar), (\bbbar A \rgr\ \bbbar \bbbar),
             (hA \rgr\  \bbbar \bbbar) 
           & \cite{ref:2hdm} \\
ffh \bbbar\toto & 91 
             & (\bbbar h \rgr\ \bbbar \toto), (\bbbar A \rgr\ \bbbar \toto),
               (hA \rgr\  \bbbar \toto) 
             & \cite{ref:2hdm}\\
ffh four-$\tau$ & 91 
                & (\toto h \rgr\ \toto \toto), (\toto A \rgr\ \toto \toto)
                & \cite{ref:2hdm}\\
\nnqq\     &161-172
           & (\hqq) (\Znn), (\hAA) (\Znn)
           & \cite{ref:pap96} \\ 
\tautauqq\ & 189-208
           & (\htt) (\Zqq), (\hqq) (\Ztt), (hA \rgr\ \toto \qqbar) 
           & \cite{ref:pap98,ref:pap99,ref:pap00} \\
\hZ\ four-jet &161-183
              & (\hqq) (\Zqq), (\hAA) (\Zqq)
              & \cite{ref:pap96,ref:pap97} \\
\hZ\ four-jet &192-208
              & (\hqq) (\Zqq), (\hAA) (\Zqq), (hA \rgr\ \bbbar \bbbar)
              & \cite{ref:pap99,ref:pap00} \\
\hA\ four-jet &161-172
              & (hA \rgr\ \bbbar \bbbar), (\hAA) A
              & \cite{ref:pap96} \\
\hA\ four-jet &192-208
              & (hA \rgr\ \bbbar \bbbar), (\hqq) (\Zqq) 
              & \cite{ref:pap99,ref:pap00} \\
four-b       &189-208
             &  (\hAA)A, (\hAA)Z, (hA \rgr\ \bbbar \bbbar)
             & \cite{ref:2hdm} \\
\hline
\multicolumn{4}{c}{b) different analyses - one final state} \\\hline
final state & \rs\ (GeV) & competing analyses & ref. \\\hline
four-jet  &91& \bbbar \bbbar, \bbbar \ccbar
          & \cite{ref:ha92} \\
multi-jet &91& three and four-jet analyses 
          & \cite{ref:2hdm} \\ 
\nnqq\    &192-208& low mass and high mass hZ analyses  
          & \cite{ref:pap00} \\
          &189-208& low mass and high mass flavour-blind analyses  
          & \cite{ref:papflbl} \\ 
          &189-208& hZ and flavour-blind analyses  
          & \cite{ref:pap98,ref:pap99,ref:pap00,ref:papflbl} \\
ll\qqbar,l=e,$\mu$    &189-208& hZ and flavour-blind analyses  
          & \cite{ref:pap98,ref:pap99,ref:pap00,ref:papflbl} \\
four-jet  &192-208& low mass and high mass hZ analyses 
              & \cite{ref:pap00} \\
          &189-208& low mass and high mass hZ flavour-blind analyses 
              & \cite{ref:papflbl} \\
          &189-208& three and four-jet hA flavour-blind analyses 
              & \cite{ref:papflbl} \\
          &189-208& cscs and WAWA analyses 
          & \cite{ref:paphphm} \\
   &189-208& \hZ, \hA, four-b, flavour-blind, \cscs\ and \wawa\ analyses 
   & \cite{ref:pap98,ref:pap99,ref:pap00,ref:2hdm,ref:papflbl,ref:paphphm} \\
$\tau \nu$ jet jet  &189-208
         & \cstn\ and \watn\ analyses 
         & \cite{ref:paphphm} \\ 
\hline
\end{tabular}
\caption[]{
  a) list of signals from the two lightest Higgs bosons h and A
  treated by a single analysis: the
  signal expectations are combined (rates added, PDFs summed with weights
  according to the rates) prior to the confidence level calculations.
  b) list of different analyses of the same final state: only one
  analysis is selected at each point in the scans, based on the best 
  expected performance for exclusion.
  In this Table,
  h and A denote any neutral Higgs boson allowed to be produced in
  the indicated production processes.}
\label{tab:overlap}
\end{center}
}
\end{table}

\begin{figure}[p]
\begin{tabular}{cc}
\hspace{-1.2cm}
\epsfig{file=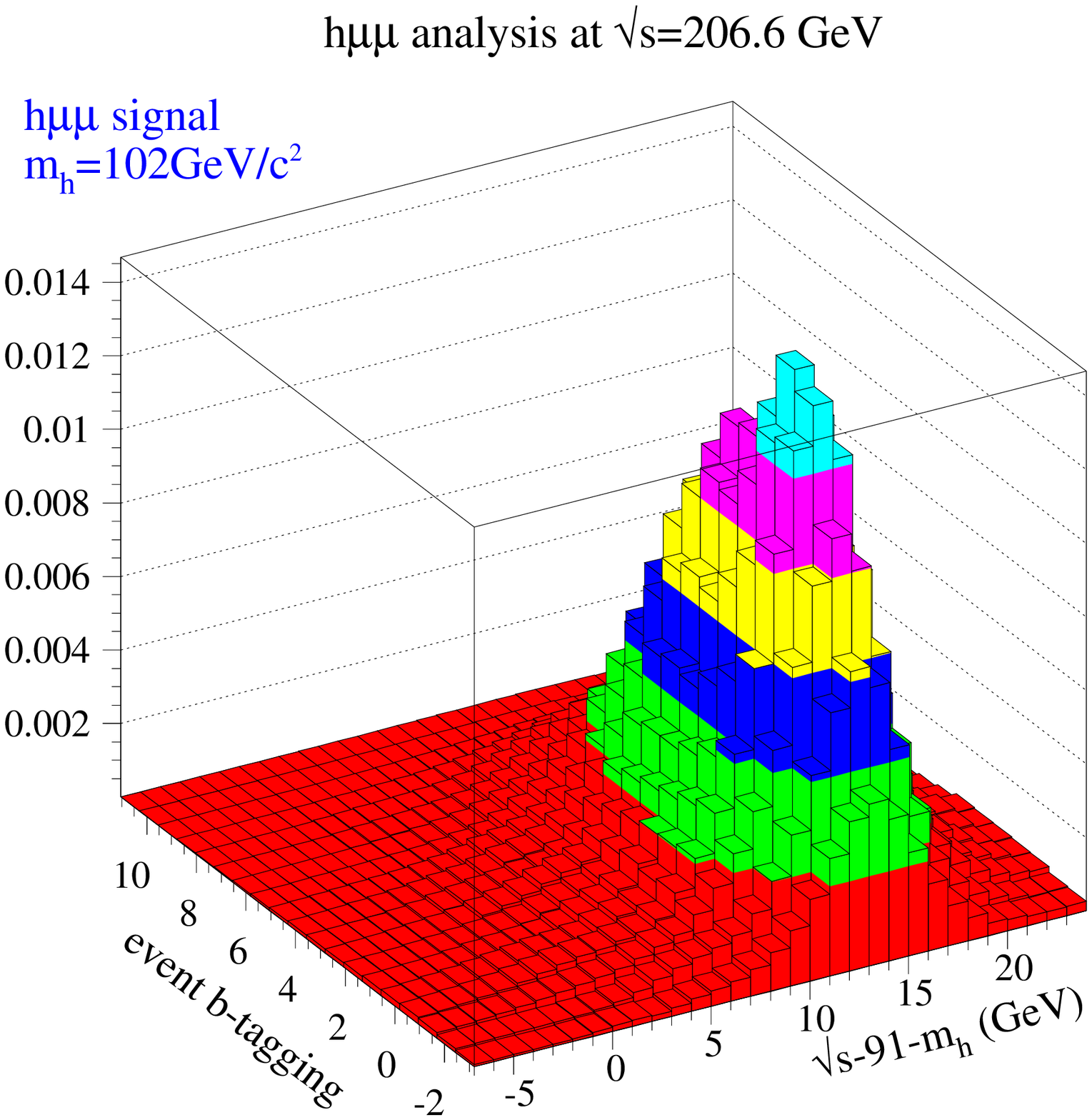,width=9.5cm} &
\hspace{-1.cm}
\epsfig{file=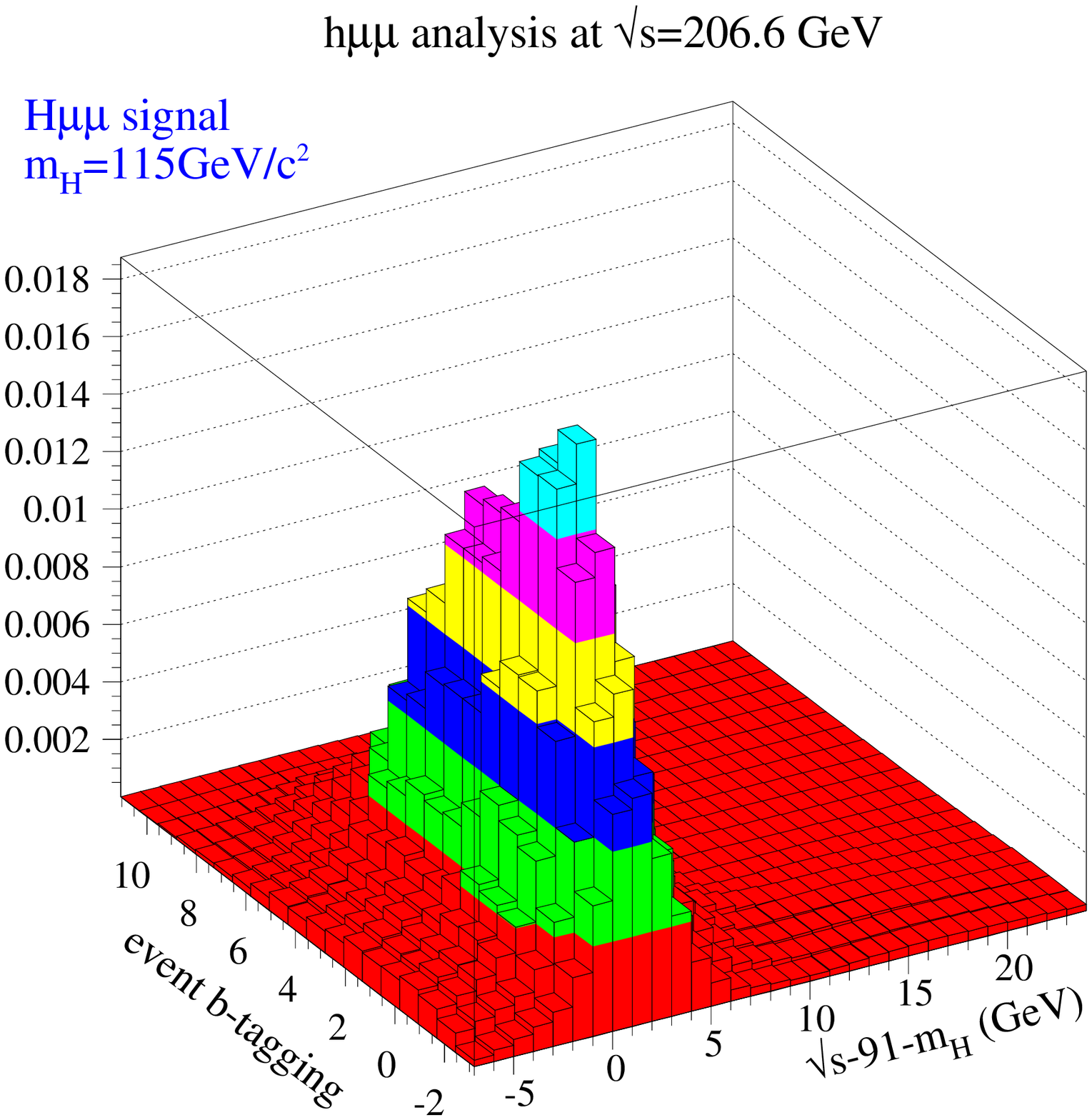,width=9.5cm} \\
\hspace{-1.2cm}
\epsfig{file=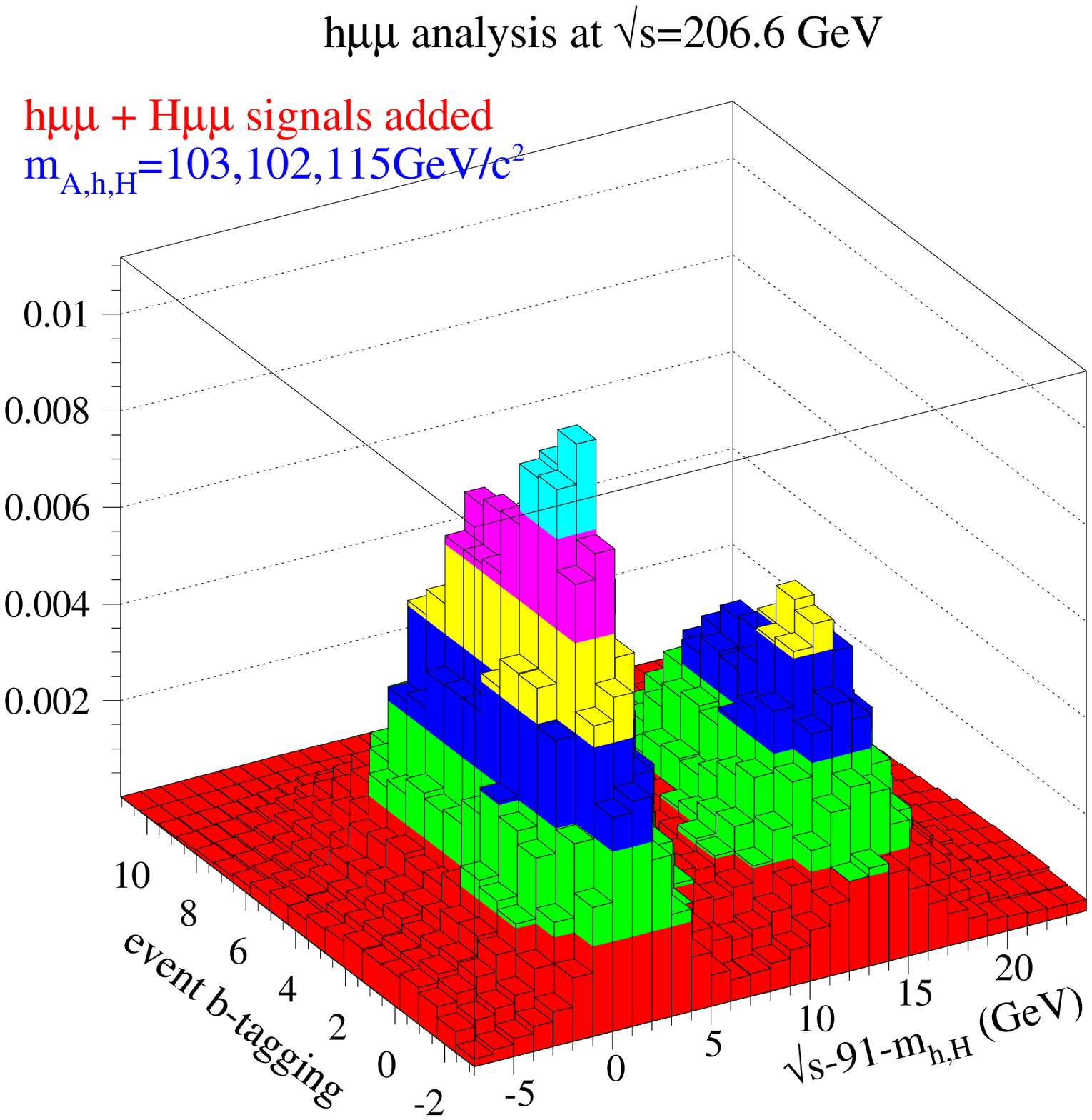,width=9.5cm} &
\end{tabular}
\caption[]{
  An example of two-dimensional PDFs from the analysis of the 
  \hZ\ $\rightarrow$ \qqbar\ \mumu\  channel at 
  \rs~=~206.6~GeV~\cite{ref:pap00}. The first discriminant variable is
  built from the reconstructed Higgs boson mass while the second
  is the event b-tagging variable. Top, left: PDF for a \hZ\ signal
  with \mh~=~102~\GeVcc. Top, right: PDF for a \HZ\ signal with
  \MH~=~115~\GeVcc. Bottom: PDF expected from the occurrence of both
  signals in a scenario 
  where the expectations for the two signals are similar
  (cross-sections 32 and 42~fb, branching fractions into
  \bbbar\ 92\% and 91\%, selection efficiencies 69\% and 66\% for \hZ\
  and \HZ, respectively) leading to a double peak in the combined PDF. 
}
\label{fig:adding_H}
\end{figure}

  The first case is that of different signals covered by the same 
analysis. The treatment of this depended upon whether the analyses
were themselves independent of the mass hypothesis for the 
Higgs bosons. 
The set of search channels (see Table~\ref{tab:channels})
contains mostly analyses of this kind.
In that case, all signals selected by one analysis were
combined into one global channel prior to the confidence level computation.
Expected rates were added together and PDFs were summed with weights
given by the expected rates of the individual signals.
As an illustration,  Table~\ref{tab:overlap}-a gives the list of these
signals and analyses on the example of the production of the lightest
Higgs bosons, h and A, through the \hZ\ and \hA\ processes. 
When extending the combination to the third Higgs boson, H,
the same procedure was followed, first for the various signals from
that boson in the \HZ\ and \HA\ processes,
and then to combine \hZ\ and \HZ\ signals or \hA\ and \HA\ signals.
The PDF combination in such a case is illustrated in Fig.~\ref{fig:adding_H}.

A different procedure was applied in the case of different 
signals covered by the same analysis whose selections do depend on the 
mass hypothesis, as most searches of Ref.~\cite{ref:papflbl} do.
Different signals are covered by these analyses only
when including signals from the third Higgs boson, H. 
In that case, in each analysis only one signal 
(from either h or H) was selected at each point in the
scanned parameter space and at each centre-of-mass energy, 
on the basis of the smallest expected \CLs\ from experiments with no signal
(that is, on the basis of the strongest average exclusion if no signal 
is present). 

The second case of non-independent channels is that of 
a large overlap in the events selected by different analyses
sensitive to the same final state. The list of such analyses 
and final states is detailed in Table~\ref{tab:overlap}-b. Again,
for each final state, 
only that analysis with the strongest expected exclusion power
was retained at each test point.
This is not optimal but ensures that the channels which are then
combined in the global confidence level computations are independent.

When the two cases just described (different signals covered by one
analysis, different analyses sensitive to the same final state)
were present 
simultaneously, the signal addition was performed before the final
analysis selection. Then if that step involved more than two analyses,
the final selection was made in successive iterations.
To quote the four-jet final state as an example, at energies above 190~GeV,
the total \hZ\ and \hA\ signals were first computed in 
each of the three four-jet analyses of Ref.~\cite{ref:pap00}
and in the four-b analysis of Ref.~\cite{ref:2hdm}.
This summed three signals in the low and high mass 
\hZ\ dedicated four-jet analyses
((\hqq) (\Zqq), (\hAA) (\Zqq) and hA \rgr\ \bbbar \bbbar),
two signals in the \hA\ dedicated four-jet analysis
(hA \rgr\ \bbbar \bbbar\ and (\hqq) (\Zqq)) 
and three signals in the four-b analysis
((\hAA)A, (\hAA)(\Zqq) and hA \rgr\ \bbbar \bbbar).
The signals due the third Higgs boson, H, were computed in the same way
and added to those from the h boson. 
Then, a choice was made between the low and 
high mass \hZ\ dedicated four-jet analyses. 
The result of this selection was compared 
with the \hA\ dedicated four-jet analysis, 
and the best of these was confronted with the four-b analysis. 
A choice was made
between the remaining analysis and the best between the various
flavour-blind multi-jet analyses, that is the
low mass and high mass hZ dedicated flavour-blind analyses, and
the three and four-jet hA dedicated flavour-blind 
analyses~\cite{ref:papflbl}.
As multi-jet flavour-blind analyses use mass-hypothesis dependent criteria,
selecting the best one implied also a choice between the h and H signals
for each of them.
The analysis retained was finally compared with the result of the 
selection between the two charged Higgs multi-jet analyses, 
the cscs and WAWA dedicated analyses~\cite{ref:paphphm}.

\section{The {\sc CP}-conserving {\sc MSSM} scenarios}\label{sec:benchmark}

In most of the parameter space of the 
{\sc CP}-conserving {\sc MSSM} scenarios, 
only \hZ\ and \hA\ productions are kinematically possible at 
{\sc LEP} energies. 
These processes have complementary cross-sections since the
hZZ and hAZ couplings are proportional to \sinab\ and \cosab, respectively,
where \tbeta\ is the ratio of the doublet vacuum expectation values 
and \alp\ is  the Higgs doublet mixing angle which enters the definition 
of the two CP-even Higgs eigenstates as a mixture of the real, neutral 
components of the initial Higgs field 
doublets~\cite{ref:mssmpheno,ref:FDmssmpheno}.
If kinematically allowed, \hZ\ production dominates at low \tbeta\
or at large \MA, while in the rest of the parameter space, it is suppressed
with respect to \hA\ pair-production. 
The third neutral Higgs boson, H, 
in some scenarios and in limited regions of the
parameter space, is light enough and can be produced with a 
large \HZ\ or \HA\ cross-section. 
As the HZZ coupling is proportional to \cosab,
and the HAZ one is proportional to \sinab, 
\HZ\ production, when allowed by kinematics, plays a role at large 
\tbeta, and \HA\ production at low \tbeta. Similarily, 
charged Higgs bosons kinematically accessible at {\sc LEP2}
energies are predicted in limited regions of the parameter space, 
typically when A is light, whatever \tbeta.
The minimal value of the mass of such charged Higgs bosons
is 60~\GeVcc\ in the scenarios under study. The coverage of the
region of the {\sc MSSM} parameter space kinematically accessible
at {\sc LEP} is then assured primarily by  the \hZ\ and \hA\  searches,
with the help of the \HZ, \HA\ and to a lesser extent \HH\ channels.

   At tree level, the production cross-sections and the 
Higgs branching fractions in the  {\sc MSSM}
depend on two free parameters, usually chosen as \tbeta\ and one Higgs 
boson mass, or, alternatively, two Higgs boson masses, e.g. \MA\ and \mh. 
Radiative corrections introduce additional parameters related
to supersymmetry breaking~\cite{ref:mssmpheno,ref:FDmssmpheno}.
Hereafter, the usual assumption that some of them
are equal at a given energy scale is made: hence, 
the SU(2) and U(1) gaugino mass parameters are assumed to be
unified at the so-called GUT scale, 
while the sfermion mass parameters or the squark trilinear
couplings are taken to be equal at the EW scale.
Within these assumptions, the parameters beyond tree level are:
the top quark mass, the Higgs mixing parameter, $\mu$, which defines
the Higgsino mass parameter at the EW scale,
the common sfermion mass parameter at the EW scale, $M_{\rm susy}$,
the SU(2) gaugino mass parameter at the EW scale, $M_2$,
the gluino mass, $m_{\tilde{g}}$,
and the common squark trilinear coupling at the EW scale, 
$A$.
The U(1) gaugino mass term at the EW scale, $M_1$, is 
related to $M_2$ through the GUT relation
$M_1 = (5/3) {\rm \tan}^2\theta_W M_2$.
The radiative corrections affect 
the Higgs boson masses and couplings,
with the largest contributions arising from 
loops involving the third generation quarks and squarks (top/stop
and, at large values of \tbeta, bottom/sbottom).
As an example, the h boson mass, which is below that of the Z boson
at tree level, increases by a few tens of \GeVcc\ in some regions
of the {\sc MSSM} parameter space due to radiative corrections.

\subsection{The benchmark scenarios} 

 In the following, eight benchmark scenarios are considered, as suggested 
in Ref.~\cite{ref:new_pres}. The values of their underlying parameters  
are quoted in Table~\ref{ta:benchmarks}.
The first three scenarios are those usually studied at {\sc LEP}. They
have been proposed to test the sensitivity of {\sc LEP} to Higgs bosons
with either masses close to the kinematic limit or 
decays difficult to detect. Similarly, 
the five other scenarios are aimed at testing the sensitivity of the 
Higgs boson searches at hadron colliders. It is thus interesting to
establish the {\sc LEP} constraints in such models too.

The first two scenarios, called the
\mbox{$ m_{\mathrm h}^{\rm max}$} scenario and the no mixing scenario,
differ only by the value of $X_t = A - \mu \cot \beta$, 
the parameter which controls the mixing in the stop sector
(through the product \mtop $X_t$). This parameter
has the largest impact on the mass of the h boson.
The \mbox{$ m_{\mathrm h}^{\rm max}$} scenario leads to the maximum
possible h mass as a function of \tbeta. 
The no mixing scenario is its counterpart with vanishing mixing, 
leading to theoretical upper bounds on \mh\ which are at least 
15~\GeVcc\ lower than in the \mbox{$ m_{\mathrm h}^{\rm max}$} scheme. 

\begin{table}[t]
\begin{center}
\begin{tabular}{l|cccccc}     \hline
scenario & $M_{\rm susy}$ & $M_2$ & $m_{\tilde{g}}$ & $\mu$ 
         & $X_t$ \\
& (\GeVcc) & (\GeVcc) & (\GeVcc) & (\GeVcc) & (\GeVcc)\\
\hline
\mbox{$ m_{\mathrm h}^{\rm max}$} &
      1000 & 200 & 800 & -200 & 2 $M_{\rm susy}$ \\
no mixing &
      1000 & 200 & 800 & -200 & 0 \\
large $\mu$ &
      400 & 400 & 200 & 1000 & -300 \\\hline
\mbox{$ m_{\mathrm h}^{\rm max}$}, $\mu >0$  &
      1000 & 200 & 800 & 200 & 2 $M_{\rm susy}$ \\
\mbox{$ m_{\mathrm h}^{\rm max}$}, $\mu >0$, $X_t<0$  &
      1000 & 200 & 800 & 200 & -2 $M_{\rm susy}$ \\
no mixing, $\mu >0$, large  $M_{\rm susy}$  &
      2000 & 200 & 800 & 200 & 0 \\
gluophobic   &
      350 & 300 & 500 & 300 & -750 \\
small $\alpha$  &
      800 & 500 & 500 & 2.5 $M_{\rm susy}$ & -1100 \\
\hline
\end{tabular}
\caption[]{
Values of the underlying parameters for the eight
representative {\sc MSSM} scenarios scanned in this paper.
Note that  $X_t = A - \mu \cot \beta$. These scenarios
have been studied for several values of the top quark mass,
\mtop~=~169.2, 174.3, 179.4 and 183.0~\GeVcc.
}
\label{ta:benchmarks}
\end{center}
\end{table}

The third scenario is called the large $\mu$ scenario to account for 
a large, positive value of $\mu$. As a consequence of the low value
of $M_{\rm susy}$ and the moderate mixing in the stop sector, this
scenario predicts at least one CP-even Higgs boson with a mass within 
kinematic reach at {\sc LEP2} in each point of the {\sc MSSM} parameter space. 
However, there are regions for which detecting such a Higgs boson is
difficult because of vanishing branching fractions into b-quarks. 
The values chosen for $\mu$ and $X_t$ are indeed such that, 
in these regions, radiative corrections lead to suppressed couplings 
to b-quarks for one or the other CP-even Higgs boson.
The dominant decays in these regions being still into hadrons,
the main analysis channels suffer from large backgrounds.
This scenario was designed to test the sensitivity of {\sc LEP} through
analyses that could not benefit from the b-tagging capabilities of the
experiments.

Among the five other benchmark scenarios, three are variants of the 
\mbox{$ m_{\mathrm h}^{\rm max}$} and no mixing scenarios. 
The sign of $\mu$ and that of the mixing parameter have been reversed in
the two scenarios derived from the {\sc LEP} \mbox{$ m_{\mathrm h}^{\rm max}$} 
scenario. The changes in the Higgs boson mass spectrum and properties
are small. 
The sign of $\mu$ has been reversed and the value of $M_{\rm susy}$
has been doubled in the scenario derived from the no mixing scenario
of {\sc LEP}. The higher $M_{\rm susy}$ scale leads to a few \GeVcc\
increase of the theoretical upper bound on \mh.
The last two scenarios have been proposed to test potentially
difficult cases for the searches at hadron colliders. Hence, the gluophobic
scenario presents regions where the main production channel at the {\sc LHC}, 
gluon fusion, is suppressed due to cancellations between the top
quark and stop quark loops in the production process.
Finally, in the small $\alpha$ scenario, important
decay channels at the Tevatron and at the {\sc LHC}, \hbb\ and \htt,
are suppressed at large \tbeta\ and moderate \mA. In these regions,
the radiatively corrected mixing angle \alp\ is low, resulting in
suppressed couplings of the ligthest CP-even Higgs boson to down-type
fermions since these couplings are proportional to $-\sin \alpha$/$\cos \beta$.

\begin{table}[t]
\begin{center}
\begin{tabular}{l|cccc}     \hline
         & \multicolumn{4}{c}{\mtop\ (\GeVcc)} \\
scenario &   169.2 & 174.3 & 179.4 & 183.0\\
\hline
\mbox{$ m_{\mathrm h}^{\rm max}$}  
          & 128.2 & 132.9 & 138.6 & 142.7 \\   
no mixing 
          & \bf{112.8} & \bf{115.5} & 118.2 & 120.3 \\        
large $\mu$ 
          & \bf{106.1} & \bf{108.0} & \bf{110.1} & \bf{111.6} \\\hline 
\mbox{$ m_{\mathrm h}^{\rm max}$}, $\mu >0$   
          & 128.4 & 134.1 & 140.1 & 144.3 \\ 
\mbox{$ m_{\mathrm h}^{\rm max}$},  $\mu >0$, $X_t<0$    
          & 124.5 & 128.8 & 134.3 & 138.2 \\   
no mixing, $\mu >0$, large  $M_{\rm susy}$ 
          & \bf{117.0} & 120.2 & 123.7 & 126.3 \\ 
gluophobic 
          & \bf{115.7} & 118.8 & 122.0 & 124.4 \\ 
small $\alpha$ 
          & 118.5 & 122.2 & 126.2 & 129.1 \\ 
\hline
\end{tabular}
\caption[]{
  Maximal value of \mh\ (in \GeVcc) 
  in the eight benchmark {\sc MSSM} scenarios 
  studied in this paper, as a function of \mtop. Radiative corrections
  include all dominant second-order loop terms~\cite{ref:FDradco_new}.
  The maximum value of \mh\ corresponds approximately to the minimum 
  value of the mass of the third Higgs boson, H.  
  Bold values indicate scenarios where this boson
  is kinematically accessible at {\sc LEP}.
}
\label{tab:mhmax}
\end{center}
\end{table}
 
\begin{figure}[htbp]
\vspace{-1.7cm}
\begin{center}
\begin{tabular}{cc}
\epsfig{figure=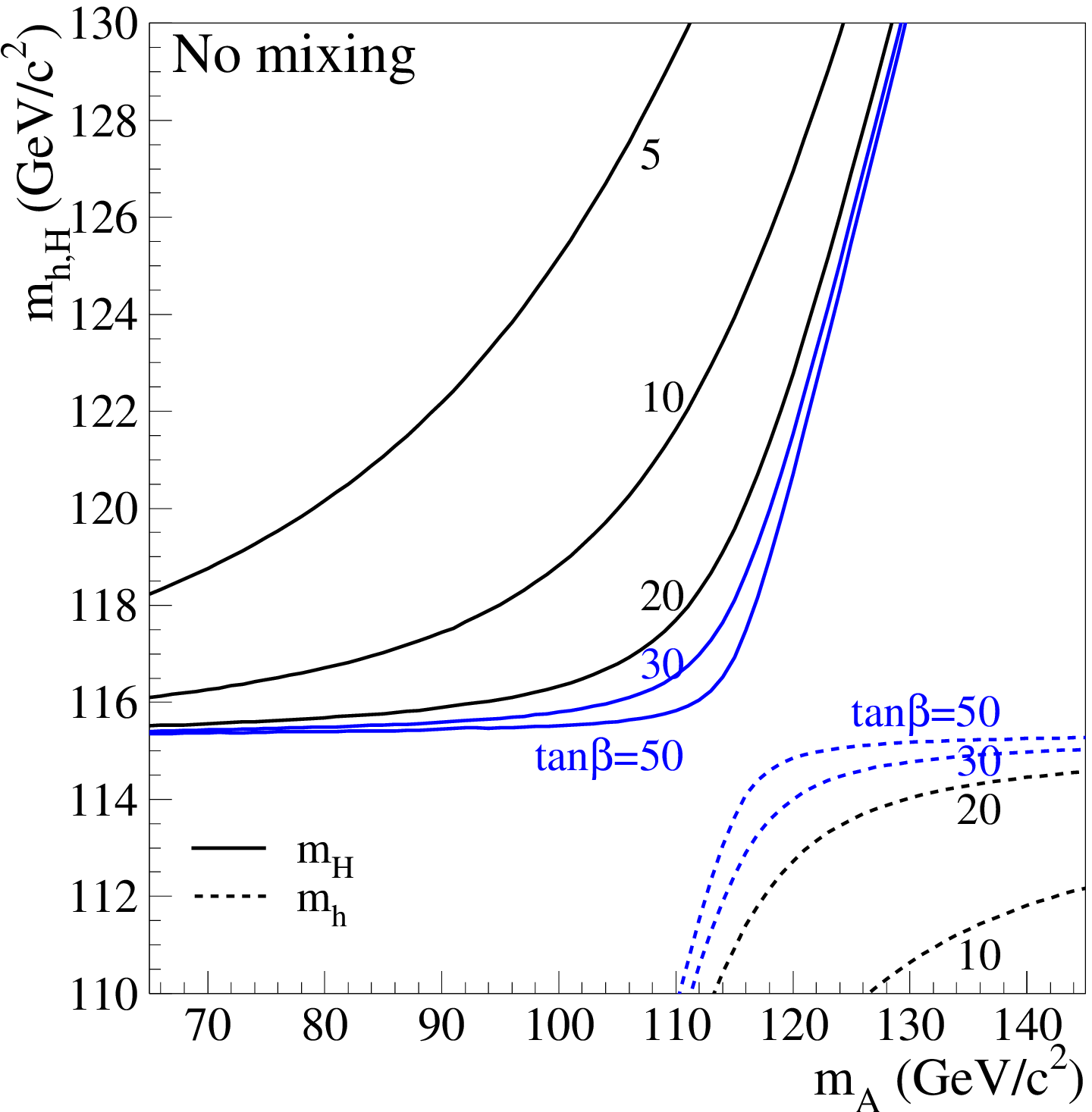,height=7.6cm} &
\epsfig{figure=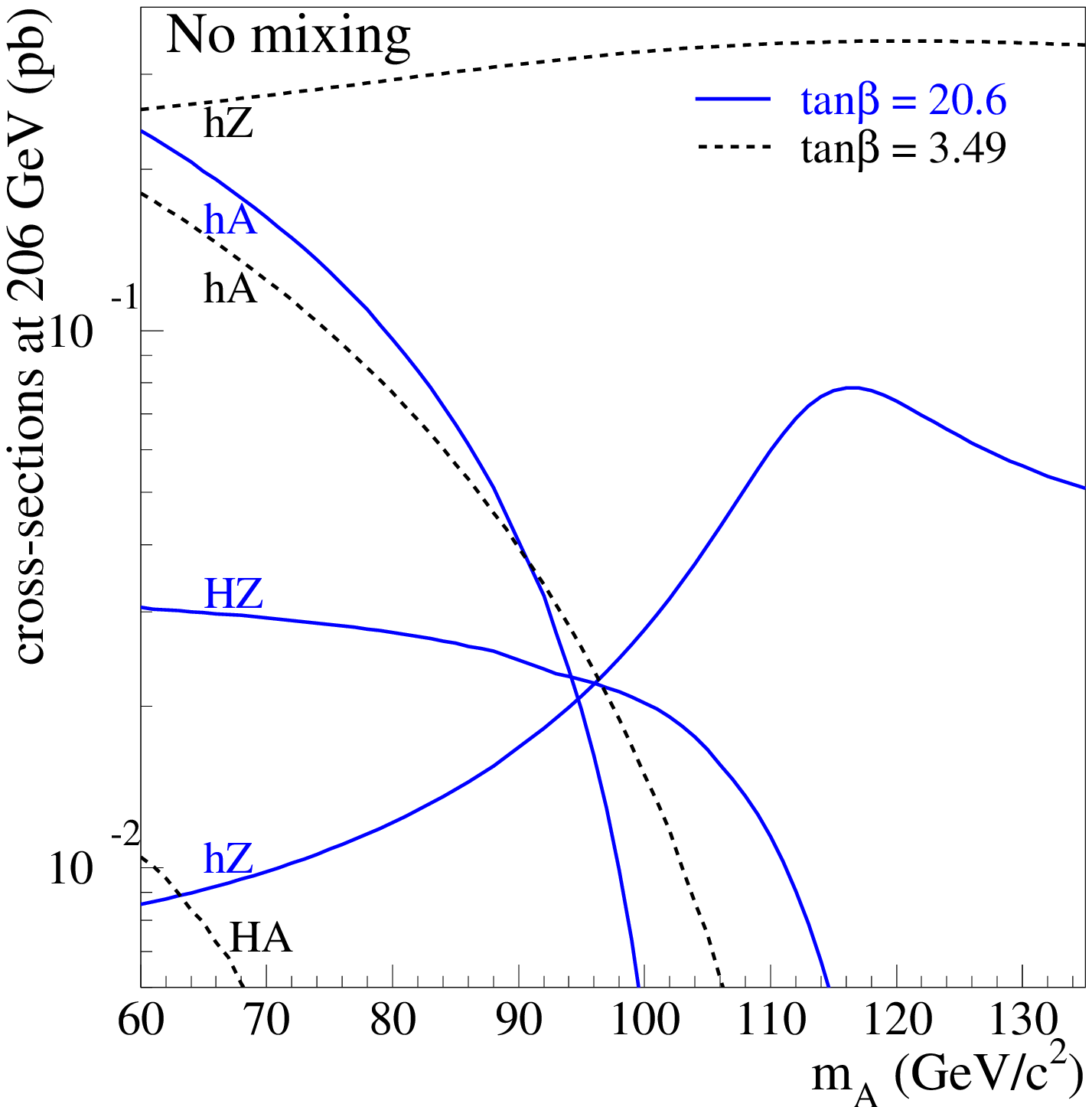,height=7.6cm}\\
\epsfig{figure=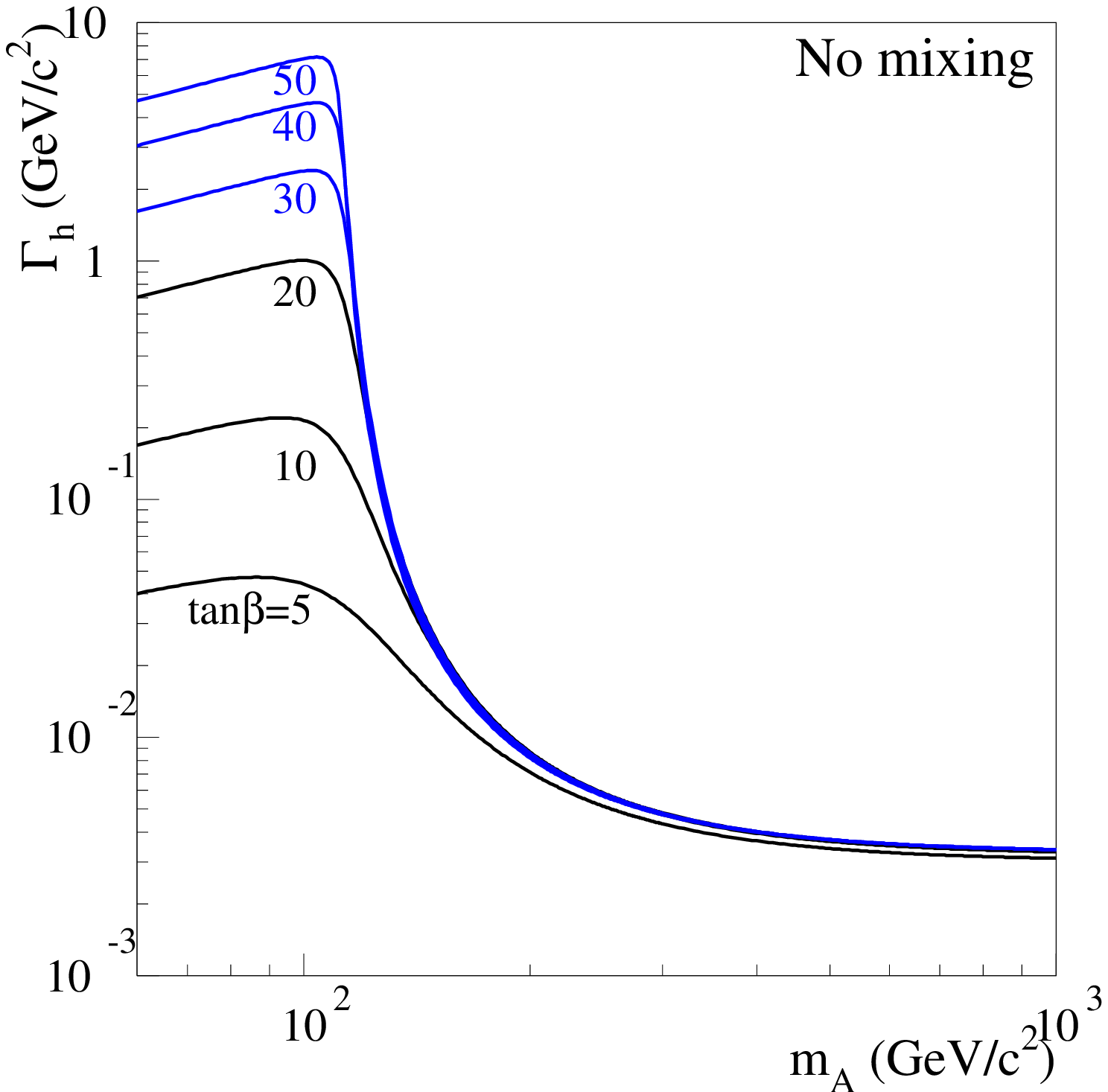,height=7.6cm} &
\epsfig{figure=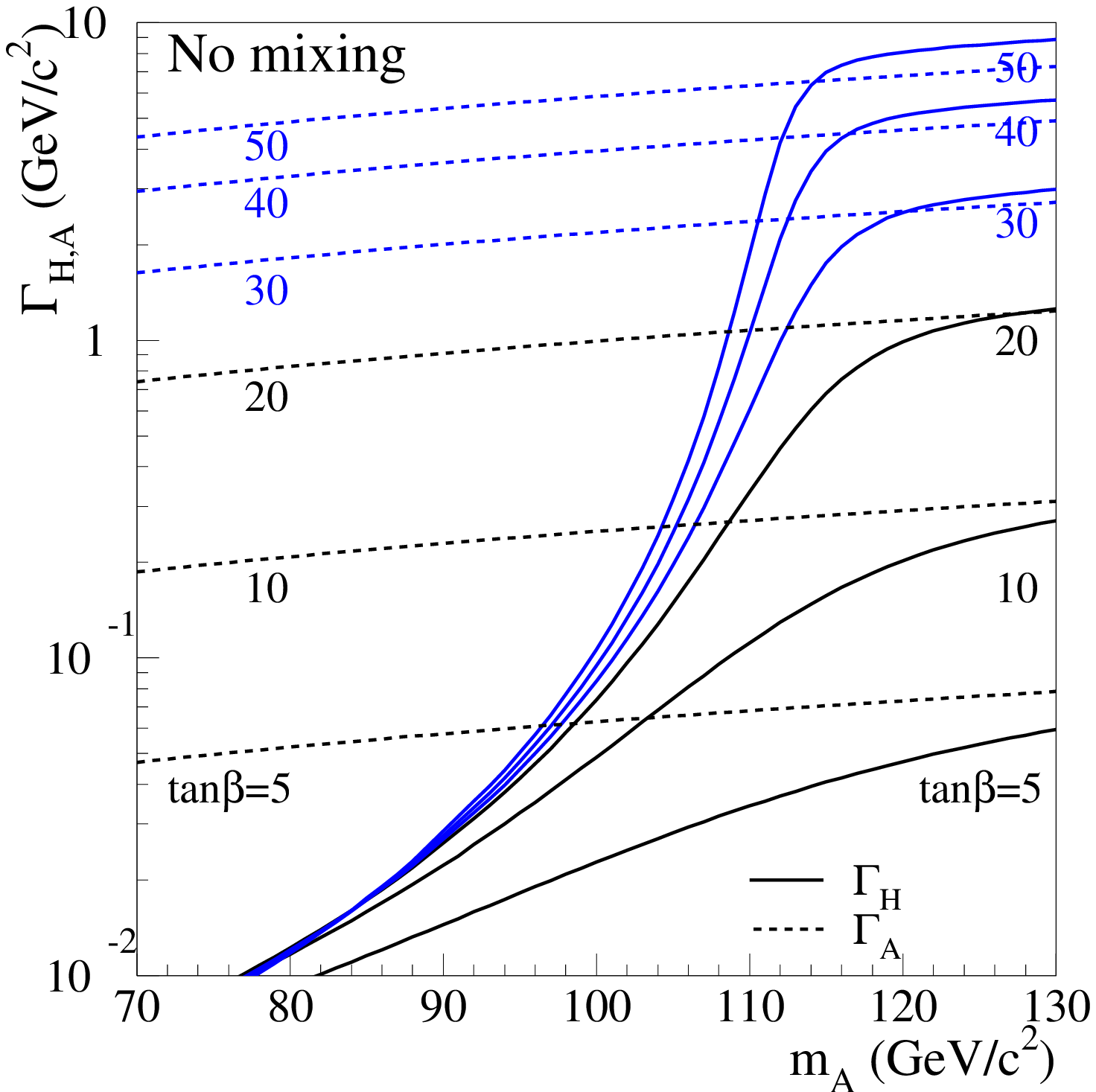,height=7.6cm}\\
\epsfig{figure=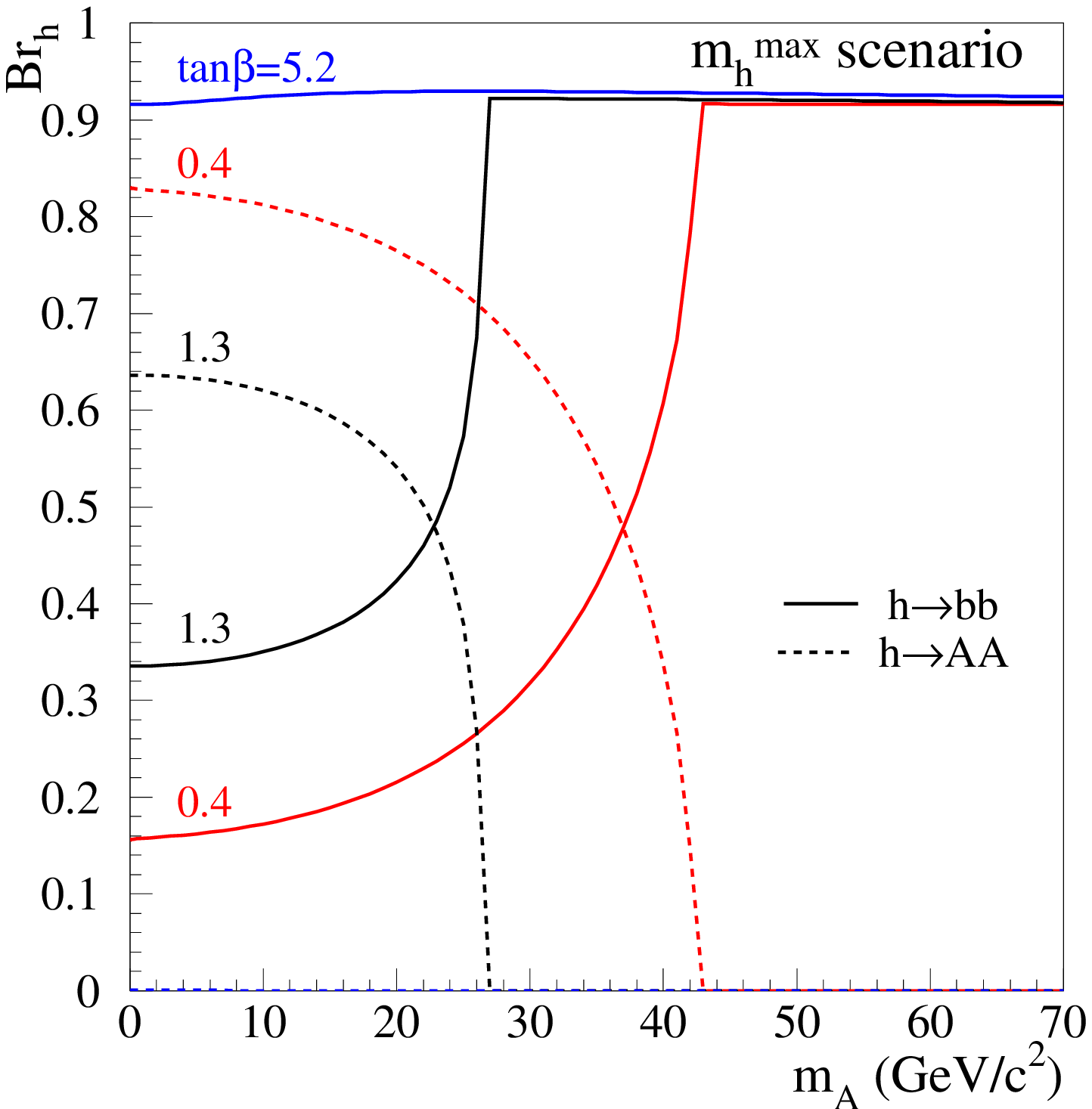,height=7.6cm} &
\epsfig{figure=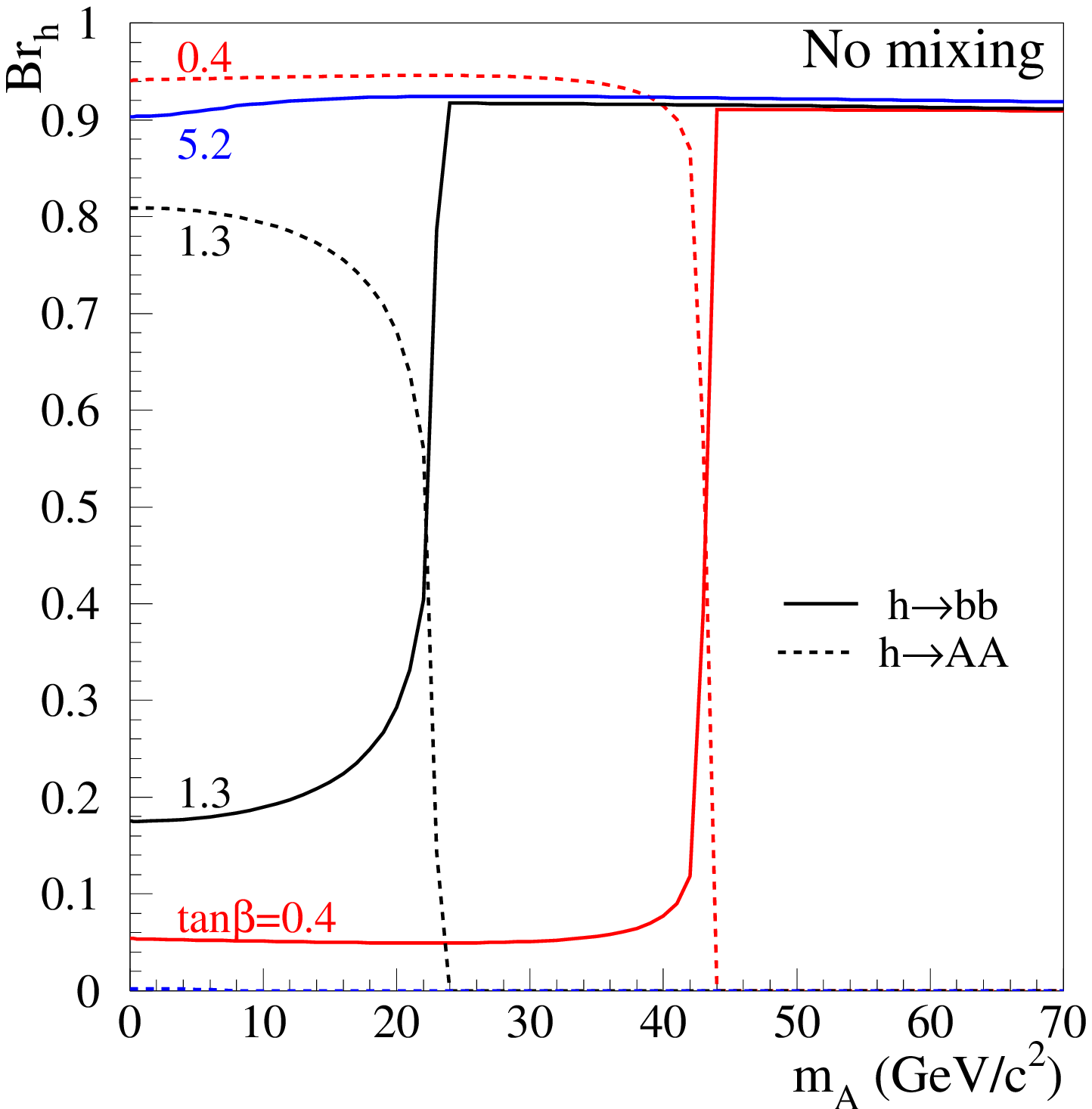,height=7.6cm} 
\end{tabular}
\caption[]{
    Properties of the three neutral Higgs bosons
    of the {\sc CP}-conserving {\sc MSSM} in the no mixing and 
    \mbox{$ m_{\mathrm h}^{\rm max}$} scenarios with  
    \mtop~=~174.3~\GeVcc. 
    Top: H and h masses and H, h and A production cross-sections
    at \rs~=~206~GeV, at various \tbeta\ values.
    Middle: h, H and A widths as a function of \mA\ and \tbeta. 
    Bottom: h branching fractions as
    a function of \mA\ at low to moderate values of \tbeta.
    Decays into \bbbar\ (solid lines) and AA (dashed lines) are compared.
    All dominant two-loop
    order radiative corrections are included~\cite{ref:FDradco_new}.}
\label{fig:heavy_neutral_new}
\end{center}
\end{figure}

In all scenarios, the radiative corrections have been computed 
in the Feynman-diagrammatic approach with all dominant two-loop 
order terms included, using version 2.0 of the {\sc FeynHiggs} 
code~\cite{ref:FDradco_new}. 
As a first illustration of the different scenarios, Table~\ref{tab:mhmax} 
gives the maximum value of \mh\ allowed by theory in each of them, 
for the four values of \mtop\ studied in this paper.
At a given \mtop\ value, the three \mbox{$ m_{\mathrm h}^{\rm max}$} 
scenarios give the highest upper bounds on \mh, the positive $\mu$
scenario leading to the maximal value. The large $\mu$ scenario presents
the lowest upper bound, followed in increasing order by the no mixing
scenario, the gluophobic one, the  no mixing scenario with positive $\mu$ 
and the small $\alpha$ scheme.
The maximum value of \mh\ increases significantly with \mtop. The effect
is most important in the three \mbox{$ m_{\mathrm h}^{\rm max}$} scenarios, 
and is much smaller in the others, especially in the large $\mu$
scheme. It must be noted that the maximum value of
\mh\ corresponds approximately to the minimum value of \mH\ 
in regions of large HZZ couplings (see Fig.\ref{fig:heavy_neutral_new}). 
Thus, 
there are a few scenarios where the H signal is expected to contribute 
to the experimental sensitivity. These are indicated in bold
characters in Table~\ref{tab:mhmax}.

 To illustrate further the Higgs boson phenomenology at {\sc LEP}, a few
properties are compared in Fig.~\ref{fig:heavy_neutral_new} in the case 
of the no mixing and \mbox{$ m_{\mathrm h}^{\rm max}$} scenarios for a top 
quark mass of 174.3~\GeVcc. 
The figures showing masses and cross-sections  underline 
the importance of the signal from the heavy scalar, H, which 
can be kinematically accessible at {\sc LEP2} energies with a large 
\HZ\ production cross-section at large \tbeta\ and moderate \mA, up
to about 100~\GeVcc. 
The width curves demonstrate that, at large \tbeta, 
neutral Higgs bosons can have 
a width exceeding the experimental resolution which is of the order
of 1 to 3~\GeVcc\ depending on the search channel.
At moderate \MA, this affects the h and A bosons and thus the \hA\
production mode, but not the \HZ\ one. At large \MA, width effects
become negligible for the h boson so that the \hZ\ production mode, 
which is the only possible dominant mode in that region, 
is not affected.
The figures showing branching fractions compare
the no mixing and \mbox{$ m_{\mathrm h}^{\rm max}$} scenarios at low \tbeta.
In both scenarios, the h branching fraction into \bbbar\ decreases
to the profit of that into AA at very low \tbeta, 
but the residual \bbbar\ branching ratio is significantly higher  
in the  \mbox{$ m_{\mathrm h}^{\rm max}$} scenario.
   
Finally, it should be noted that our previous {\sc MSSM} 
interpretation of Ref.~\cite{ref:pap00} relied on partial two-loop 
order radiative corrections~\cite{ref:FDradco_old}.
In the present paper, these
have been updated to include all dominant two-loop order 
corrections~\cite{ref:FDradco_new}.
This leads to significant changes in the Higgs boson  masses and properties. 
The main effect is an increase of the maximum (resp. minimum) 
allowed value of the h (resp. H) boson mass at fixed \tbeta.
As a consequence, the experimental sensitivity in \tbeta\ and that in \mH\ 
are expected to decrease. 
A review of the changes induced by the more complete corrections 
on the experimental sensitivity of {\sc DELPHI} is given in
Ref.~\cite{ref:pap03,ref:pap04}
in the framework of the three {\sc LEP} scenarios, keeping
identical experimental inputs.

\subsection{Scan procedure} \label{sec:cpc_procedure}
 
  In each scenario, a scan was performed over the {\sc MSSM} 
parameters \tbeta\ and \MA. The range in \MA\ spans from 0.02~\GeVcc\ 
up to 1~TeV/$c^2$.
Values of \MA\ leading to unphysical negative mass squared values 
were removed from the scans. Such points are rather rare, except in 
the large $\mu$, gluophobic and small $\alpha$ scenarios (see
section~\ref{sec:results}).
The range in \tbeta\ extends from the minimal value allowed in each 
scenario\footnote{The minimal value of \tbeta\ is around 0.7 in the large
$\mu$ scenario and in the no mixing scenario with positive $\mu$ 
and 0.4 in all other schemes. Lower \tbeta\ values
give rise to unphysical negative mass squared values in the Higgs sector.}
up to 50, a value chosen in the vicinity of the ratio of 
the top- and b-quark masses,
above which the Higgs-bottom Yukawa coupling is expected to become
unreliable 
(see e.g.~\cite{ref:mssmpheno}).
The scan steps were 1~\GeVcc\ in \MA\ and 0.1 in \tbeta\ in the regions 
where \mh\ varies rapidly with these parameters. At low \MA, where the
decay modes change rapidly with the Higgs boson mass, values
tested were 0.02, 0.1, 0.25, 0.5, 1.5 and 3~\GeVcc.

  At each point of the parameter space, the neutral and charged Higgs
cross-sections and their branching fractions were taken from 
databases provided by the {\sc LEP} Higgs working group~\cite{ref:hwg}, 
on the basis of the
theoretical calculations in Ref.~\cite{ref:FDradco_new}, completed by
that in Ref.~\cite{ref:hphm_br} for the
charged Higgs boson branching fractions.
The signals from the third Higgs boson, H, were included in the
channel combination
at each point where \mH\
was found to be below 120~\GeVcc, the ultimate sensitivity of {\sc LEP}.
The signal expectations in each channel were then derived from the 
theoretical cross-sections and branching fractions,
the experimental luminosity and the efficiencies. If necessary, a correction
was applied to account for different branching fractions of the Higgs bosons 
between the test point and the simulation (e.g.~for the 
\hZ\ and \HZ\ processes, 
the simulation was done in the {\sc SM} framework).

\begin{figure}[hp]
\begin{tabular}{cc}
\hspace{-1.6cm} \epsfig{file=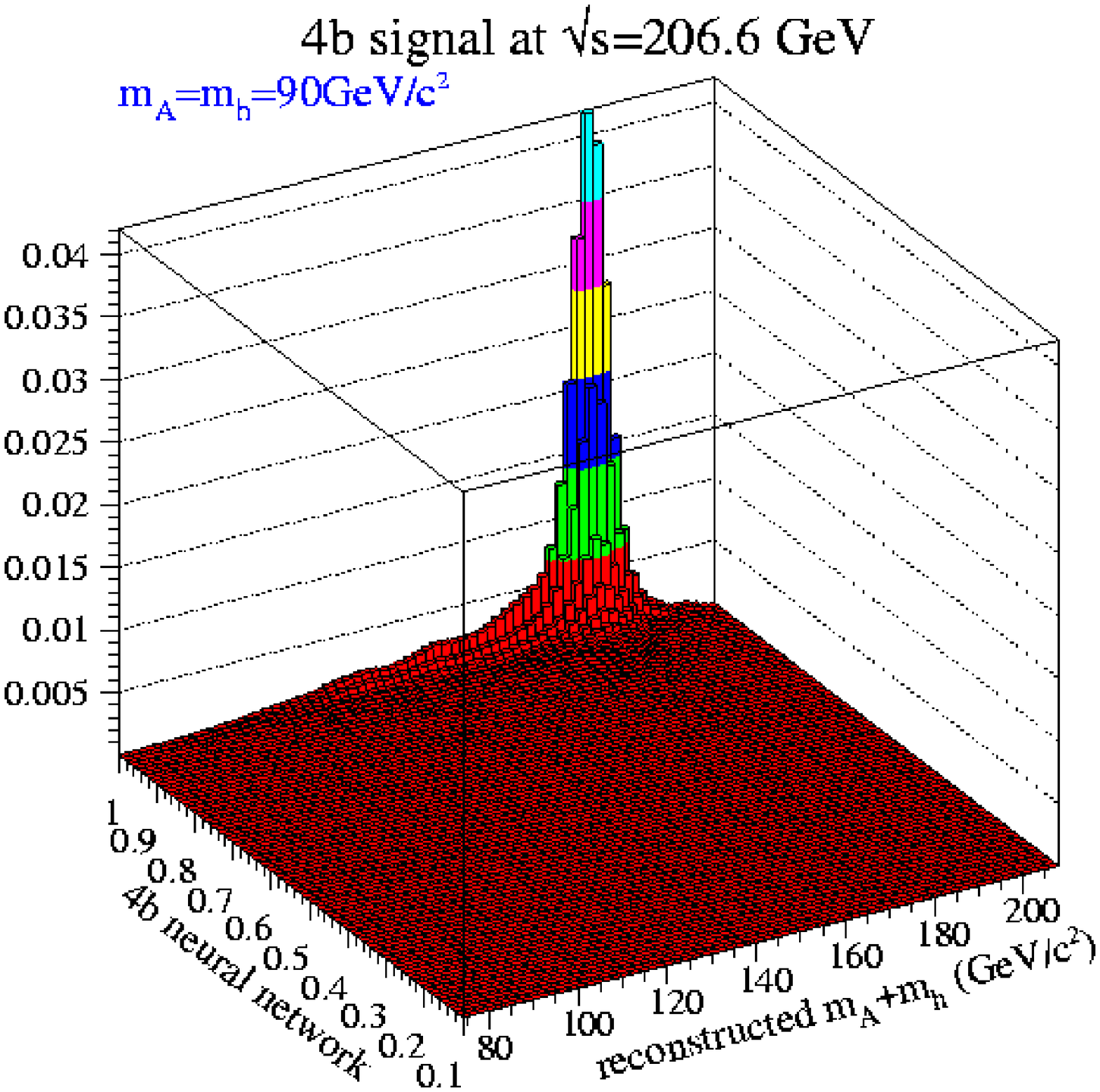,width=9.cm} &
\hspace{-0.6cm} \epsfig{file=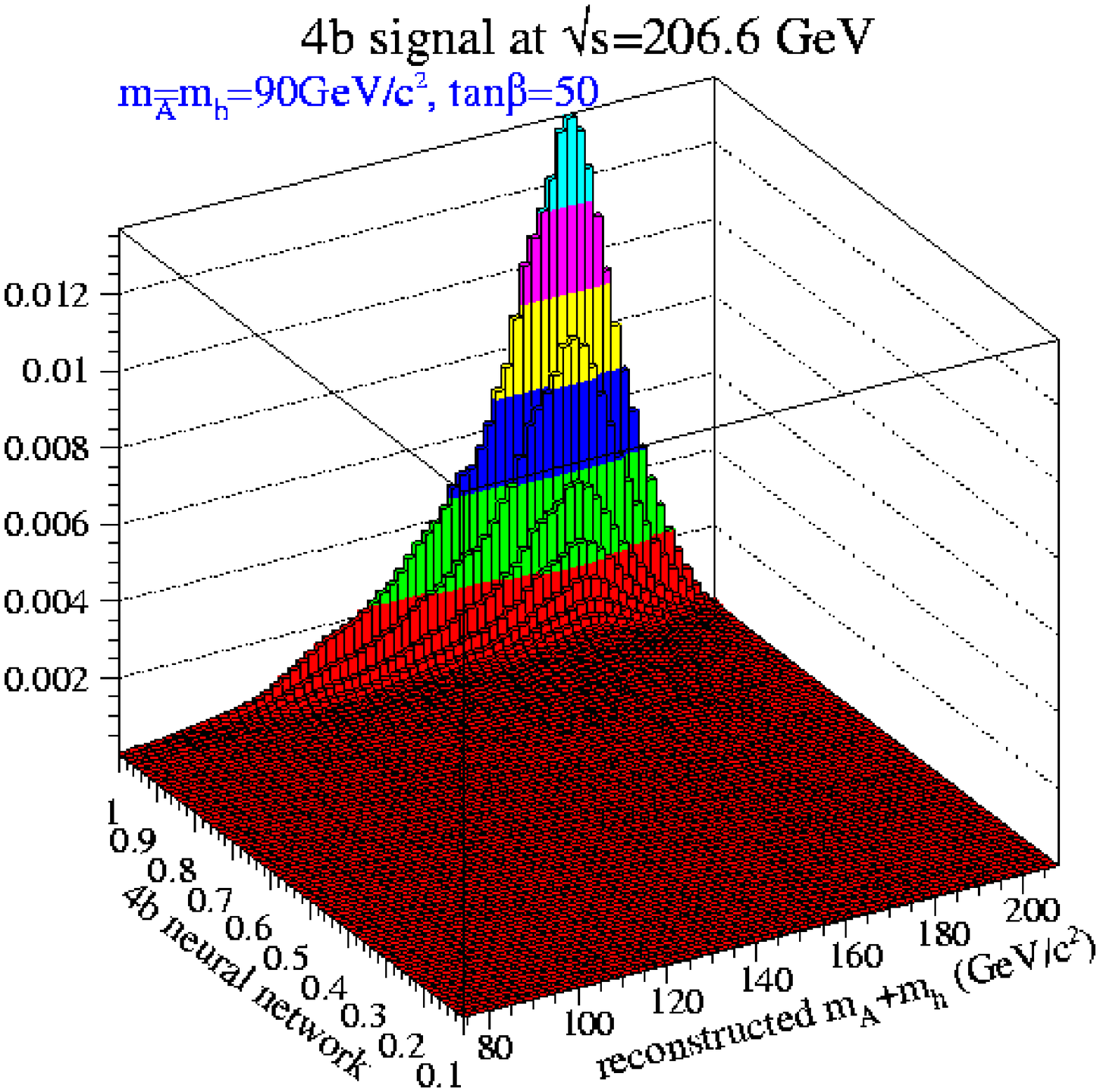,width=9.cm} \\
\hspace{-1.6cm} \epsfig{file=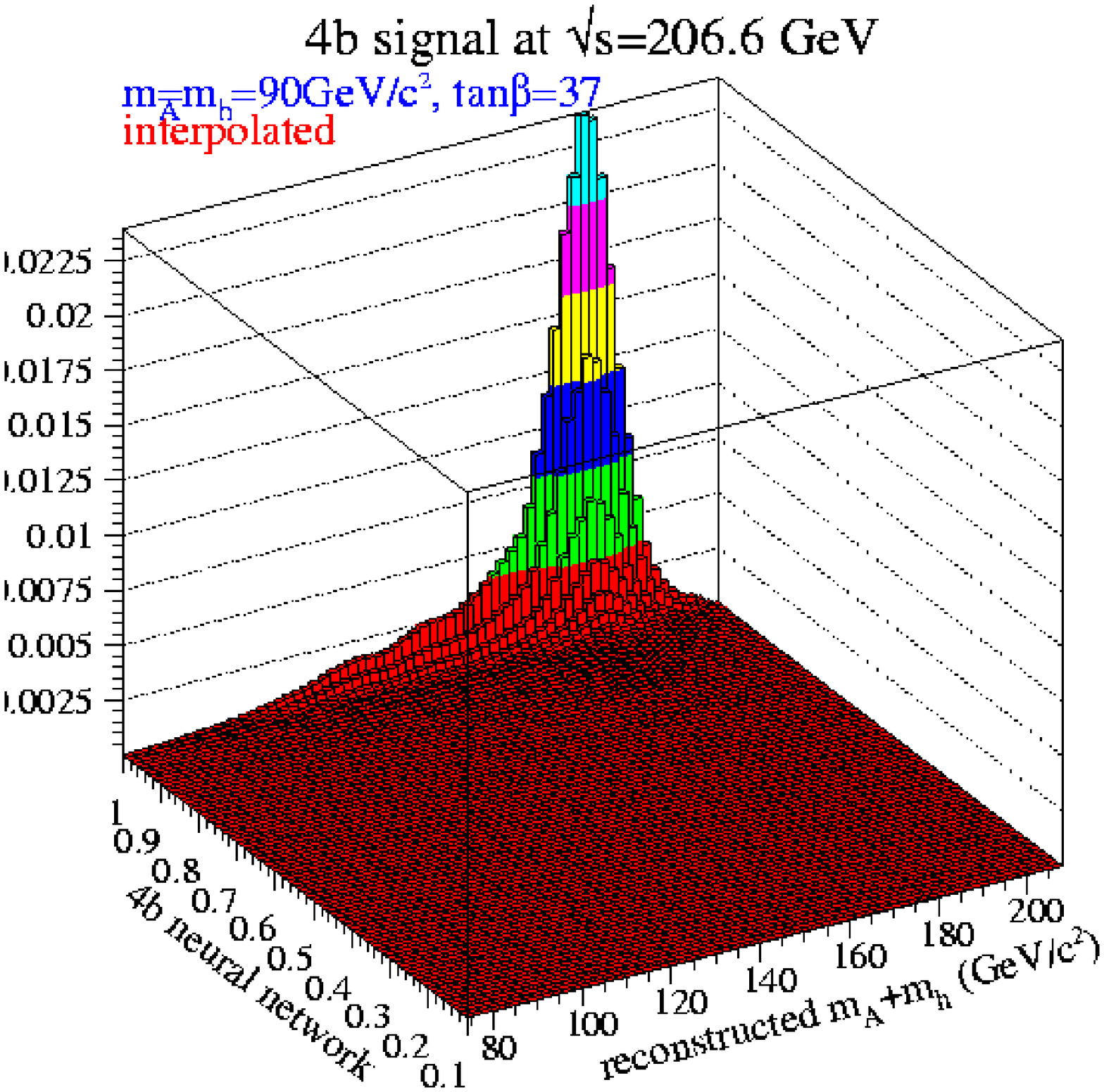,width=9.cm} &
\end{tabular}
\caption[]{
  Two-dimensional PDFs used in the analysis of the 
  \hA\ \rgr\ \bbbar \bbbar\  channel at 
  \rs~=~206.6~GeV~\cite{ref:pap00}. The first discriminant variable is
  the sum of the reconstructed Higgs boson masses while the second
  is a neural network output. 
  Top, left: PDF for a \hA\ signal with \mA~=~\mh~=~90~\GeVcc\ and 
  h and A widths below 1~\GeVcc. 
  Top, right: PDF for a \hA\ signal with \mA~=~\mh~=~90~\GeVcc\ and 
  \tbeta~=~50. The Higgs boson widths in that case are 5 and 9~\GeVcc\
  for A and h, respectively.
  Bottom: PDF linearily interpolated in \tbeta\ at a value of 37. 
}
\label{fig:inter_tb}
\end{figure}

  As stated in the previous section, neutral Higgs bosons can
have non-negligible widths at large \tbeta\ when \mA\ is above a few tens
of~\GeVcc. 
In this region, the experimental sensitivity is dominated 
by the {\sc LEP2} \hA\ analyses dedicated to standard {\sc MSSM}
final states. To account for width effects in these channels,
efficiencies derived from simulations with
h and A widths below 1~\GeVcc\ (see e.g. Ref.~\cite{ref:pap00}) were 
applied for \tbeta~$<$ 30 only. Above that value, efficiencies
were linearly interpolated in \tbeta\ between the efficiencies from 
these simulations and those from simulations at \tbeta~=~50 where
the Higgs boson widths exceed the experimental resolution. 
As the Higgs boson widths grow approximately linearly with 
\tbeta\ above 30, a linear interpolation is valid. 
The same holds for the discriminant information, for which 
the same interpolation software was used as discussed in
section~\ref{sec:pdf}
for the PDF interpolation in mass or centre-of-mass energy. 
The effect of the Higgs boson widths on the PDFs of the \hA\ 
signals and the interpolation in \tbeta\ of these PDFs 
are illustrated in Fig.~\ref{fig:inter_tb}. Note that
the \hZ\ and \HZ\ channels at large \tbeta\ are not affected
by such an effect since in most of the regions where they possibly
contribute, their widths are below the experimental resolution,
as shown in Fig.~\ref{fig:heavy_neutral_new}.

\section{
Results in {\sc CP}-conserving {\sc MSSM} scenarios}\label{sec:results}

The regions of the {\sc MSSM} parameter space excluded at 
95\% {\sc CL} or more by combining the searches of 
Table~\ref{tab:channels} are hereafter discussed in turn 
for each scenario. The exclusion is dominated by the
searches for neutral Higgs bosons in standard {\sc MSSM} final states. 
The searches for neutral Higgs bosons decaying into hadrons of any flavour
and the charged Higgs boson searches complete the exclusion in
restricted regions of the parameter space. In addition,
the limit on the Z partial width that would be due to 
new physics~\cite{ref:width}, $\Gamma^{\rm{new}} < 6.6$~\MeVcc\
is used as an external constraint on the \hA\ process at {\sc LEP1}.
A detailed account of the impact of
these auxiliary constraints can be found in Ref.~\cite{ref:pap04,ref:pap05}.

\subsection{The \mbox{$ m_{\mathrm h}^{\rm max}$} scenario}\label{sec:mhmax}

\begin{figure}[htbp]
\begin{center}
\begin{tabular}{cc}
\hspace{-1.4cm}
\epsfig{figure=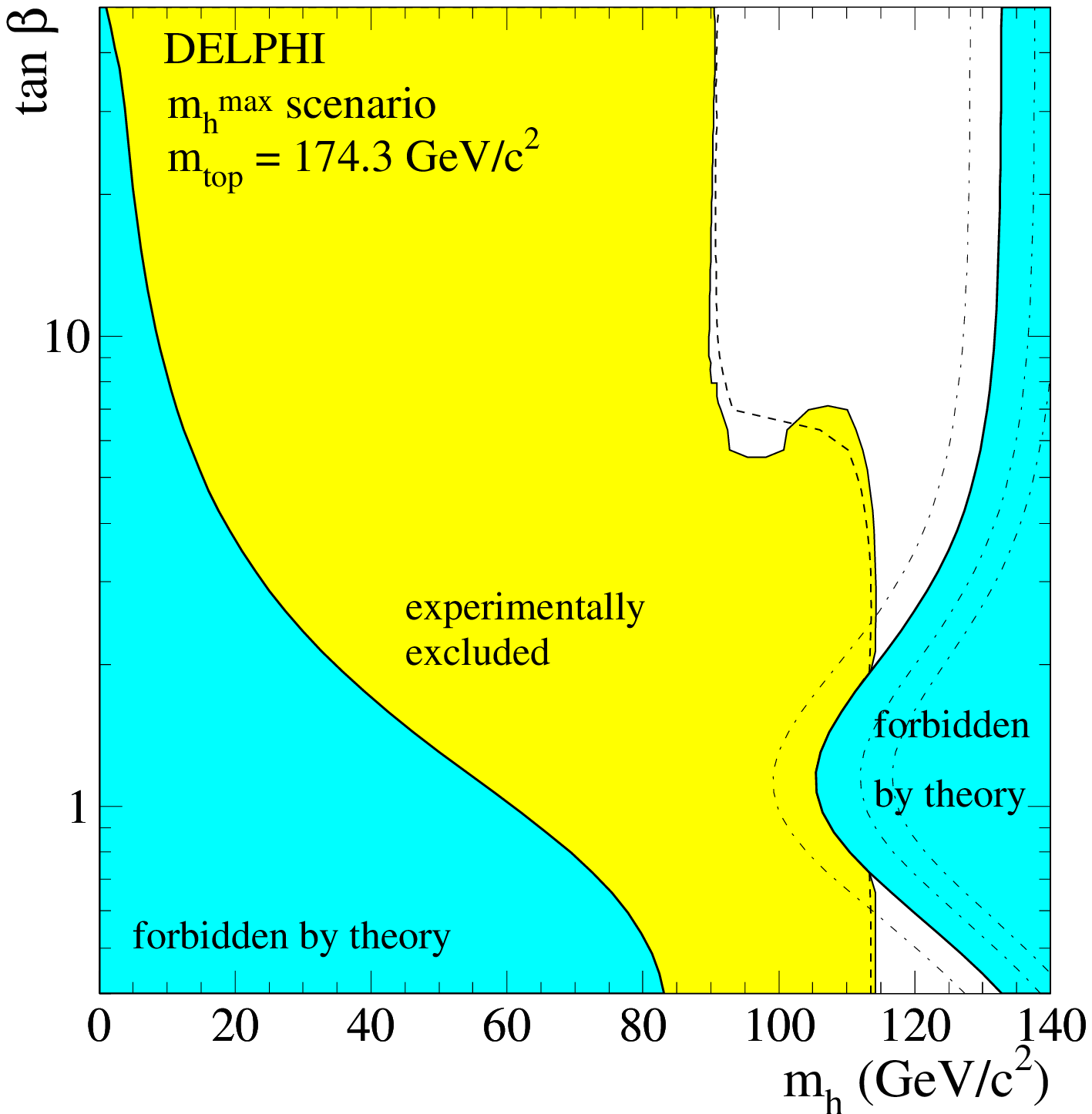,height=9cm} &
\hspace{-0.5cm}
\epsfig{figure=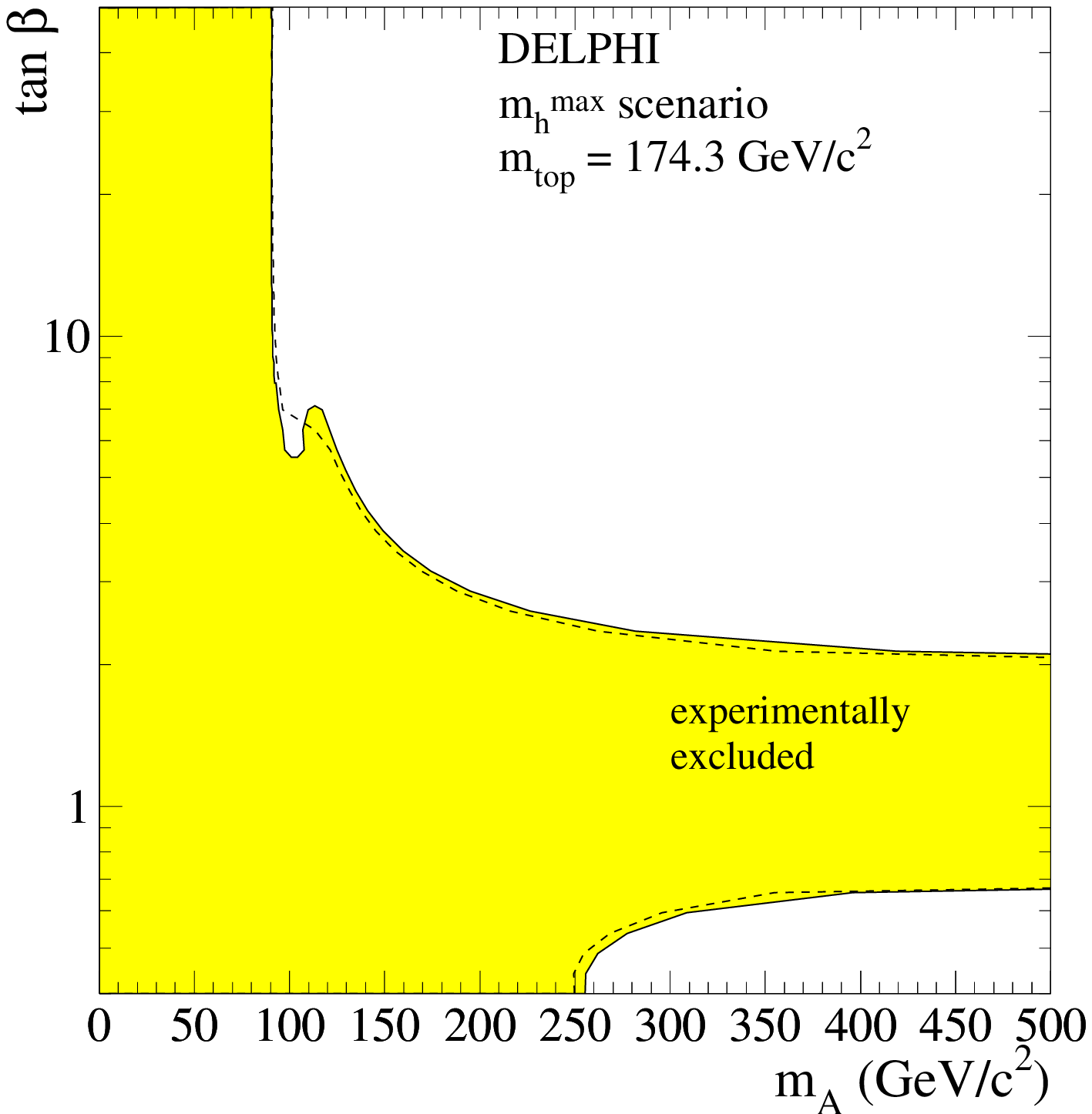,height=9cm} \\
\hspace{-1.4cm}
\epsfig{figure=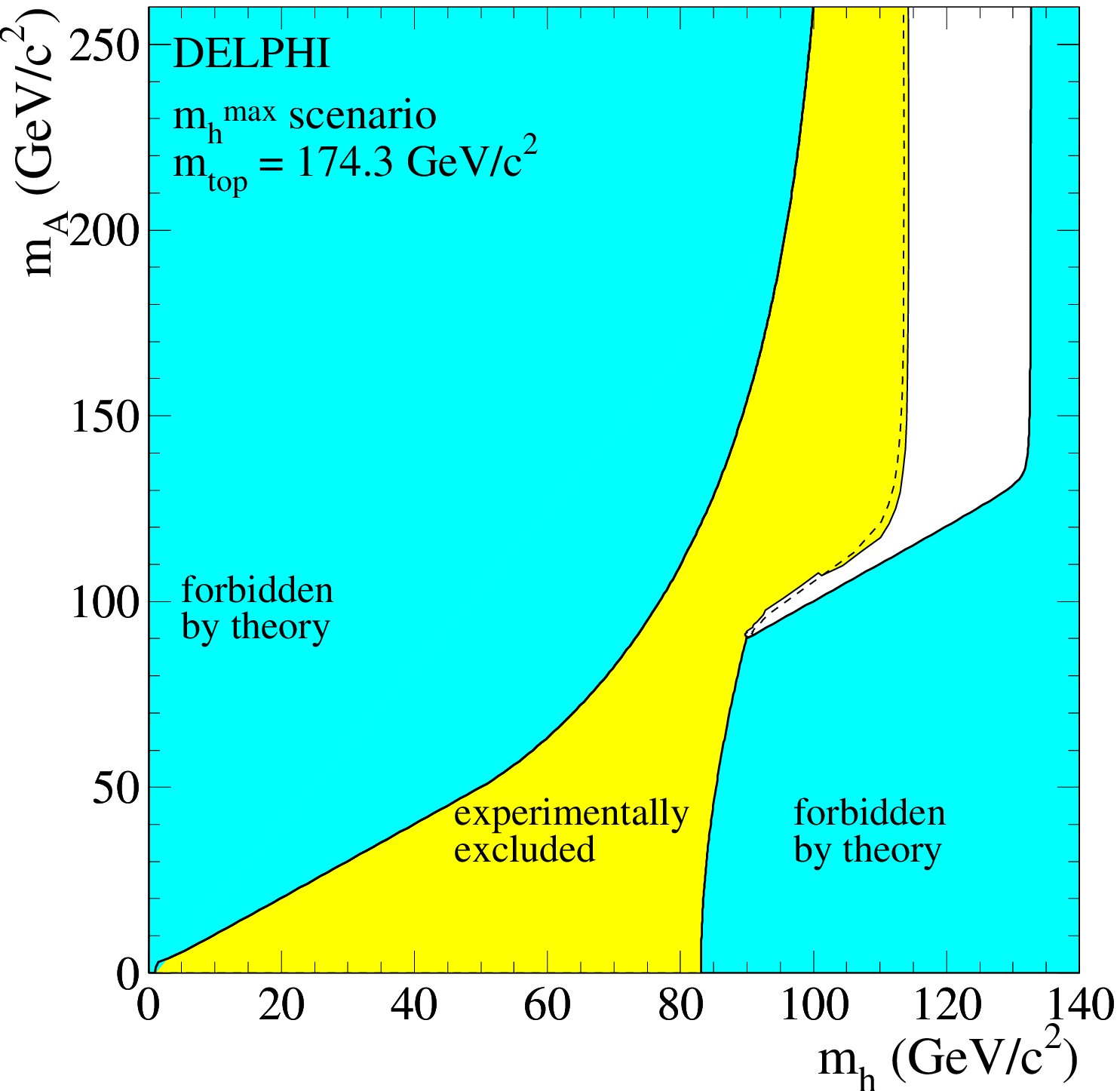,height=9cm} &
\hspace{-0.5cm}
\epsfig{figure=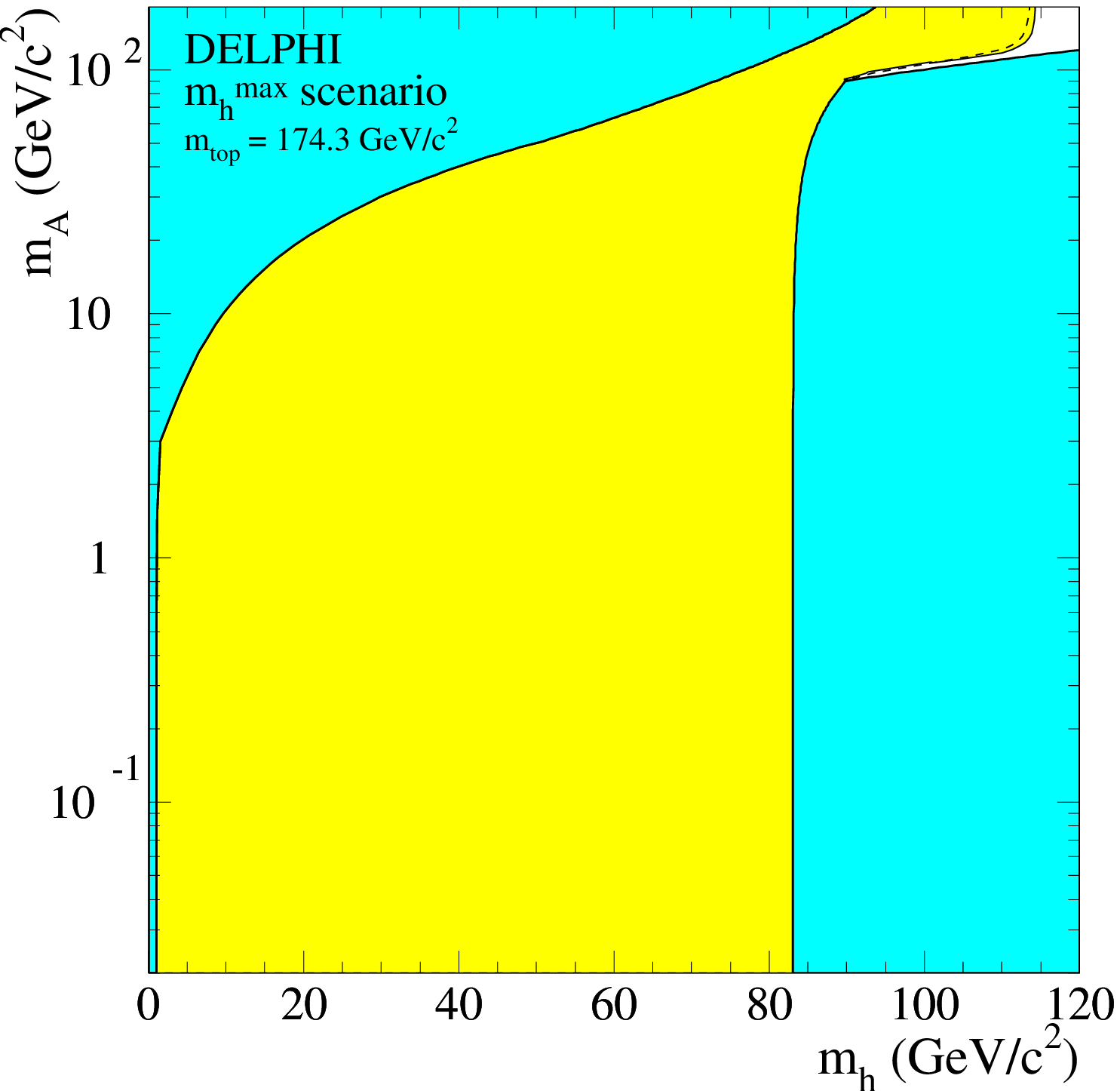,height=9cm} 
\end{tabular}
\caption[]{
   {\sc MSSM} \mbox{$ m_{\mathrm h}^{\rm max}$} scenario for
   a top mass of 174.3~\GeVcc: regions
   excluded at 95\% {\sc CL} by combining the results of the Higgs 
   boson searches in the whole {\sc DELPHI} data sample  
   (light-grey). 
   The dashed curves show the median expected limits.
   The medium-grey areas are the regions not allowed by theory. 
   The dash-dotted 
   lines in the top left-hand plot are the theoretical upper 
   bounds for a top mass of 169.2, 179.4 and 183.0~\GeVcc\
   (from left to right).
   }
\label{fig:limit_max}
\end{center}
\end{figure}

  The excluded regions in the \mbox{$ m_{\mathrm h}^{\rm max}$} scenario
are presented in the (\mh, \tbeta), (\MA, \tbeta) and (\mh, \MA) planes 
in Fig.~\ref{fig:limit_max} for a top mass value of 174.3~\GeVcc.
Basically,
the exclusion is made by the results in the \hZ\ (\hA) channels in the
low (large) \tbeta\ region while they both contribute at intermediate
values. The searches for the heavy scalar, H, brings no additional
sensitivity since H is not kinematically accessible
in this scenario (see Table~\ref{tab:mhmax}).
The above results establish the following 
95\% {\sc CL} lower limits on \mh\ and \MA\ for \mtop~=~174.3~\GeVcc:

\[ \mh > 89.7~\GeVcc \hspace{1cm}
   \MA > 90.4~\GeVcc  \]

\noindent
for any value of \tbeta\ between 0.4 and 50.
The expected median limits are 90.6~\GeVcc\ for \mh\ and 
90.8~\GeVcc\ for \MA. 
The observed limit in \MA\ (\mh) is reached at \tbeta\ around 20 (10),
in a region where both the \hZ\ and \hA\ processes contribute.
For \mtop~=~174.3~\GeVcc\, the range in \tbeta\ between  
0.7 and 1.9 (expected [0.7-1.9]) is excluded 
for any value of \mA\ between 0.02 and 1000~\GeVcc.
These limits and exclusions, as well as those for all the 
{\sc CP}-conserving scenarios, are summarized in Table~\ref{tab:limits}.

The \mtop\ dependence of the above limits was also studied, as reported
in Table~\ref{tab:limits}. The mass limits remain unchanged when 
varying \mtop, for \mh\ is insensitive to \mtop\ in the region
of large \tbeta\ and intermediate \mA\ where the limits are set. 
On the other hand, the excluded 
range in \tbeta\ is governed by the maximal value of \mh, which is 
reached at large \mA\ where \mh\ is very sensitive to \mtop, as 
illustrated in the top left-hand plot in Fig.~\ref{fig:limit_max}: 
hence the variation of the limits
in \tbeta\ as reported in Table~\ref{tab:limits} and Fig.~\ref{fig:mtop}.
An exclusion in \tbeta\ exists for a top mass up to 179.4~\GeVcc\ 
which is about three standard deviations higher than
the current average \mtop\ measurement. The exclusion would vanish for
a top mass as high as 183.0~\GeVcc.

\subsection{The \mbox{$ m_{\mathrm h}^{\rm max}$} scenario but  
with $\mu$ positive and either sign for $X_t$}

\begin{figure}[htbp]
\begin{center}
\begin{tabular}{cc}
\hspace{-1.4cm}
\epsfig{figure=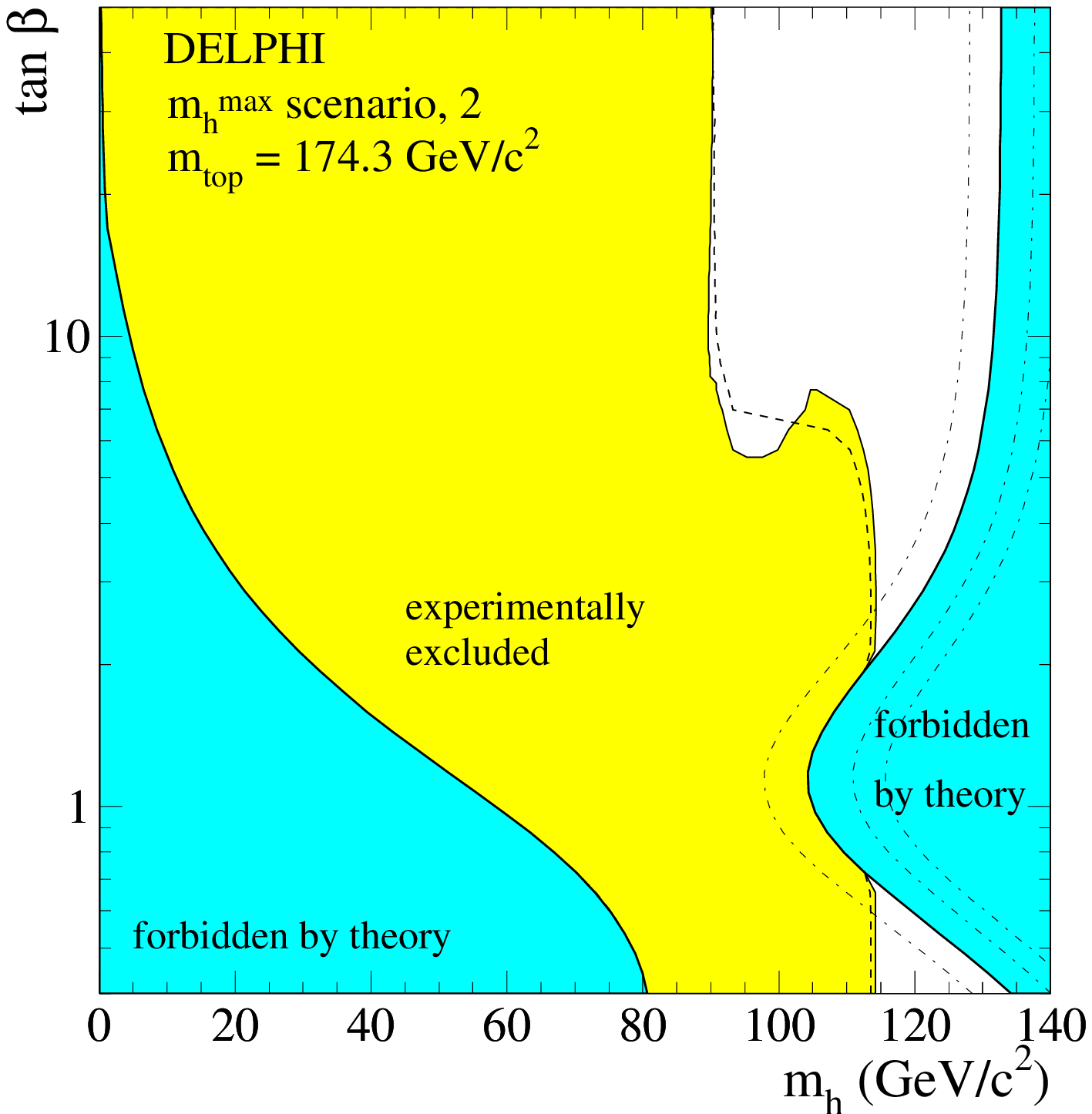,height=9cm} &
\hspace{-0.5cm}
\epsfig{figure=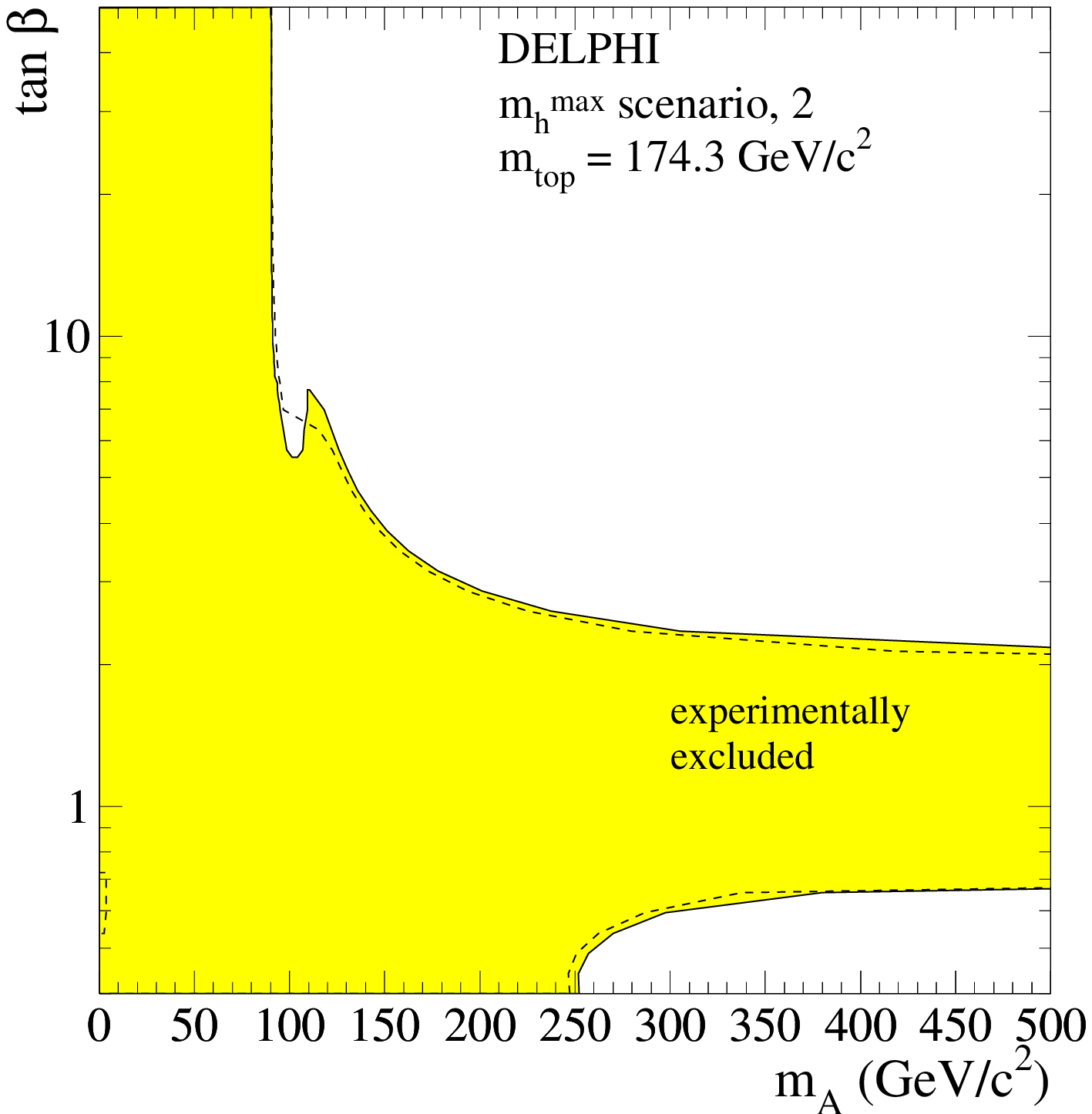,height=9cm} \\
\hspace{-1.4cm}
\epsfig{figure=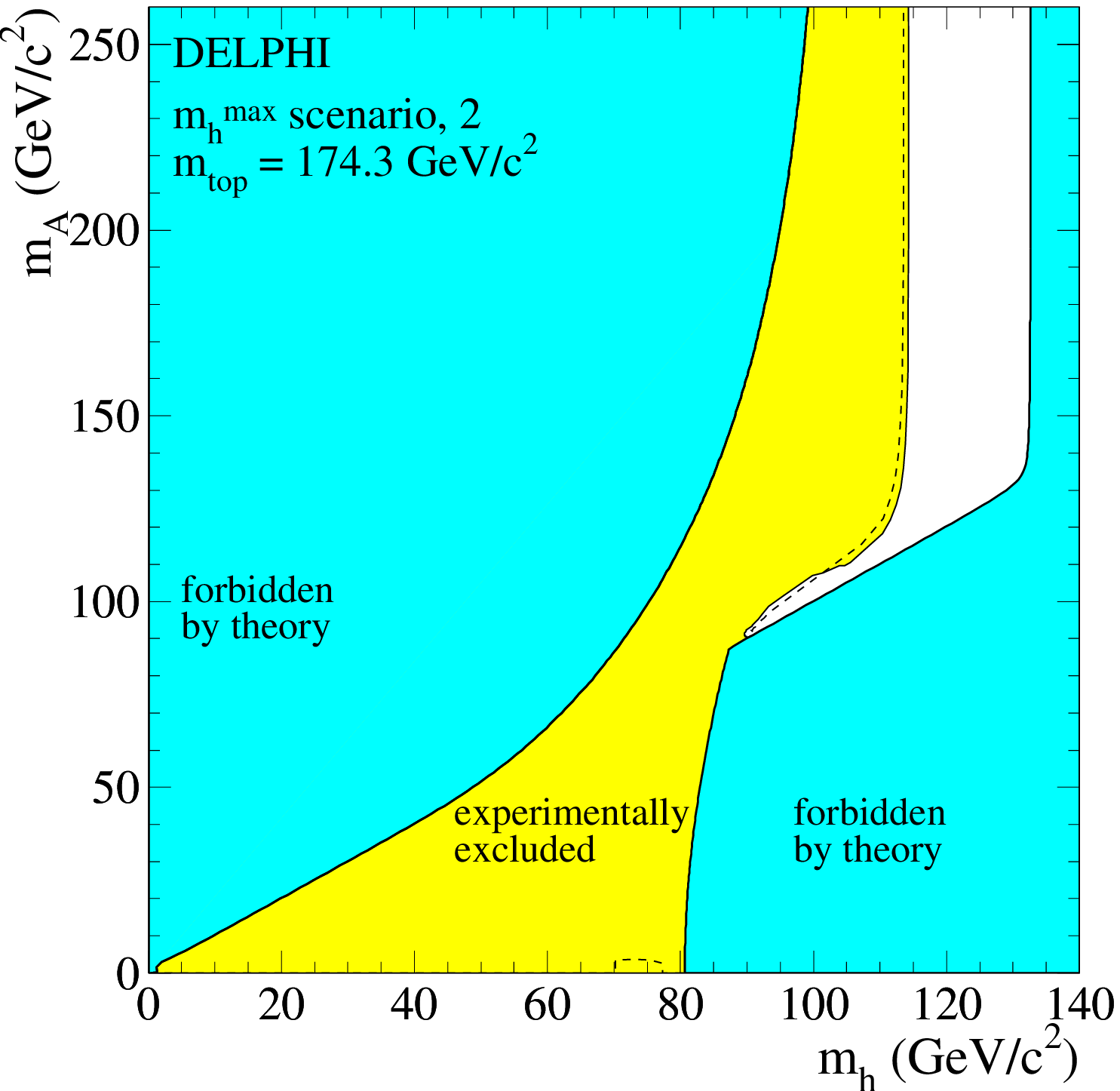,height=9cm} &
\hspace{-0.5cm}
\epsfig{figure=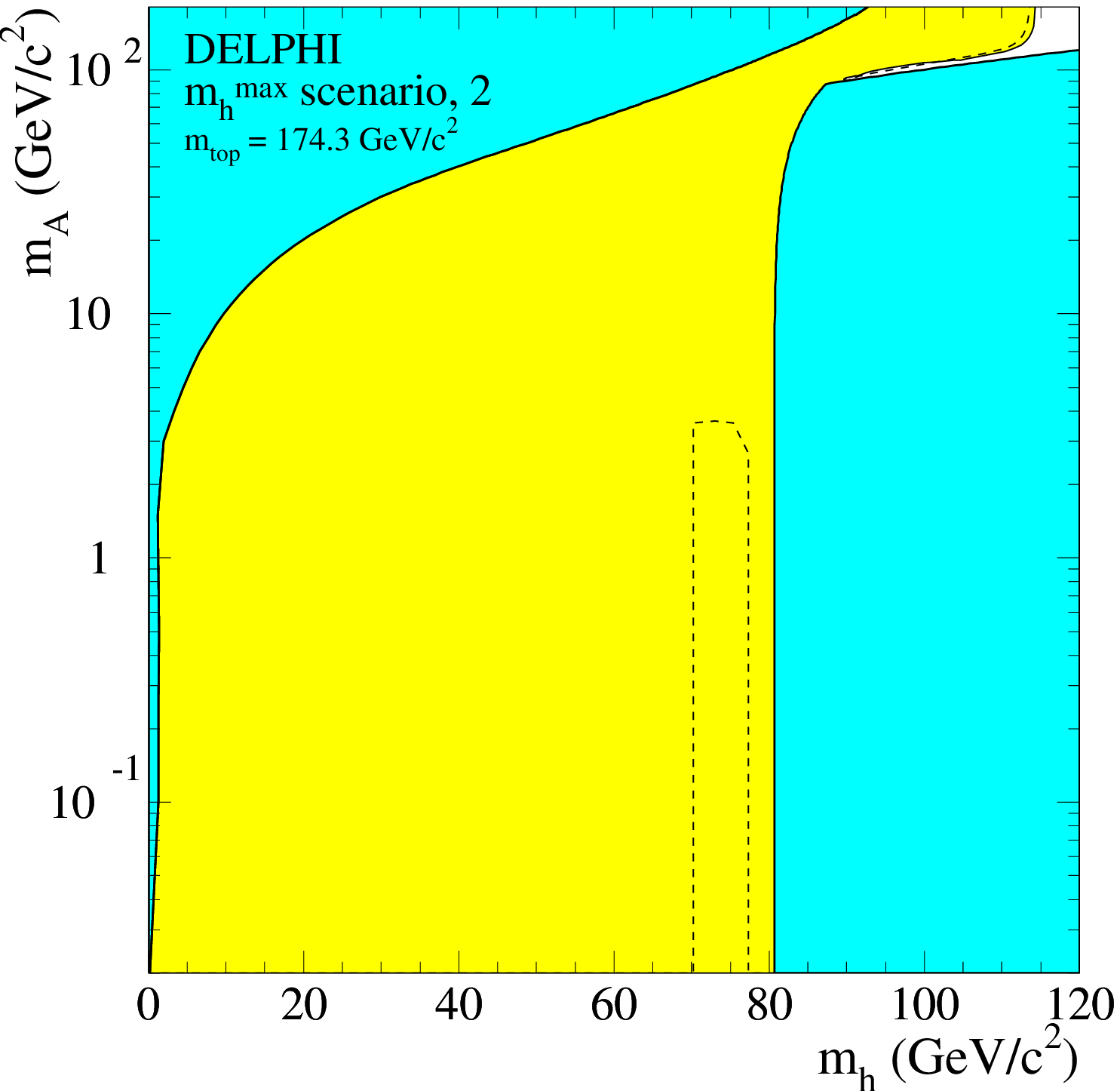,height=9cm} 
\end{tabular}
\caption[]{
   {\sc MSSM} \mbox{$ m_{\mathrm h}^{\rm max}$} scenario with
   positive $\mu$ for a top mass of 174.3~\GeVcc: regions
   excluded at 95\% {\sc CL} by combining the results of the Higgs 
   boson searches in the whole {\sc DELPHI} data sample  
   (light-grey). 
   The dashed curves show the median expected limits.
   The medium-grey areas are the regions not allowed by theory.
   The dash-dotted 
   lines in the top left-hand plot are the theoretical upper 
   bounds for a top mass of 169.2, 179.4 and 183.0~\GeVcc\
   (from left to right).
   }
\label{fig:limit_rev}
\end{center}
\end{figure}

\begin{figure}[htbp]
\begin{center}
\begin{tabular}{cc}
\hspace{-1.4cm}
\epsfig{figure=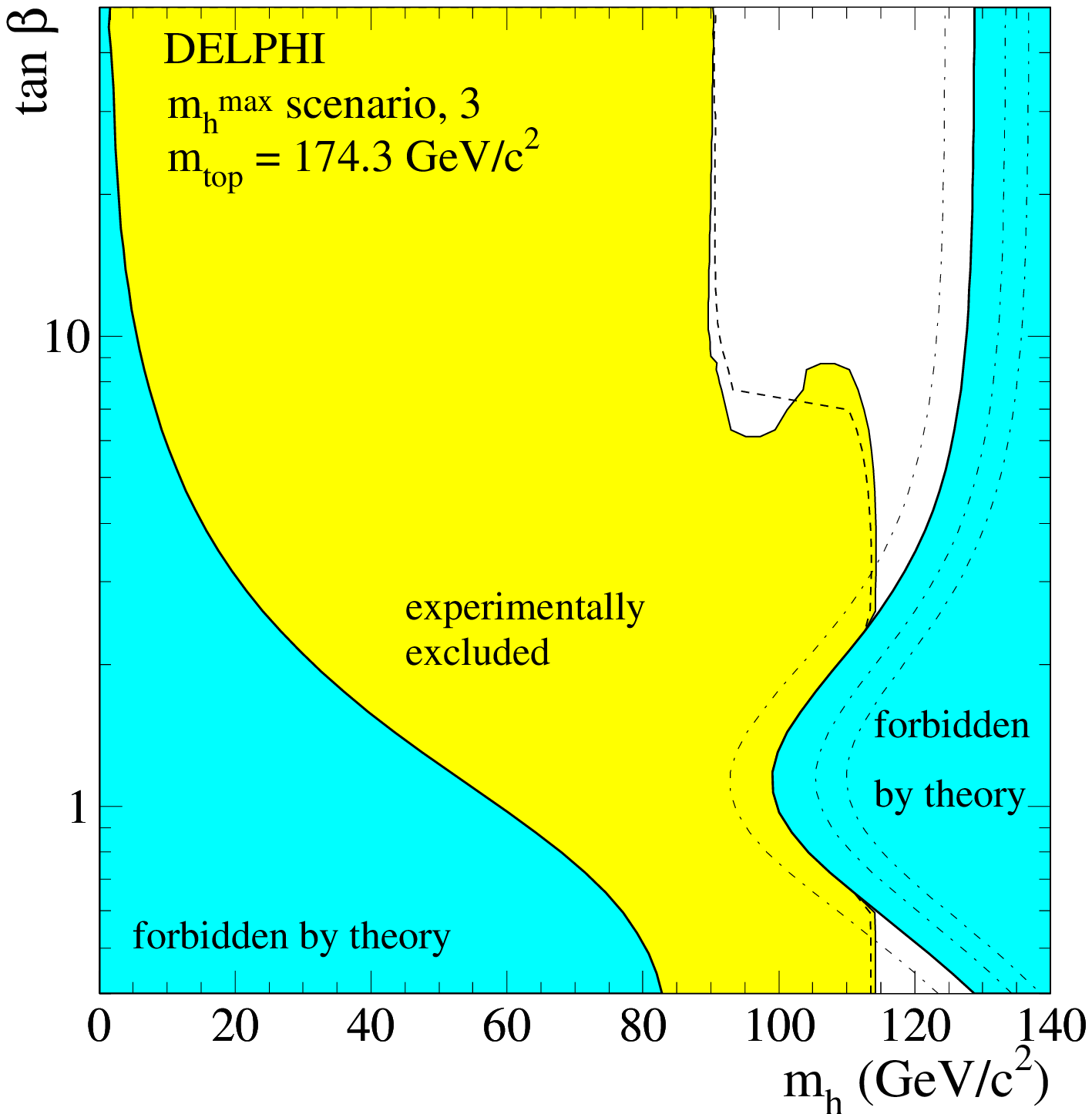,height=9cm} &
\hspace{-0.5cm}
\epsfig{figure=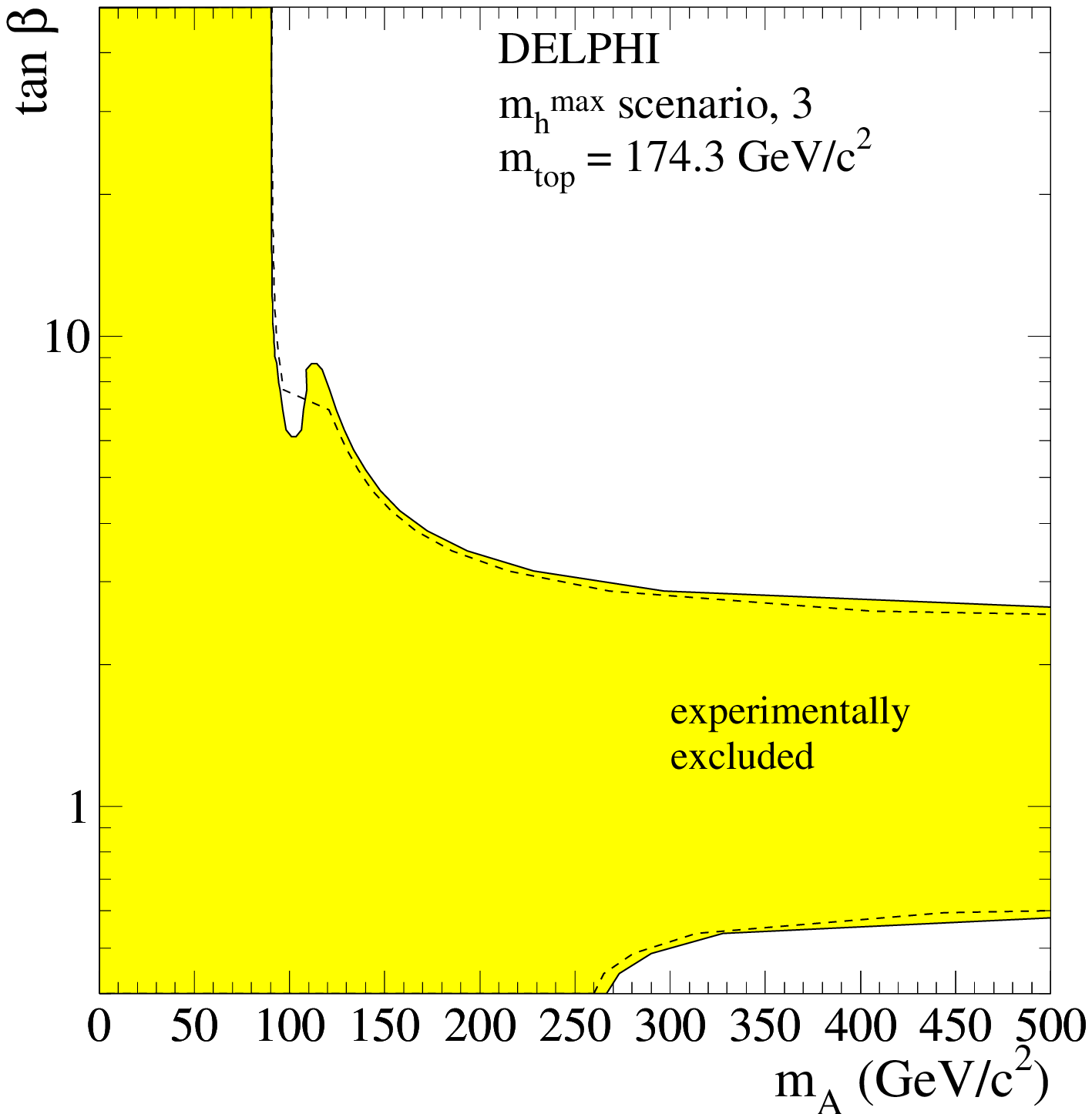,height=9cm} \\
\hspace{-1.4cm}
\epsfig{figure=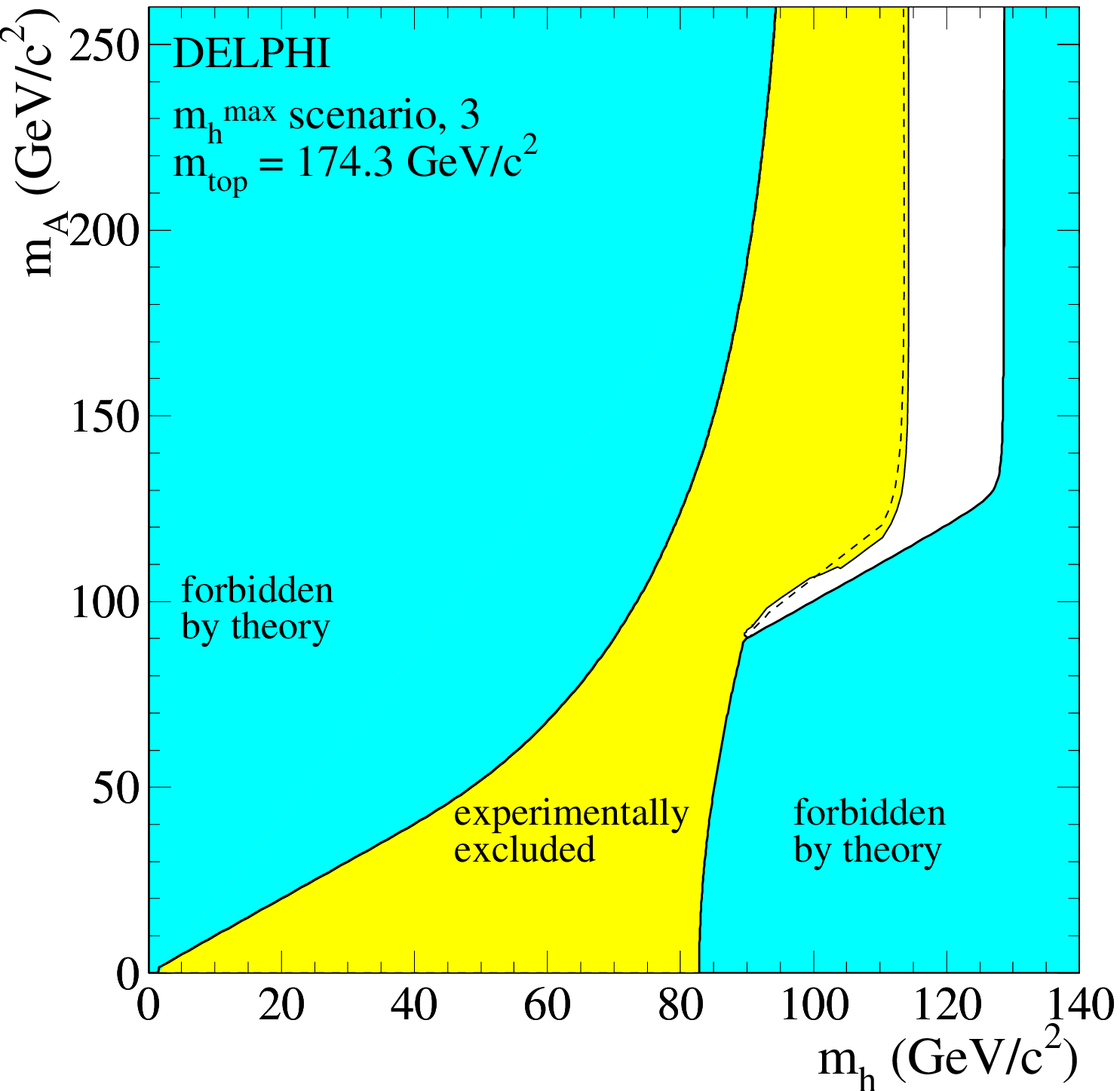,height=9cm} &
\hspace{-0.5cm}
\epsfig{figure=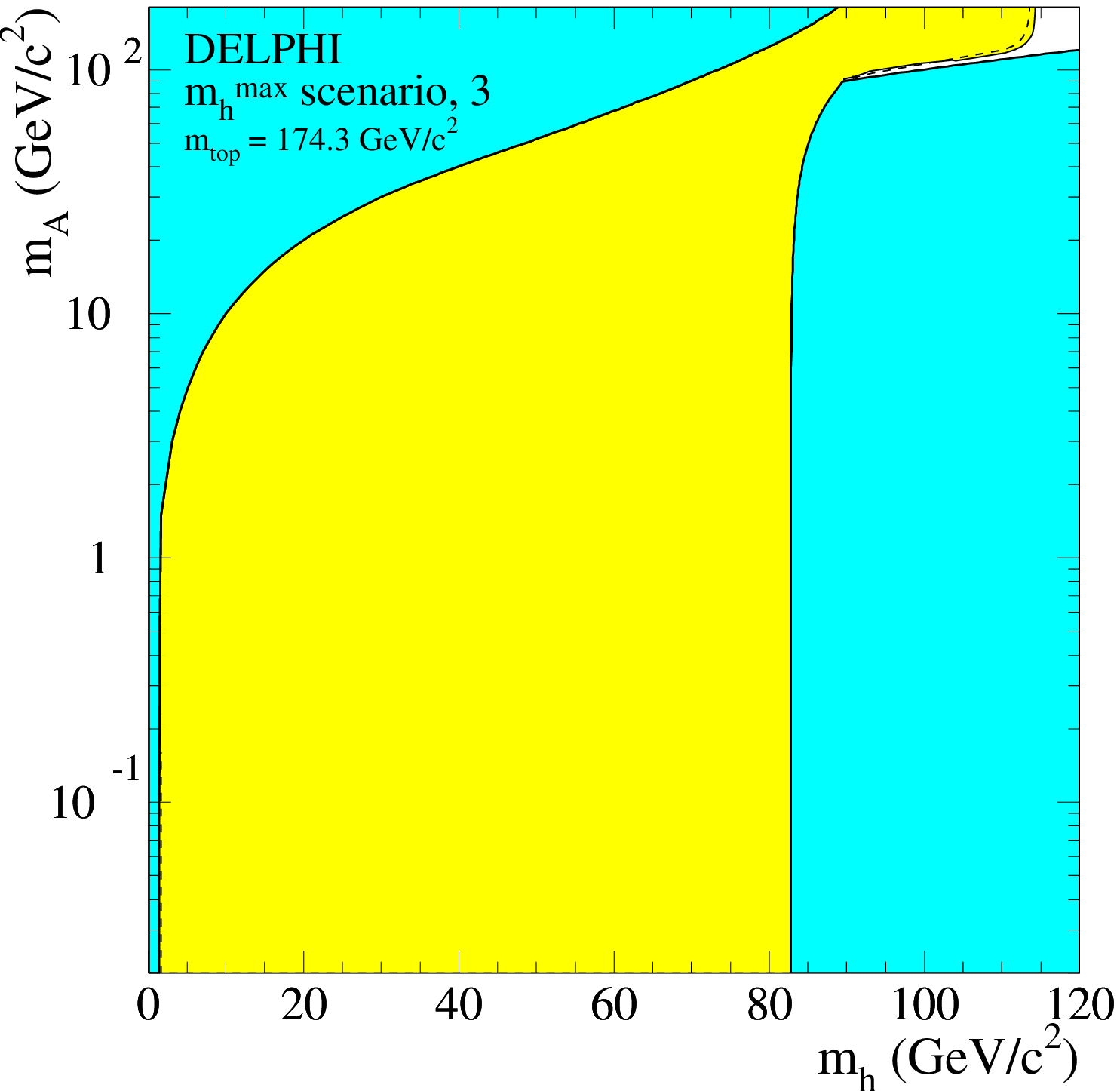,height=9cm} 
\end{tabular}
\caption[]{
   {\sc MSSM} \mbox{$ m_{\mathrm h}^{\rm max}$} scenario with
   positive $\mu$ and negative $X_t$ for a top mass of 174.3~\GeVcc: 
   regions
   excluded at 95\% {\sc CL} by combining the results of the Higgs 
   boson searches in the whole {\sc DELPHI} data sample  
   (light-grey). 
   The dashed curves show the median expected limits.
   The medium-grey areas are the regions not allowed by theory. 
   The dash-dotted 
   lines in the top left-hand plot are the theoretical upper 
   bounds for a top mass of 169.2, 179.4 and 183.0~\GeVcc\
   (from left to right).
   }
\label{fig:limit_maxl}
\end{center}
\end{figure}

  The excluded regions 
for a top mass value of 174.3~\GeVcc\ are presented 
in Fig.~\ref{fig:limit_rev} for the 
\mbox{$ m_{\mathrm h}^{\rm max}$} scenario with positive $\mu$,
keeping $X_t$ positive as in the original 
\mbox{$ m_{\mathrm h}^{\rm max}$} scenario,
and in Fig.~\ref{fig:limit_maxl} for the 
\mbox{$ m_{\mathrm h}^{\rm max}$} scenario with positive $\mu$
and negative $X_t$. 
The results are quite similar to those in the original 
\mbox{$ m_{\mathrm h}^{\rm max}$} scenario.
Mass limits are within 200~\MeVcc\ of those in the previous section
and do not vary significantly with \mtop,
as reported in Table~\ref{tab:limits}.
 
To compare observed and median limits,
the 95\% {\sc CL} lower limits on \mh\ and \MA\  in the
\mbox{$ m_{\mathrm h}^{\rm max}$} scenario with positive $\mu$ 
for \mtop~=~174.3~\GeVcc\ are:

\[ \mh > 89.6~\GeVcc \hspace{1cm}
   \MA > 90.3~\GeVcc  \]

\noindent
for any value of \tbeta\ between 0.4 and 50.
The expected median limits are 90.3~\GeVcc\ for \mh\ and 
90.4~\GeVcc\ for \MA. 
The 95\% {\sc CL} lower limits on \mh\ and \MA\  in the
\mbox{$ m_{\mathrm h}^{\rm max}$} scenario with positive $\mu$ and
negative $X_t$ for \mtop~=~174.3~\GeVcc\ are:

\[ \mh > 89.6~\GeVcc \hspace{1cm}
   \MA > 90.4~\GeVcc  \]

\noindent
for any value of \tbeta\ between 0.4 and 50.
The expected median limits are 90.4~\GeVcc\ for \mh\ and 
90.6~\GeVcc\ for \MA. 

The excluded ranges in \tbeta\ are different in the three 
\mbox{$ m_{\mathrm h}^{\rm max}$} scenarios,
since they have different theoretical upper bounds on \mh.
For \mtop~=~174.3~\GeVcc\, the excluded range in the
\mbox{$ m_{\mathrm h}^{\rm max}$} scenario with positive $\mu$ lies
between 0.7 and 2.0 (expected [0.8-2.0]),
while in the
\mbox{$ m_{\mathrm h}^{\rm max}$} scenario with positive $\mu$ and
negative $X_t$ it spans
from 0.6 to 2.5 (expected [0.6-2.4]).
These limits are valid for any value
of \mA\ between 0.02 and 1000~\GeVcc. Note that despite the higher
maximal value of \mh\ in the 
\mbox{$ m_{\mathrm h}^{\rm max}$} scenario with positive $\mu$,
the most conservative limits in \tbeta\ are still derived in the original
\mbox{$ m_{\mathrm h}^{\rm max}$} scenario (see Section~\ref{sec:mhmax}), 
reflecting the
differences in the theoretical upper bounds at \tbeta\ around 1
(see top left-hand plots in 
Fig.~\ref{fig:limit_max},~\ref{fig:limit_rev} and ~\ref{fig:limit_maxl}).
The \mtop\ dependence of the above limits is presented
in Table~\ref{tab:limits} and Fig.~\ref{fig:mtop}.
For a top mass as high as 183~\GeVcc, there would be no longer
any exclusion in \tbeta\ in the 
\mbox{$ m_{\mathrm h}^{\rm max}$} scenario with positive $\mu$,
while there is still one in the scenario with
positive $\mu$ and negative $X_t$ due to the lower
maximal value of \mh\ in that scenario.

\subsection{The no mixing scenario}

\begin{figure}[htbp]
\begin{center}
\begin{tabular}{cc}
\hspace{-1.4cm}
\epsfig{figure=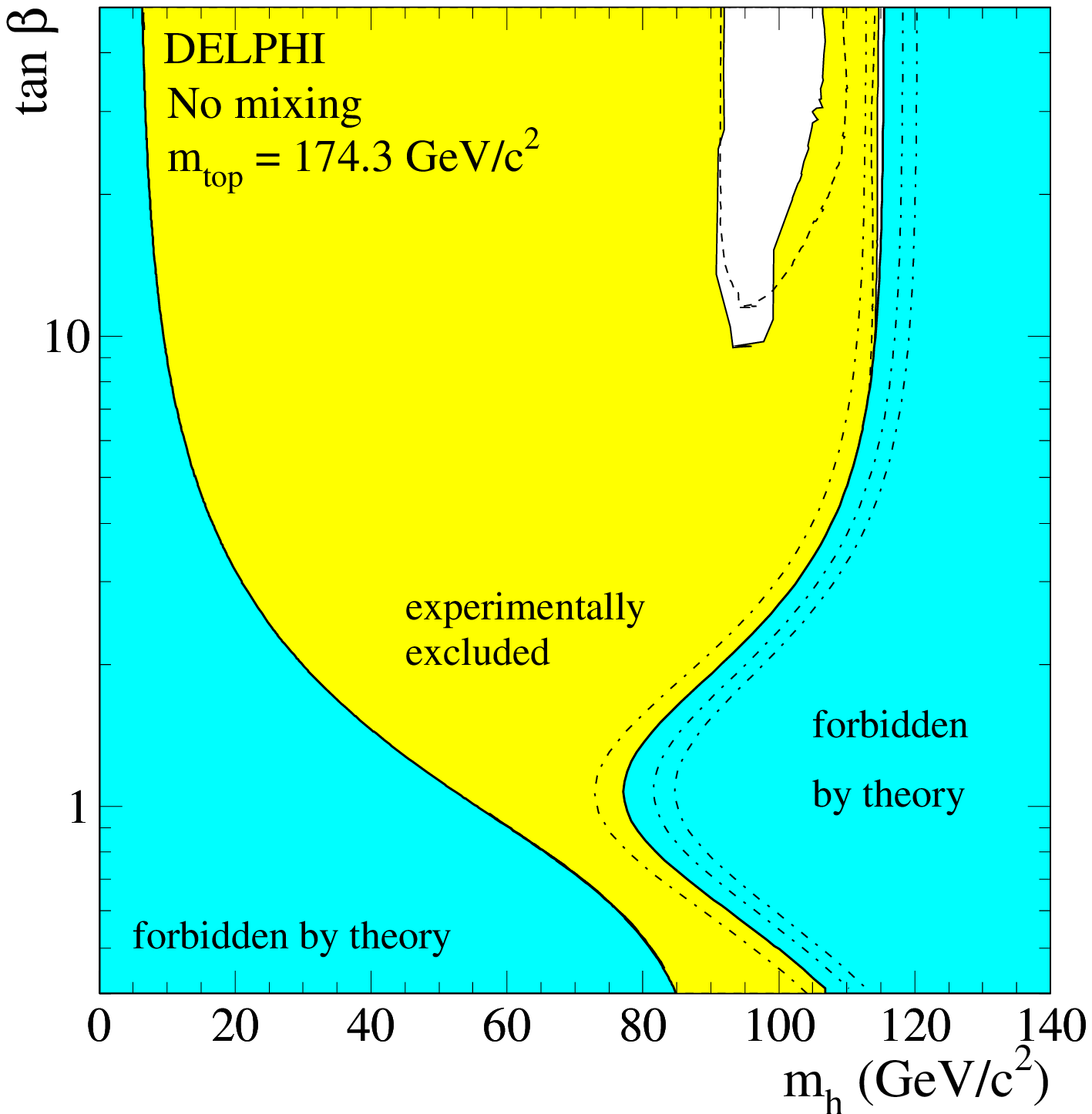,height=9cm} &
\hspace{-0.5cm}
\epsfig{figure=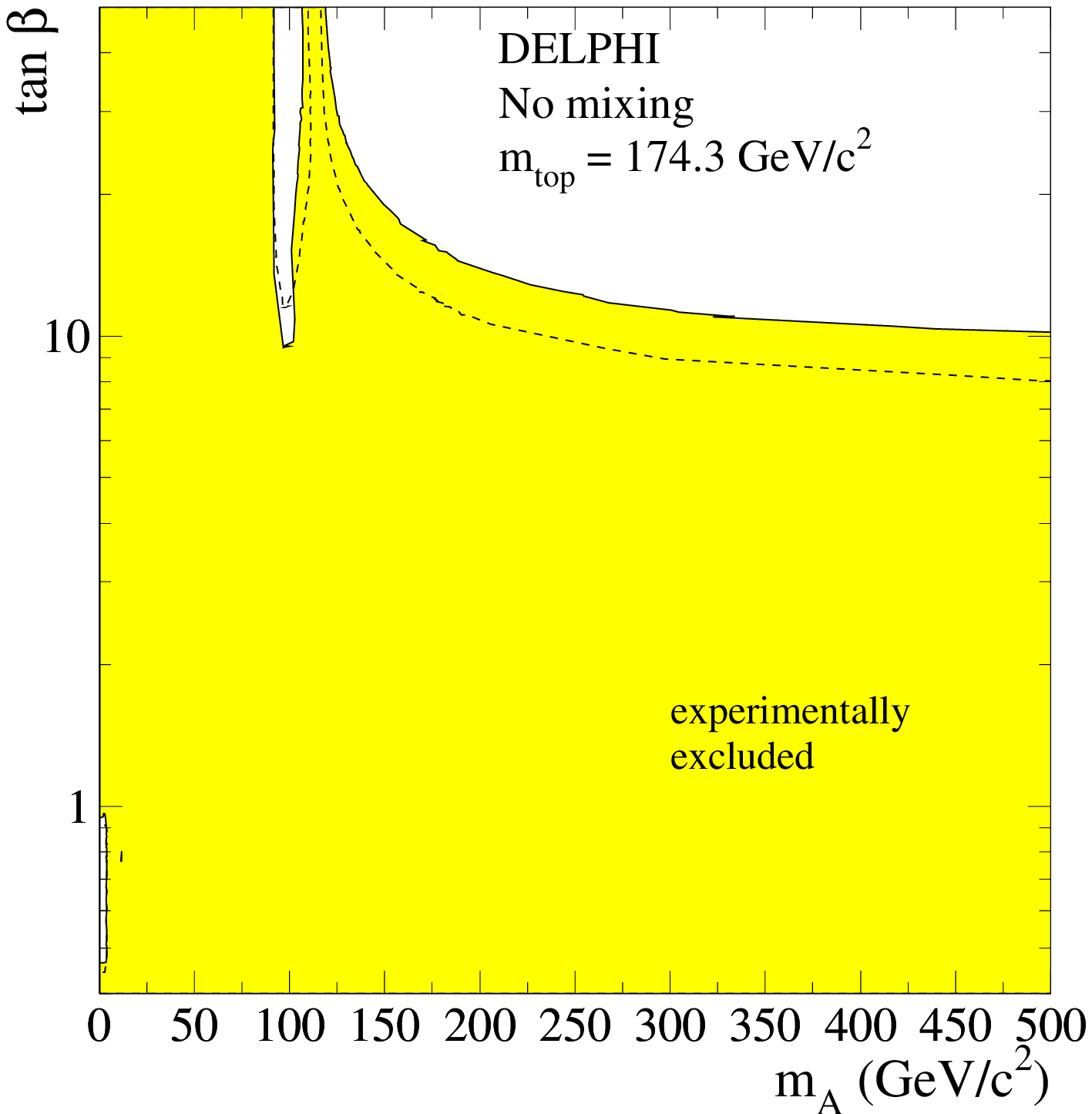,height=9cm} \\
\hspace{-1.4cm}
\epsfig{figure=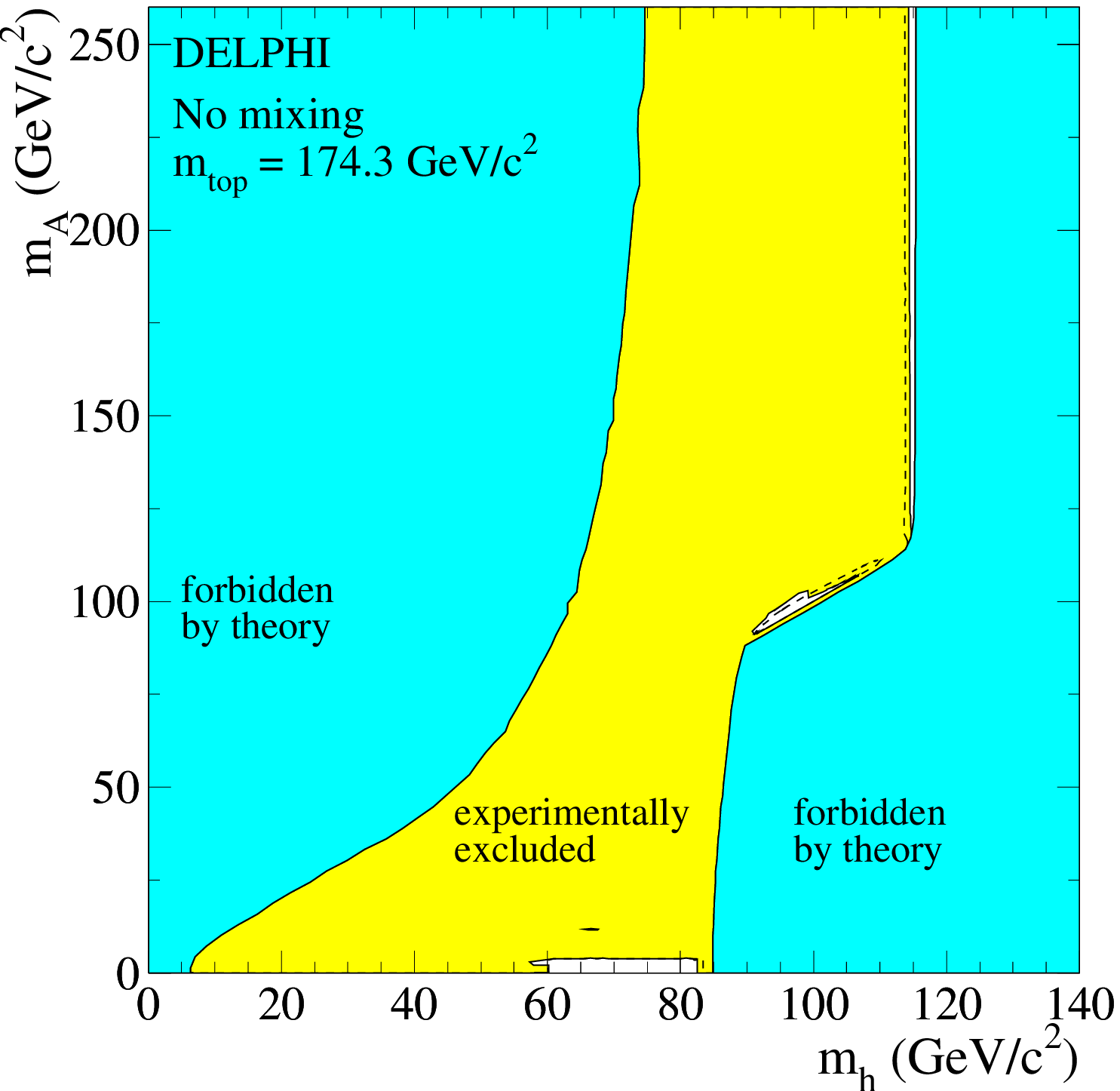,height=9cm} &
\hspace{-0.5cm}
\epsfig{figure=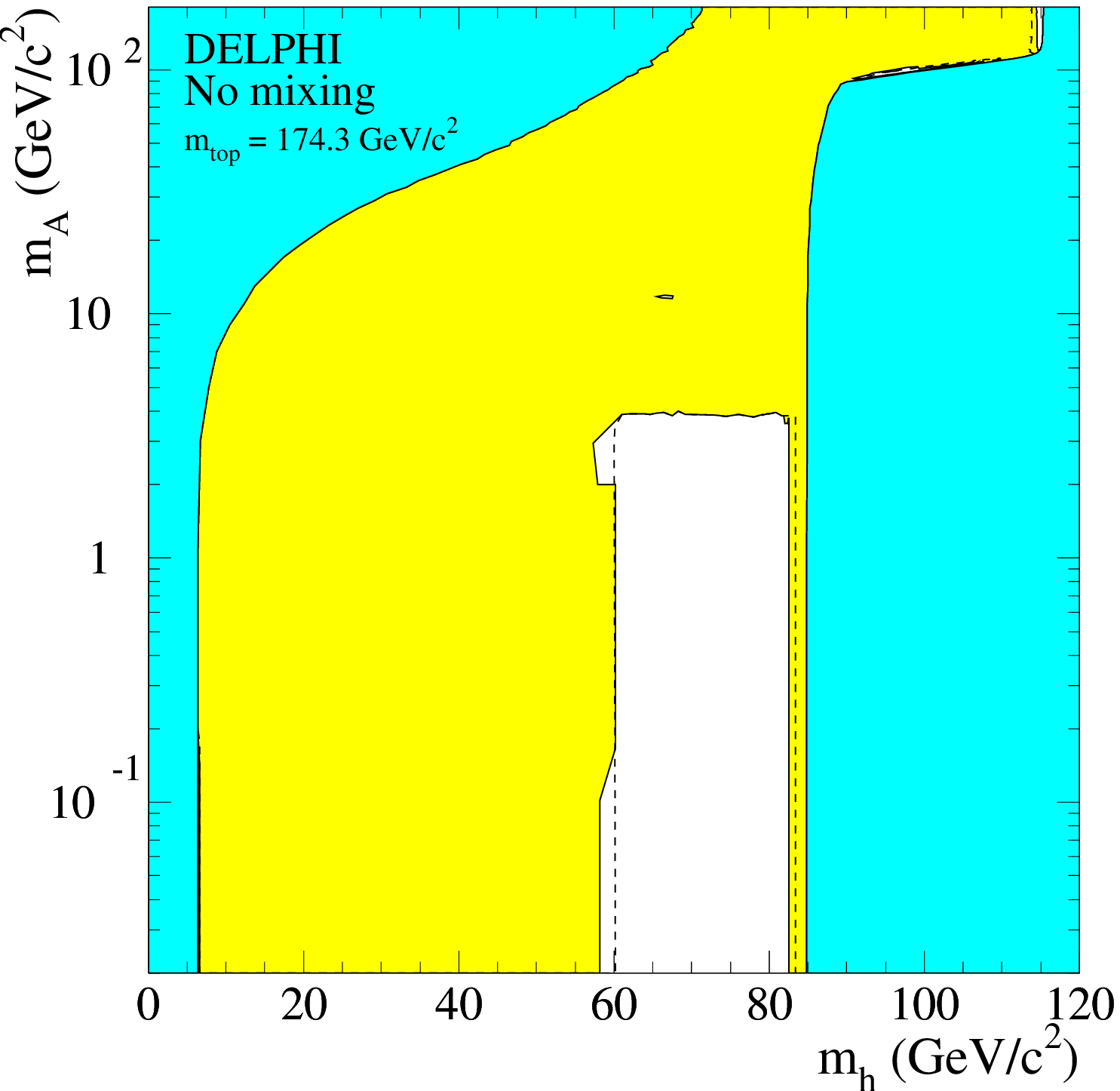,height=9cm}
\end{tabular}
\caption[]{
    {\sc MSSM} no mixing scenario for a top mass of 174.3~\GeVcc: 
    regions excluded at 95\% {\sc CL} by combining the results of the Higgs 
    boson searches in the whole {\sc DELPHI} data sample  
    (light-grey). 
    Among the three unexcluded regions at low \mA, 
    the strip at low \mh\ is fully excluded by the limit on 
    the Z partial width that would be due to new 
    physics~\cite{ref:width}.
    The dashed curves show the median expected limits.
    The medium-grey areas are the regions not allowed by theory. 
    The dash-dotted 
    lines in the top left-hand plot are the theoretical upper 
    bounds for a top mass of 169.2, 179.4 and 183.0~\GeVcc\
    (from left to right).
   }
\label{fig:limit_no}
\end{center}
\end{figure}

  The excluded regions in the no mixing scenario are presented 
in Fig.~\ref{fig:limit_no} for a top mass value of 174.3~\GeVcc.
In this scenario, if the top is not too heavy,
the heavy scalar, H, is kinematically accessible at 
large \tbeta\ and moderate \mA, the region where the mass limits in 
\mA\ and \mh\ are set. 
Thus, allowing for its production increases the sensitivity of the 
searches. 

The zoom at low \mA\ in the (\mh, \MA) projection shows that
the direct searches leave three unexcluded regions below 12~\GeVcc\ in \mA. 
The thin strip along the theoretical lower bound on \mh\ at
very low \mA\ (hardly visible in the figure)
is excluded by the limit on the Z partial width 
that would be due to new physics~\cite{ref:width}, 
$\Gamma^{\rm{new}} < 6.6$~\MeVcc, which, when applied to the \hA\ process, 
translates into an excluded region that encompasses that area.
This is not the case for the two other unexcluded regions. 
These have \tbeta\ below 1.0 and \mh\ between 59 and 82~\GeVcc.
In that region, \MA\ is below the kinematic threshold 
\mh~=~2\MA, the decay \hAA\ opens and supplants the \hbb\ mode,
as can be seen in Fig.~\ref{fig:heavy_neutral_new}.
Our {\sc LEP2} \hAA\ searches, covering A masses above the \ccbar\ 
threshold (see Table~\ref{tab:channels}), have no sensitivity
below 4~\GeVcc\ in \mA.
Similarly, charged Higgs bosons, although kinematically accessible
with a mass between 57 and 82~\GeVcc, have a large branching fraction into
W$^*$A in this region. 
As our charged Higgs boson searches in these channels assume
\MA\ above 12~\GeVcc\ (see Table~\ref{tab:channels}), the overall
experimental sensitivity in these regions remains weak and no exclusion
at 95\% {\sc CL} can be derived, in agreement with the expected
performance. The largest value of \CLs\ is 7\% in the
unexcluded region around 12~\GeVcc\  in \mA\ and 33\% in
the unexcluded hole below  4~\GeVcc.
Note that the nearby region with \mh\ from 82~\GeVcc\ to the
theoretical upper bound on \mh\ is excluded at 95\% {\sc CL}
by the charged Higgs boson searches through their fermionic decays 
which dominate the W$^*$A mode there.

The above results establish the following 95\% {\sc CL} lower limits 
on \mh\ and \MA\ for \mtop~=~174.3~\GeVcc:

\[ \mh > 90.7~\GeVcc \hspace{1cm}
   \MA > 91.2~\GeVcc  \]

\noindent
for any value of \tbeta\ between 1.0 and 50.
The expected median limits are 91.1~\GeVcc\ for both \mh\ and  
\MA. The observed limits in \MA\ and \mh\ are  reached at \tbeta\ 
around 15, in a region where both the \hZ\ and \hA\ processes 
contribute. For \mtop~=~174.3~\GeVcc, two ranges in \tbeta\ are excluded
for any value of \mA\ between 0.02 and 1000~\GeVcc,
the largest interval  being 
between 1.0 and 9.7 (expected [0.9-7.7]). 

The \mtop\ dependence of the above limits was studied, as shown
in Table~\ref{tab:limits} and Fig.~\ref{fig:mtop}. 
In this scenario, both the mass limits and 
the excluded range in \tbeta\ change when varying \mtop. Indeed,
as already mentioned,  
the mass limits in \mA\ and \mh\ rely on the searches for H, whose
mass is very sensitive to \mtop\ in the region where the limits are 
set. Similarly, the  maximal value of \mh, which governs the limits
in \tbeta, is reached at large \mA\ where \mh\ is very sensitive to 
\mtop\ (see Table~\ref{tab:mhmax}). Note that for
a top mass of 169~\GeVcc, \MH\ decreases by
3~\GeVcc\ in the region where the mass limits are set, making the H signal 
more within the sensitivity of {\sc LEP2}: the whole parameter space
of the no mixing scenario is then accessible and found to be excluded
at 95\% {\sc CL}, apart from two holes at 
\tbeta\ below 1.0, one at \mA\ around 12~\GeVcc, which is excluded at
92\% {\sc CL},
and a larger one below 4~\GeVcc, which is disfavoured at
69\% {\sc CL} only.

\subsection{The no mixing scenario but
with positive $\mu$ and large $M_{\rm susy}$}

\begin{figure}[htbp]
\begin{center}
\begin{tabular}{cc}
\hspace{-1.4cm}
\epsfig{figure=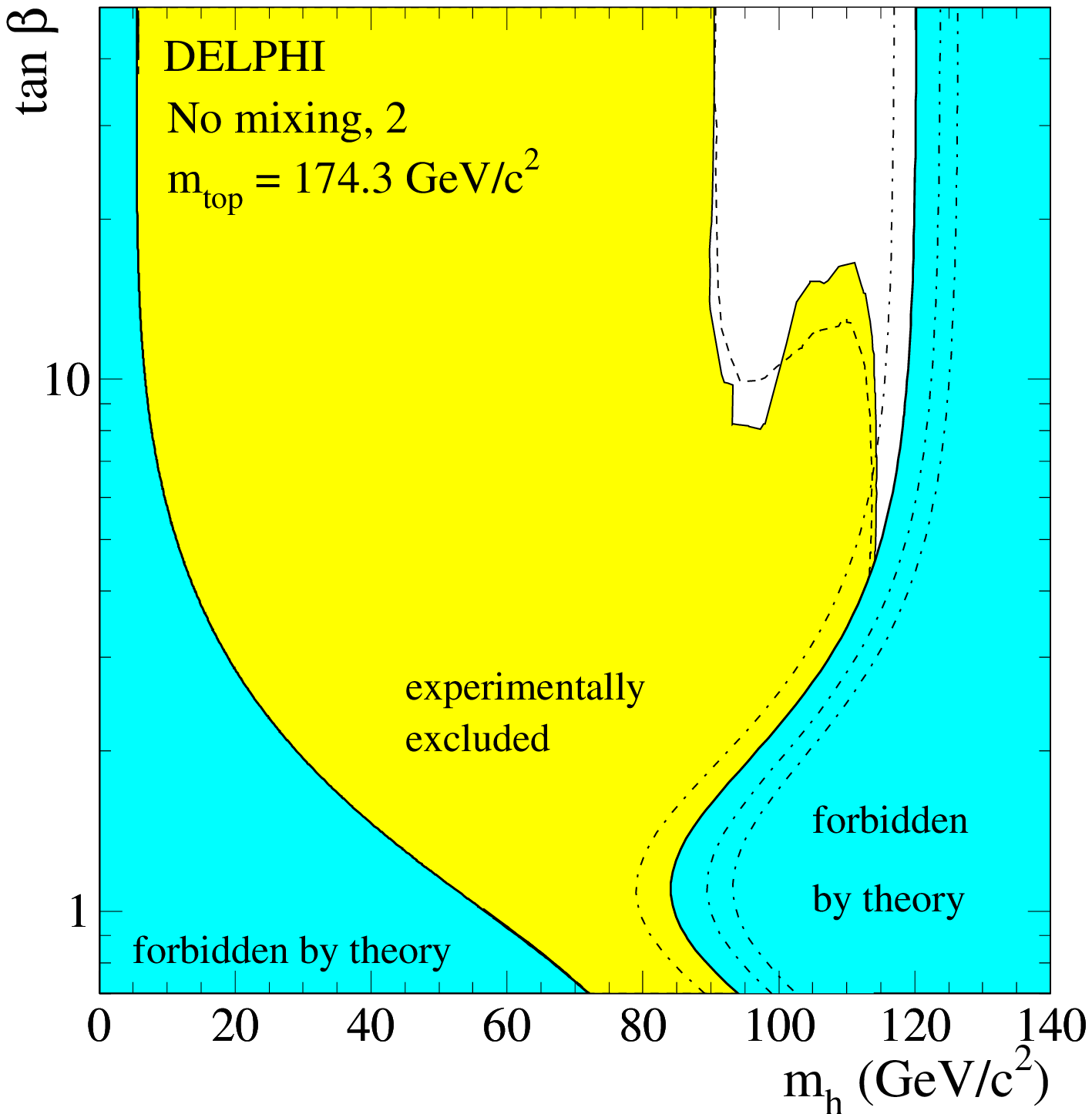,height=9cm} &
\hspace{-0.5cm}
\epsfig{figure=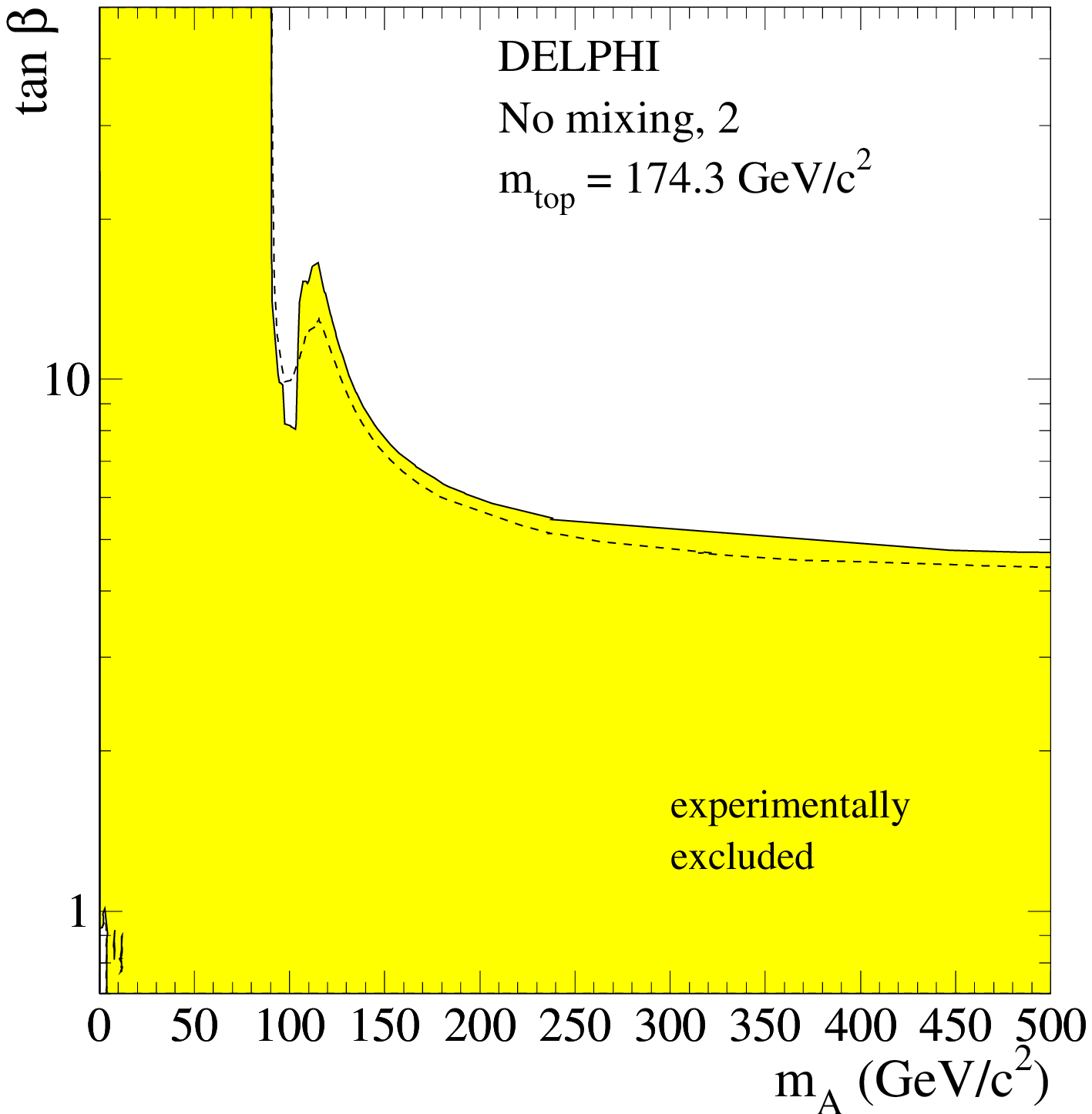,height=9cm} \\
\hspace{-1.4cm}
\epsfig{figure=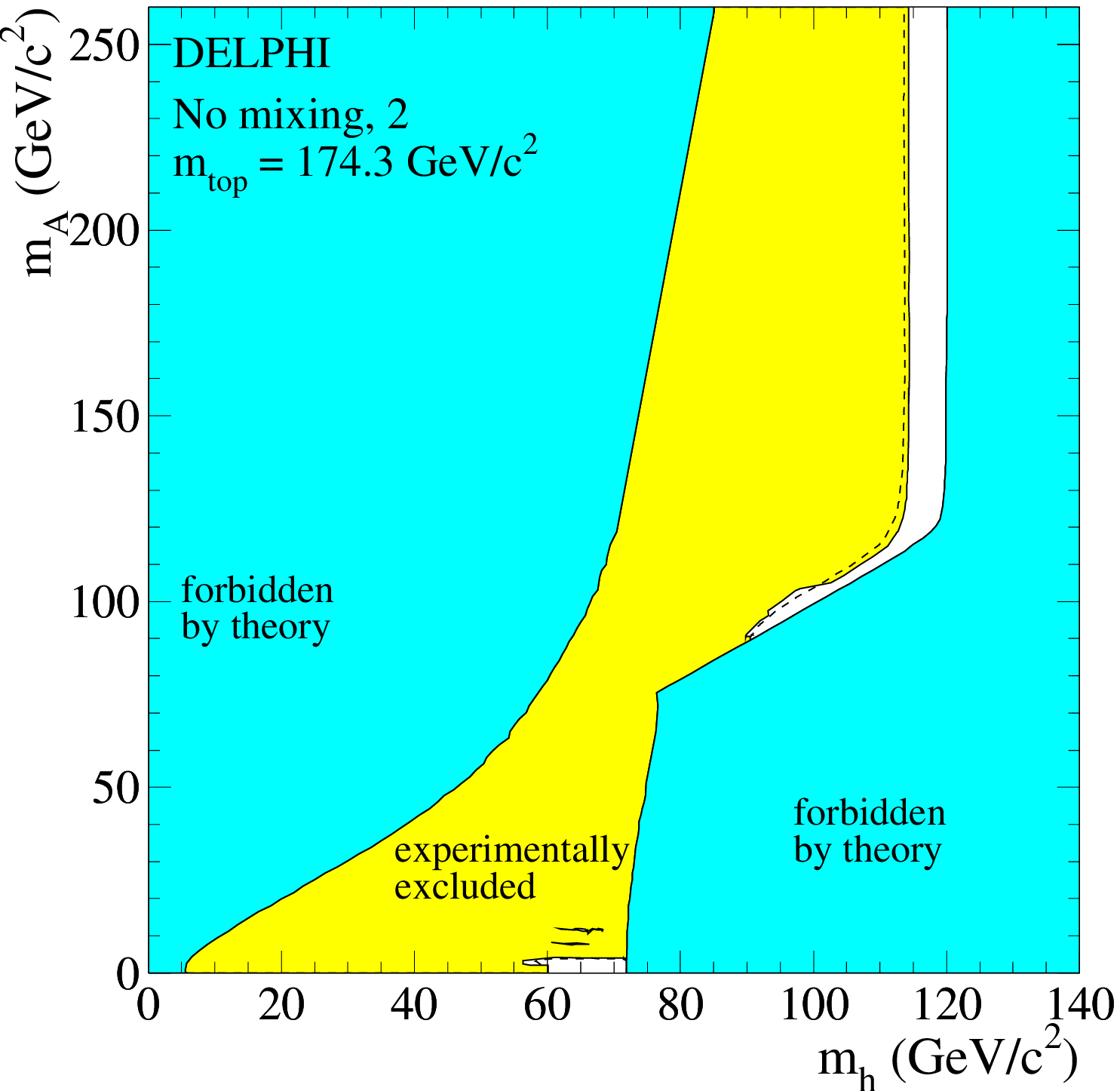,height=9cm} &
\hspace{-0.5cm}
\epsfig{figure=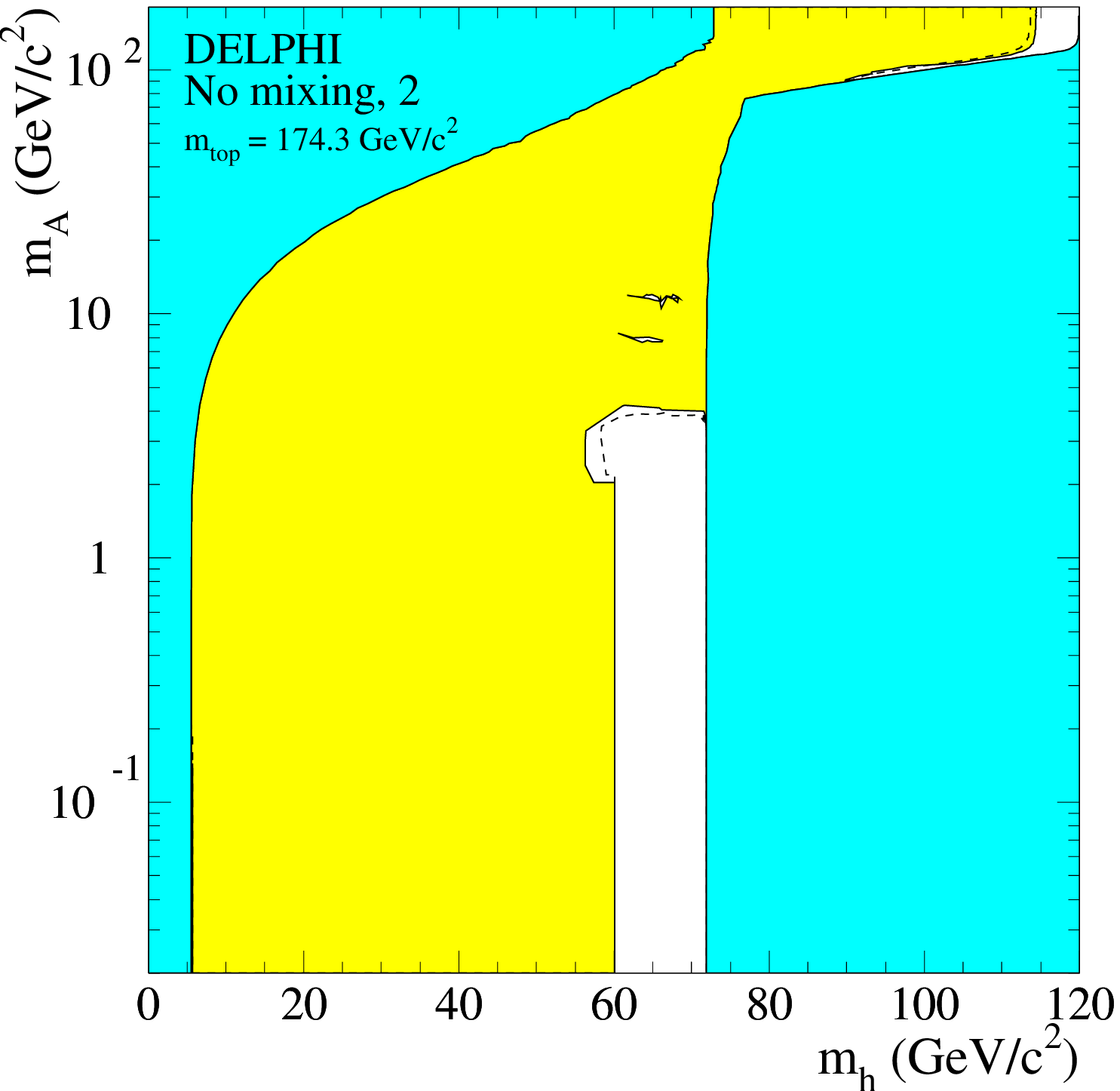,height=9cm} 
\end{tabular}
\caption[]{
    {\sc MSSM} no mixing scenario with positive $\mu$ and 
    large M$_{\rm susy}$ for a top mass of 174.3~\GeVcc: 
    regions excluded at 
    95\% {\sc CL} by combining the results of the Higgs 
    boson searches in the whole {\sc DELPHI} data sample  
   (light-grey). 
   Among the four unexcluded regions at low \mA, 
   the strip at low \mh\ is fully excluded by the limit on 
   the Z partial width that would be due to new 
   physics~\cite{ref:width}.
   The dashed curves show the median expected limits.
   The medium-grey areas are the regions not allowed by theory. 
   The dash-dotted 
   lines in the top left-hand plot are the theoretical upper 
   bounds for a top mass of 169.2, 179.4 and 183.0~\GeVcc\
   (from left to right).
   }
\label{fig:limit_nol}
\end{center}
\end{figure}

  The excluded regions in the no mixing scenario with positive $\mu$ and 
large M$_{\rm susy}$  are presented 
in Fig.~\ref{fig:limit_nol} for a top mass value of 174.3~\GeVcc.
The larger  $M_{\rm susy}$ makes the impact of the H signal, and hence
the exclusion limits, weaker than in the previous scenario. On the
other hand, the results in the low mass region, at \mA\ below 12~\GeVcc,
are similar to those in the no mixing scenario. The direct searches 
leave a tiny unexcluded strip at low \mh\ and very low \mA\
which is excluded by the
limit on $\Gamma^{\rm{new}}$. Three other regions, at \mh\ between
56 and 72~\GeVcc, remain unexcluded even when charged Higgs boson
searches are included, due to the large branching fraction into W$^*$A 
decays, which are not covered by these searches at such low A masses.
The holes around 8 and 12~\GeVcc\ in \mA\ are however excluded
at 93\% and 91\% {\sc CL}, respectively, while the larger area below
4~\GeVcc\ in \mA\ is disfavoured at 60\% {\sc CL} only.

The above results establish the following 95\% {\sc CL} lower limits 
on \mh\ and \MA\ for \mtop~=~174.3~\GeVcc:

\[ \mh > 89.8~\GeVcc \hspace{1cm}
   \MA > 90.6~\GeVcc  \]

\noindent
for any value of \tbeta\ between 1.0 and 50. 
The expected median limits are 90.5~\GeVcc\ for \mh\ and 90.6~\GeVcc\
\MA.  For \mtop~=~174.3~\GeVcc\ 
the range in \tbeta\ 
between 1.0 and 4.5 (expected [1.0-4.3]) is excluded for any value
of \mA\ between 0.02 and 1000~\GeVcc .

The \mtop\ dependence of the above limits is presented
in Table~\ref{tab:limits} and Fig.~\ref{fig:mtop}. 
The mass limits vary only slightly with \mtop, since in the
region where these are set, \mh\ is insensitive to \mtop\ while 
\mH, although sensitive to \mtop, is very close to the kinematic 
limit.
Contrary to the case of the no mixing scenario,
the parameter space of this scenario
does not become fully accessible for a top mass of 169~\GeVcc, 
due to too high an upper (resp. lower) bound 
on \mh\ (resp. \mH). The exclusion is thus much weaker than in the
no mixing scheme but stronger than in the 
\mbox{$ m_{\mathrm h}^{\rm max}$} scenarios.

\subsection{The large $\mu$ scenario}

\begin{figure}[htbp]
\begin{center}
\begin{tabular}{cc}
\hspace{-1.4cm}
\epsfig{figure=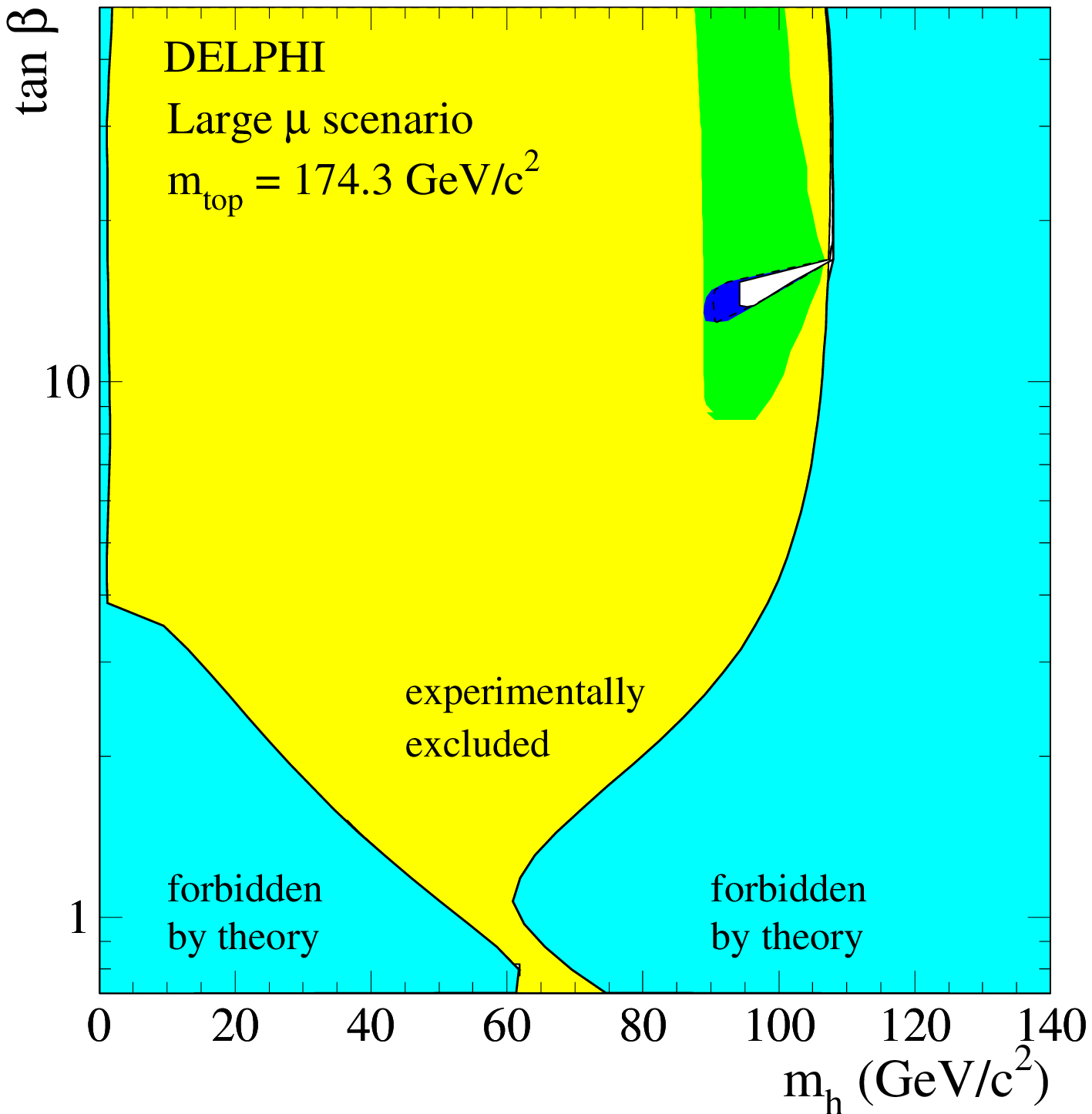,height=9cm} &
\hspace{-0.8cm}
\epsfig{figure=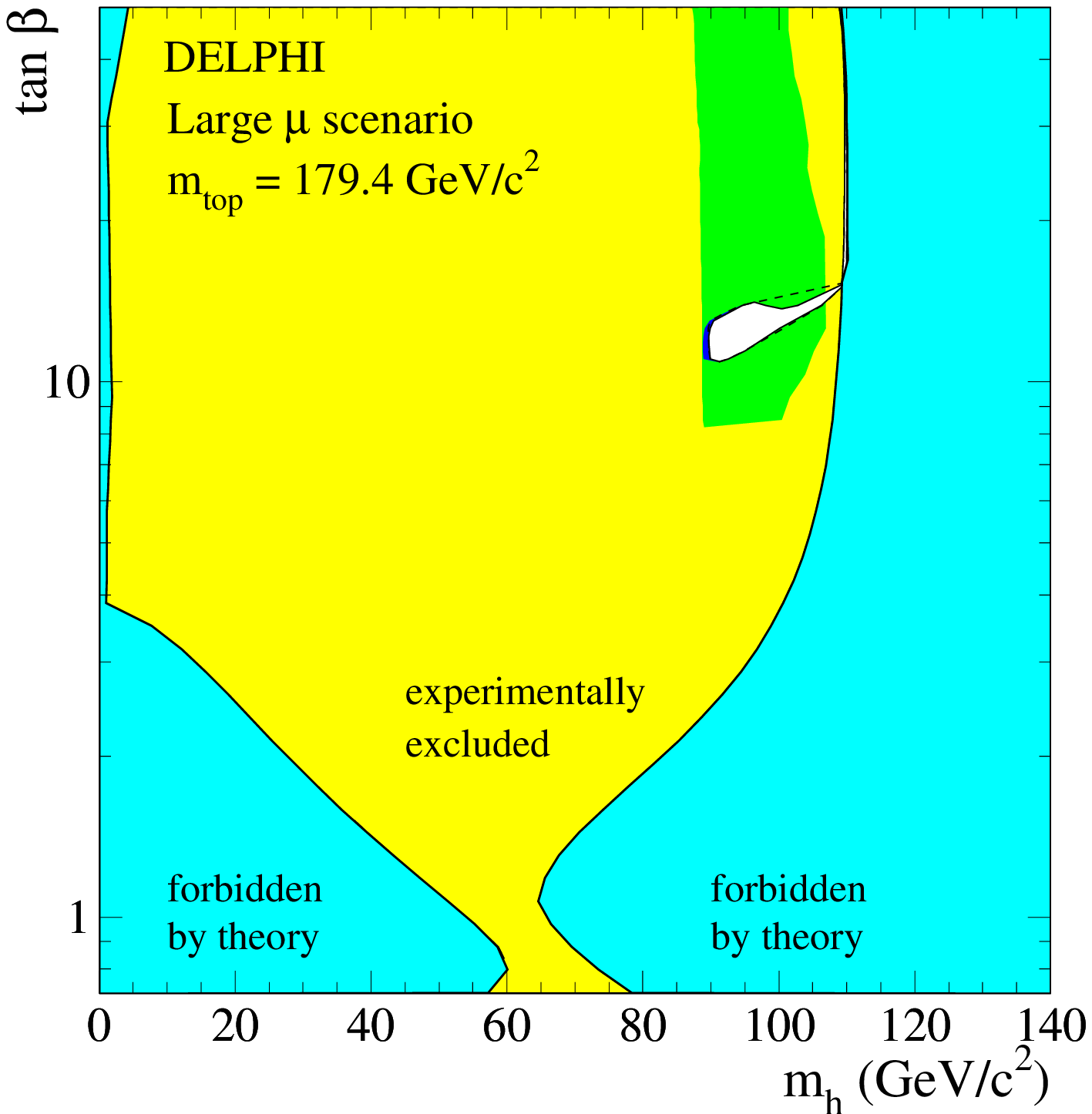,height=9cm} \\
\hspace{-1.4cm}
\epsfig{figure=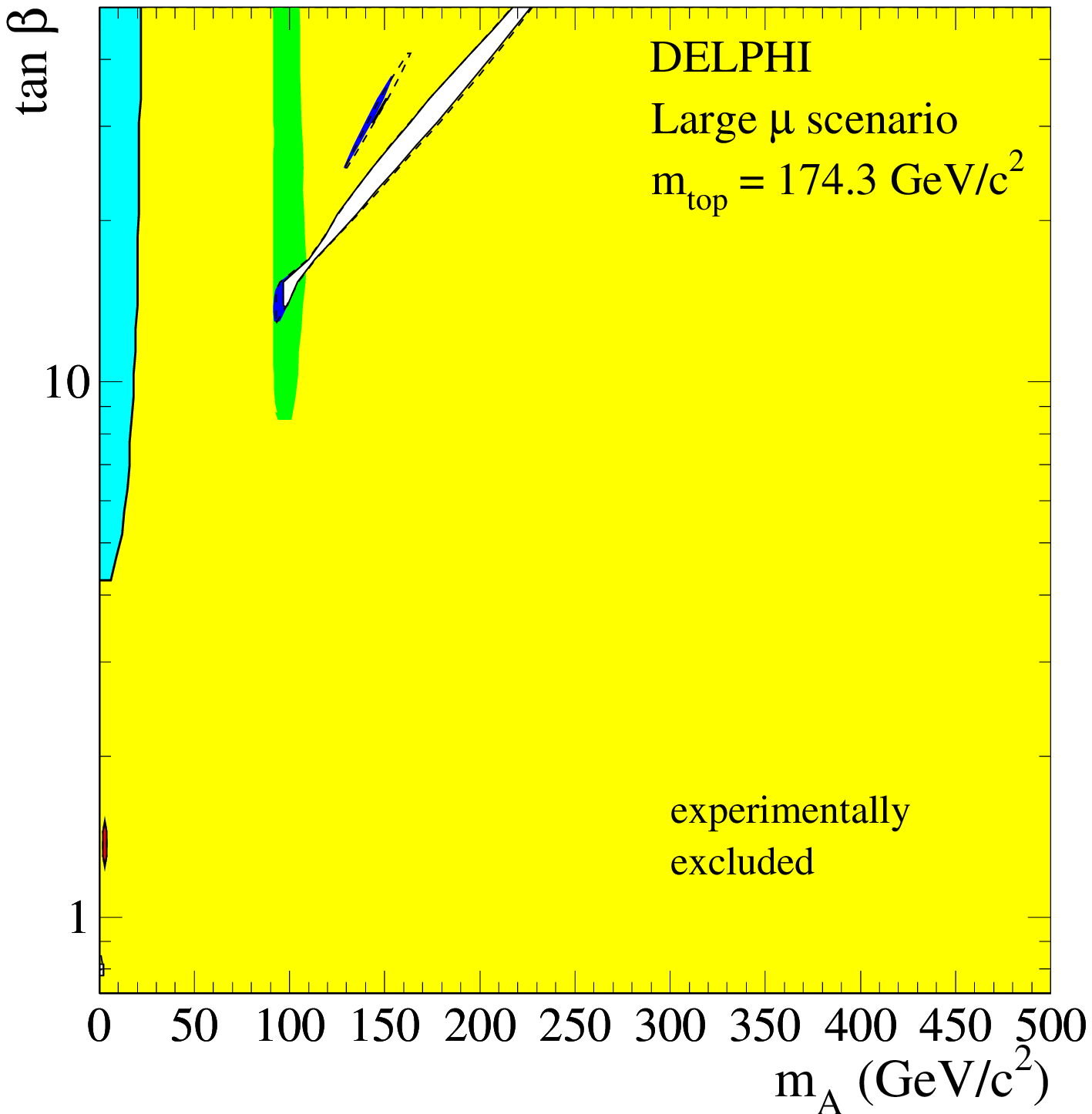,height=9cm} &
\hspace{-0.8cm}
\epsfig{figure=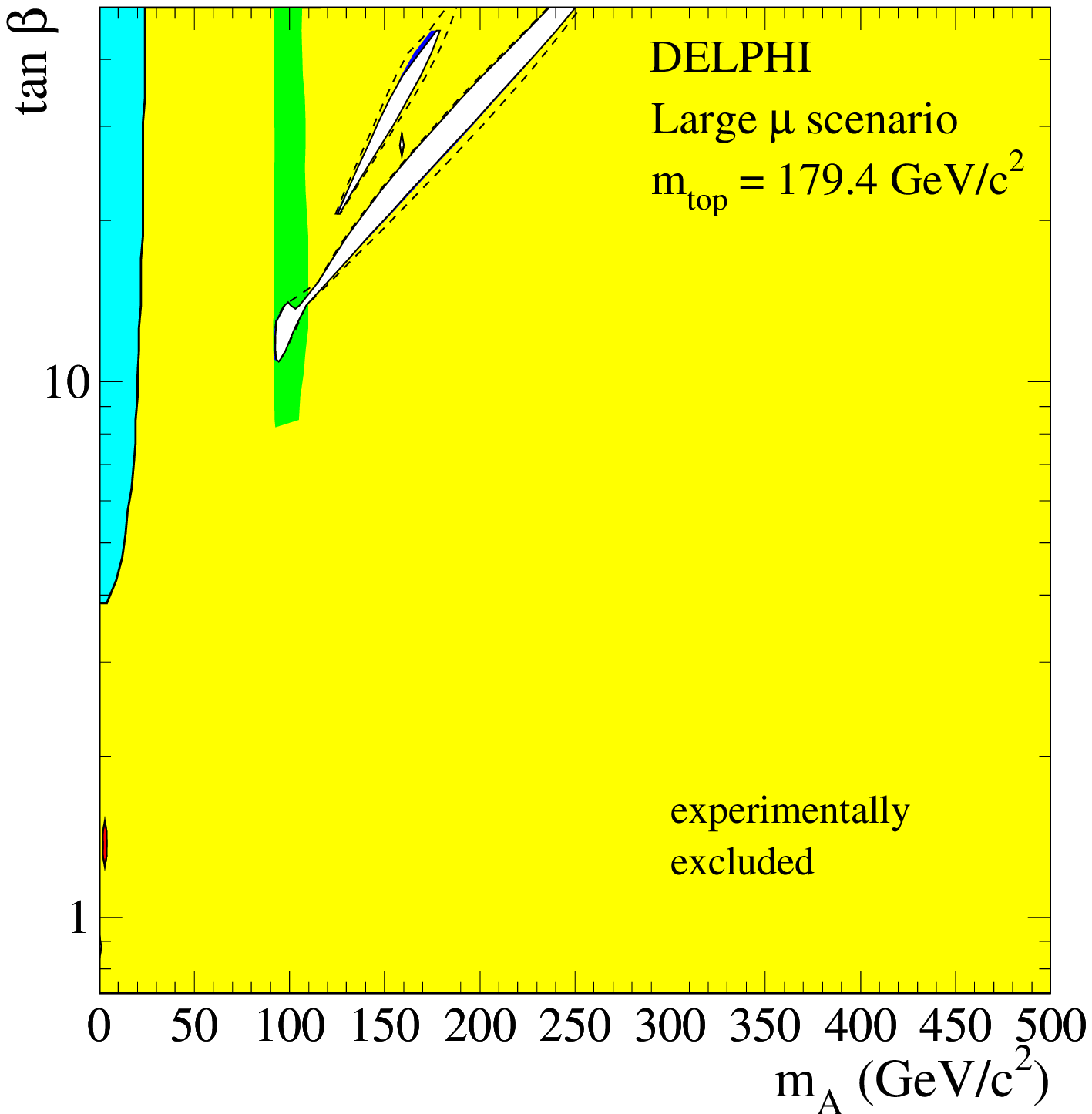,height=9cm} \\
\end{tabular}
\caption[]{
    {\sc MSSM} large $\mu$ scenario: regions excluded at 
   95\% {\sc CL} by combining the results of the Higgs boson
   searches in the whole {\sc DELPHI} data sample  
   (light-grey area and embedded domains in medium- and dark-grey). Results
   are shown for two values of the top mass, 174.3 and
   179.4~\GeVcc. The domains embedded in the light-grey area 
   at large \tbeta\ are excluded by the searches for the heavy scalar Higgs 
   boson, H (medium-grey or green) 
   and by the flavour-blind searches (dark-grey or dark-blue). 
   Of the two unexcluded holes at low \mA,
   the one at \tbeta\ above 1 
   is excluded by the limit on the Z partial width~\cite{ref:width}
   that would be due to new physics.
   The dashed curves show the  median expected limits.
   The medium-grey areas with bold contours are the regions 
   not allowed by theory. 
   Note in particular the large region forbidden
   at low \MA\ in the (\MA, \tbeta) projections, which is due to
   points leading to unphysical h masses.}
\label{fig:limit_mu}
\end{center}
\end{figure}

  The excluded regions in the large $\mu$ scenario are presented in the 
(\mh, \tbeta) and (\MA, \tbeta) planes in Fig.~\ref{fig:limit_mu} for
values of the top quark mass of 174.3 and 179.4~\GeVcc. 
In these figures, the contribution of the H signal and that of the
searches for neutral Higgs bosons decaying into hadrons of any flavour
are highlighted.

A large fraction of the allowed domain is excluded by the searches for
the h, A and H Higgs bosons into standard {\sc MSSM} final states. 
In particular, since the theoretical upper bound on the h boson 
mass in this scenario is low (around 110.0~\GeVcc, see 
Table~\ref{tab:mhmax}), the sensitivity of the \hZ\ channels is high 
even at large \tbeta, which explains why the excluded region reaches the 
theoretically forbidden area for large values of \tbeta. 
As the value of the upper bound on \mh\ is also the
theoretical lower bound on \MH\ at large \tbeta, 
allowing for the production of H translates into a 
significant gain in exclusion, namely at \tbeta\ above 8. 
The searches for neutral Higgs bosons decaying
into hadrons of any flavour bring an additional exclusion in
regions left unexcluded by the standard searches at
\tbeta\ around 14. At moderate \mA, \hZ\ and \hA\ productions 
are low due to weak hZZ couplings for \hZ\
and to kinematics for \hA. On the other hand, \HZ\ production is large 
but H is decoupled from \bbbar. At larger \mA, \hA\ and \HZ\ 
productions are kinematically forbidden, \hZ\ production is large but
the h$\rightarrow$\bbbar\ branching fraction vanishes. 
In both cases, the Higgs boson whose production is allowed (H or h) has
a large branching fraction into hadrons and a mass close to the sensitivity
of our searches for a neutral Higgs boson decaying into hadrons and 
fully coupled to the Z. This explains why these searches lead to an 
additional but only partial exclusion in these regions.
Note that increasing the top quark mass from 174.3 to 179.4~\GeVcc\ leads to 
a larger unexcluded area. There are indeed more points with vanishing
h or H branching fractions into \bbbar\ and, as \mh\ and \mH\ increase with
\mtop, the impact of the searches for hadronically decaying Higgs 
bosons also becomes weaker.
However, when combining the four {\sc LEP}
experiments, the sensitivity of these
searches increases and becomes high enough to cover almost entirely these
regions of vanishing branching fractions into \bbbar~\cite{ref:hwg}.

Below 3~\GeVcc\ in \mA, the direct searches leave
two unexcluded holes at \tbeta\ around 1.
The one at \tbeta\ above 1 is fully excluded by the limit 
on $\Gamma^{\rm{new}}$ for either value of \mtop.
The hole at \tbeta\ below 1 remains unexcluded. 
The largest value of \CLs\ in this area is 12\% for \mtop~=~174.3~\GeVc\
and 6\% for \mtop~=~179.4~\GeVc.

The above results establish the following 95\% {\sc CL} lower limits 
on \mh\ and \MA\ for \mtop~=~174.3~\GeVcc:

\[ \mh > 94.2~\GeVcc \hspace{1cm}
   \MA > 96.6~\GeVcc  \]

\noindent
for any value of \tbeta\ between 0.9 and 50. 
The expected median limits are 90.3~\GeVcc\ for \mh\ and 92.8~\GeVcc\
for \MA. 
The observed limits in \MA\ and \mh\ are  reached at \tbeta\ 
around 14, in a region where the \hZ, \HZ\ and \hA\ processes contribute. 
For \mtop~=~174.3~\GeVcc, two ranges in \tbeta\ are excluded
for any value of \mA\ between 0.02 and 1000~\GeVcc,
the largest interval being 
between 0.9 and 13.7 (expected [0.9-12.9]).

The \mtop\ dependence of the above limits is presented
in Table~\ref{tab:limits} and Fig.~\ref{fig:mtop}. 
Except for \mtop~=~174.3~\GeVcc, 
the mass limits vary only slightly with \mtop\ and are in
agreement with the expected ones. 
The difference at \mtop~=~174.3~\GeVcc\ has been traced back to the deficit
in data with respect to background expectations which was observed in the 
flavour-blind searches applied to the Higgsstrahlung 
process~\cite{ref:papflbl} when testing  
masses above 100~\GeVcc, which corresponds to the range of \mH\ values
in the region where the mass limits are obtained in the large $\mu$ scenario.
In this region, the set of independent channels which are selected to be 
statistically combined (see Sec.~\ref{sec:overlap}) 
varies strongly from one top mass value to 
the other, due to still large H branching fractions into \bbbar\
at \mtop~=~169.2~\GeVcc\ and to \mH\ values  increasing with 
\mtop\ (see Table~\ref{tab:mhmax}). At \mtop~=~174.3~\GeVcc, the weight of 
the flavour-blind \HZ\ searches is maximal and the deficit in data of 
these searches
translates into a difference between the observed and median limits.

\subsection{The gluophobic scenario}

\begin{figure}[htbp]
\begin{center}
\begin{tabular}{cc}
\hspace{-1.4cm}
\epsfig{figure=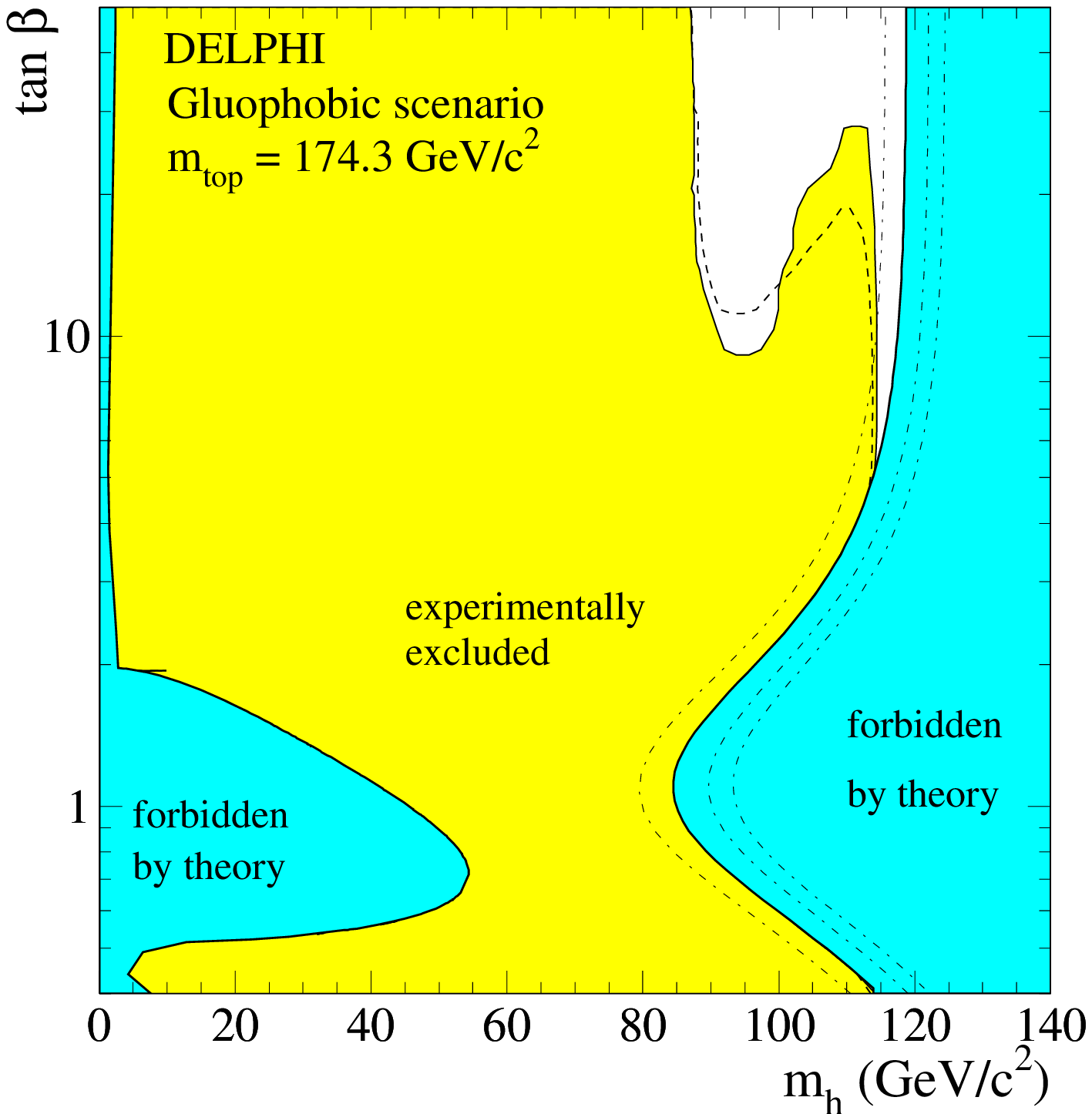,height=9cm} &
\hspace{-0.5cm}
\epsfig{figure=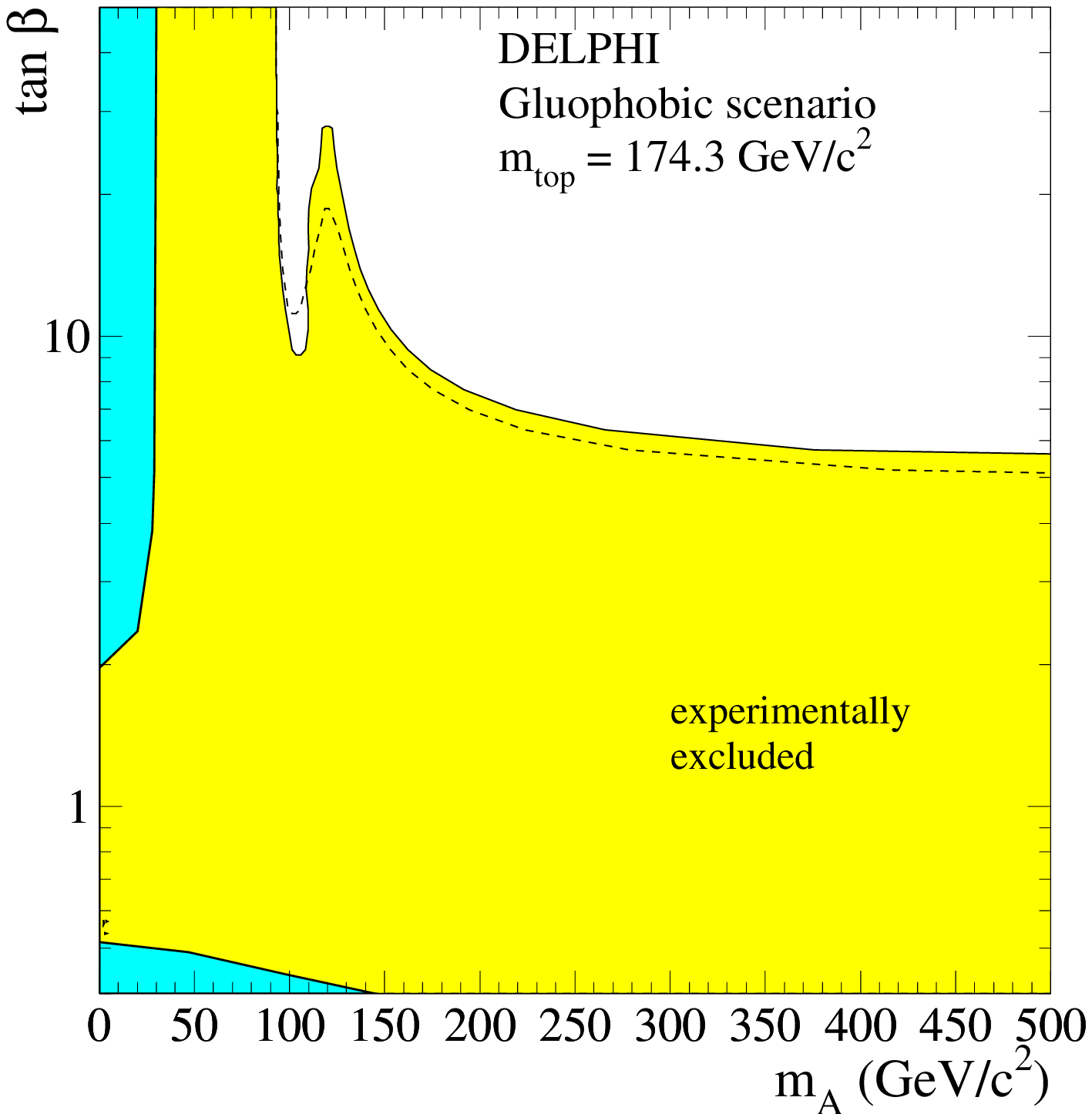,height=9cm} \\
\hspace{-1.4cm}
\epsfig{figure=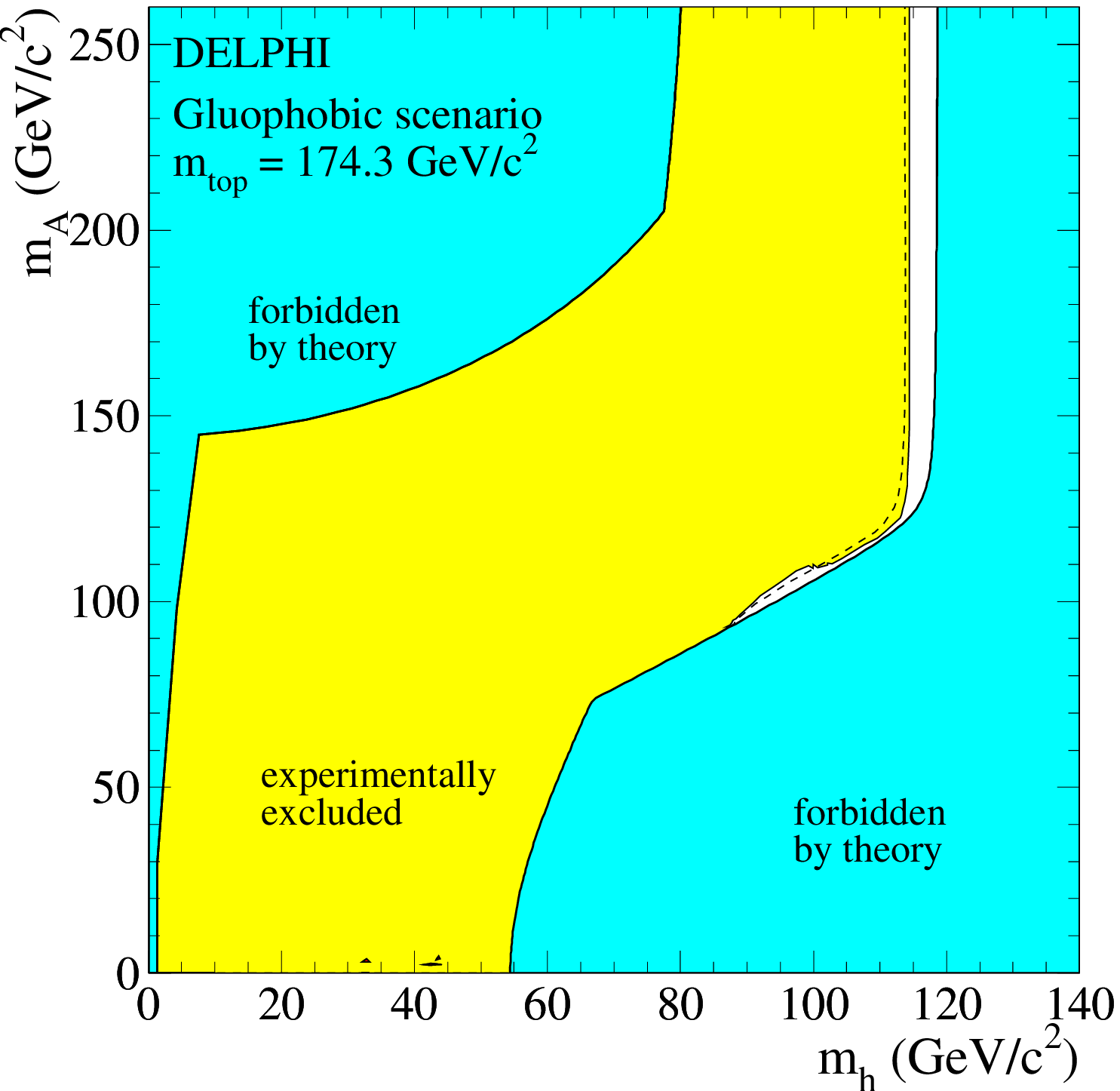,height=9cm} &
\hspace{-0.5cm}
\epsfig{figure=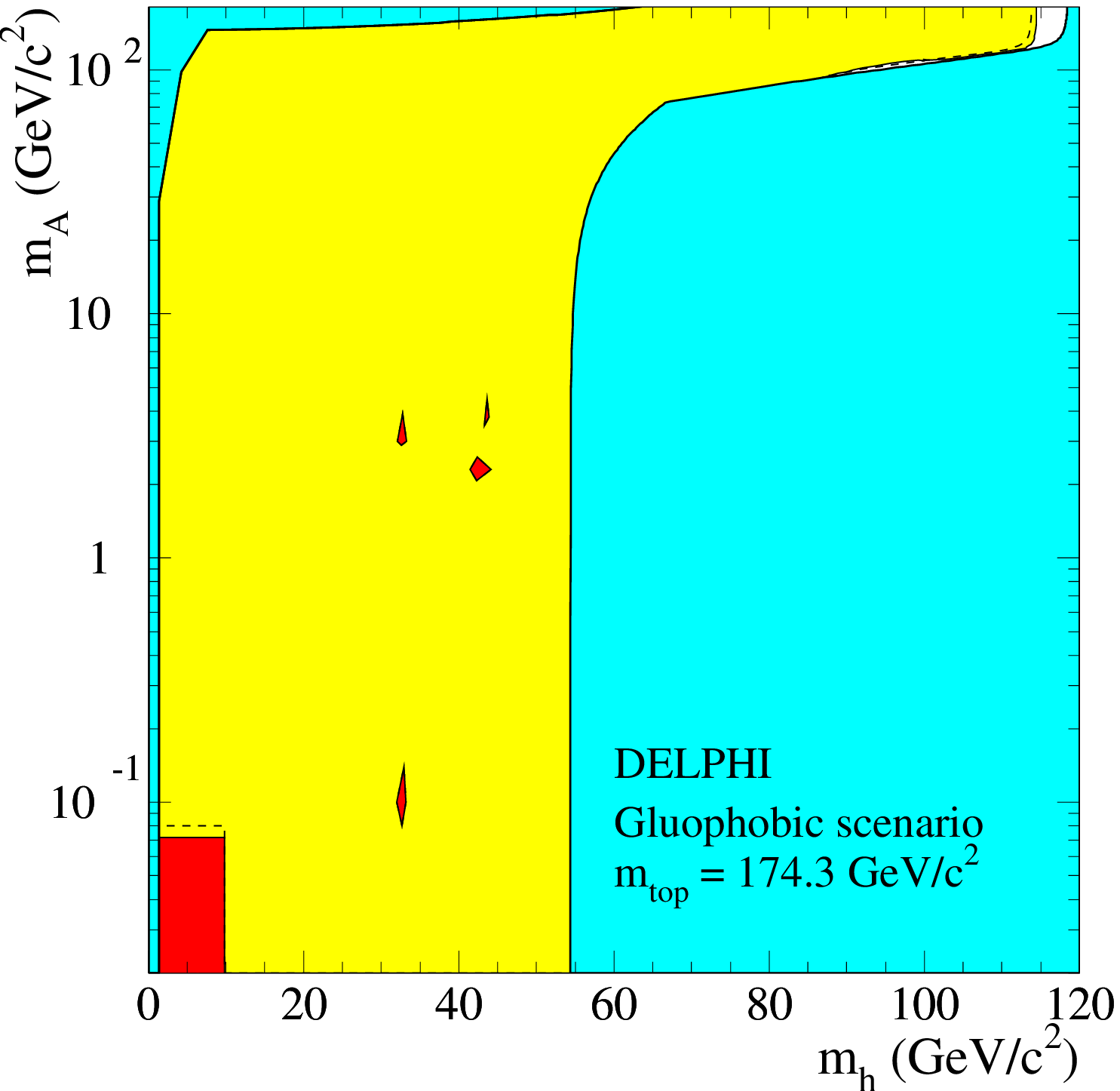,height=9cm} 
\end{tabular}
\caption[]{
    {\sc MSSM} gluophobic scenario for a top mass of 174.3~\GeVcc: 
    regions excluded at 
    95\% {\sc CL} by combining the results of the Higgs 
    boson searches in the whole {\sc DELPHI} data sample  
   (light-grey). 
   The unexcluded holes at low \mA\ are fully excluded by the limit on 
   the Z partial width~\cite{ref:width} that would be due to new 
   physics (dark-grey).
   The dashed curves show the median expected limits.
   The medium-grey areas are the regions not allowed by theory. 
   Note in particular the large forbidden region 
   in the (\MA, \tbeta) projection, which is due to
   points leading to unphysical h masses.
   The dash-dotted 
   lines in the top left-hand plot are the theoretical upper 
   bounds for a top mass of 169.2, 179.4 and 183.0~\GeVcc\
   (from left to right).
   }
\label{fig:limit_gluo}
\end{center}
\end{figure}

 For the gluophobic scenario the excluded regions  
in the (\mh, \tbeta), (\MA, \tbeta) and (\mh, \MA) planes 
are presented
in Fig.~\ref{fig:limit_gluo} for a top mass value of 174.3~\GeVcc.
Although this scenario was designed to test Higgs boson searches 
at hadron colliders, with a phenomenology very different from 
that of {\sc LEP}, results are similar to those derived in the
previous scenarios. The exclusion 
is defined by the results in the \hZ\ (\hA) channels in the
low (large) \tbeta\ region while they both contribute at intermediate
values. The direct searches leave several unexcluded holes 
below 4~\GeVcc\ in \mA\ and at \tbeta\ below 2, which are all excluded by
the limit on $\Gamma^{\rm{new}}$.

The above results establish the following 95\% {\sc CL} lower limits 
on \mh\ and \MA\ for \mtop~=~174.3~\GeVcc:

\[ \mh > 87.0~\GeVcc \hspace{1cm}
   \MA > 92.9~\GeVcc  \]

\noindent
for any value of \tbeta\ between 0.4 and 50. 
The expected median limits are 87.0~\GeVcc\ for \mh\ and 
93.0~\GeVcc\ for \MA. 
The observed limits in \MA\ and \mh\ are  reached at \tbeta\ 
around 50, in a region where only the \hA\ process contributes. 
Contrary to the other scenarios, the h and A bosons are
not degenerate in mass at large \tbeta, which reflects in the 
significant difference between the h and A mass limits.
For \mtop~=~174.3~\GeVcc,
the range in \tbeta\  
between 0.4 and 5.2 (expected [0.4-4.8]) is excluded
for any value of \mA\ between 0.02 and 1000~\GeVcc.

The \mtop\ dependence of the above limits is shown
in Table~\ref{tab:limits} and Fig.~\ref{fig:mtop}. 
As already mentioned, the h and A bosons are not degenerate at large
\tbeta\ and moderate \mA, the region where the mass limits are
set. As a consequence, the value of \mh\ at fixed \mA\ and \tbeta\ 
is observed to vary significantly with \mtop\ in that region. This
is the main reason for the variations of the mass limits with 
\mtop, an additional effect being the variations
of \mH\ which is kinematically accessible at low \mtop\
in this scenario (see Table~\ref{tab:mhmax}).
On the other hand, the variation of the excluded range in \tbeta\ 
is due, as in the other scenarios, to the change in the 
maximal value of \mh\ which is very sensitive to \mtop. 

\subsection{The small $\alpha$ scenario}

\begin{figure}[htbp]
\begin{center}
\begin{tabular}{cc}
\hspace{-1.4cm}
\epsfig{figure=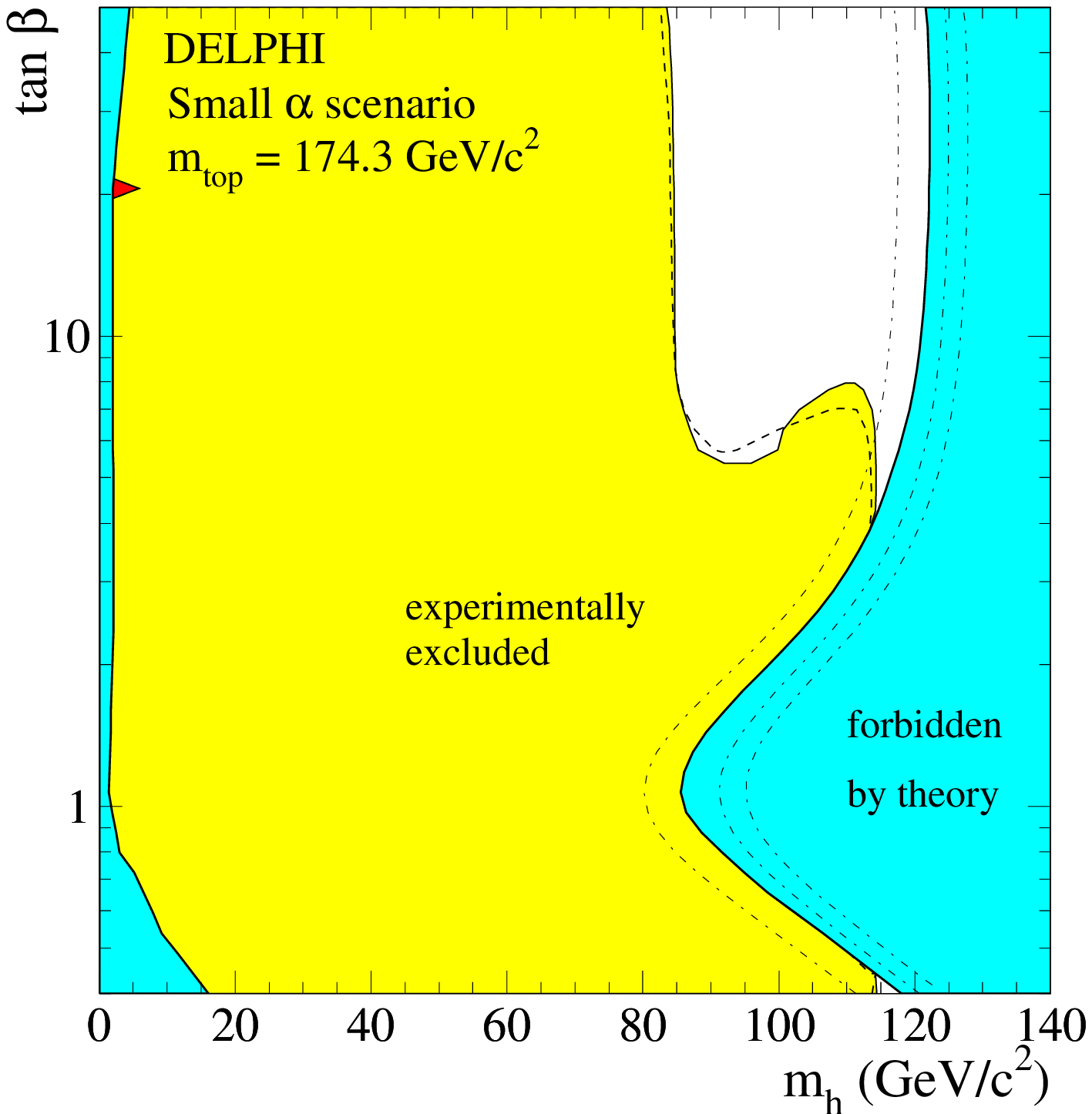,height=9cm} &
\hspace{-0.8cm}
\epsfig{figure=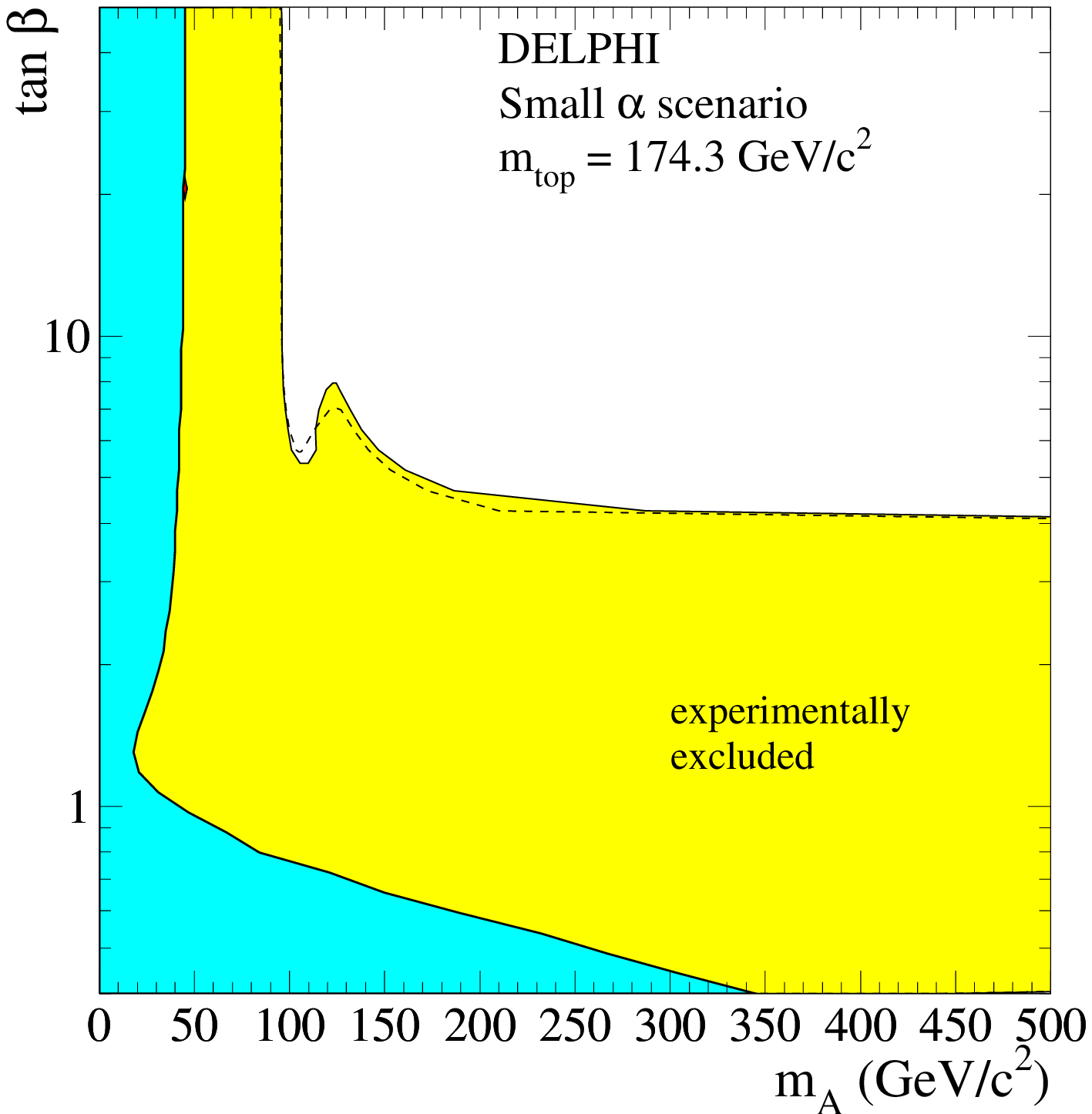,height=9cm} \\
\hspace{-1.4cm}
\epsfig{figure=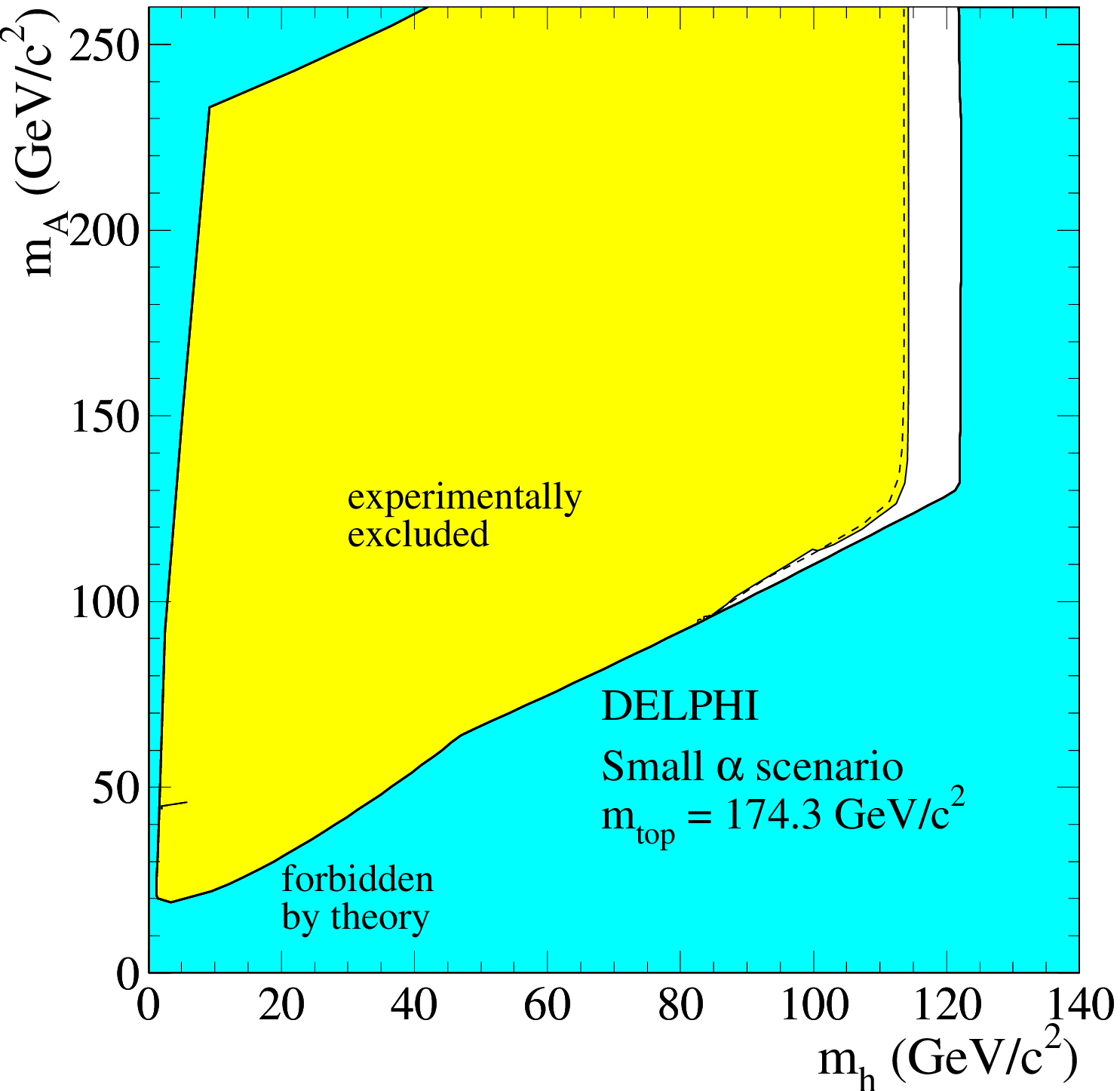,height=9cm} &
\end{tabular}
\caption[]{
    {\sc MSSM} small $\alpha$ scenario for a top mass of 174.3~\GeVcc: 
    regions excluded at 
    95\% {\sc CL} by combining the results of the Higgs 
    boson searches in the whole {\sc DELPHI} data sample  
   (light-grey). 
   There is one unexcluded hole at low \mh\ and \tbeta\ around 20
   which is excluded by the limit on 
   the Z partial width~\cite{ref:width} that would be due to new 
   physics (dark-grey).
   The dashed curves show the median expected limits.
   The medium-grey areas are the regions not allowed by theory. 
   Note in particular the large forbidden region 
   in the (\MA, \tbeta) projection, which is due to
   points leading to unphysical h masses.
   The dash-dotted 
   lines in the top left-hand plot are the theoretical upper 
   bounds for a top mass of 169.2, 179.4 and 183.0~\GeVcc\
   (from left to right).
   }
\label{fig:limit_small}
\end{center}
\end{figure}
 
The excluded regions in the small $\alpha$ scenario
are presented in Fig.~\ref{fig:limit_small} 
for a top mass value of 174.3~\GeVcc.
The small $\alpha$ scheme is the second example of a scenario aiming at
testing potentially difficult cases for the Higgs boson searches at hadron 
colliders. As mentioned in section~\ref{sec:benchmark}, 
this scenario presents regions of the parameter space where the 
\hbb\ and \htt\ decays vanish, which could be a problem at {\sc LEP} too.
The results in  Fig.~\ref{fig:limit_small}, similar to
those derived in the previous scenarios, show that this is not the case.
At large \tbeta, in the region accessible at {\sc LEP}, 
the \hbb\ branching fraction, although reduced, 
remains high enough (e.g. above 70\% in the region where the
mass limits are set) to ensure a good sensitivity.
At low \mh, the direct searches leave 
one unexcluded island that is fully excluded by the limit on 
$\Gamma^{\rm{new}}$.

The above results establish the following 95\% {\sc CL} lower limits 
on \mh\ and \MA\ for \mtop~=~174.3~\GeVcc:

\[ \mh > 83.5~\GeVcc \hspace{1cm}
   \MA > 95.8~\GeVcc  \]

\noindent
for any value of \tbeta\ between 0.4 and 50. 
The expected median limits are 82.6~\GeVcc\ for \mh\ and 
95.0~\GeVcc\ for \MA. 
The observed limits in \MA\ and \mh\ are reached at \tbeta\ 
around 50, in a region where only the \hA\ process contributes. 
As in the previous scenario, the h and A bosons are
not degenerate in mass at large \tbeta, which reflects in the 
significant difference between the h and A mass limits.
For \mtop~=~174.3~\GeVcc,
the range in \tbeta\
between 0.4 and 4.0 (expected [0.5-3.9]) is excluded
for any value of \mA\ between 0.02 and 1000~\GeVcc .

\begin{figure}[hp]
\begin{center}
\begin{tabular}{cc}
\hspace{-1.5cm}
\epsfig{figure=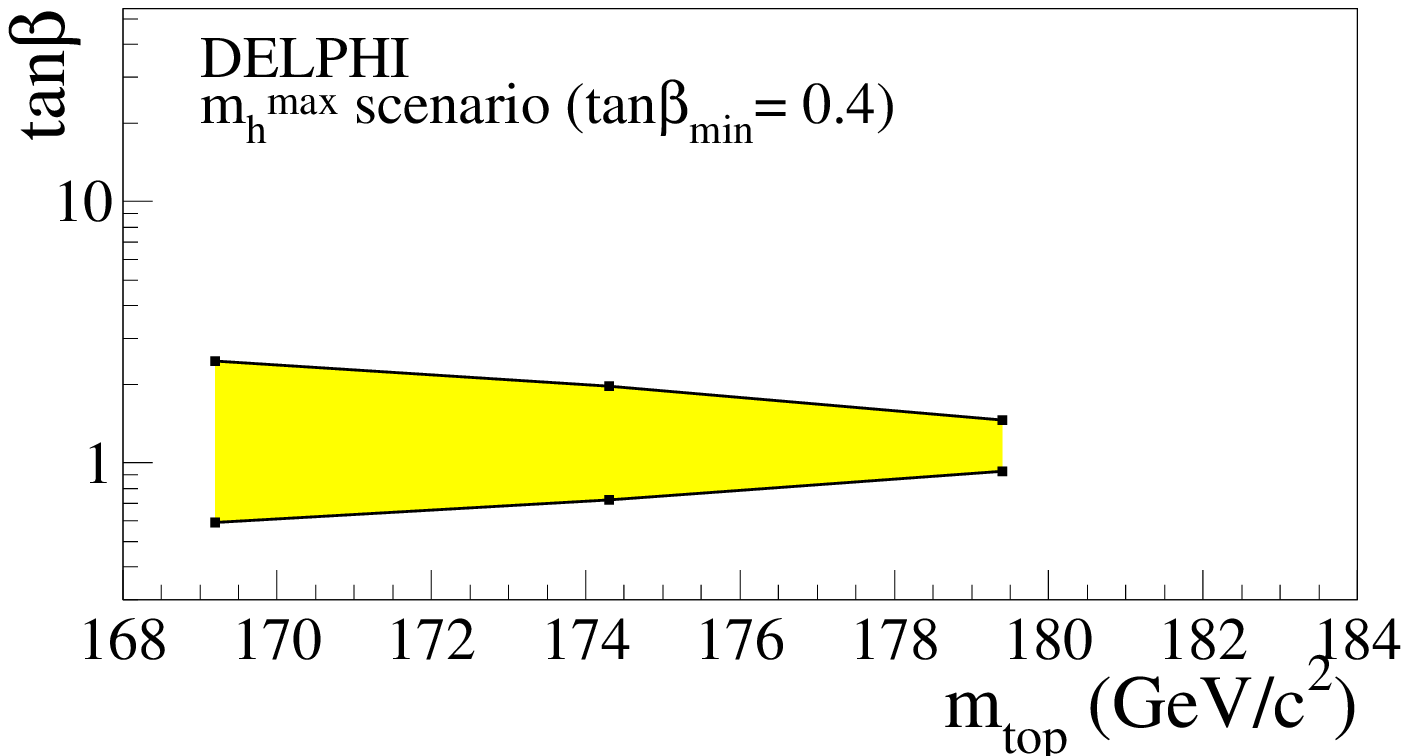,width=9cm} &
\hspace{-0.8cm}
\epsfig{figure=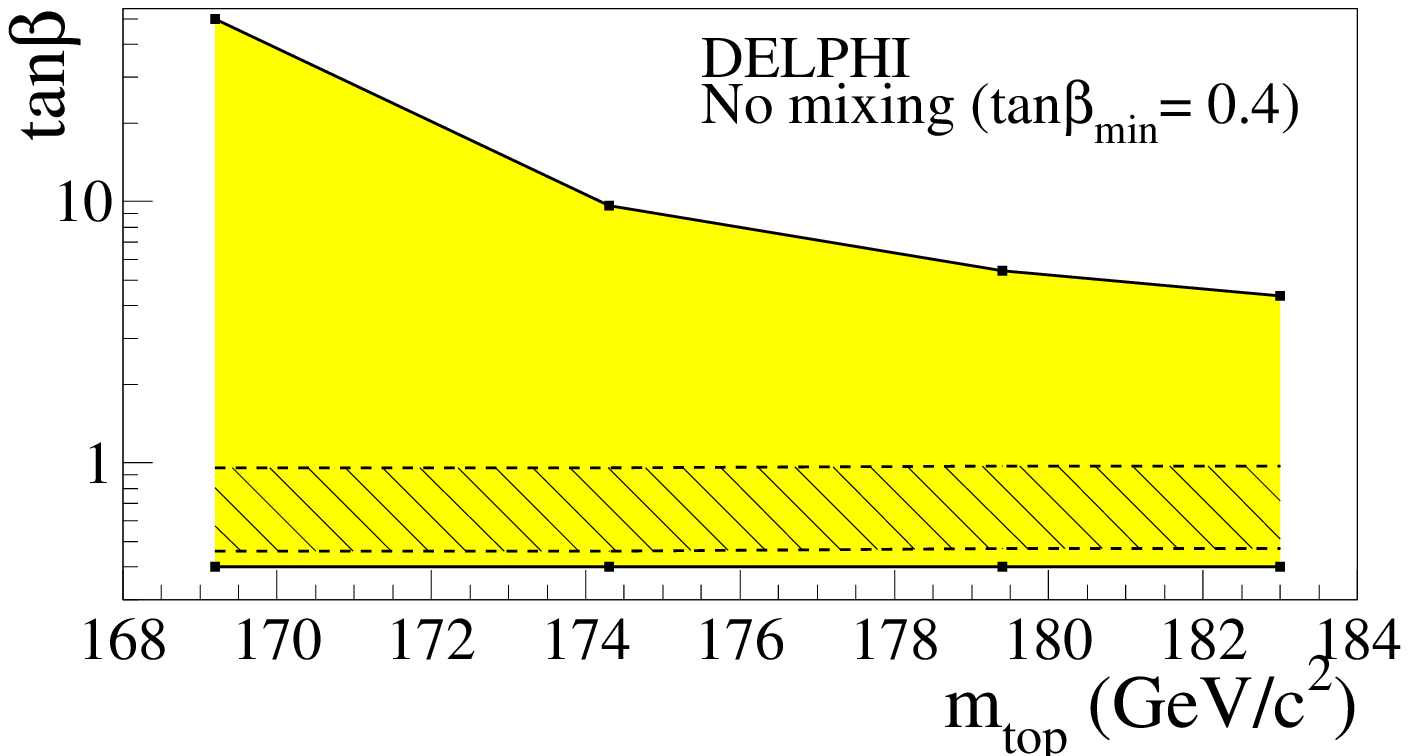,width=9cm} \\
\hspace{-1.5cm}
\epsfig{figure=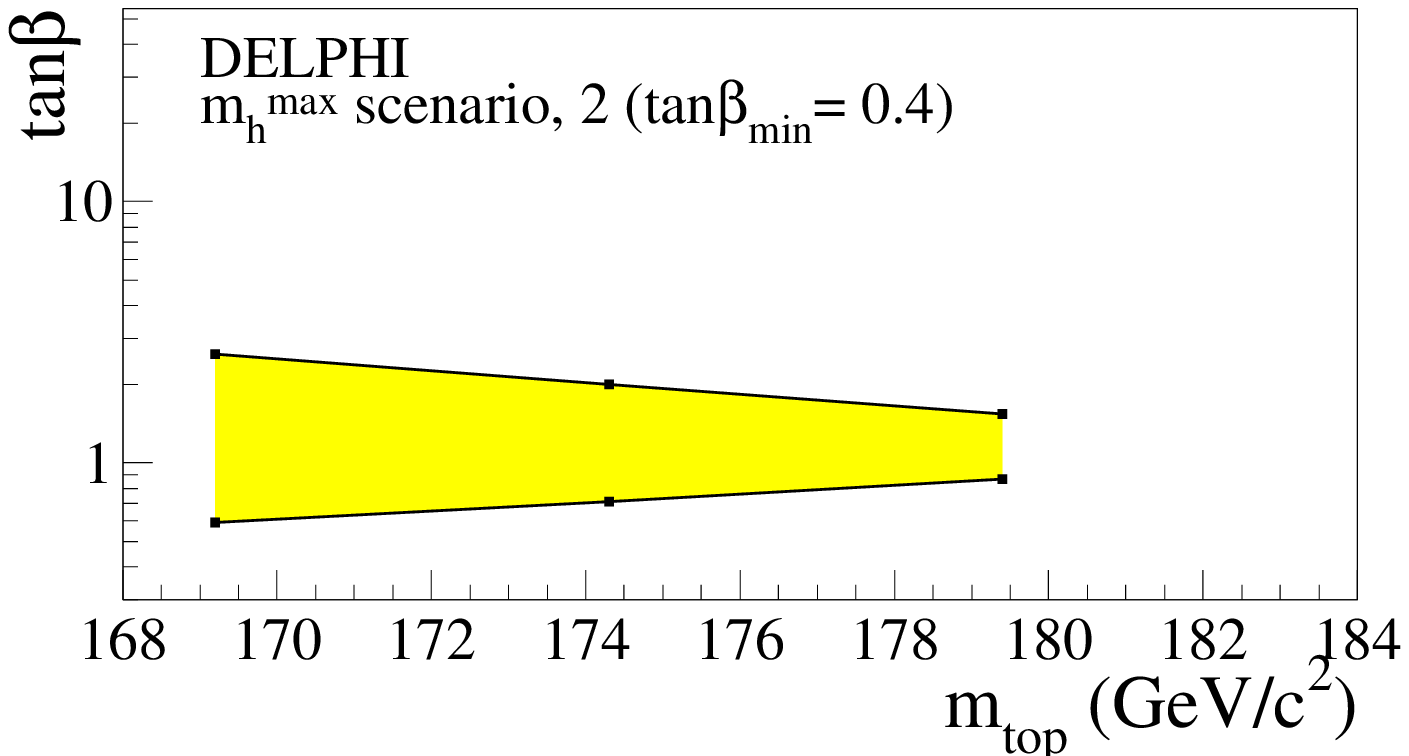,width=9cm} &
\hspace{-0.8cm}
\epsfig{figure=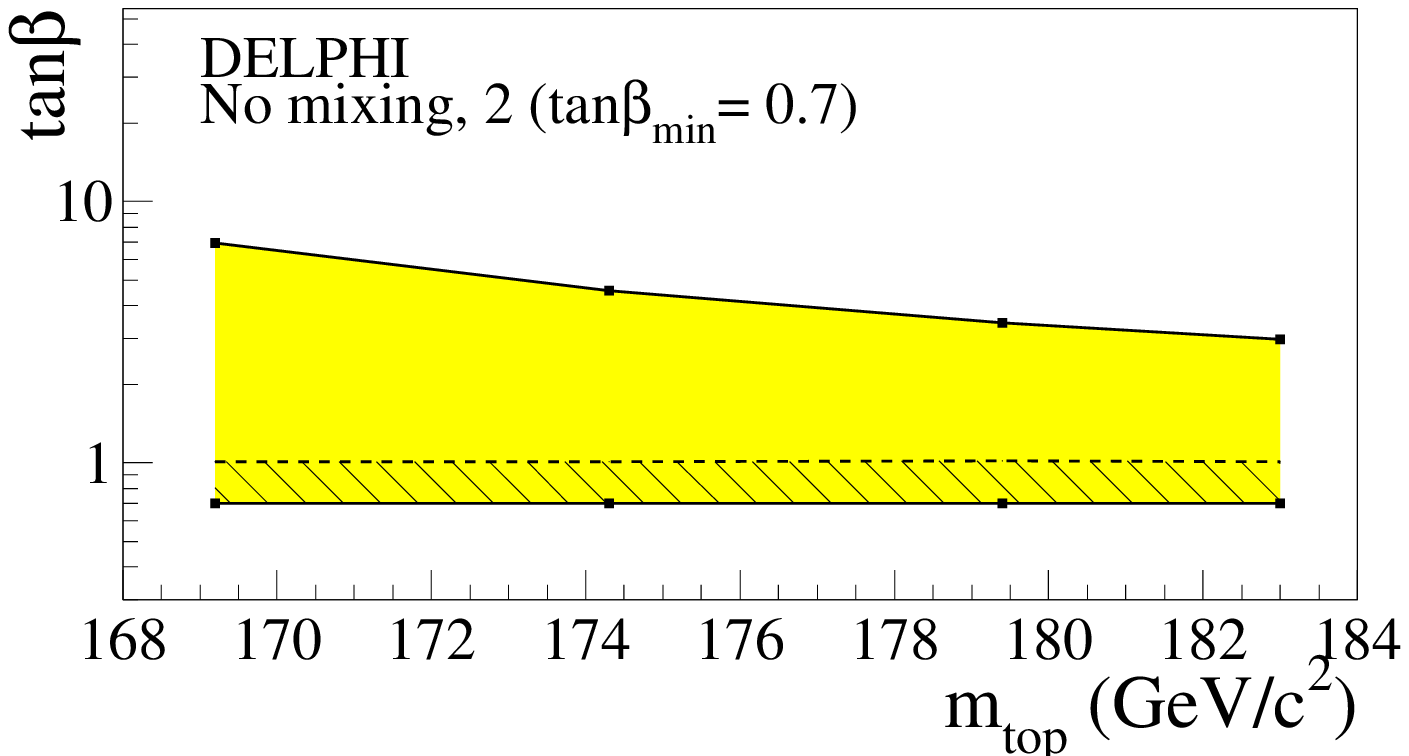,width=9cm} \\
\hspace{-1.5cm}
\epsfig{figure=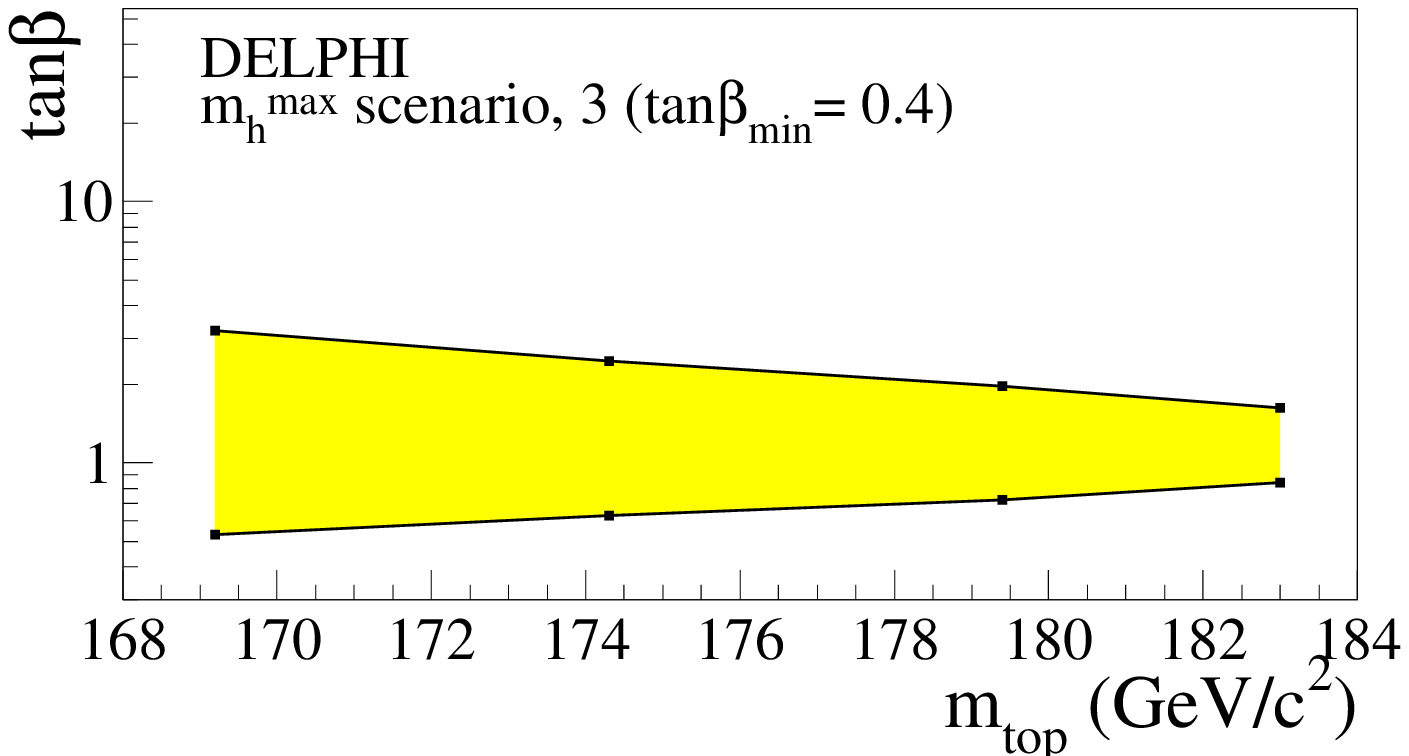,width=9cm} & 
\hspace{-0.8cm}
\epsfig{figure=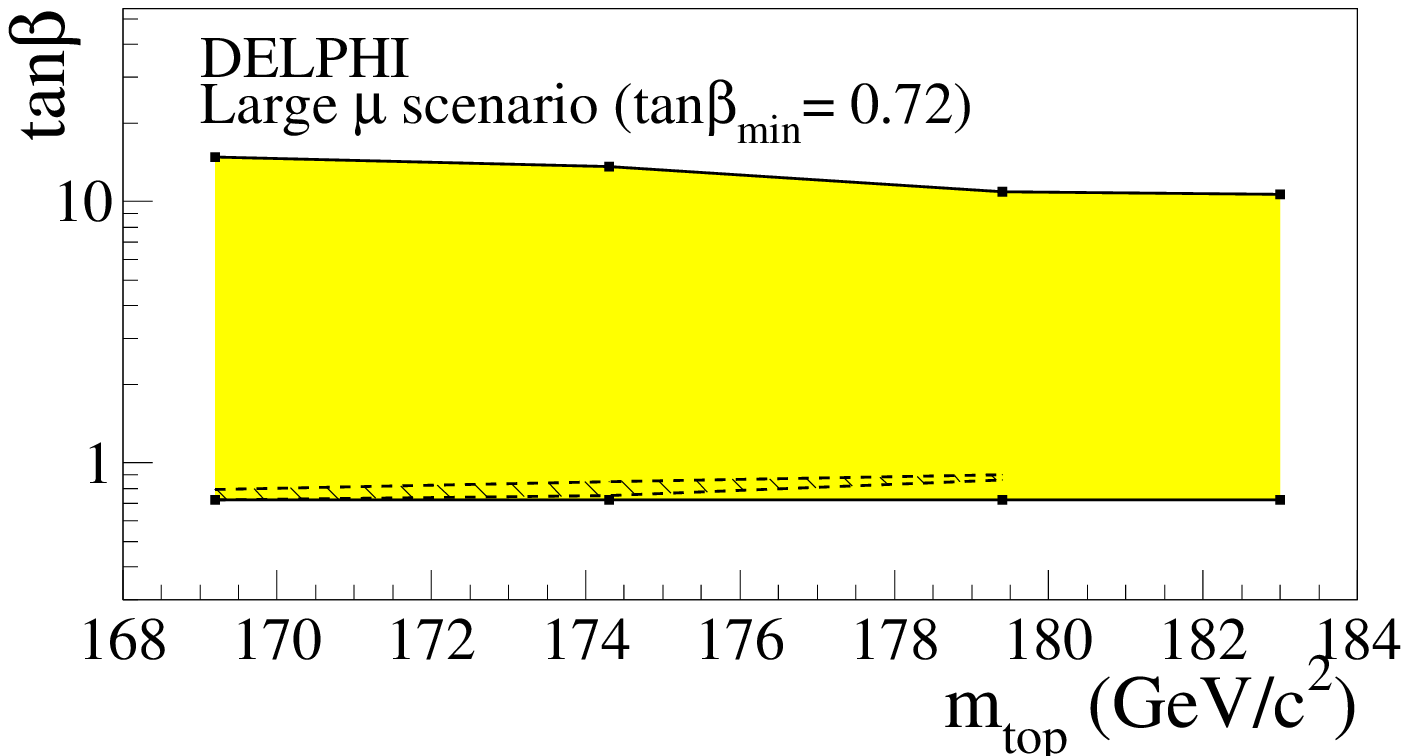,width=9cm}\\
\hspace{-1.5cm}
\epsfig{figure=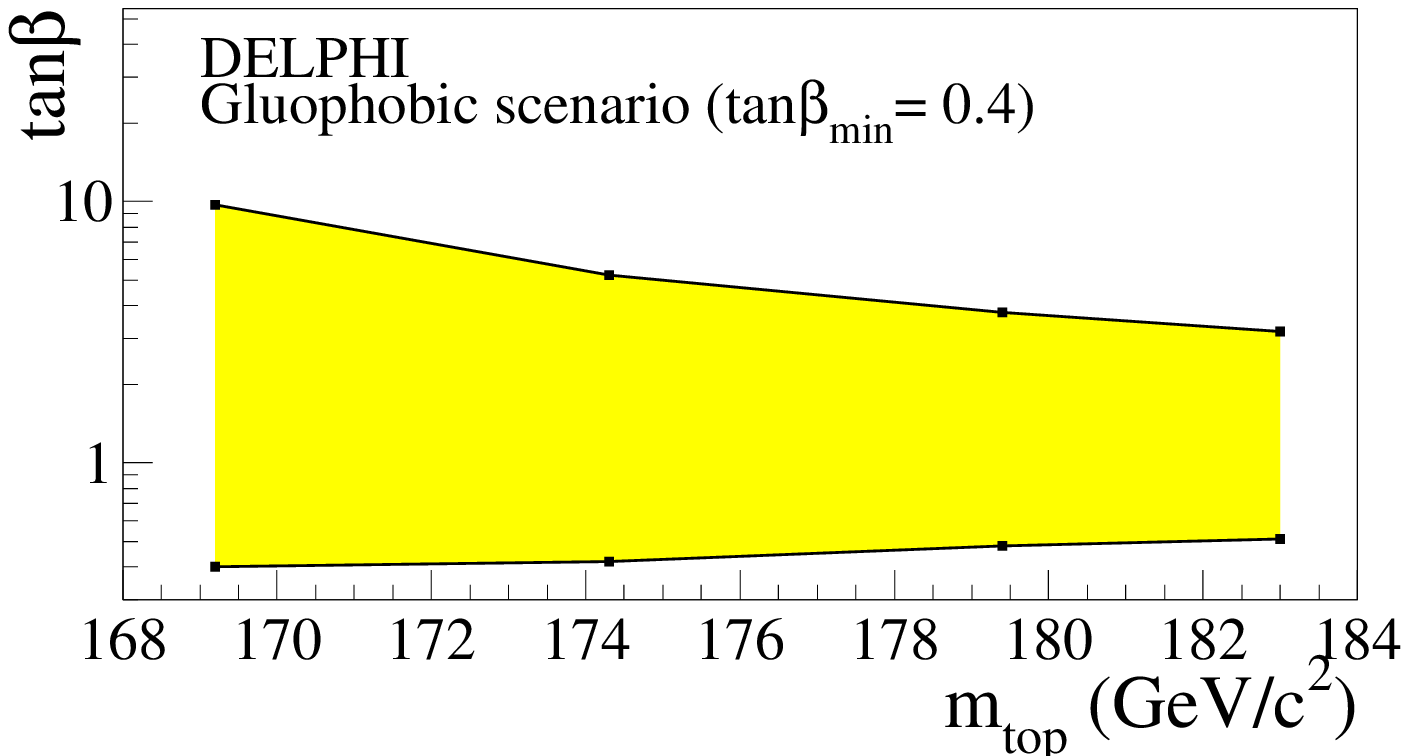,width=9cm} &
\hspace{-0.8cm}
\epsfig{figure=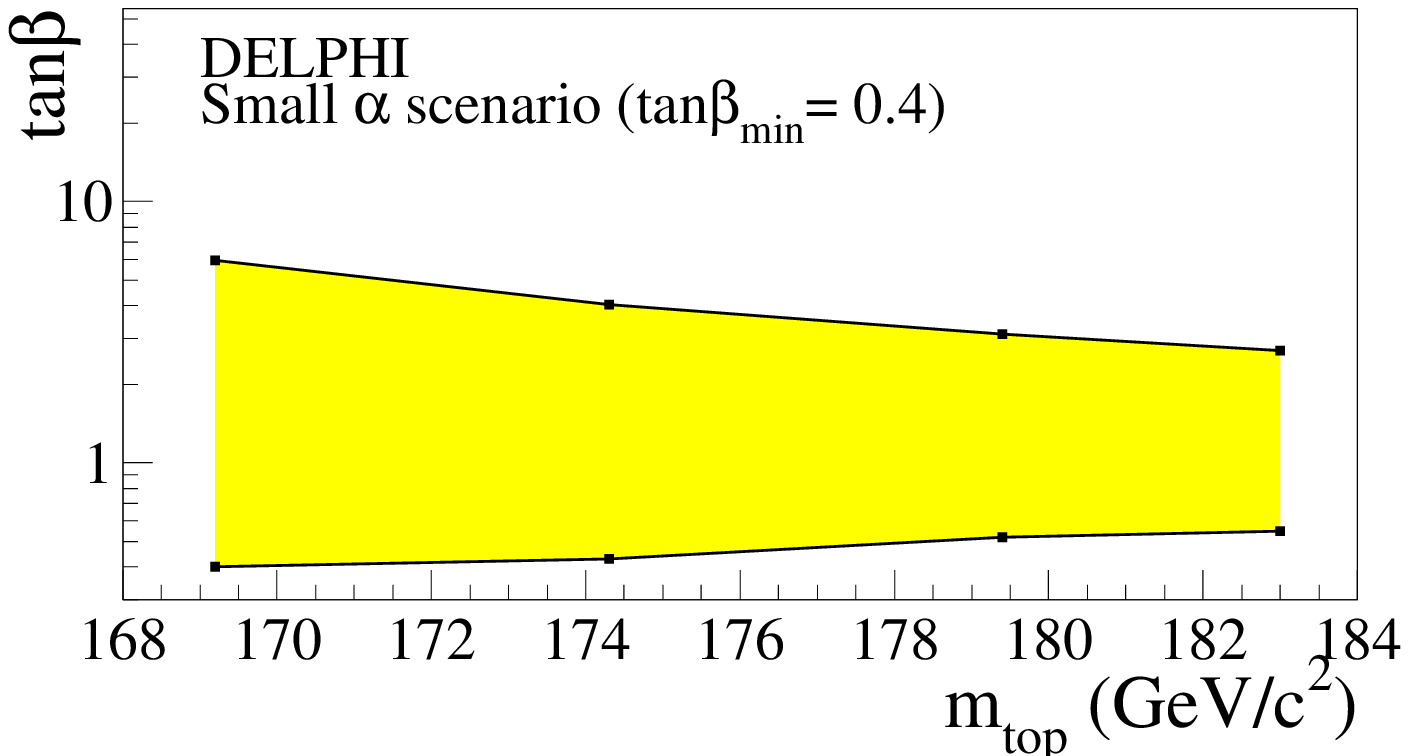,width=9cm} 
\end{tabular}
\caption[]{
      Variation with \mtop\ of the ranges in \tbeta\
      excluded by {\sc DELPHI}
      in the {\sc CP}-conserving  {\sc MSSM} benchmark scenarios.
      Note that each bound in \tbeta\ is a limit (either upper
      or lower) at 95\% {\sc CL}. These bounds hold for the
      whole interval of \MA\ between 0.02 and 1000~\GeVcc, except
      in the hatched intervals, where the exclusion is valid above
      12~\GeVcc\ in the two no mixing scenarios and above
      2.4~\GeVcc\ in the large $\mu$ scenario.
}
\label{fig:mtop}
\end{center}
\end{figure}

The \mtop\ dependence of the above limits is shown
in Table~\ref{tab:limits} and Fig.~\ref{fig:mtop}. 
As in the previous scenario,
the value of \mh\ at fixed \mA\ and \tbeta\ 
varies significantly with \mtop\ in the region where
the mass limits are set, which explains the variations of
the latter. The H signal, being kinematically inaccessible for
most values of \mtop\ (see Table~\ref{tab:mhmax}) 
plays no role in this scenario.
Finally, the variation of the 
excluded range in \tbeta\ is due to the change in the 
maximal value of \mh\ which is very sensitive to \mtop. 

\begin{table}[hp]
\begin{center}
\begin{tabular}{c|c|cccc}     \hline
  &  & \multicolumn{4}{c}{\mtop\ (\GeVcc)} \\
scenario & limits &  169.2 & 174.3 & 179.4 & 183.0\\
\hline

\mbox{$ m_{\mathrm h}^{\rm max}$} &
    \mh\  & 89.7 & 89.7 & 89.7 & 89.6\\
&   \mA\  & 90.4 & 90.4 & 90.4 & 90.4\\
&   \tbeta\       & 0.59 - 2.46 & 0.72 - 1.96 & 0.93 - 1.46 & none
\\\hline

\mbox{$ m_{\mathrm h}^{\rm max}$}  &
    \mh\  & 89.6 & 89.6 & 89.5 &  89.6 \\
$\mu >0$ &
    \mA\  & 90.3 & 90.3 & 90.3 & 90.3 \\
&   \tbeta\       & 0.59 - 2.61 & 0.71 - 2.00 & 0.87 - 1.54 & none
\\\hline

\mbox{$ m_{\mathrm h}^{\rm max}$} &
    \mh\  & 89.6 & 89.6 & 89.5 & 89.6 \\
$\mu >0, {\rm X_t}<0$ &
    \mA\  & 90.5 & 90.4 & 90.4 & 90.4 \\
&   \tbeta\       & 0.53 - 3.20 & 0.63 - 2.46 & 0.72 - 1.96 & 0.84 - 1.63
\\\hline

no mixing &
    \mh\ & 112.8 & 90.7 & 90.0 & 89.9 \\
&   \mA\ & 1000. & 91.2 & 90.8 & 90.5 \\
&   \tbeta\      & 0.40 - 50.0 &  0.40 - 9.70 & 0.40 - 5.40   & 0.40 - 4.40 \\
& \tbeta, \mA $>$12.0
                 & 0.46 - 0.96 &  0.46 - 0.96 &  0.47 - 0.97 & 0.47 - 0.97
\\\hline

no mixing &
    \mh\ & 89.9 & 89.8 & 89.7 & 89.8 \\
$\mu >0$&
    \mA\ & 90.8 & 90.6 & 90.4 & 90.3 \\
large M$_{\rm SUSY}$&
     \tbeta\ & 0.70 - 6.95  & 0.70 - 4.55  & 0.70 - 3.43 & 0.70 - 2.97 \\
& \tbeta, \mA $>$12.0
                 & 0.70 - 1.01 &  0.70 - 1.01 &  0.70 - 1.02 & 0.70 - 1.01
\\\hline

Large $\mu$ &
    \mh\  & 90.2 & 94.2 & 89.7 & 89.3 \\
&   \mA\  & 92.5 & 96.6 & 92.6 & 92.5 \\
&   \tbeta\       & 0.72 - 14.79 & 0.72 - 13.68 & 0.72 - 10.91 & 0.72 - 10.63 \\
& \tbeta, \mA $>$2.4
                 & 0.72 - 0.79 &  0.75 - 0.85 &  0.86 - 0.90 & none
\\\hline

Gluophobic &
    \mh\  & 87.8 & 87.0 & 86.4 & 86.2 \\
&   \mA\  & 93.0 & 92.9 & 93.2 & 93.5 \\
&   \tbeta\       & 0.40 - 9.70 & 0.42 - 5.22 & 0.48 - 3.76 & 0.51 - 3.19
\\\hline

Small $\alpha$ &
    \mh\  & 84.3 & 83.5 & 82.5 &  82.0 \\
&   \mA\  & 95.0 & 95.8 & 96.5 & 97.2 \\
&   \tbeta\       & 0.40 - 5.97 & 0.43 - 4.03 & 0.52 - 3.12 & 0.55 - 2.69

\\\hline
\end{tabular}
\caption[]{95\% {\sc CL} lower bounds on \mh\ and \mA\ in \GeVcc\ and
  excluded ranges in \tbeta\ obtained in the different {\sc MSSM}
  {\sc CP}-conserving benchmark scenarios, as a function of \mtop. 
  Except for the two no mixing and the large $\mu$ scenarios,
  the exclusions in mass are valid for all values of \tbeta\ between 0.4
  and 50, and the exclusions in \tbeta\ hold for all values of \mA\ between
  0.02 and 1000~\GeVcc.
  In the three other scenarios, part of the interval in \tbeta\
  is excluded only for \mA\ above a few \GeVcc\ threshold: 
  this sub-interval is indicated 
  in a fourth line together with the threshold in \mA. 
  As a consequence, the mass bounds in these scenarios are valid only for
  values of \tbeta\ outside the quoted sub-interval.
}
\label{tab:limits}
\end{center}
\end{table}

\subsection{Summary}

The lower bounds in mass and excluded ranges in \tbeta\ obtained
in the eight {\sc CP}-conserving benchmark scenarios presented
in the previous sections are summarized in Table~\ref{tab:limits}.
The variation with \mtop\ of the excluded ranges in \tbeta\ is 
further illustrated in Fig.~\ref{fig:mtop}. 
All lower bounds in mass are at the 95\% {\sc CL}, as well as each
individual (either lower or upper) bound in \tbeta.
In all scenarios, the radiative corrections on the Higgs boson
masses and couplings have been computed 
in the Feynman-diagrammatic approach with all dominant two-loop 
order terms included, using version 2.0 of the {\sc FeynHiggs} 
code~\cite{ref:FDradco_new}.

\newpage
\section{The {\sc CP}-violating {\sc MSSM} scenarios}\label{sec:cpv}

In most of the parameter space of the {\sc CP}-violating {\sc MSSM} 
scenarios studied in this paper, only the two lightest neutral
Higgs bosons, H$_1$ and H$_2$ are kinematically accessible at 
{\sc LEP} energies. If their couplings to the Z boson are not 
strongly suppressed by {\sc CP}-violation,
the main production processes are 
the \HuZ, \HdZ\ and \HuHd\ processes, with \HuZ\ dominating at low
\tbeta, \HuHd\ at large \tbeta\ and \HdZ\ contributing over the whole
range of \tbeta\ values allowed in each scenario.
In restricted areas of the parameter space, the second pair-production
process, \HuHt, can add a non-negligible signal and has 
also been considered in the searches. On the other hand, in most
scenarios, charged Higgs bosons have a mass above 100~\GeVcc, and thus
have not been included.

   As already mentioned, {\sc CP} violation in the {\sc MSSM} Higgs sector
is introduced through radiative corrections. Besides the two parameters
used to define the scenarios at tree level, chosen as \tbeta\ and 
\mHp, 
radiative corrections introduce additional parameters. 
As in the {\sc CP}-conserving case, these are primarily
\mtop\ and the set of parameters related to supersymmetry breaking: 
$\mu$,  $M_{\rm susy}$, $M_2$, $m_{\tilde{g}}$ and $A$, as defined in
section~\ref{sec:benchmark}~\cite{ref:mssmpheno,ref:FDmssmpheno}.  
In addition, {\sc CP} violation introduces phases. 
The unification assumptions made for the supersymmetry breaking 
parameters, and the global symmetries
that govern the dimension-four operators of the {\sc MSSM} Lagrangian,
can be used to reduce the number of {\sc CP}-violating phases to 
only two~\cite{ref:cpv_mssm}. In the scenarios studied hereafter, these
phases are taken as the phase of the gluino mass, arg($m_{\tilde{g}}$) and
the phase of the common stop and sbottom trilinear coupling, 
arg($A$).

\subsection{The benchmark scenarios} \label{sec:cpv_bench}

The dominant {\sc CP}-violating effects on the neutral Higgs boson
masses and couplings to gauge bosons are proportional to 
\[ \frac{m_{top}^4}{v^2} \frac{Im(\mu A)}{M_{susy}^2} \]
where $v^2$ is the quadratic sum of the vacuum expectation values of the two
Higgs field doublets~\cite{ref:cpv_bench}.
Sizeable effects are thus expected for moderate values of $M_{\rm susy}$,
large values of $\mu$ and phases arg($A$) around 90$^\circ$. A
strong dependence on the value of \mtop\ is also to be expected.

Along these lines, Ref.~\cite{ref:cpv_bench} proposed a benchmark 
scenario with maximal {\sc CP}-violation, the {\sc CPX} scenario, as 
an appropriate scheme for direct searches at {\sc LEP} and other 
colliders. The values of
its underlying parameters are quoted in Table~\ref{ta:cpv_scen}.
As expected from the above discussion, the value of $M_{\rm susy}$,  
a few hundred \GeVcc, is moderate, $\mu$ and $|A|$ take 
large values, 2 and 1~\TeVcc\ respectively, and the {\sc CP}-violating
phase arg($A$) is set at 90$^\circ$. 
Although the gluino-mass phase has a small impact on the 
{\sc CP}-violating effects, these appear to be reinforced 
at 90$^\circ$~\cite{ref:cpv_mssm}, a value which was thus 
retained for arg($m_{\tilde{g}}$).
The values listed in Table~\ref{ta:cpv_scen}
fulfill the existing constraints from measurements of the
electron and neutron electric dipole moments, by making the first
two generations of squarks sufficiently heavy, with masses above 
1~\TeVcc. In the following,
the CPX scenario has been studied for four values of the top quark mass,
\mtop~=~169.2, 174.3, 179.4 and 183.0~\GeVcc.

\begin{table}[t]
{\small
\begin{center}
\begin{tabular}{l|ccccccc}     \hline
scenario & $M_{\rm susy}$ & $M_2$ & $|m_{\tilde{g}}|$ & $\mu$ 
         & $|A|$ & arg($m_{\tilde{g}}$)=arg($A$) \\
& (\GeVcc) & (\GeVcc) & (\GeVcc) & (\GeVcc) & (\GeVcc) & 
  (degrees) \\
\hline
{\sc CPX}  &
      500 & 200 & 1000 & 2000 & 1000  & 90  \\
phase study  &
      500 & 200 & 1000 & 2000 & 1000 & 0,30,60,135,180\\
 $\mu$ study  &
      500 & 200 & 1000 & 500,1000,4000 & 1000 & 90\\
 $M_{\rm susy}=$1 TeV/c$^2$ &
      1000 & 200 & 1000 & 2000 & 1000 & 90\\
 $M_{\rm susy}=$1 TeV/c$^2$, scaled &
      1000 & 200 & 2000 & 4000 & 2000 & 90\\
\hline
\end{tabular}
\caption[]{
Values of the underlying parameters for the representative 
{\sc CP}-violating {\sc MSSM} scenarios scanned in this paper,
namely the {\sc CPX} scenario and its ten variants.
}
\label{ta:cpv_scen}
\end{center}
}
\end{table}

In addition to the {\sc CPX} scenario, a few variants have also been 
considered in order to study the dependence of the {\sc CP}-violation 
effects on the values of phases, $\mu$ and $M_{\rm susy}$. The
values tested are quoted in Table~\ref{ta:cpv_scen}. 
The two {\sc CP}-violating phases, still taken to be equal, were varied 
from 0 to 180$^\circ$, keeping all other parameters as in the
{\sc CPX} scenario. Values of $\mu$ below and above 2~\TeVcc\ were
studied in the same way. Finally, the value of  $M_{\rm susy}$
was increased from 500~\GeVcc\ to 1~\TeVcc, keeping the phases
at 90$^\circ$, and either keeping all
other parameters to their {\sc CPX} values, or
scaling the other parameters in such a way that the relation 
between $|m_{\tilde{g}}|$, $|A|$ and $\mu$ is 
as in the {\sc CPX} scenario. In the following,
these ten variants have been studied for \mtop~=~174.3~\GeVcc\ only.

In all scenarios, theoretical databases provided by the
{\sc LEP} Higgs working group were used~\cite{ref:hwg}. In these,
radiative corrections have been computed in two different approaches,
the Feynman-diagrammatic approach of Ref.~\cite{ref:FDradco_new},
already selected 
in the {\sc CP}-conserving case (see Section~\ref{sec:benchmark}), 
and the renormalization group approach of Ref.~\cite{ref:RGradco}.
As in Section~\ref{sec:benchmark}, the Feynman-diagrammatic calculations
use version 2.0 of the {\sc FeynHiggs} code.
The renormalization group corrections rely on 
the {\sc CP}-violating version {\sc CPH} of the {\sc SUBHPOLE} 
code.\footnote{Since this work, updated versions of the two codes,  
{\sc CPsuperH}~\cite{ref:CPsuperH} and
{\sc FeynHiggs 2.5}~\cite{ref:FeynHiggs2.5} have been made available.
In both cases, the changes concern the Higgs boson decays and have
no substantial impact on the phenomenology at {\sc LEP}.}
Contrary to the {\sc CP}-conserving case, where the calculations
in the Feynman-diagrammatic approach were the most complete
due to the inclusion of all dominant two-loop
order terms, in the case of {\sc CP}-violation
neither of the two calculations can be preferred on theoretical grounds.
Both contain one and two-loop corrections, but
the {\sc CPH} code has a more complete phase dependence at the two-loop
order while {\sc FeynHiggs} contains more corrections at the one-loop order
with the full complex phase dependence and more corrections at the two-loop
order but without the full phase dependence.
This may result in large differences when convoluted with the 
experimental inputs.
We thus present our results in the two frameworks separately. 
A comparison between the two calculations in the {\sc CP}-conserving case
can be found in Ref.~\cite{ref:FDRGcomp}.

\begin{figure}[htbp]
\vspace{-1.7cm}
\begin{center}
\begin{tabular}{cc}
\epsfig{figure=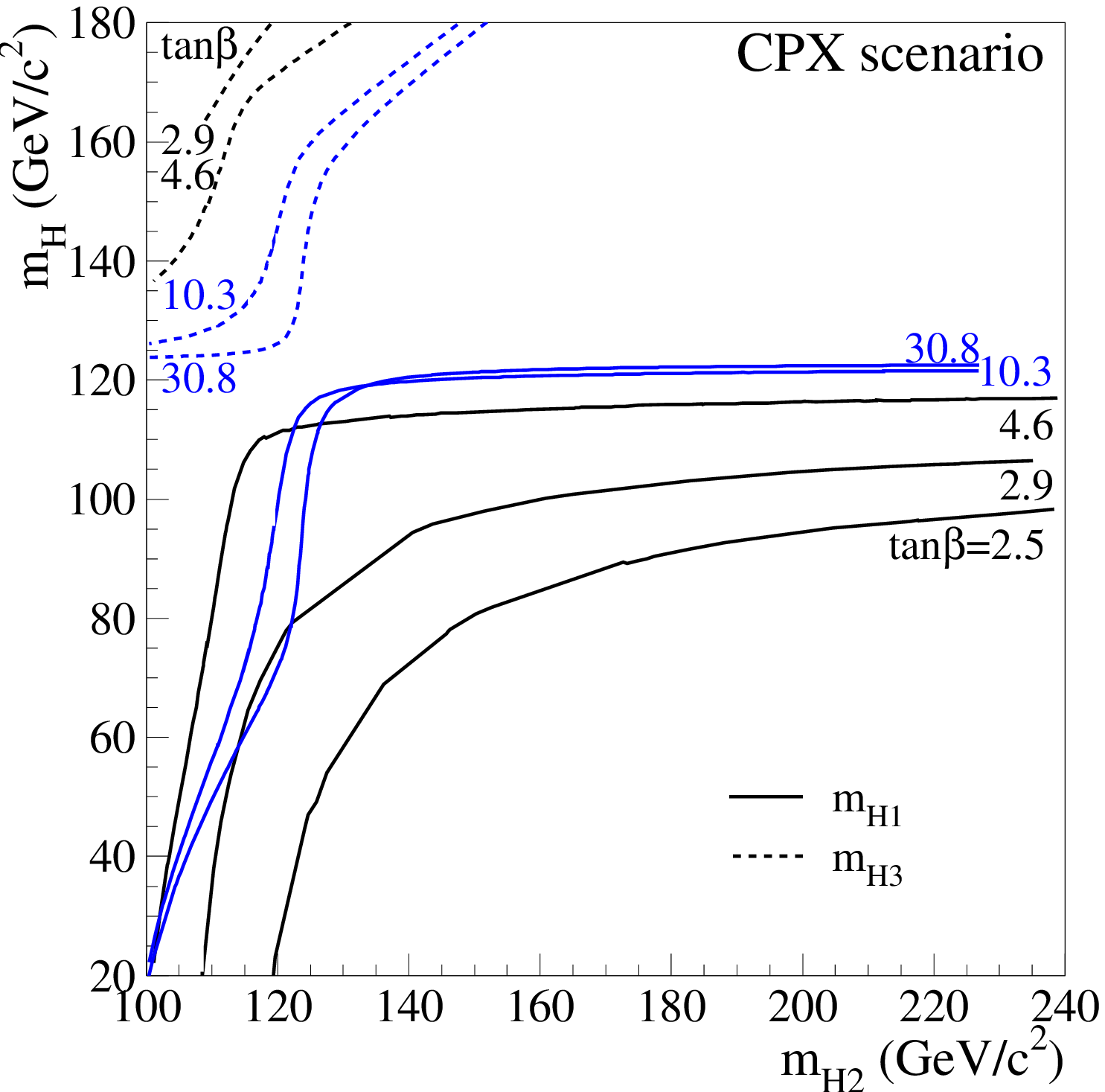,height=7.6cm} &
\epsfig{figure=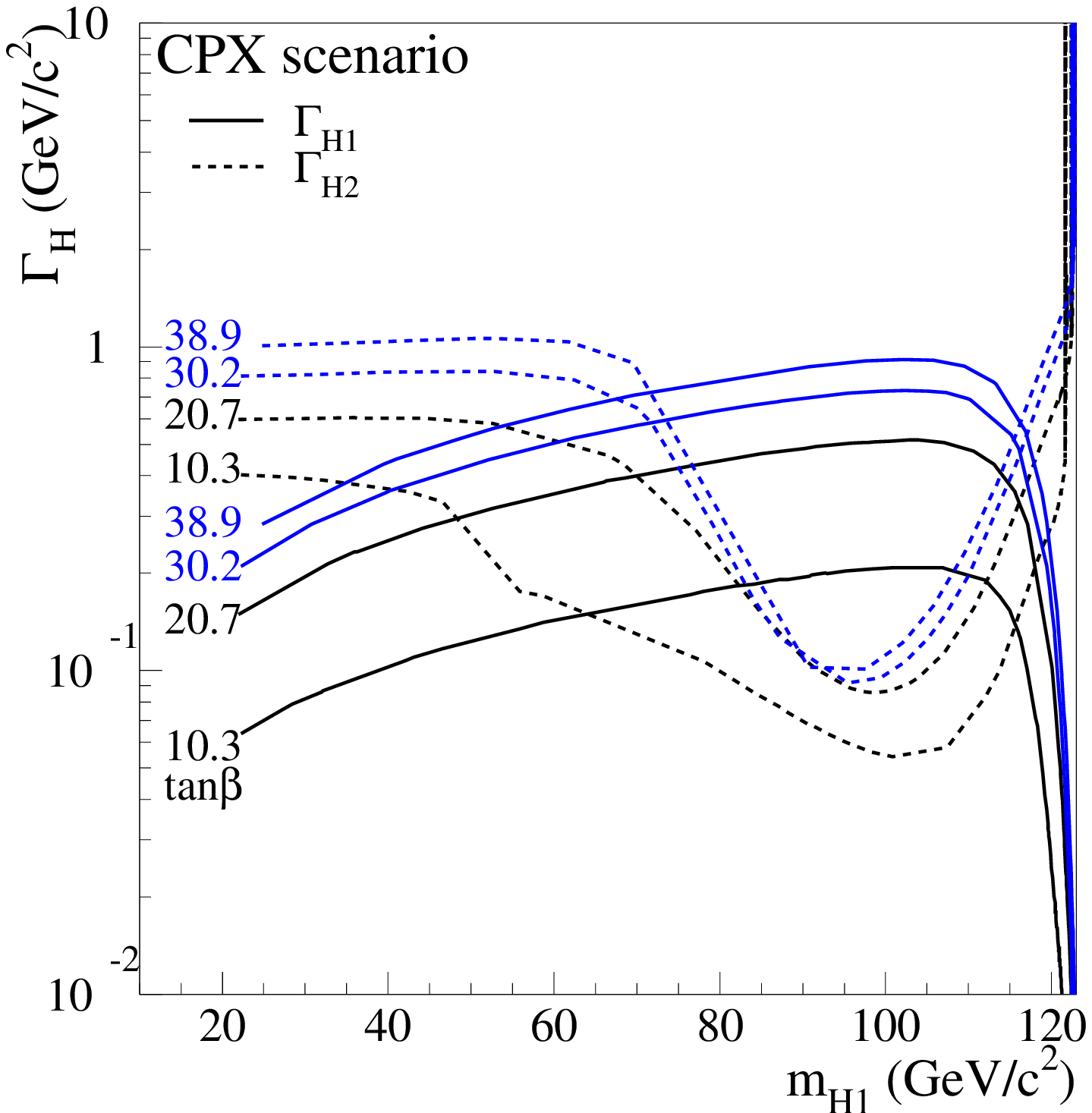,height=7.6cm}\\
\epsfig{figure=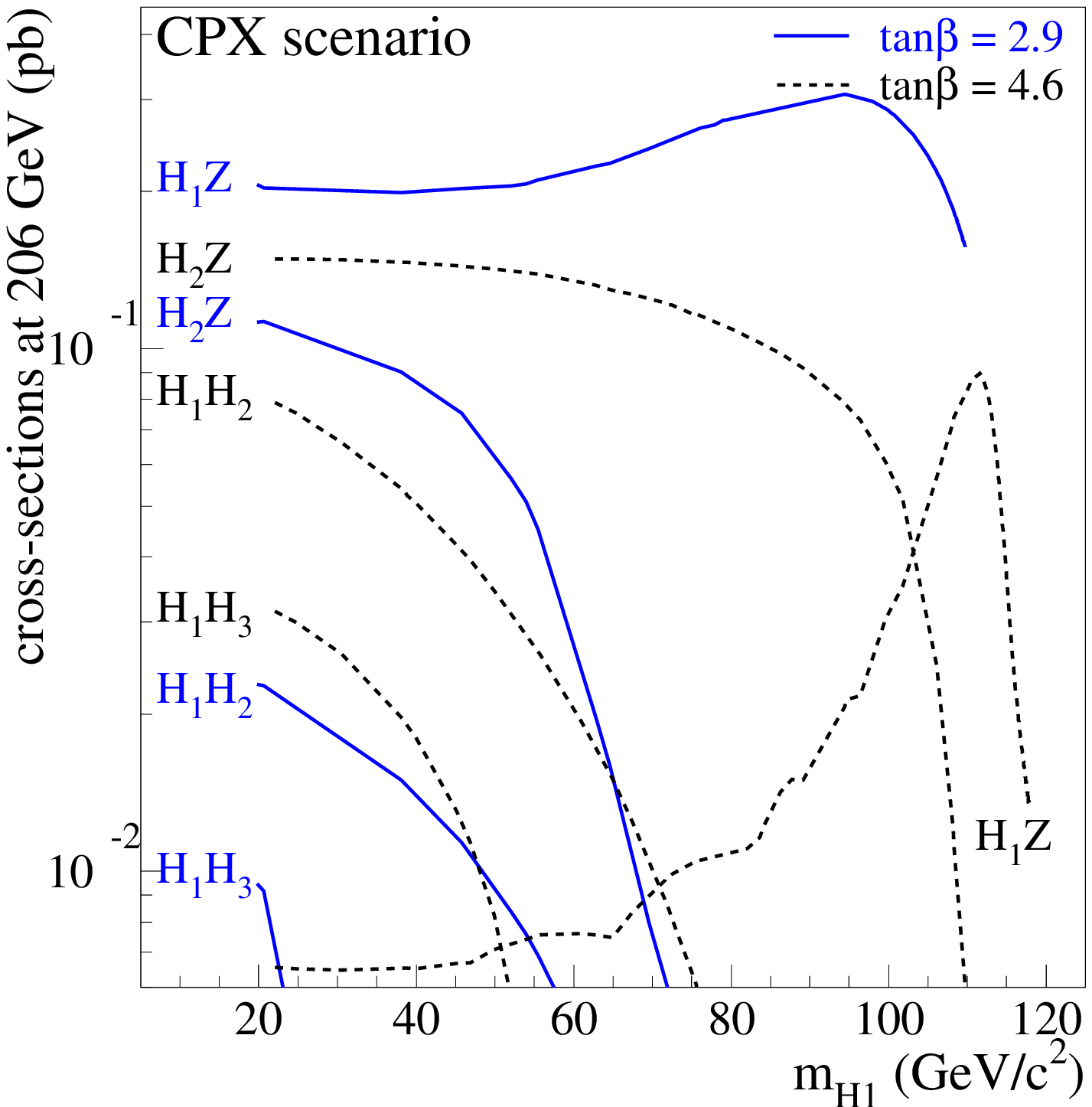,height=7.6cm} &
\epsfig{figure=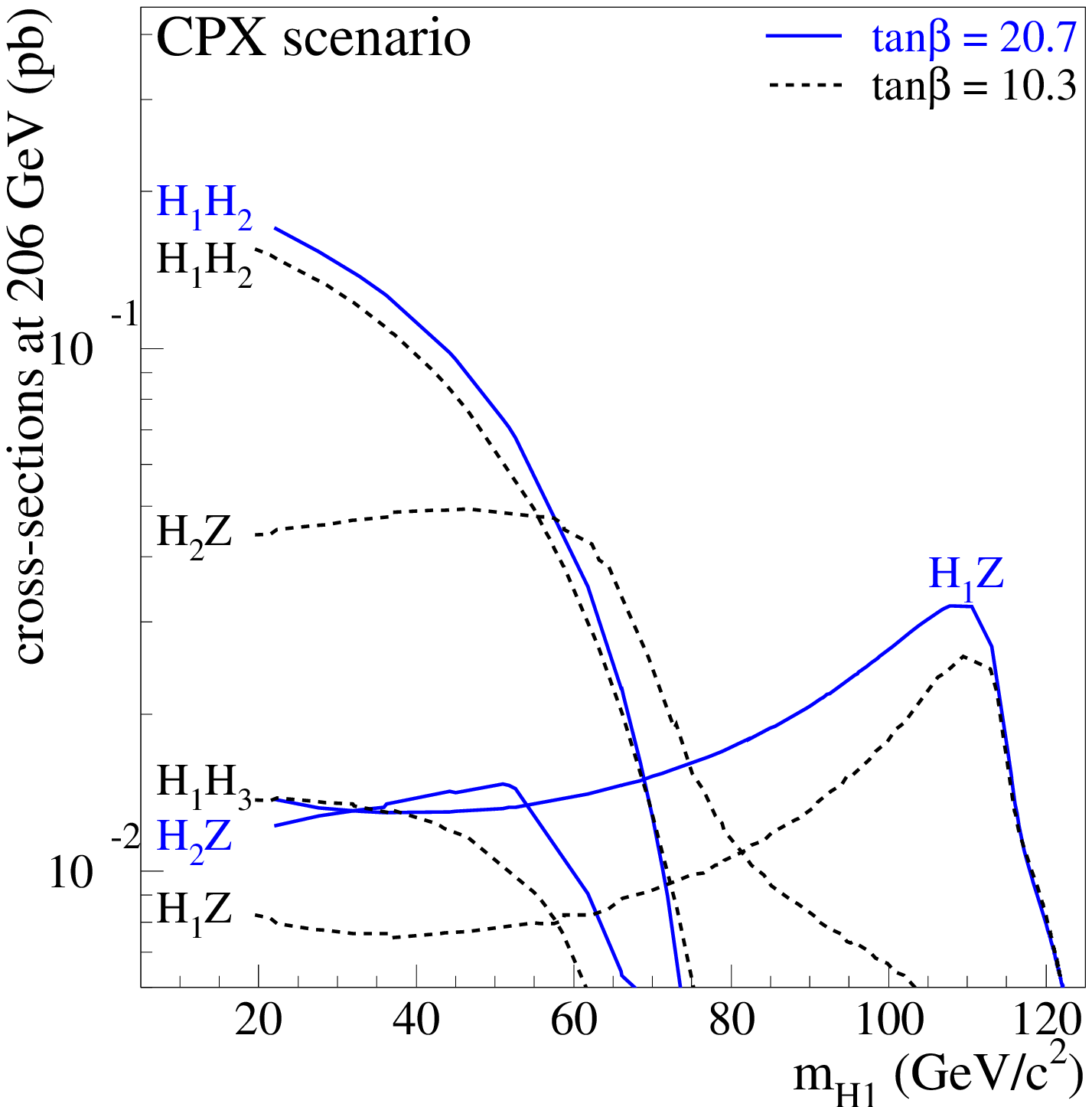,height=7.6cm}\\
\epsfig{figure=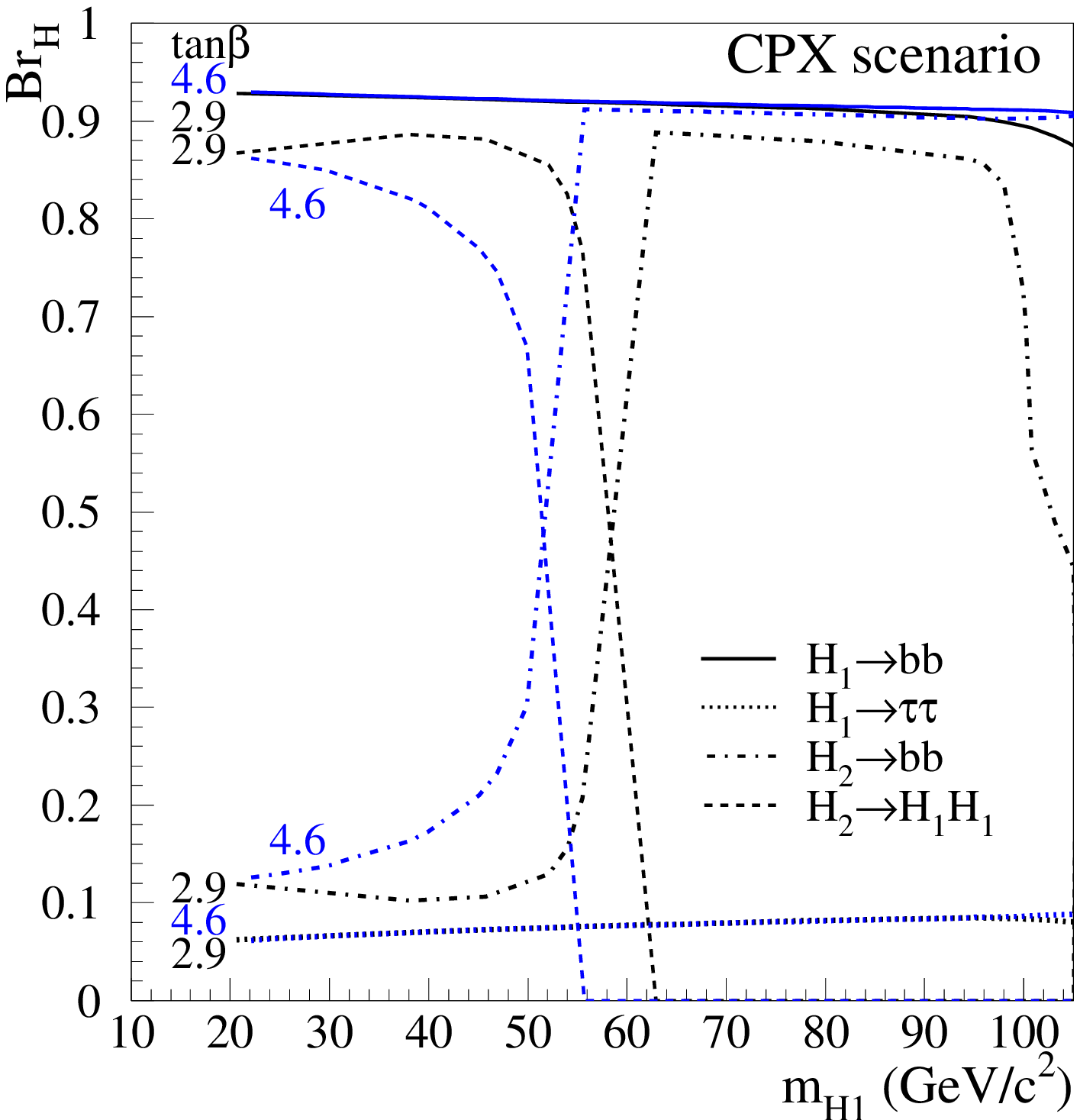,height=7.6cm} &
\epsfig{figure=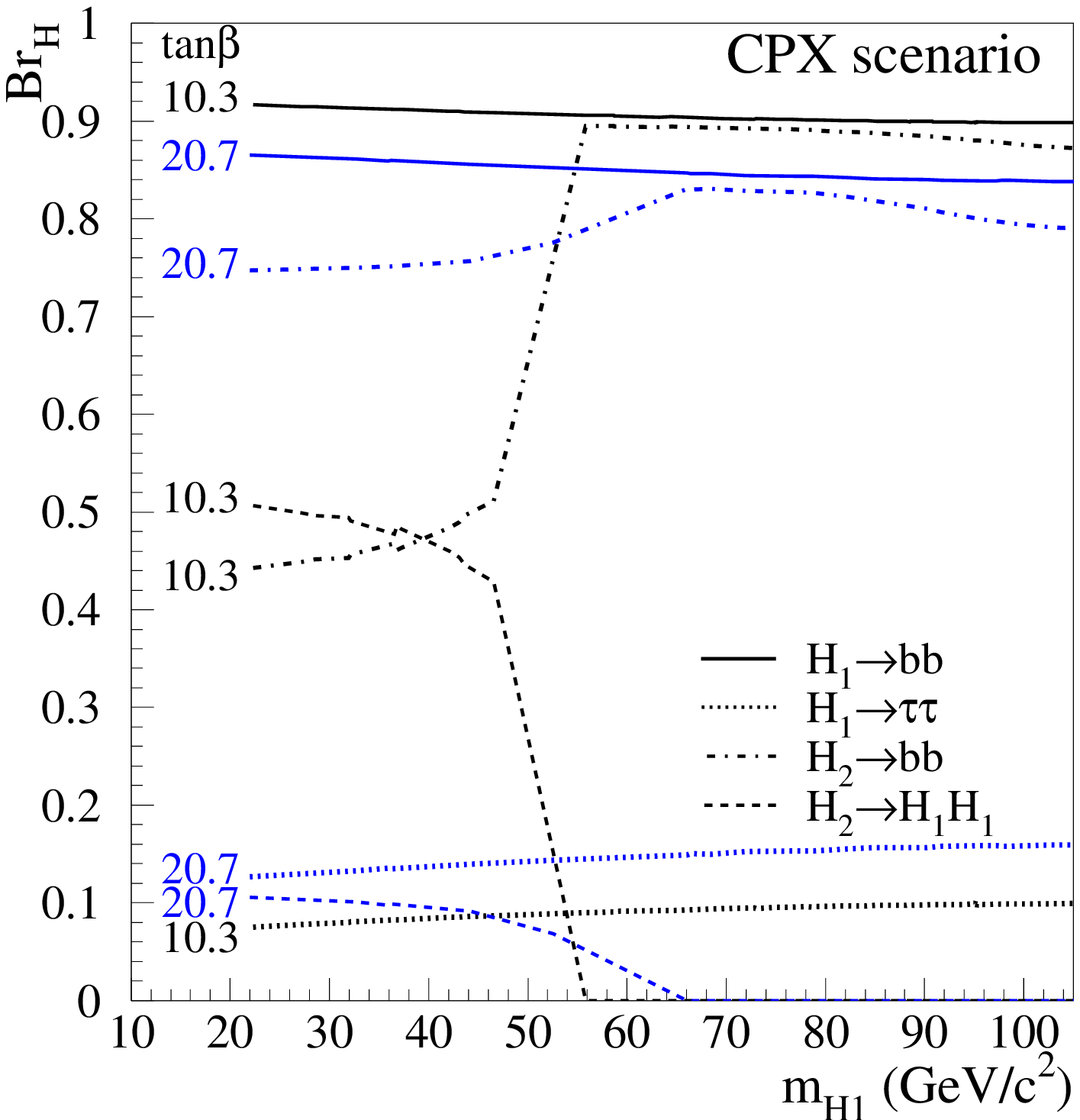,height=7.6cm} 
\end{tabular}
\caption[]{
    Properties of the three neutral Higgs bosons
    of the {\sc CP}-violating {\sc MSSM} in the 
    {\sc CPX} scenario with \mtop~=~174.3~\GeVcc. 
    Top: H$_1$, H$_2$ and H$_3$ masses and H$_1$, H$_2$ widths at 
    various \tbeta\ values.
    Middle: production cross-sections at \rs~=~206~GeV, 
    as a function of \mHu\ and \tbeta. 
    Bottom: H$_1$ and H$_2$ dominant branching fractions as
    a function of \mHu\ and \tbeta. H$_1$ and H$_2$ decays into \bbbar\ 
    (solid and dash-dotted lines) are compared with H$_1$ decays into 
    \toto\ (dotted lines) and 
    H$_2$ decays into H$_1$H$_1$ (dashed lines).
    The radiative corrections are as in Ref.~\cite{ref:RGradco}.}
\label{fig:cpv_neutral}
\end{center}
\end{figure}

 The phenomenology of the three neutral Higgs bosons of the 
{\sc CP}-violating {\sc MSSM} is illustrated in 
Fig.~\ref{fig:cpv_neutral} in the case 
of the {\sc CPX} scenario for a top quark mass of 174.3~\GeVcc, with 
radiative corrections computed in the renormalization group approach.
The top figures show that the two lightest scalars, H$_1$ and H$_2$,
are likely to be kinematically accessible at {\sc LEP2} in 
wide regions of the parameter space, in which their widths remain
lower than 1~\GeVcc, that is below the experimental resolution.
The cross-section curves show that at low and large \tbeta, the
dominant production processes are the \HuZ\ and \HuHd\ processses, 
respectively, as in the {\sc CP}-conserving case 
(see Fig.~\ref{fig:heavy_neutral_new}). On the other hand, 
at intermediate \tbeta\ and moderate \mHu,
the \HuZ\ cross-section is significantly weakened,
as a result of the suppressed H$_1$ZZ coupling due to {\sc CP}-violation.
In the same region, the \HdZ\ process compensates only partly for this loss.
Finally, the figures showing branching fractions compare the dominant 
H$_1$ and H$_2$ branching fractions for different values of \tbeta.
For all values of \tbeta, decays into \bbbar\ and \toto\ saturate
the width of the lightest Higgs boson, H$_1$, in the mass range above
the \bbbar\ threshold up to the maximal sensitivity of {\sc LEP}.
In the same mass range, the second lightest Higgs boson, H$_2$, decays
predominantly into \bbbar\ at large \tbeta\ only. At low and intermediate
\tbeta, the cascade decay \Hcas\ dominates over the \bbbar\ final state
at masses up to 50~\GeVcc\ or so. 
A loss in experimental sensitivity can thus be anticipated
in regions where the \HuZ\ cross-section is negligible and 
the H$_2$Z signals are not significant with respect to background,
due to too weak \HdZ\ cross-sections or H$_2$ branching fractions
into fermions.

\subsection{Scan procedure}
The scan procedure is similar to that described in 
Section~\ref{sec:cpc_procedure} for the {\sc CP}-conserving scenarios. 
The only changes are the following. 
The scan was performed over the {\sc MSSM} parameters \tbeta\ and 
\mHp. The range in \mHp\ spans from 4~\GeVcc\ 
up to 1~TeV/$c^2$. Values of \mHp\ below about 100~\GeVcc\ were
noticed to give unphysical negative mass squared values in most
scenarios and thus were removed from the scans.
The range in \tbeta\ extends from the minimal value allowed in each 
scenario up to 40, a value above which the Higgs-bottom Yukawa coupling 
calculation becomes unreliable in the {\sc CP}-violating {\sc MSSM} 
scenarios.
Theoretical points were generated randomly in both \tbeta\ and \mHp\
with a granularity which is sufficient to map the general features
of the exclusion regions.

 The signal expectations in each channel were then computed as outlined 
in Section~\ref{sec:cpc_procedure}, except for the width effects. 
In the {\sc CP}-violating {\sc MSSM} scans, the widths remain well below 
the experimental resolution for \tbeta\ up to 40 and \mHu\ below
120~\GeVcc\ (see Fig.~\ref{fig:cpv_neutral}). Signal efficiencies and
PDFs were thus exclusively determined from simulations with Higgs boson
widths below  1~\GeVcc.

\section{Results in {\sc CP}-violating {\sc MSSM} scenarios}\label{sec:resucpv}
The regions of the {\sc MSSM} parameter space excluded at 
95\% {\sc CL} or more by combining the neutral Higgs boson searches of 
Table~\ref{tab:channels} are hereafter discussed in turn 
for each scenario. 
The additional constraint from the limit on the Z partial width that would 
be due to new physics~\cite{ref:width} 
(used as described in Section 3.2 of~\cite{ref:hwg})
brings no gain in sensitivity in any of the scenarios tested.
Results are presented only in the (\mHu, \tbeta)
plane, which is the only one relevant at {\sc LEP} since the
minimal values of \mHd\ and \tbeta\ in most scenarios are such
that the region accessible at {\sc LEP} is much reduced in the other
projections.

\subsection{Dependence on the phases}

\begin{figure}[htbp]
\vspace{-1.7cm}
\begin{center}
\begin{tabular}{cc}
\hspace{-1.cm}
\epsfig{figure=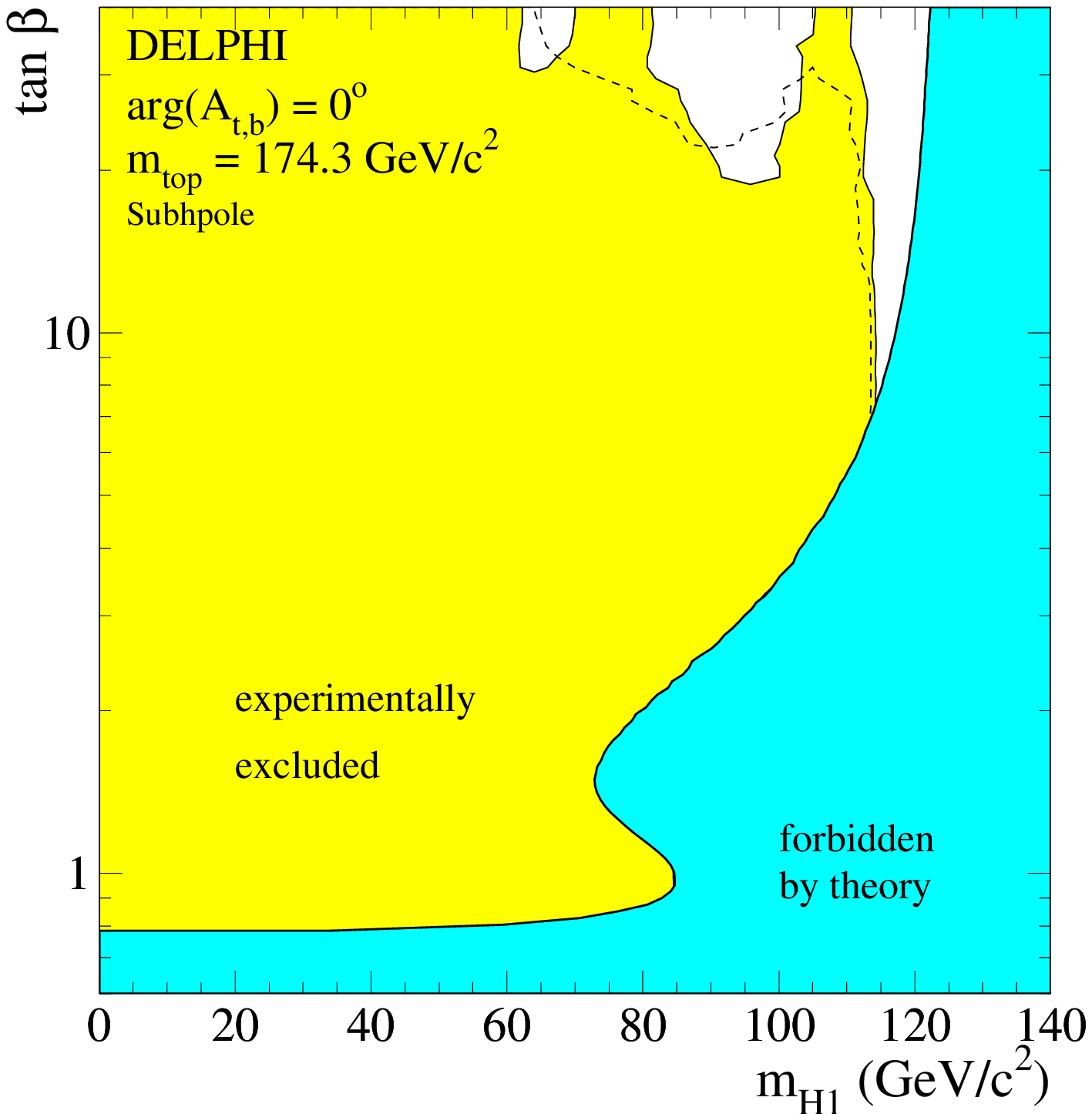,height=8cm} &
\hspace{-0.8cm}
\epsfig{figure=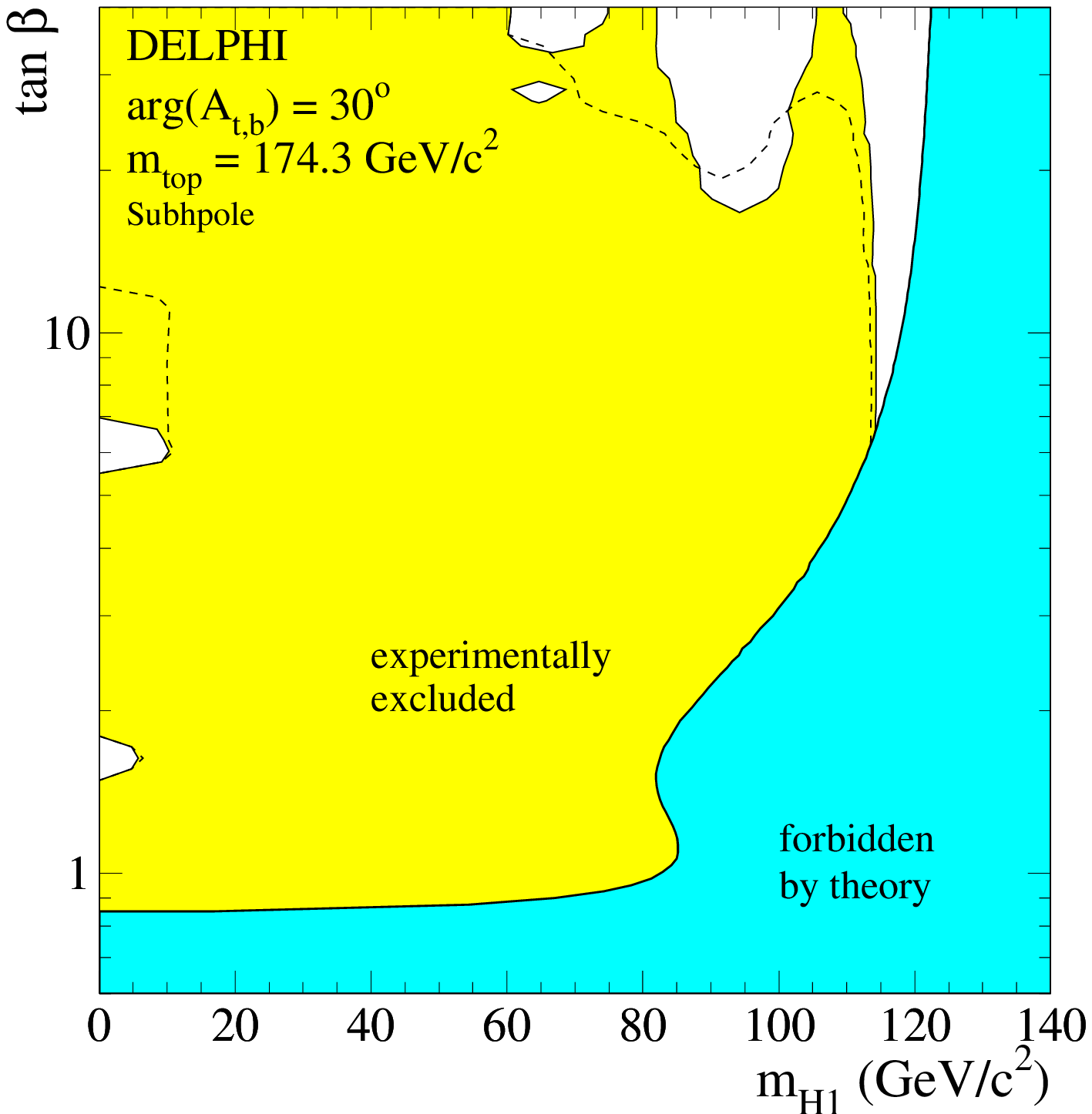,height=8cm} \\
\hspace{-1.cm}
\epsfig{figure=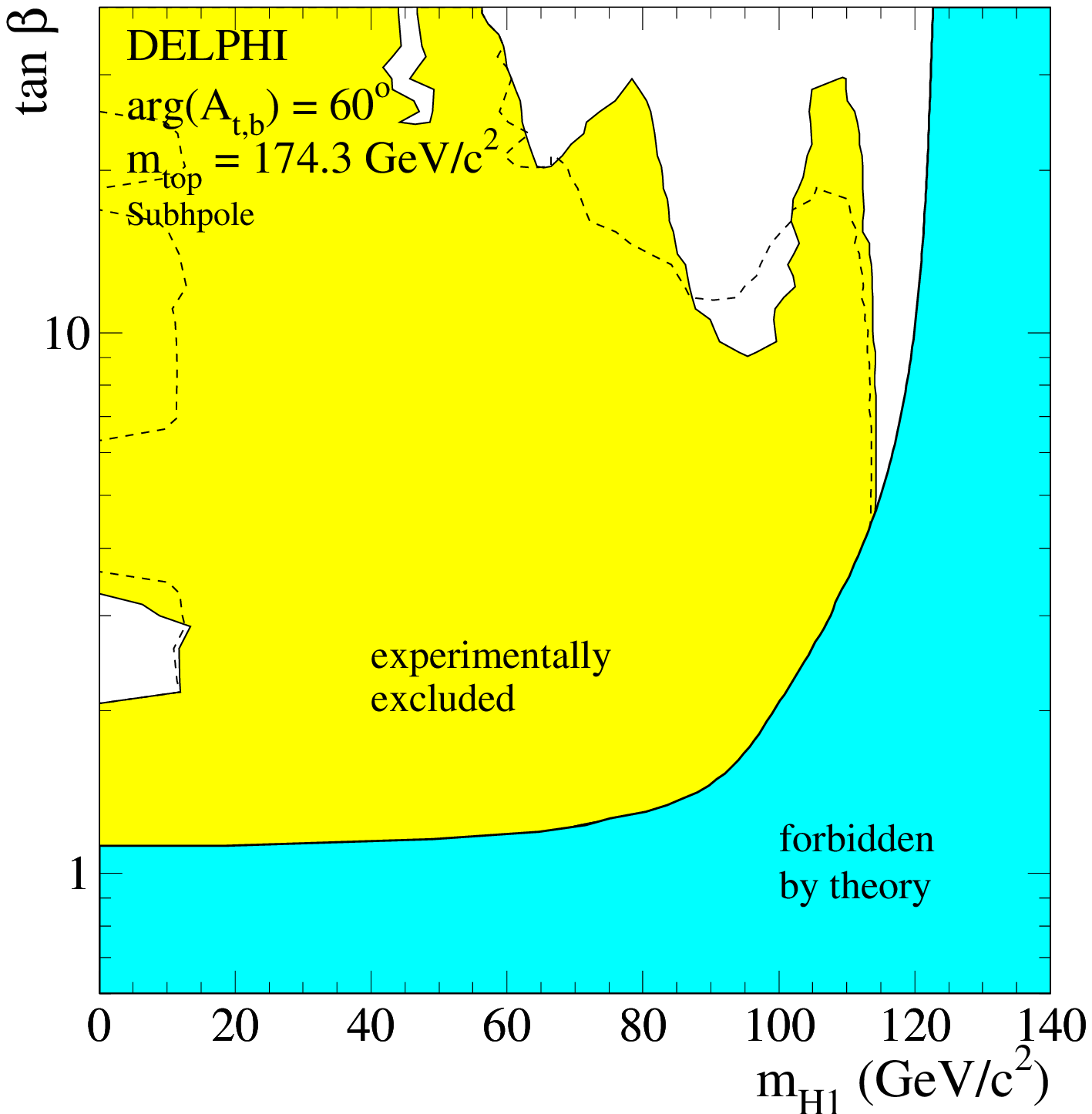,height=8cm} &
\hspace{-0.8cm}
\epsfig{figure=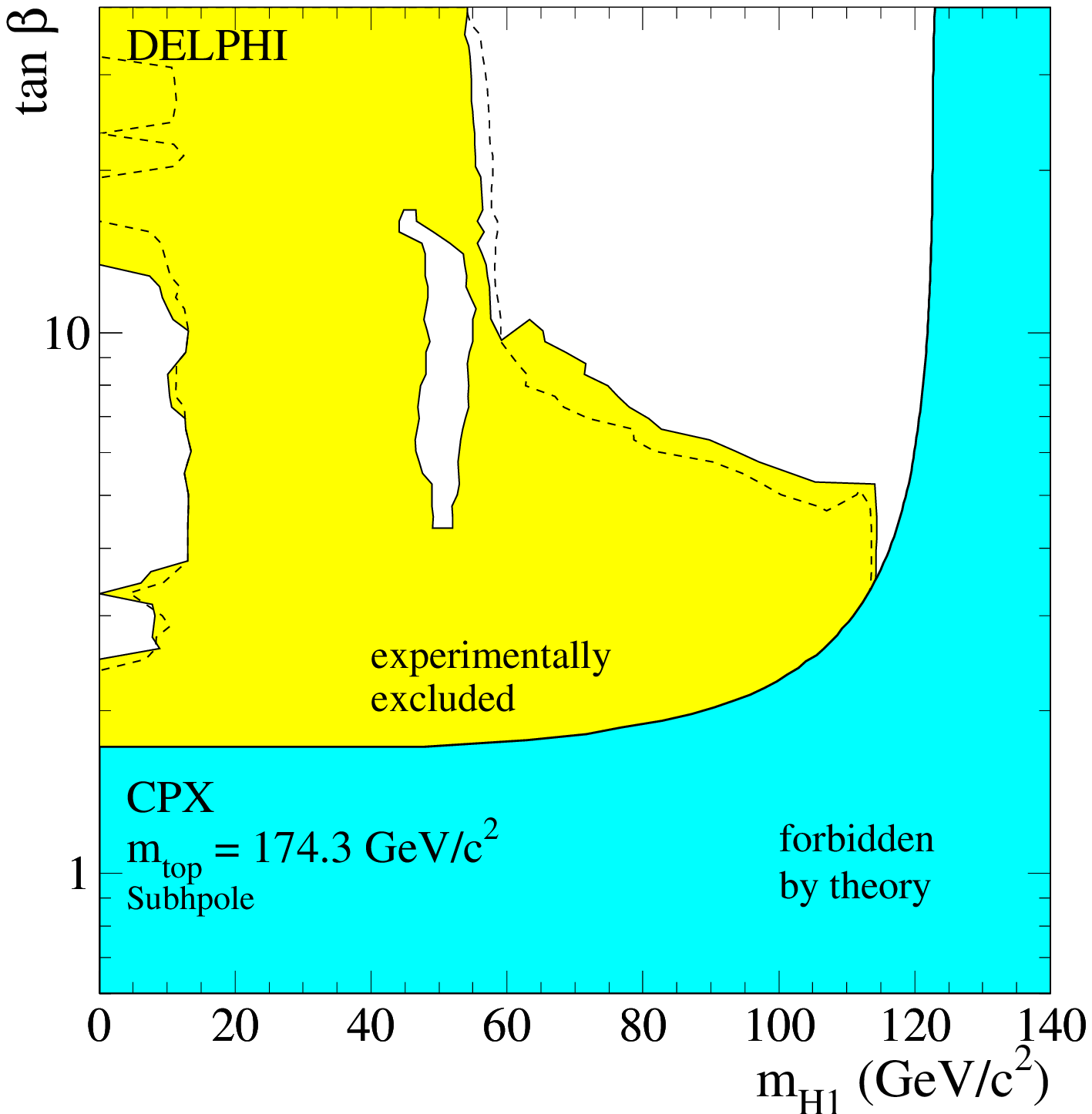,height=8cm} \\
\hspace{-1.cm}
\epsfig{figure=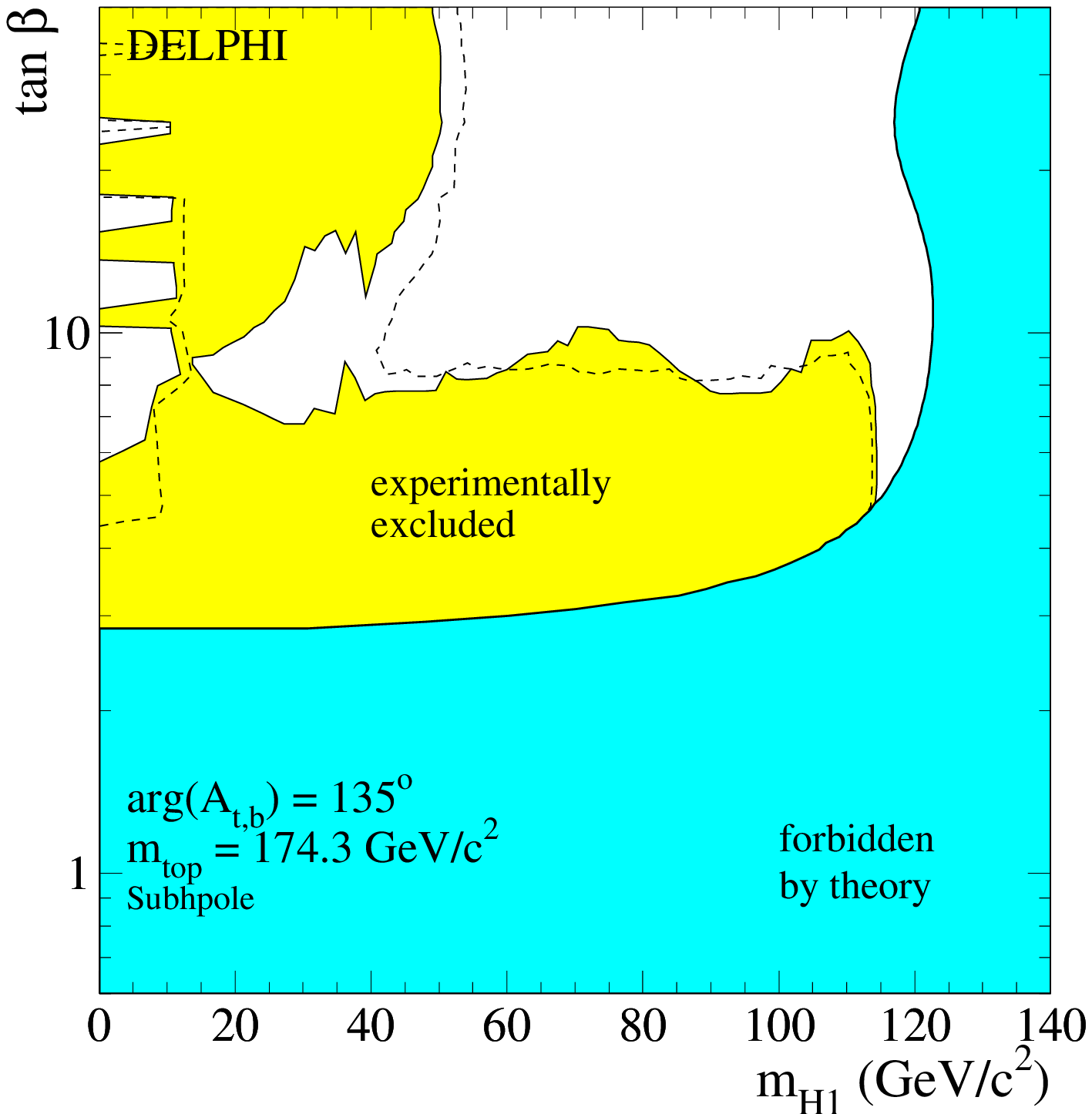,height=8cm} &
\hspace{-0.8cm}
\epsfig{figure=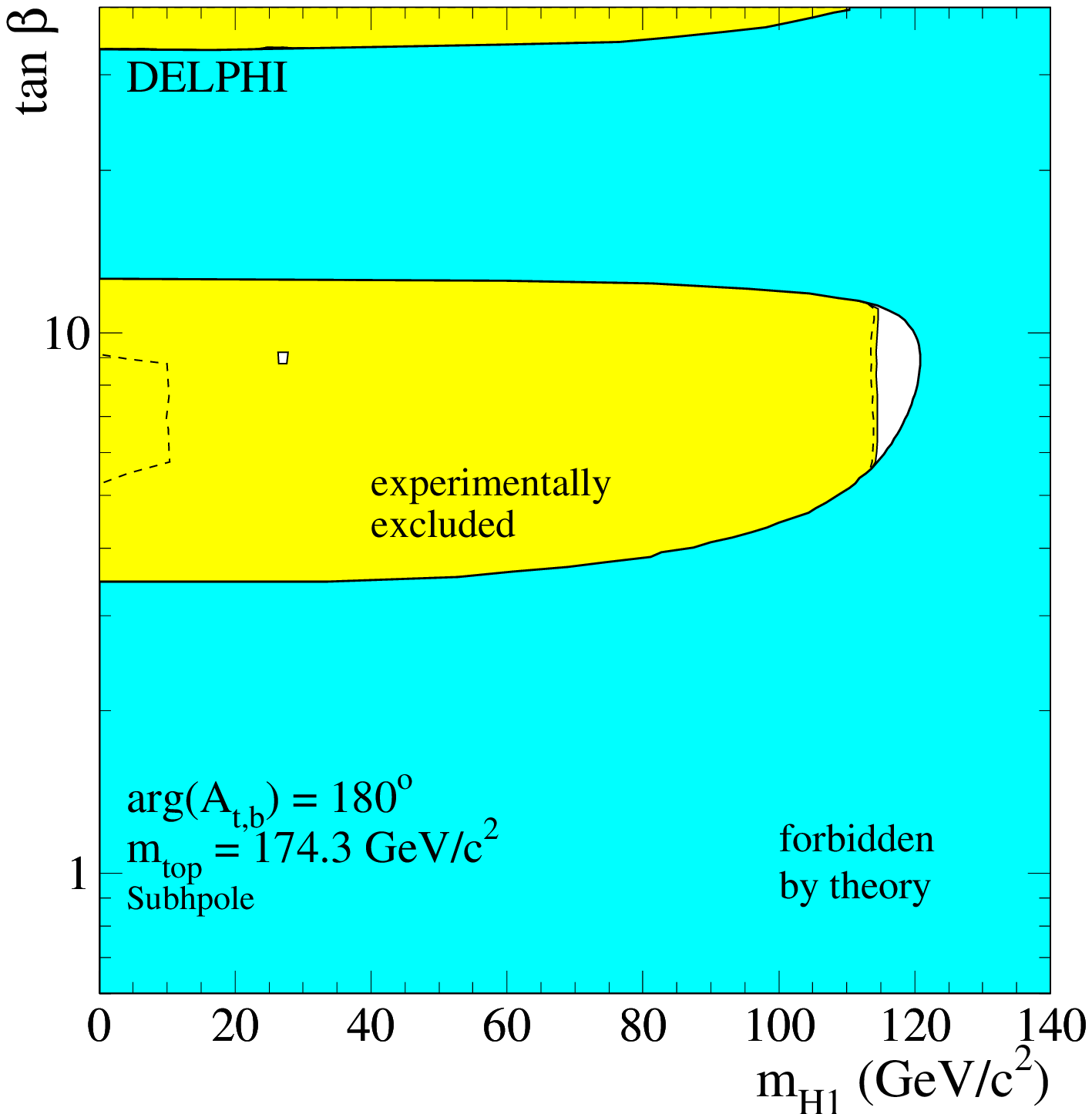,height=8cm} 
\end{tabular}
\caption[]{
    {\sc CP}-violating {\sc MSSM} scenarios with corrections 
    as in Ref.~\cite{ref:RGradco} for different values of the 
    phases: regions excluded at 
    95\% {\sc CL} by combining the results of the neutral Higgs 
    boson searches in the whole {\sc DELPHI} data sample  
    (light-grey). 
   The dashed curves show the median expected limits.
   The medium-grey areas are the regions not allowed by theory.
   The {\sc CPX} scenario corresponds to phases of 90$^{\circ}$.
   }
\label{fig:limit_phases_sub}
\end{center}
\end{figure}

\begin{figure}[htbp]
\vspace{-1.7cm}
\begin{center}
\begin{tabular}{cc}
\hspace{-1.cm}
\epsfig{figure=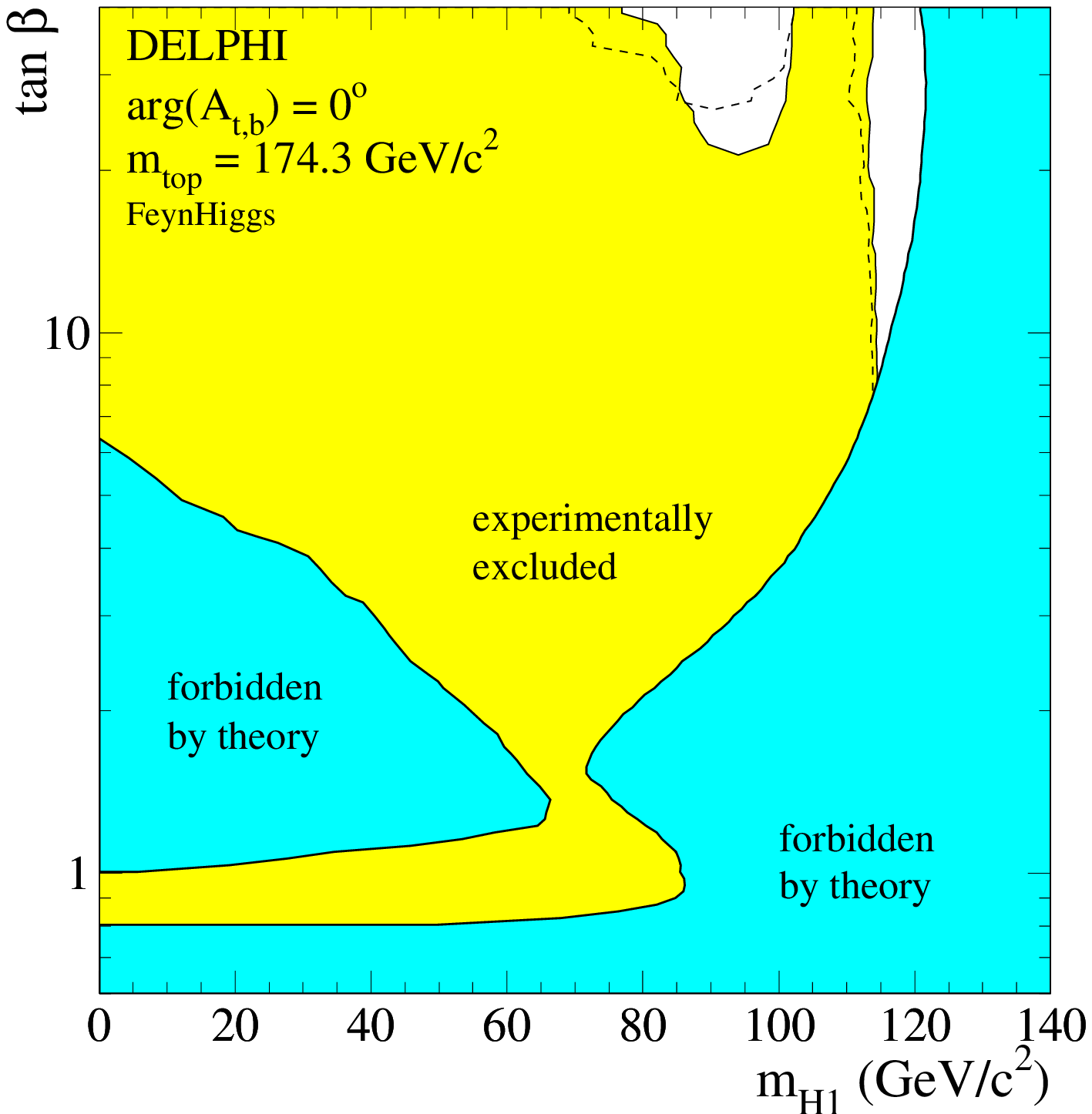,height=8cm} &
\hspace{-0.8cm}
\epsfig{figure=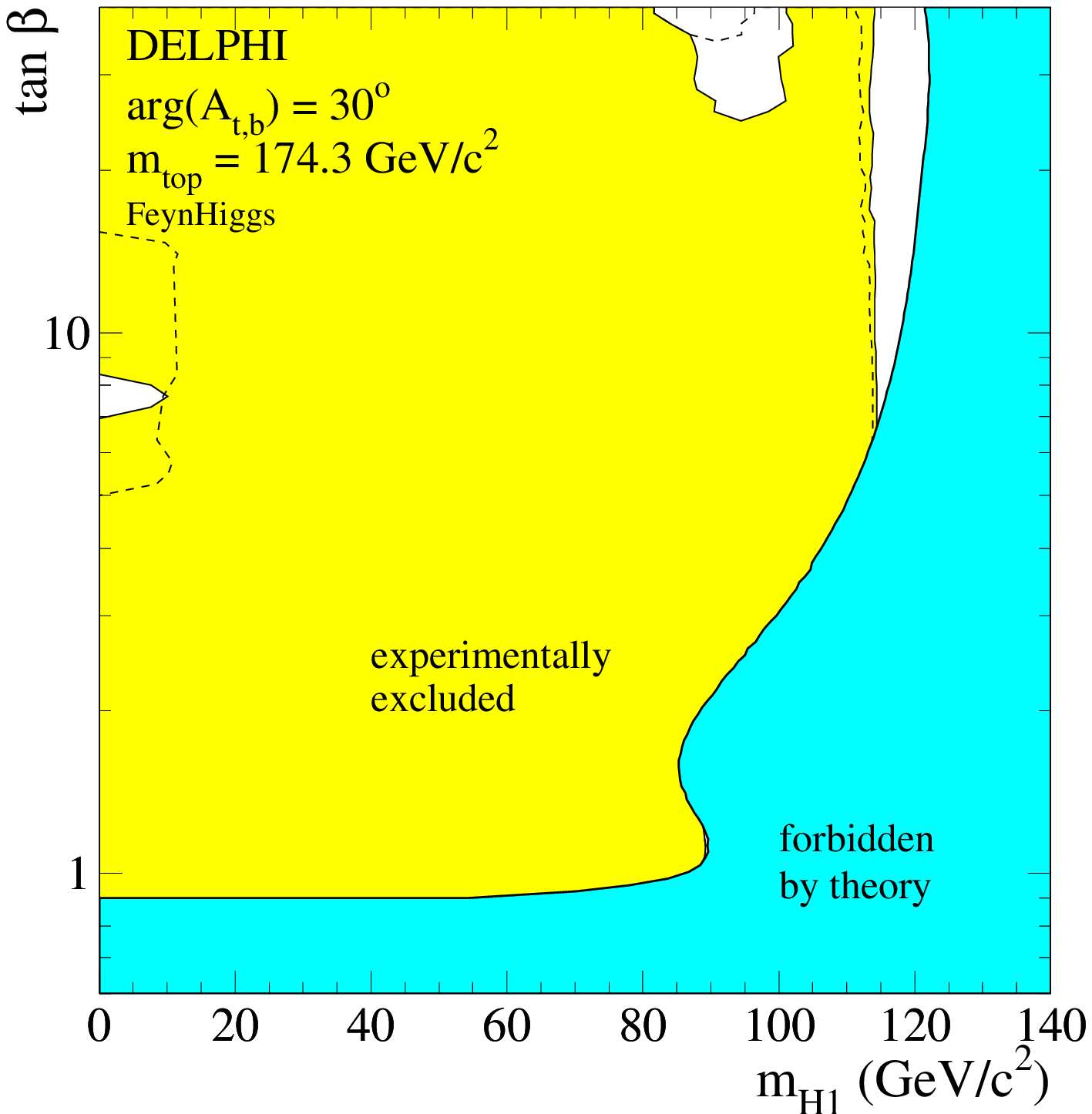,height=8cm} \\
\hspace{-1.cm}
\epsfig{figure=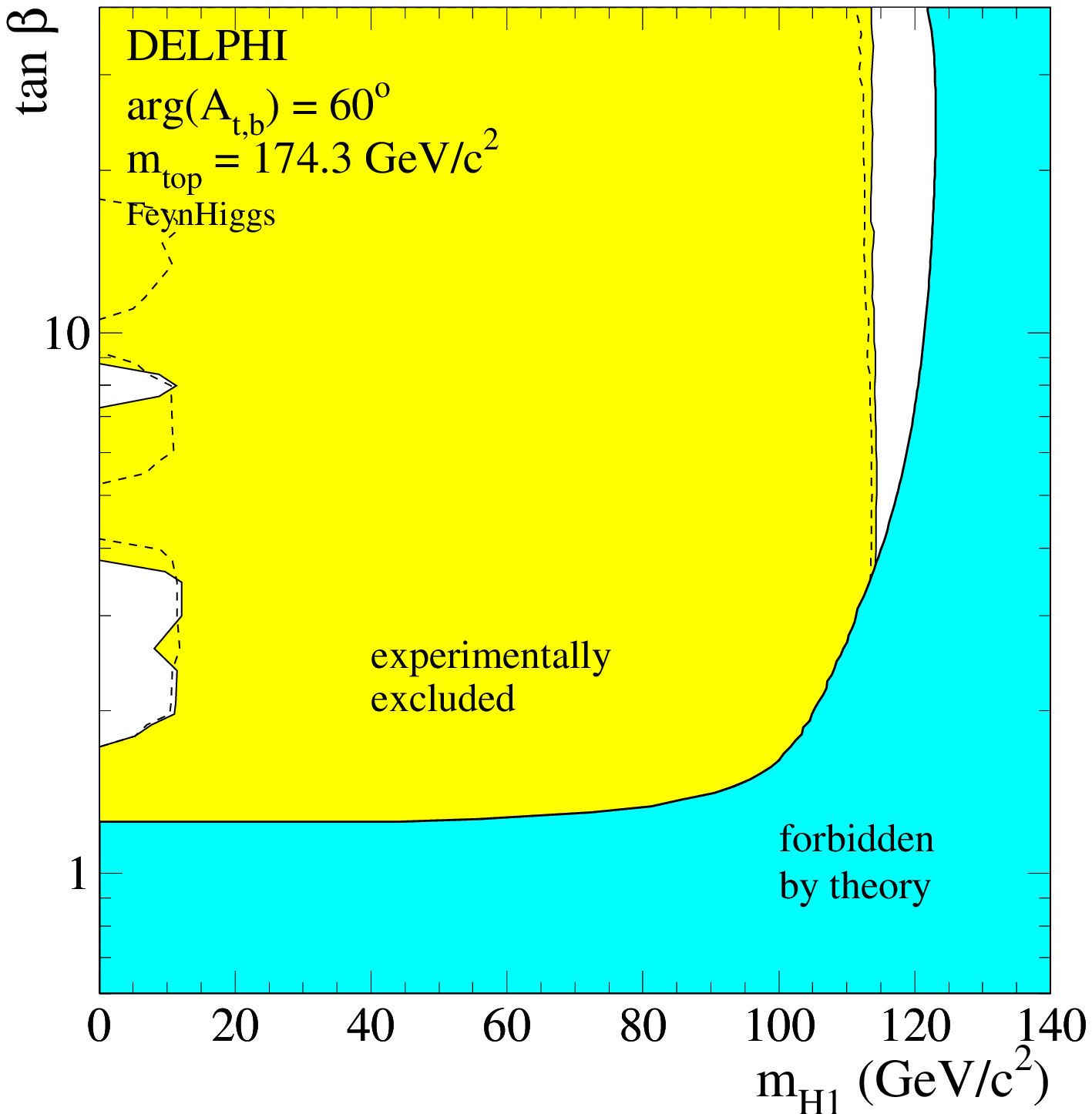,height=8cm} &
\hspace{-0.8cm}
\epsfig{figure=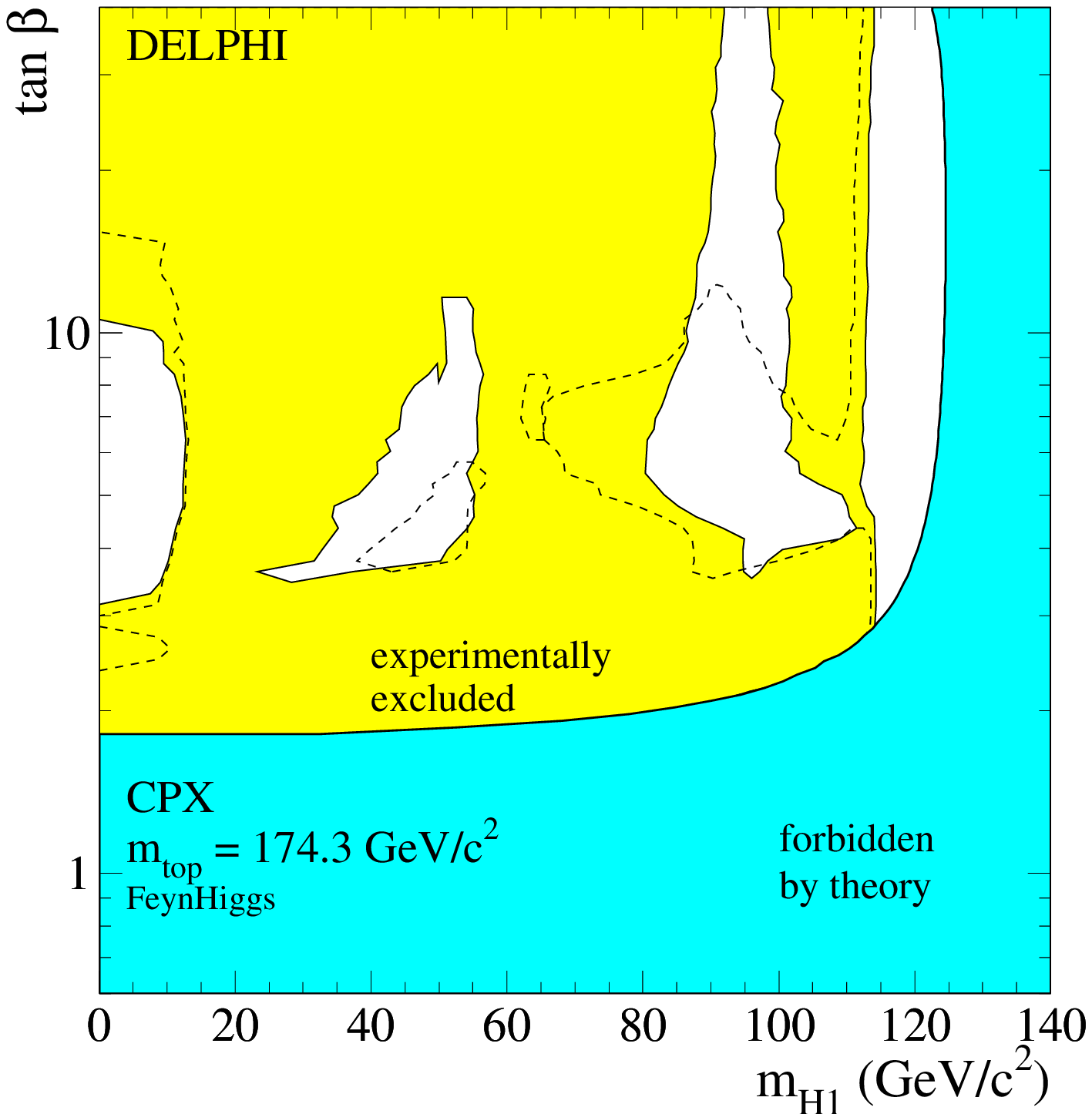,height=8cm} \\
\hspace{-1.cm}
\epsfig{figure=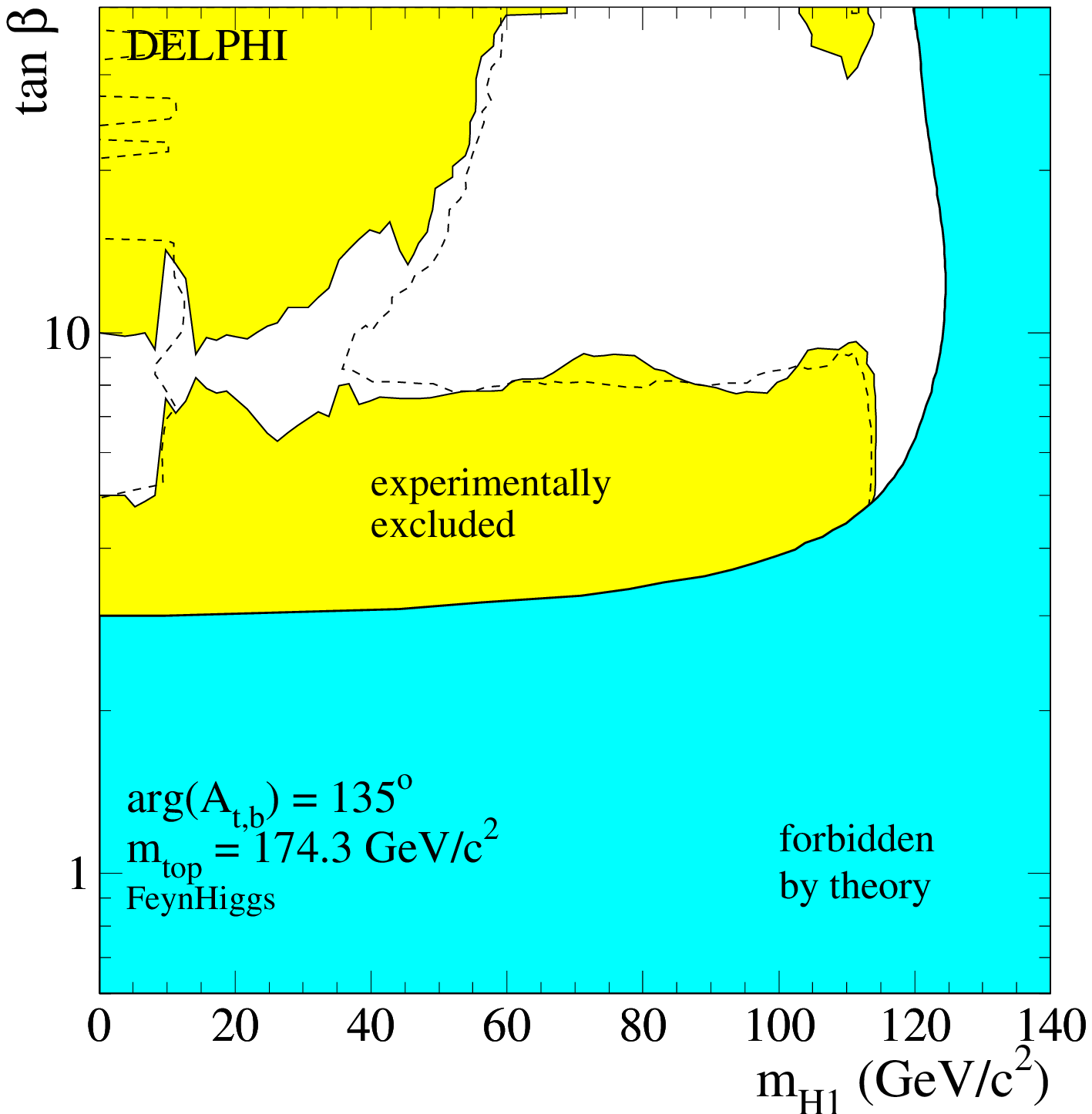,height=8cm} &
\hspace{-0.8cm}
\epsfig{figure=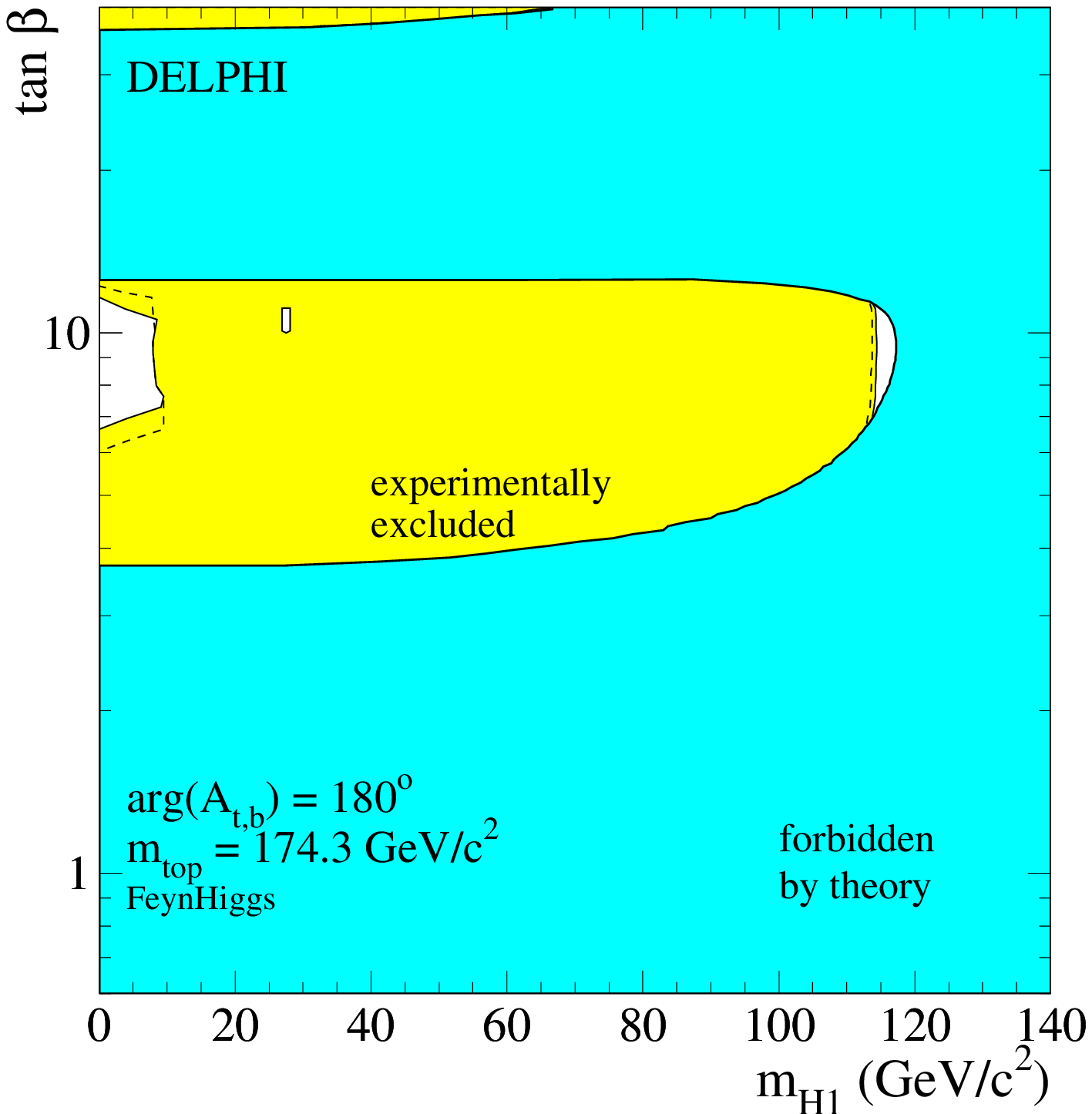,height=8cm} 
\end{tabular}
\caption[]{
    {\sc CP}-violating {\sc MSSM} scenarios with corrections 
    as in Ref.~\cite{ref:FDradco_new} for different values of the phases: 
    regions excluded at 
    95\% {\sc CL} by combining the results of the neutral Higgs 
    boson searches in the whole {\sc DELPHI} data sample  
    (light-grey). 
   The dashed curves show the median expected limits.
   The medium-grey areas are the regions not allowed by theory.
   The {\sc CPX} scenario corresponds to phases of 90$^{\circ}$.
   }
\label{fig:limit_phases_fey}
\end{center}
\end{figure}

The excluded regions in the (\mHu, \tbeta) plane for the {\sc CPX}
scenario and its variants with different phase values are
presented in Fig.~\ref{fig:limit_phases_sub} for the renormalization
group approach~\cite{ref:RGradco} and in Fig.~\ref{fig:limit_phases_fey} 
for the Feynman-diagrammatic calculations~\cite{ref:FDradco_new}. 
The top mass value is 174.3~\GeVcc\ in all plots.


Going from 0$^{\circ}$ to 180$^{\circ}$, the excluded domain varies
significantly. The qualitative trend, valid in the two theoretical
approaches, is as follows.
The extreme values (0$^{\circ}$ and 180$^{\circ}$) correspond to
scenarios with no {\sc CP} violation, and hence to a large 
excluded region.
Moreover, at  180$^{\circ}$, the theoretically allowed
region is reduced, especially at large \tbeta\ due
to unphysical values of the Higgs-bottom Yukawa coupling.
At phases between 60$^{\circ}$ and 135$^{\circ}$,
losses in sensitivity are observed
at large \tbeta\ and \mHu\ above 50~\GeVcc, as well as 
in the intermediate \tbeta\ range for \mHu\ below 60~\GeVcc.
This is the consequence of the strong suppression of the 
H$_1$ZZ coupling due to {\sc CP}-violation, as already encountered 
in Fig.~\ref{fig:cpv_neutral} in the case of the {\sc CPX}
scenario. More generally~\cite{ref:cpv_mssm}, 
the H$_1$ZZ coupling decreases slowly (by a few tens of \%) 
with phases below about 75$^{\circ}$ and is strongly
suppressed (by three to four orders of magnitude) for phases around 
90$^{\circ}$. For phases above 100$^{\circ}$, 
the coupling is partially restored, mostly at low \tbeta.
This explains the evolution of the upper bound of the experimentally
excluded area as a function of phases in 
Figs.~\ref{fig:limit_phases_sub} and~\ref{fig:limit_phases_fey}. 
The changes are moderate for phases up to 60$^{\circ}$ and 
significant for the 90$^{\circ}$ and 135$^{\circ}$ phases, 
where the experimental
sensitivity relies mainly on the \HdZ\ process at low \tbeta\
and on the \HuHd\ production at large \tbeta, both giving large
signals at moderate \mHu\ only, typically below 60~\GeVcc\
(see Fig.~\ref{fig:cpv_neutral}). 

At the 90$^{\circ}$ and 135$^{\circ}$ phases, there are also unexcluded 
areas at masses lower than 60~\GeVcc\ in the intermediate \tbeta\ range, 
between about 4 and 16. These are related to 
weakened sensitivities in the \HdZ\ or \HuHd\ searches.
To take the {\sc CPX} scenario as an example,
at masses below 15~\GeVcc, the dominant final state
is the (\Hcas)Z channel.
The lack of experimental searches at {\sc LEP2} for such final states
with \mHu\ below the \bbbar\ threshold
(see Tab.~\ref{tab:channels}) explains the unexcluded area
which is observed at these masses, in agreement with the expected 
sensitivity. The largest value of \CLs\ in this region is 52\% in the 
renormalisation group framework and 50\% in the
Feynman-diagrammatic approach.
Still in the {\sc CPX} scenario,
the hole at \mHu\ around 50~\GeVcc\ arises in the region where
the decays \Hcas\ and \Hdbb\ become approximately equal,
leading to a loss of significance of the H$_2$ signals,
as pointed out in Section~\ref{sec:cpv_bench} 
(see Fig.~\ref{fig:cpv_neutral}).
The largest value of \CLs\ in this region is 17\% (expected 4\%) in the 
renormalisation group framework and 37\% (expected 11\%) in the
Feynman-diagrammatic approach. In both frameworks, 
these \CLs\ values are observed at \tbeta$\sim$4,  
\mHu$\sim$50~\GeVcc\ and \mHd$\sim$105/107~\GeVcc.
The observed exclusion in this region is weaker than expected,
which is due to a slight excess of data over the expected 
background. The value of 1-\CLb\ at the point of weakest 
exclusion is indeed 15\% (corresponding to a 1.4 sigma deviation)
in the renormalisation group framework  
and 12\% (1.5 sigma deviation) in the Feynman-diagrammatic approach.
Conversely, the largest deviation in the whole hole 
has a value of 1-\CLb\
of 3.3\% (2.1 sigma deviation) in the two approaches. This value
is observed at \tbeta$\sim$16, \mHu$\sim$45~\GeVcc\ and 
\mHd$\sim$107~\GeVcc\ in the renormalisation group framework
and \tbeta$\sim$11, \mHu$\sim$52~\GeVcc\ and 
\mHd$\sim$111~\GeVcc\ in the Feynman-diagrammatic approach.
At this point, the \CLs\ values are 5.4\% (expected 0.1\%)
and  6.5\% (expected 0.2\%) in the two frameworks, respectively.
The combined {\sc LEP} data show also deviations in this
region~\cite{ref:hwg}.

Finally, differences between the two theoretical frameworks are visible 
mainly at large \tbeta, where the Feynman-diagrammatic
calculations predict significantly higher \HuZ\ residual cross-sections
(e.g. a factor about 4 in the {\sc CPX} scenario
for \mHu\ between 40 and 80~\GeVcc),
leading to a better experimental sensitivity. 
Differences in the phase dependence of the results 
are also visible, which are likely to reflect the 
different phase treatment between the two calculations.

\subsection{Dependence on $\mu$ and $M_{\rm susy}$}

\begin{figure}[htbp]
\vspace{-1.7cm}
\begin{center}
\begin{tabular}{cc}
\hspace{-1.cm}
\epsfig{figure=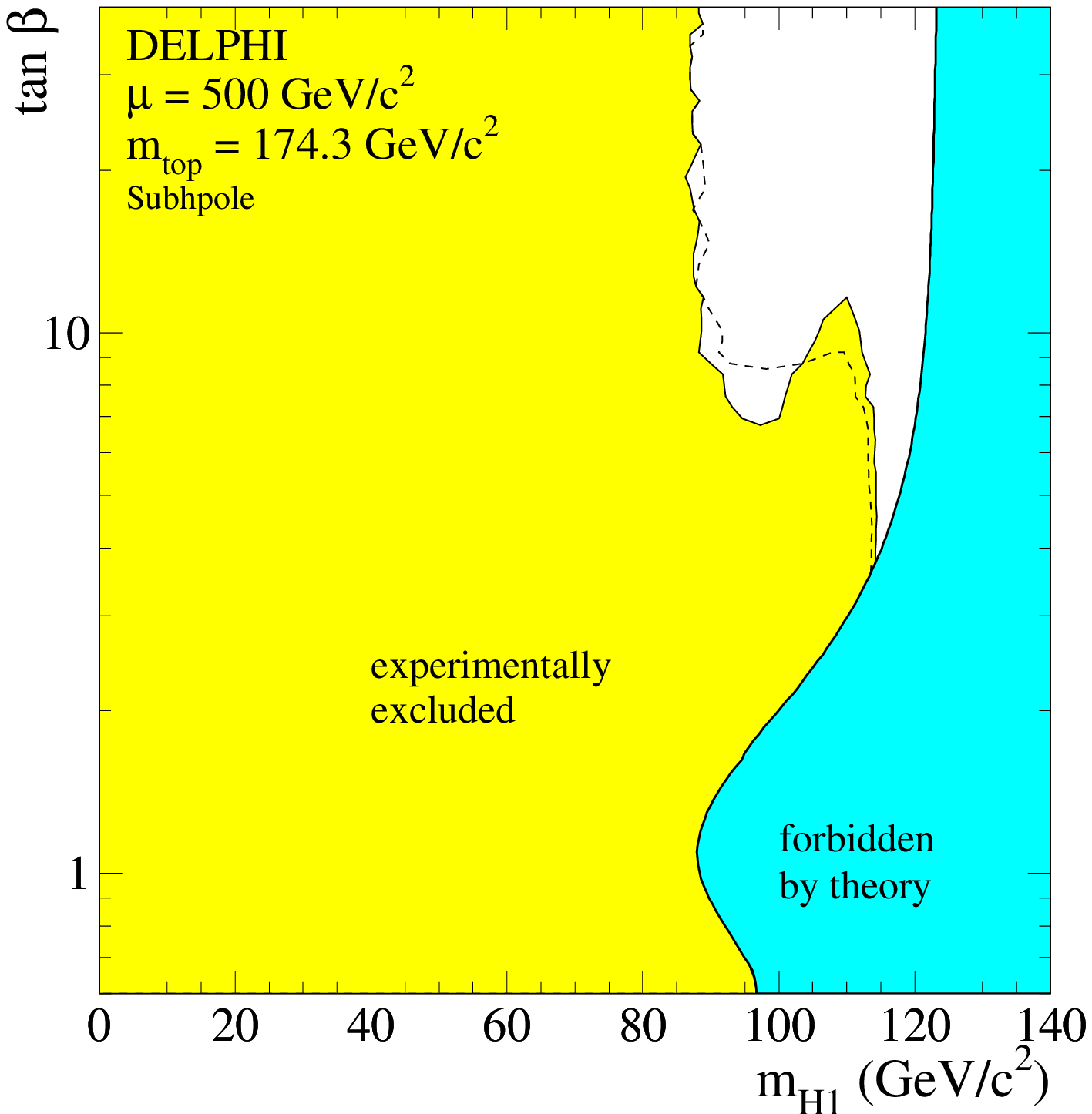,height=8cm} &
\hspace{-0.8cm}
\epsfig{figure=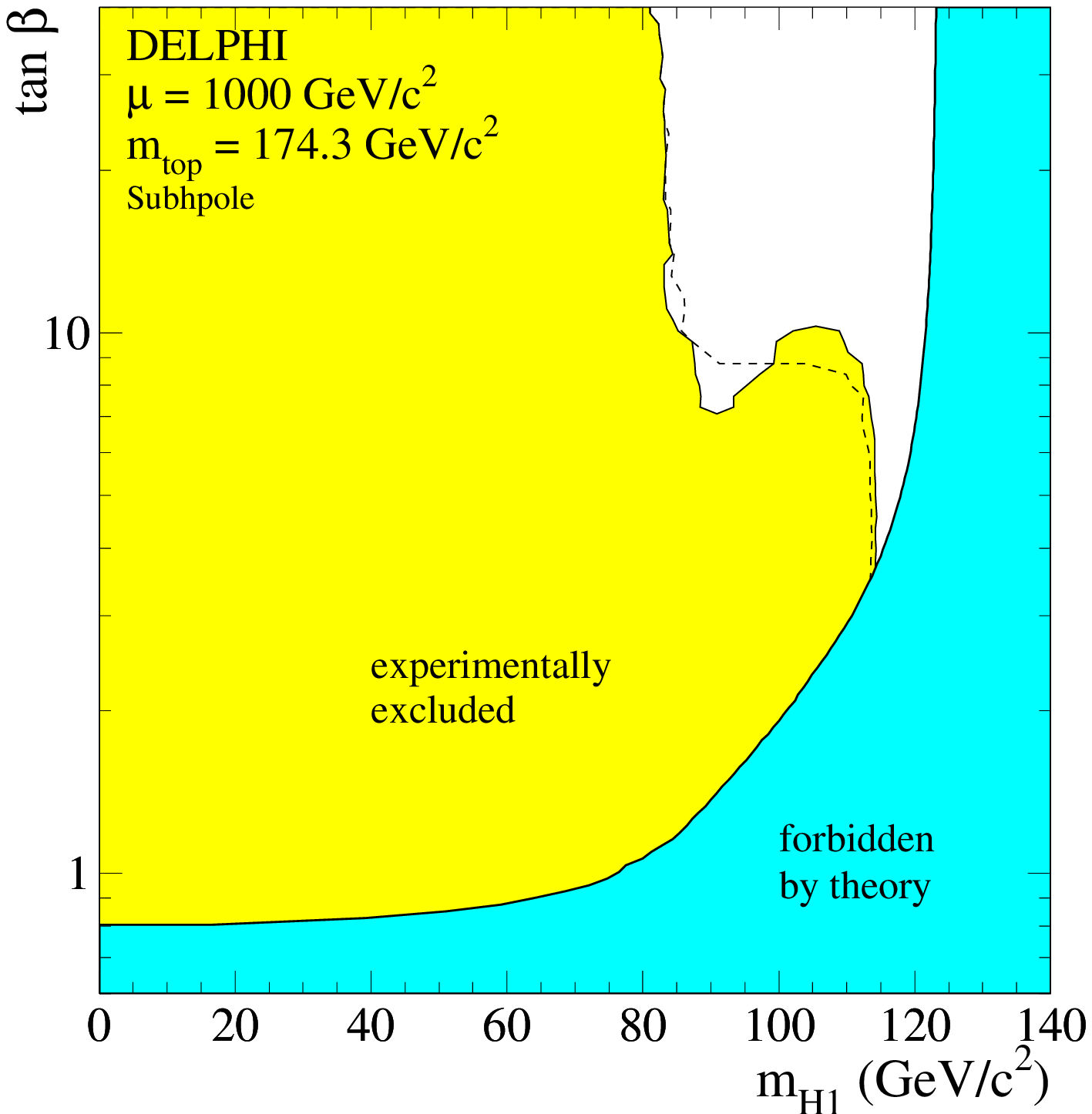,height=8cm} \\
\hspace{-1.cm}
\epsfig{figure=cpv_90_174_tb_mh1_sub.eps,height=8cm} &
\hspace{-0.8cm}
\epsfig{figure=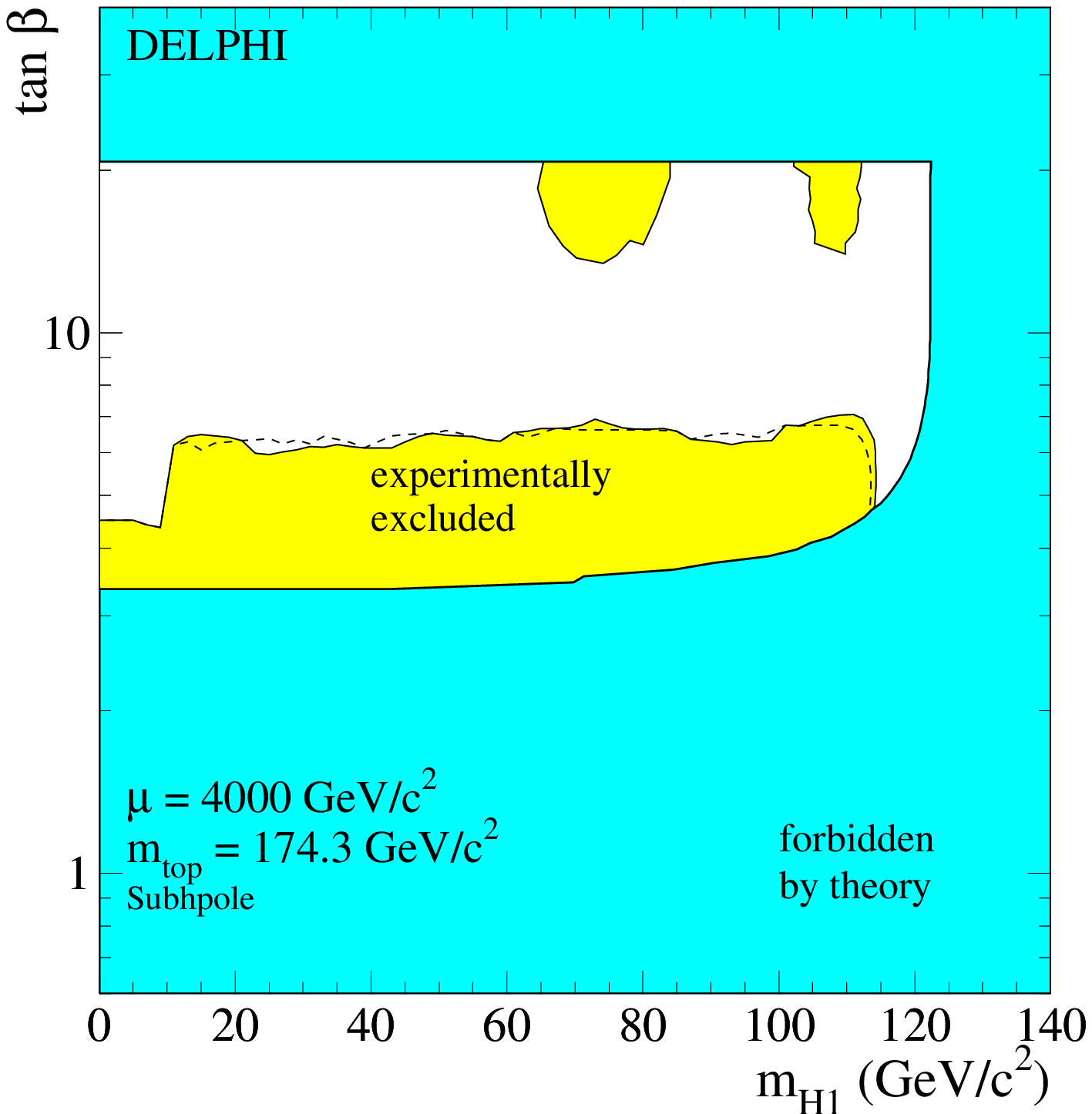,height=8cm} \\
\hspace{-1.cm} 
\epsfig{figure=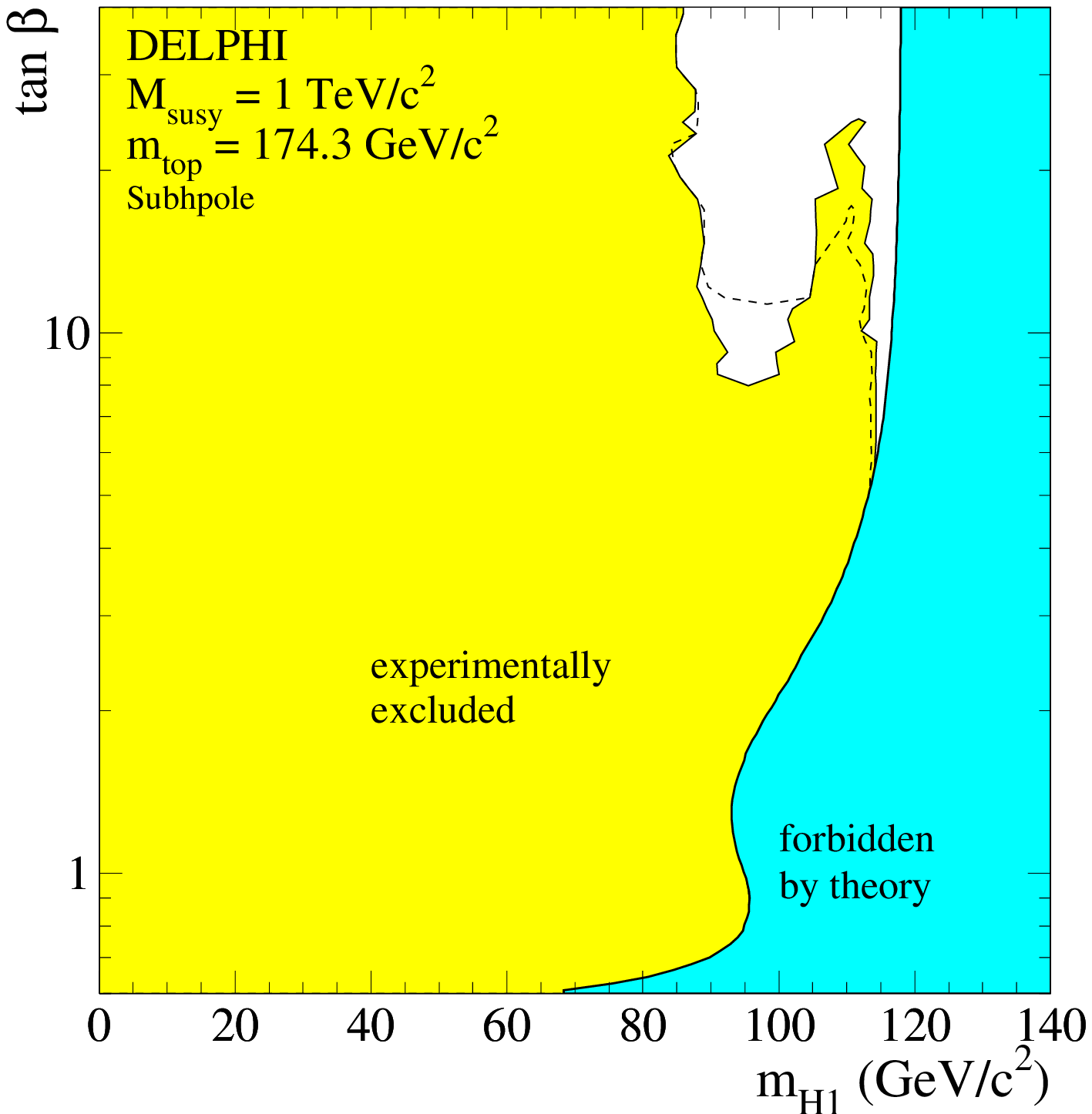,height=8cm} &
\hspace{-0.8cm}
\epsfig{figure=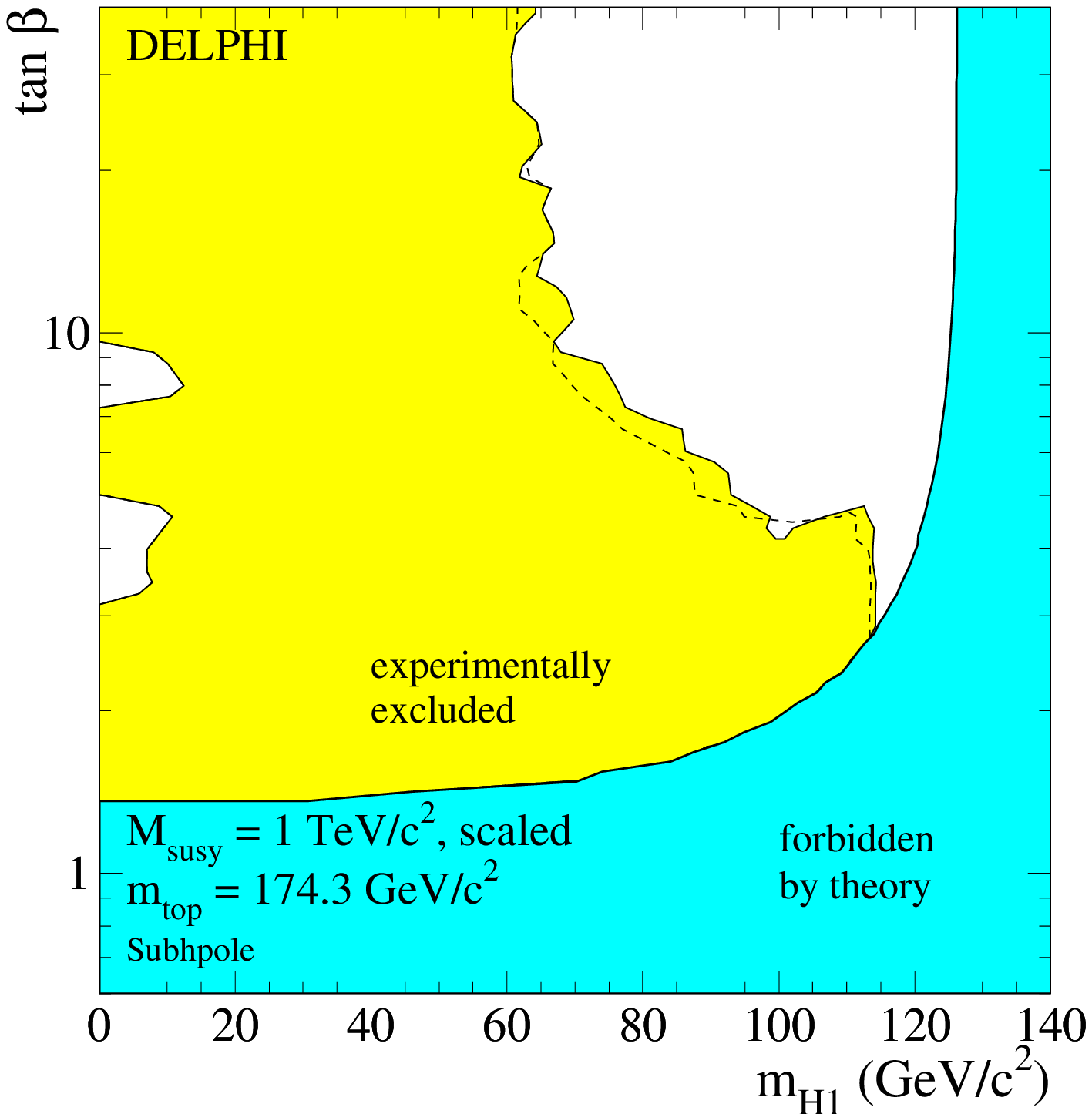,height=8cm}
\end{tabular}
\caption[]{
    {\sc CP}-violating {\sc MSSM} scenarios  with corrections 
    as in Ref.~\cite{ref:RGradco} for different values of 
    $\mu$ and $M_{\rm susy}$: regions excluded at 
    95\% {\sc CL} by combining the results of the neutral Higgs 
    boson searches in the whole {\sc DELPHI} data sample  
    (light-grey). 
   The dashed curves show the median expected limits.
   The medium-grey areas are the regions not allowed by theory.
   The {\sc CPX} scenario corresponds to $\mu$~=~2000~\GeVcc\ and
   $M_{\rm susy}$~=~500~\GeVcc.
   }
\label{fig:limit_mu_sub}
\end{center}
\end{figure}

\begin{figure}[htbp]
\vspace{-1.7cm}
\begin{center}
\begin{tabular}{cc}
\hspace{-1.cm}
\epsfig{figure=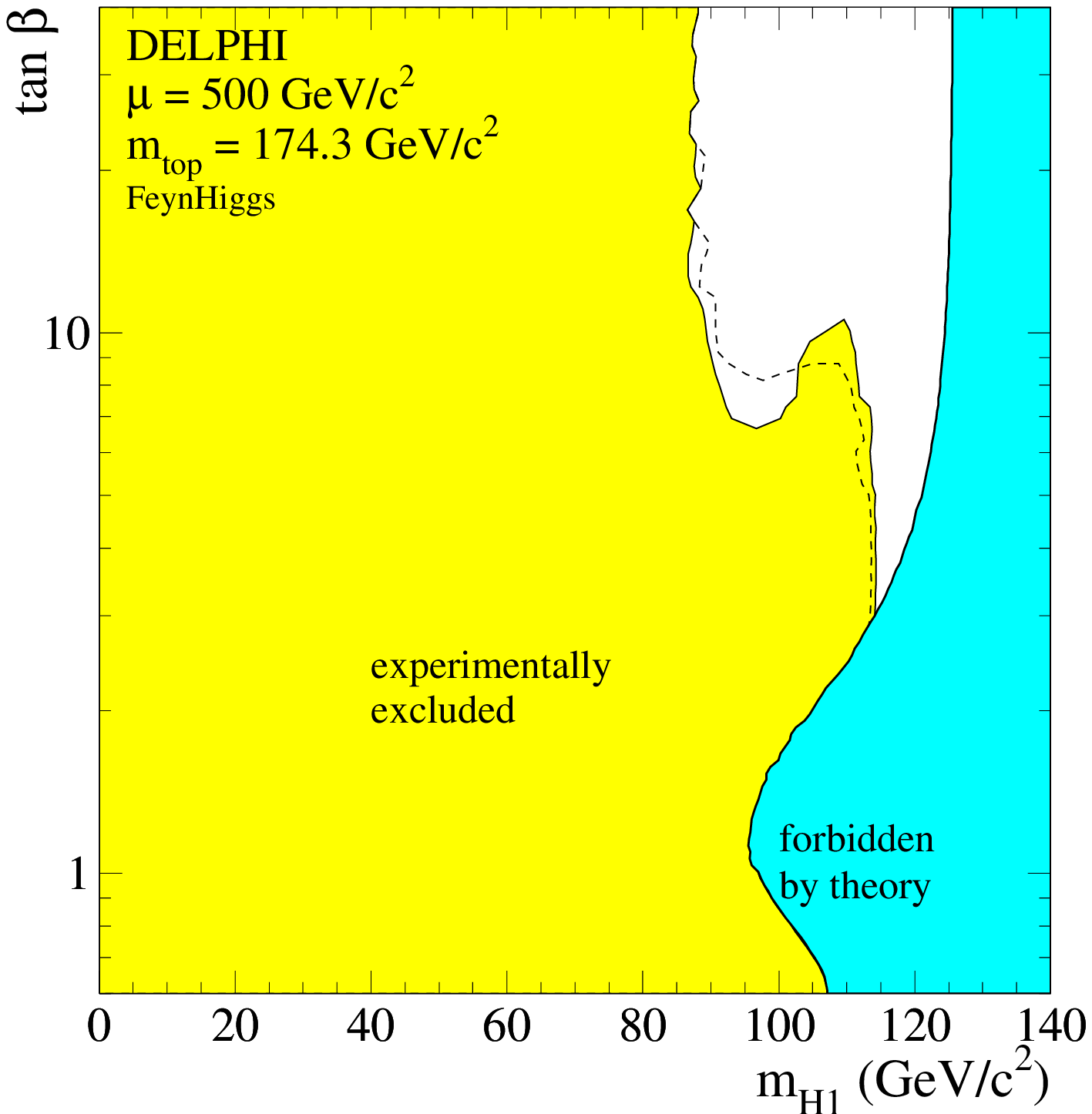,height=8cm} &
\hspace{-0.8cm}
\epsfig{figure=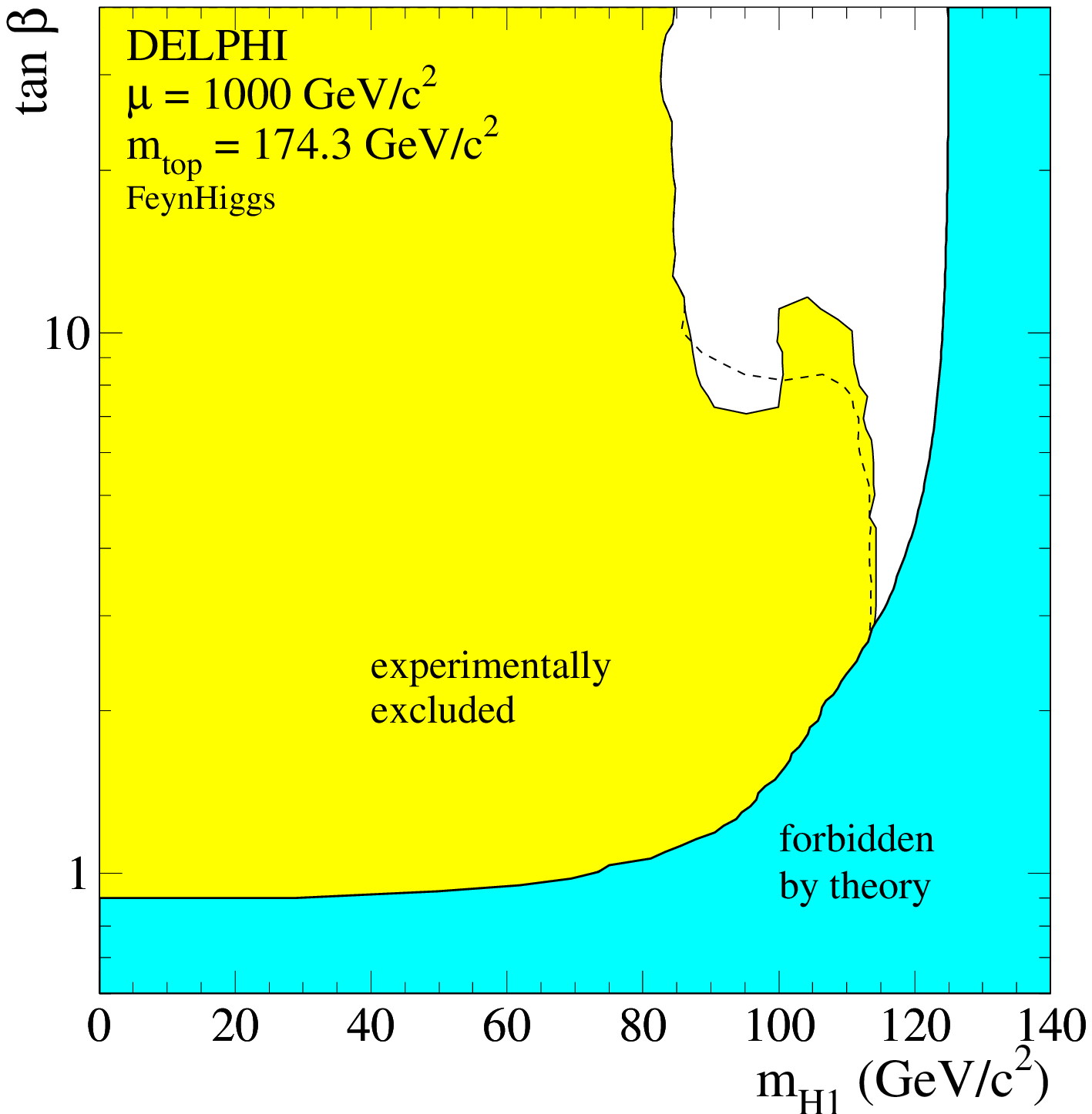,height=8cm} \\
\hspace{-1.cm}
\epsfig{figure=cpv_90_174_tb_mh1_fey.eps,height=8cm} &
\hspace{-0.8cm}
\epsfig{figure=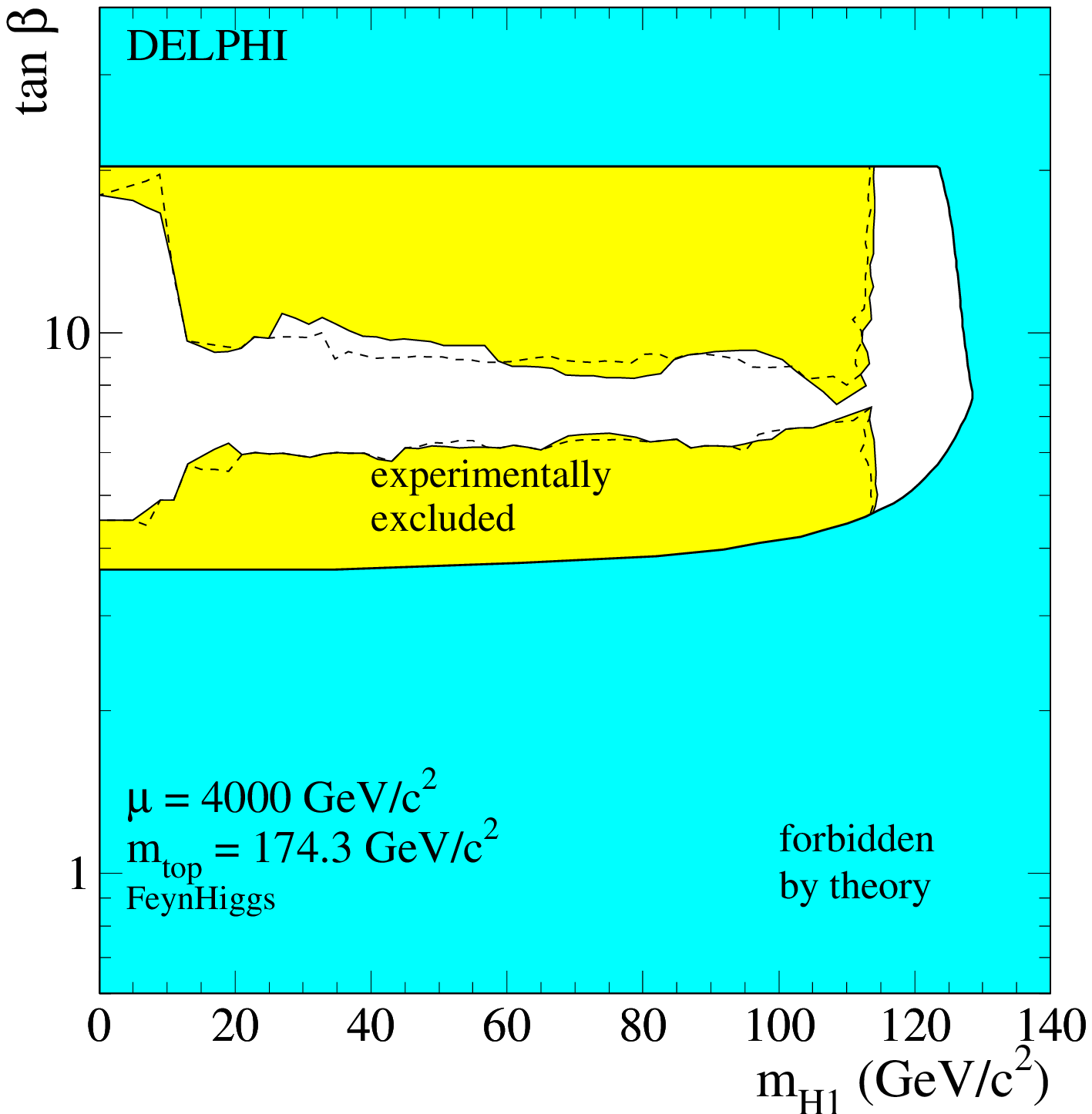,height=8cm} \\
\hspace{-1.cm} 
\epsfig{figure=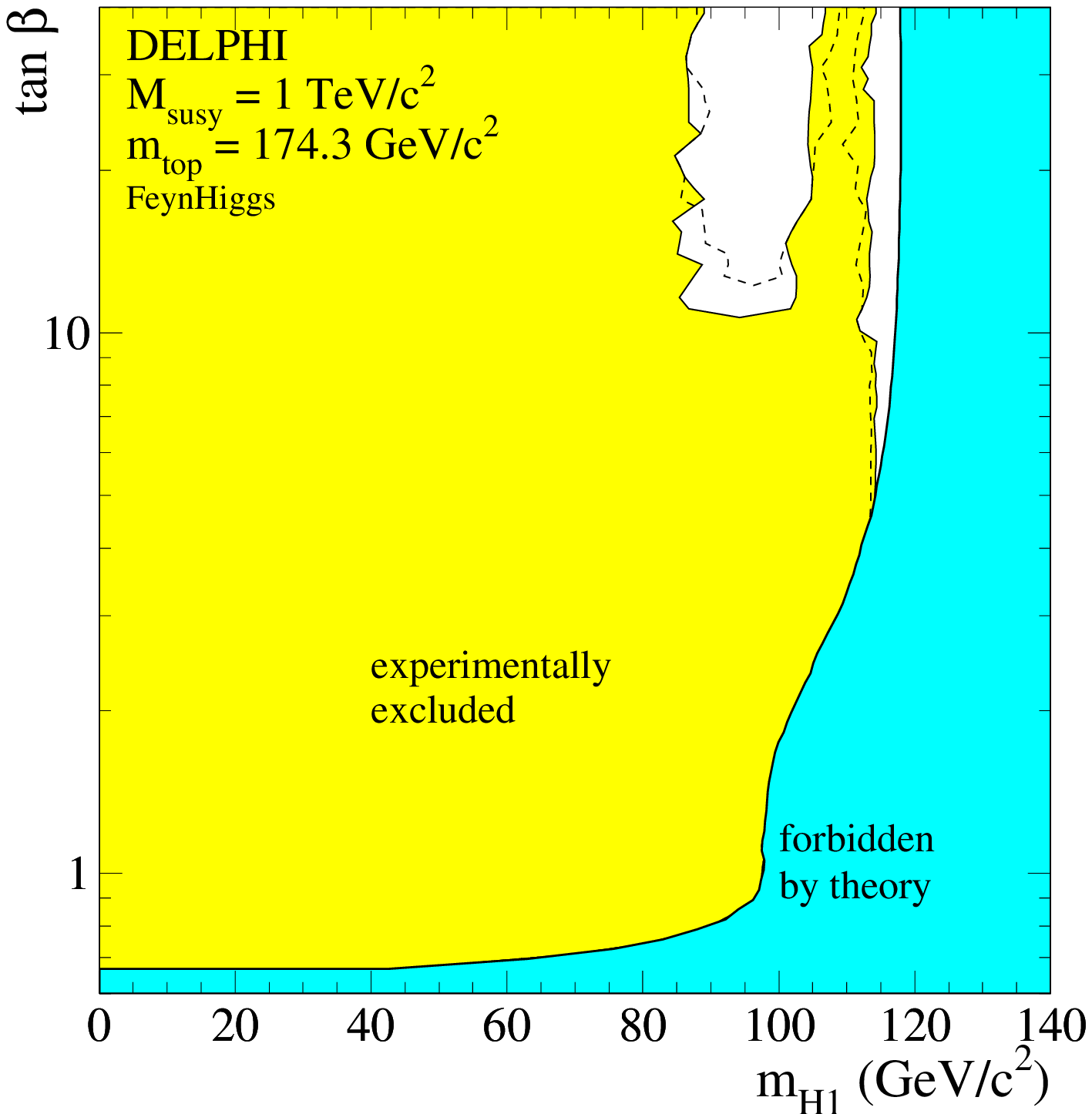,height=8cm} &
\hspace{-0.8cm}
\epsfig{figure=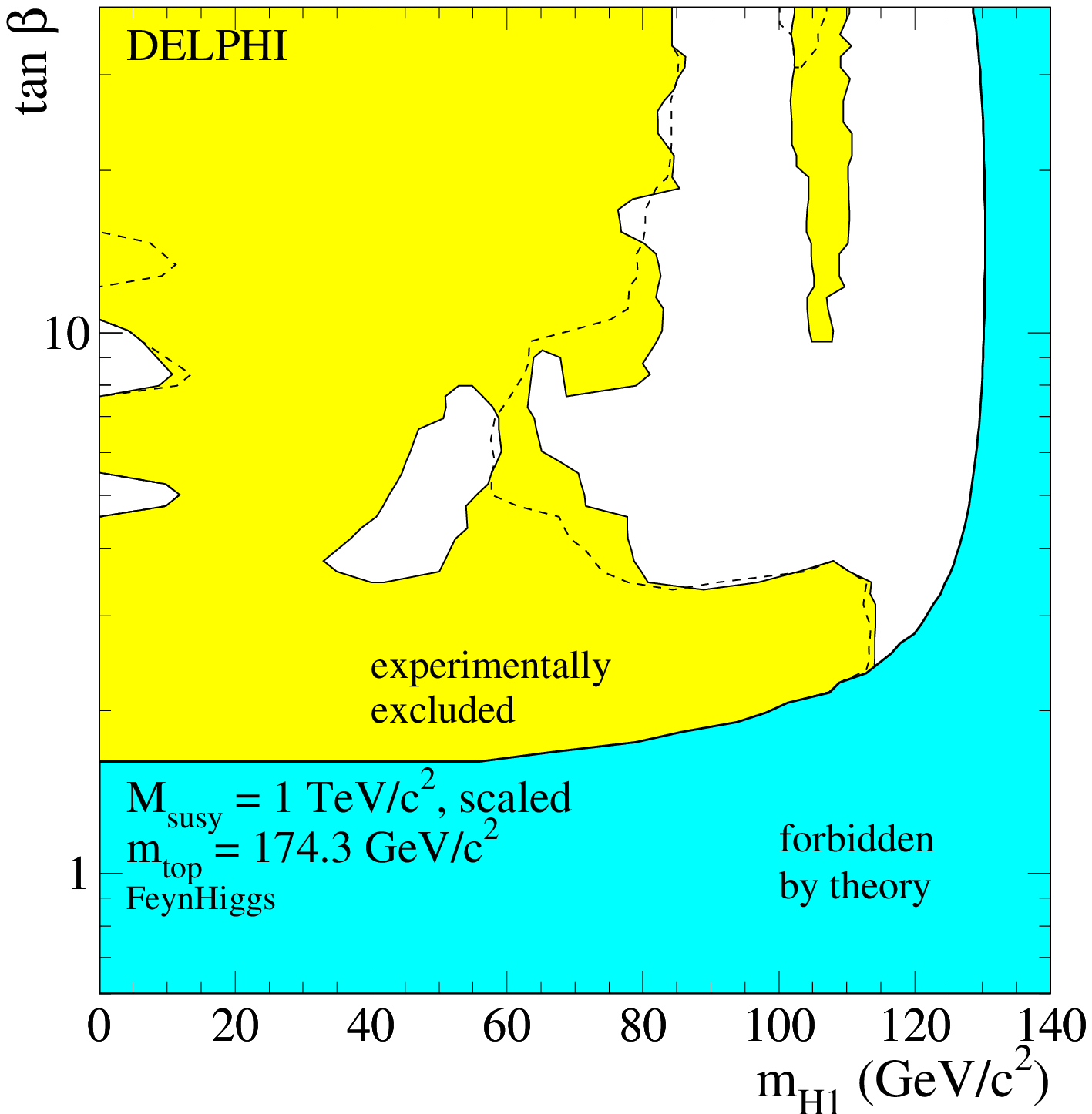,height=8cm}
\end{tabular}
\caption[]{
    {\sc CP}-violating {\sc MSSM} scenarios  with corrections 
    as in Ref.~\cite{ref:FDradco_new}
    for different values of $\mu$ and $M_{\rm susy}$: 
    regions excluded at 
    95\% {\sc CL} by combining the results of the neutral Higgs 
    boson searches in the whole {\sc DELPHI} data sample  
    (light-grey). 
   The dashed curves show the median expected limits.
   The medium-grey areas are the regions not allowed by theory.
   The {\sc CPX} scenario corresponds to $\mu$~=~2000~\GeVcc\ and
   $M_{\rm susy}$~=~500~\GeVcc.
   }
\label{fig:limit_mu_fey}
\end{center}
\end{figure}

The excluded regions in the (\mHu, \tbeta) plane for the {\sc CPX}
scenario and its variants with different values of 
$\mu$ and $M_{\rm susy}$ are
presented in Fig.~\ref{fig:limit_mu_sub} for the renormalization
group approach~\cite{ref:RGradco} and in Fig.~\ref{fig:limit_mu_fey} 
for the Feynman-diagrammatic calculations~\cite{ref:FDradco_new}. 
In all plots, the common {\sc CP}-violating phase is 90$^{\circ}$ and
the top mass value is 174.3~\GeVcc.

The dependence of the results on the value of $\mu$ 
is as expected from the scaling of the
dominant {\sc CP}-violating effects with $Im(\mu A)$
(see Section~\ref{sec:cpv_bench}).
The exclusion is almost entirely 
restored for values of $\mu$ lower than 2~\TeVcc,
the value in the {\sc CPX} scenario, and gets weaker at 4~\TeVcc.
In the first two variants, despite the {\sc CP}-violating phase being at
90$^{\circ}$, there are always two production processes
with significant rates in every point of the kinematically accessible 
parameter space. 
In the variant at 4~\TeVcc, due to the large value of $\mu$ and the
{\sc CP}-violating phase at 90$^{\circ}$, 
the \HuHd\ and \HdZ\ processes are suppressed for all values of \tbeta, 
as well as the \HuZ\ process at intermediate and large \tbeta\ values.
In the Feynman-diagrammatic approach, the \HuZ\ cross-sections are
partly restored at large \tbeta\ which explains the difference
between the results in the two theoretical frameworks in that region.
Note also that the theoretically allowed region is much reduced at large 
\tbeta\ in this scenario due to unphysical values of the bottom Yukawa
coupling.

The dependence on the value of $M_{\rm susy}$ is presented in the two
bottom plots of Figs.~\ref{fig:limit_mu_sub} and~\ref{fig:limit_mu_fey}.
The first scenario corresponds to setting $M_{\rm susy}$ at 1~\TeVcc,
twice its value in the {\sc CPX} scenario. As the dominant 
{\sc CP}-violating effects are proportional to $M_{\rm susy}^{-2}$,
the exclusion is restored in this variant. The reason is as in
the case of the two variants with low values of $\mu$, i.e.
there are always two production processes
with significant rates in every point of the kinematically accessible 
parameter space. 
In the second scenario,  $M_{\rm susy}$ is still set at  1~\TeVcc\ but
the values of $|m_{\tilde{g}}|$, $\mu$ and $|A|$ are also scaled
by a factor 2, leaving the {\sc CP}-violating effects almost unchanged
(see Section~\ref{sec:cpv_bench}) with respect to the {\sc CPX} scenario. 
This explains why the exclusion region 
in this variant is close to that in the {\sc CPX} scenario.
The few differences between the excluded regions in these two scenarios 
are due to different cross-sections for some of the processes
which contribute most to the experimental sensitivity, that is the
\HuHd\ and \HdZ\ processes at masses below 60~\GeVcc, 
and the \HuZ\ process at higher masses. 
As an example, the better
coverage of the low mass region at intermediate \tbeta\ values
in the scaled variant is explained by slightly higher \HuHd\ cross-sections.

\subsection{Dependence on \bf{\mtop}}

\begin{figure}[htbp]
\begin{center}
\begin{tabular}{cc}
\hspace{-1.2cm}
\epsfig{figure=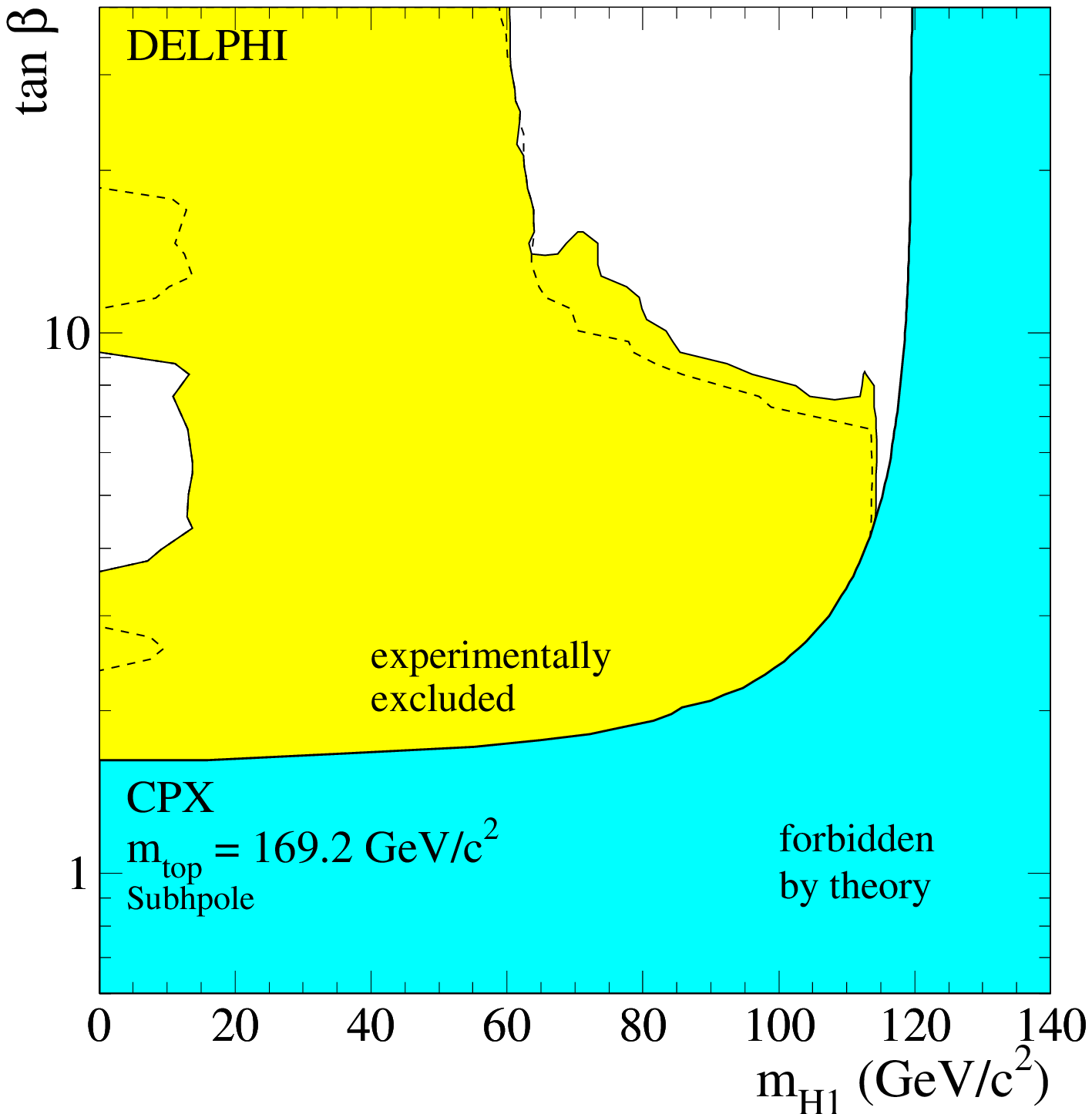,height=9cm} &
\hspace{-0.8cm}
\epsfig{figure=cpv_90_174_tb_mh1_sub.eps,height=9cm} \\
\hspace{-1.2cm}
\epsfig{figure=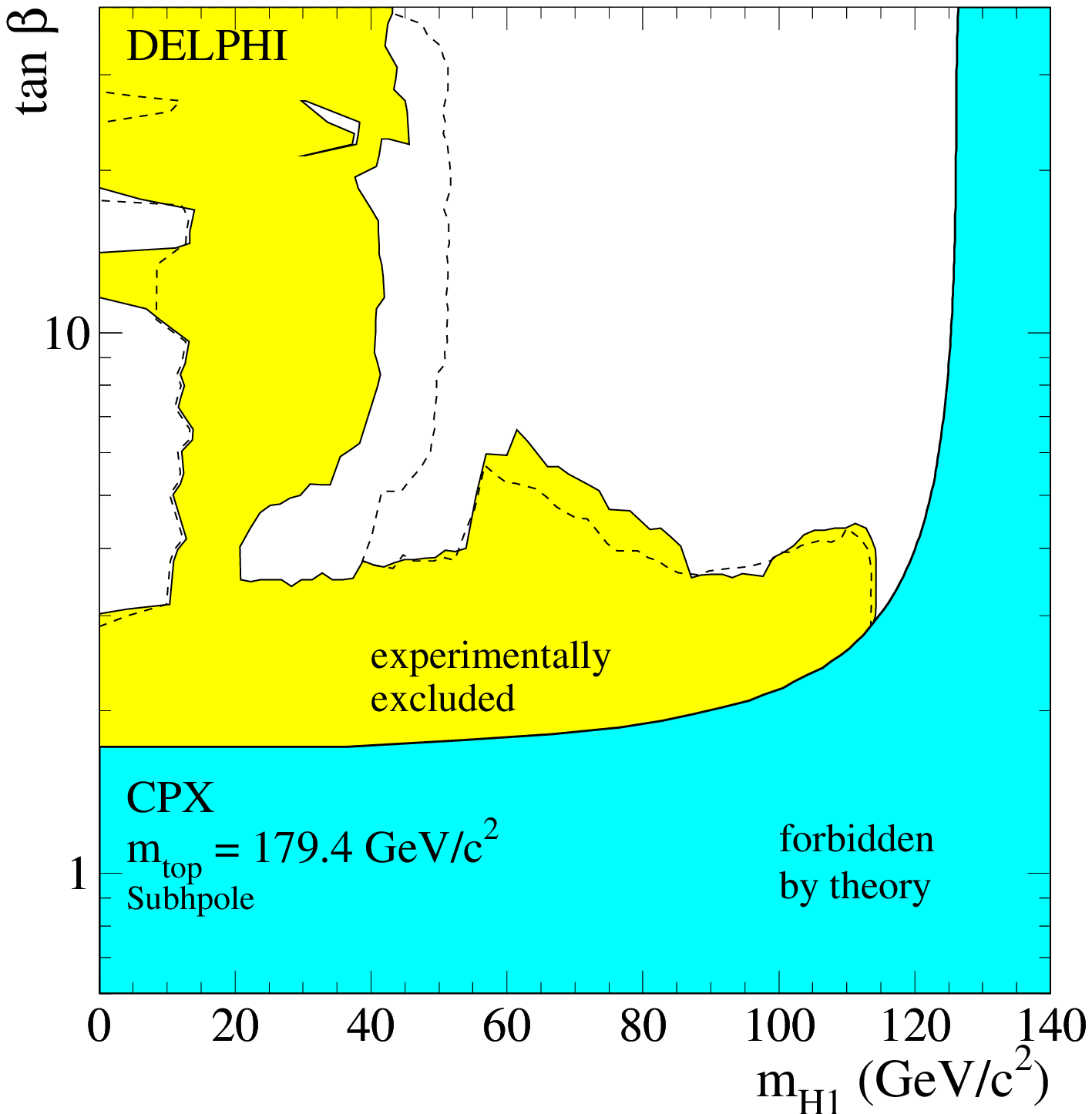,height=9cm} &
\hspace{-0.8cm}
\epsfig{figure=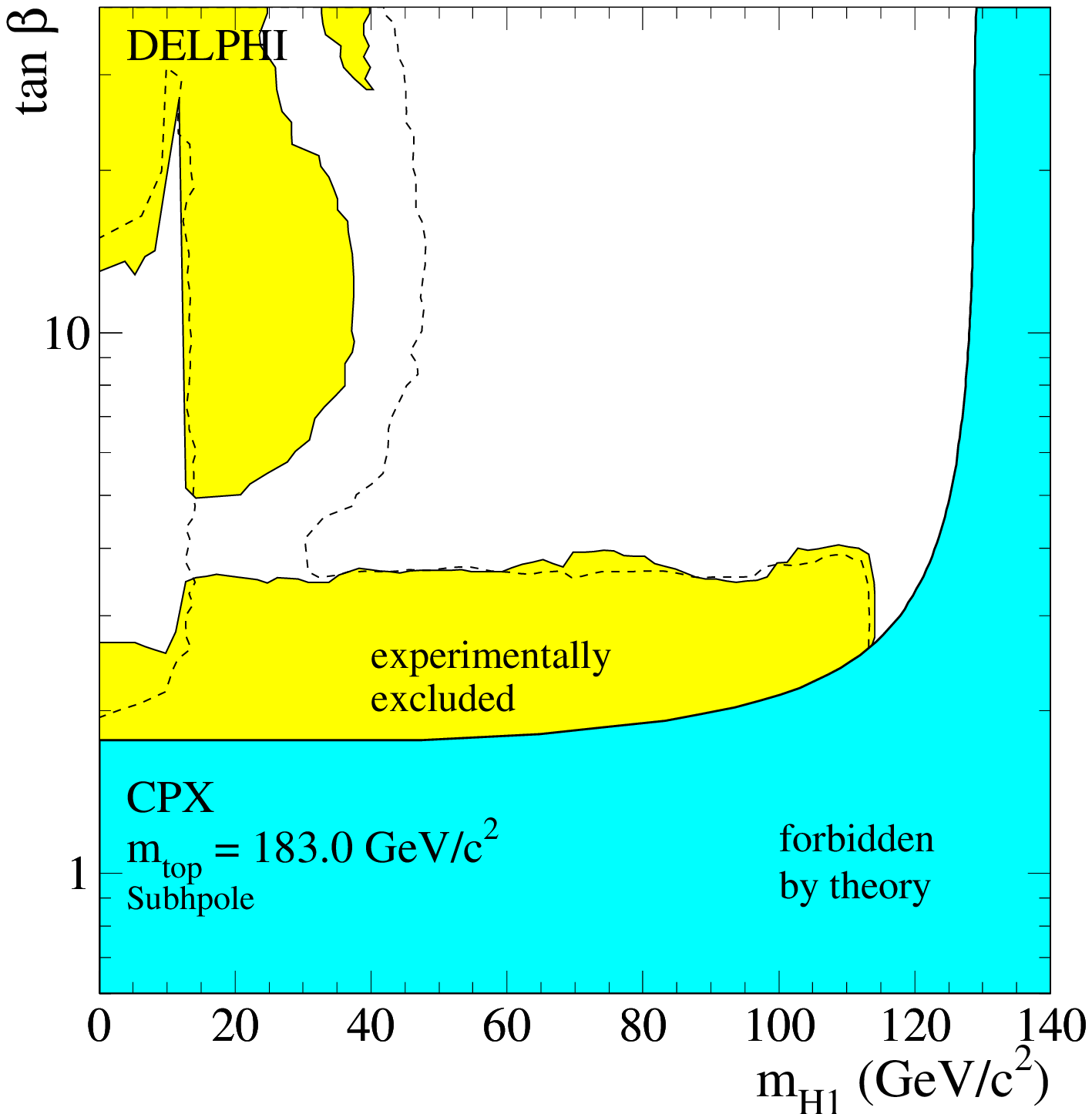,height=9cm} \\
\hspace{-1.cm}
\end{tabular}
\caption[]{
    {\sc CP}-violating {\sc MSSM} scenarios with corrections 
    as in Ref.~\cite{ref:RGradco} for different values of \mtop: 
    regions excluded at 
    95\% {\sc CL} by combining the results of the neutral Higgs 
    boson searches in the whole {\sc DELPHI} data sample  
    (light-grey). 
   The dashed curves show the median expected limits.
   The medium-grey areas are the regions not allowed by theory.
   }
\label{fig:limit_mtop_sub}
\end{center}
\end{figure}

\begin{figure}[htbp]
\begin{center}
\begin{tabular}{cc}
\hspace{-1.2cm}
\epsfig{figure=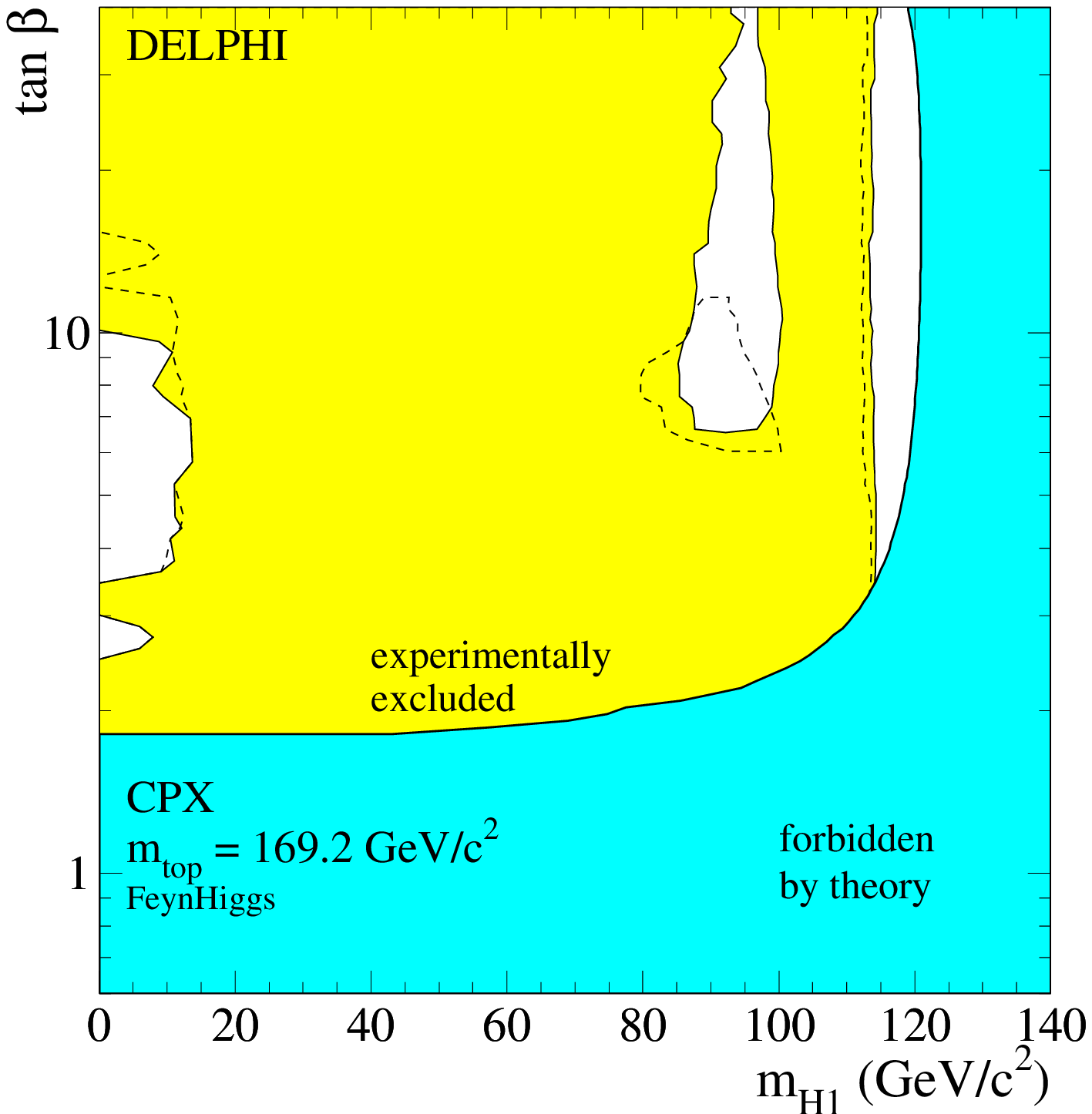,height=9cm} &
\hspace{-0.8cm}
\epsfig{figure=cpv_90_174_tb_mh1_fey.eps,height=9cm} \\
\hspace{-1.2cm}
\epsfig{figure=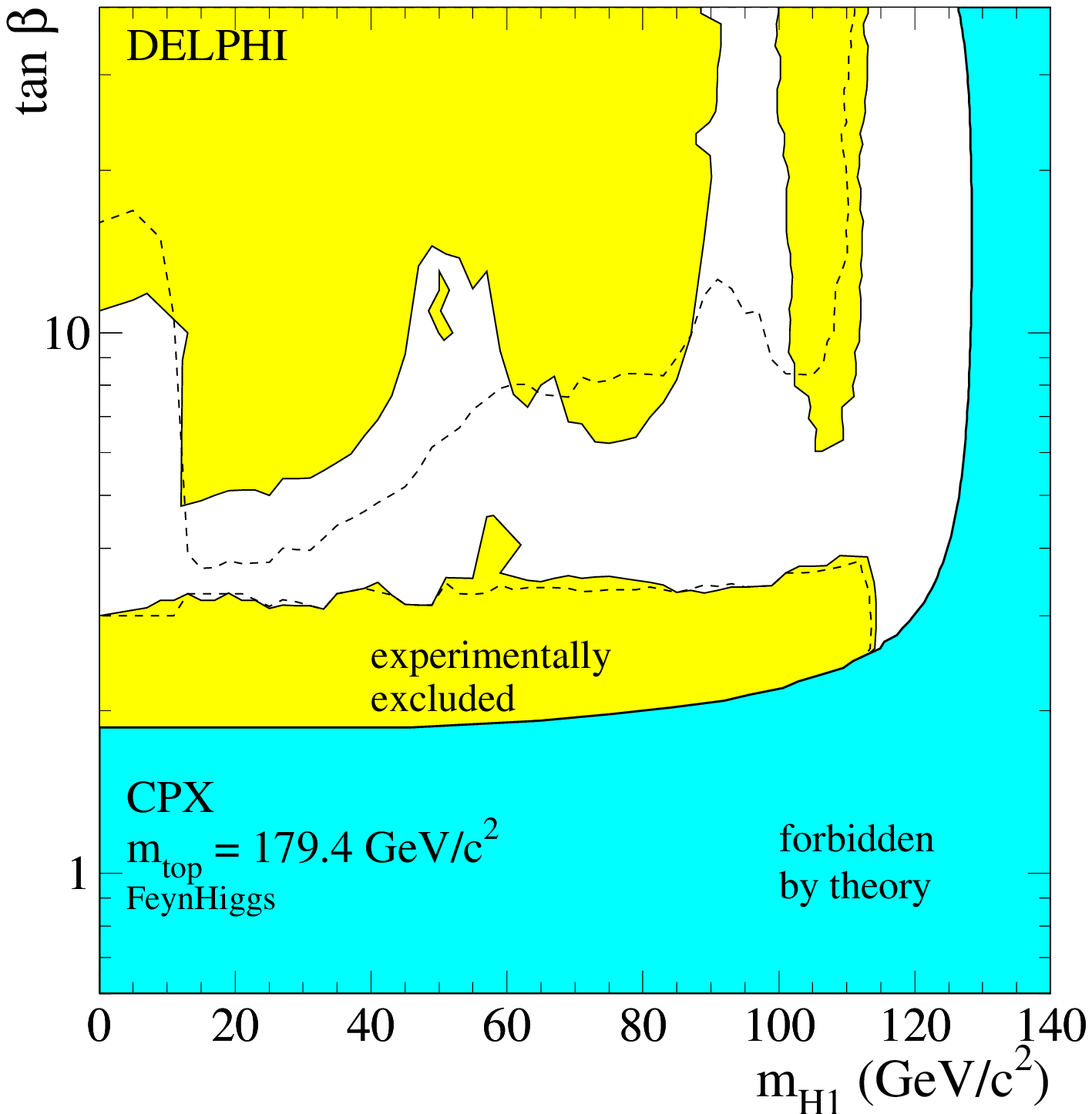,height=9cm} &
\hspace{-0.8cm}
\epsfig{figure=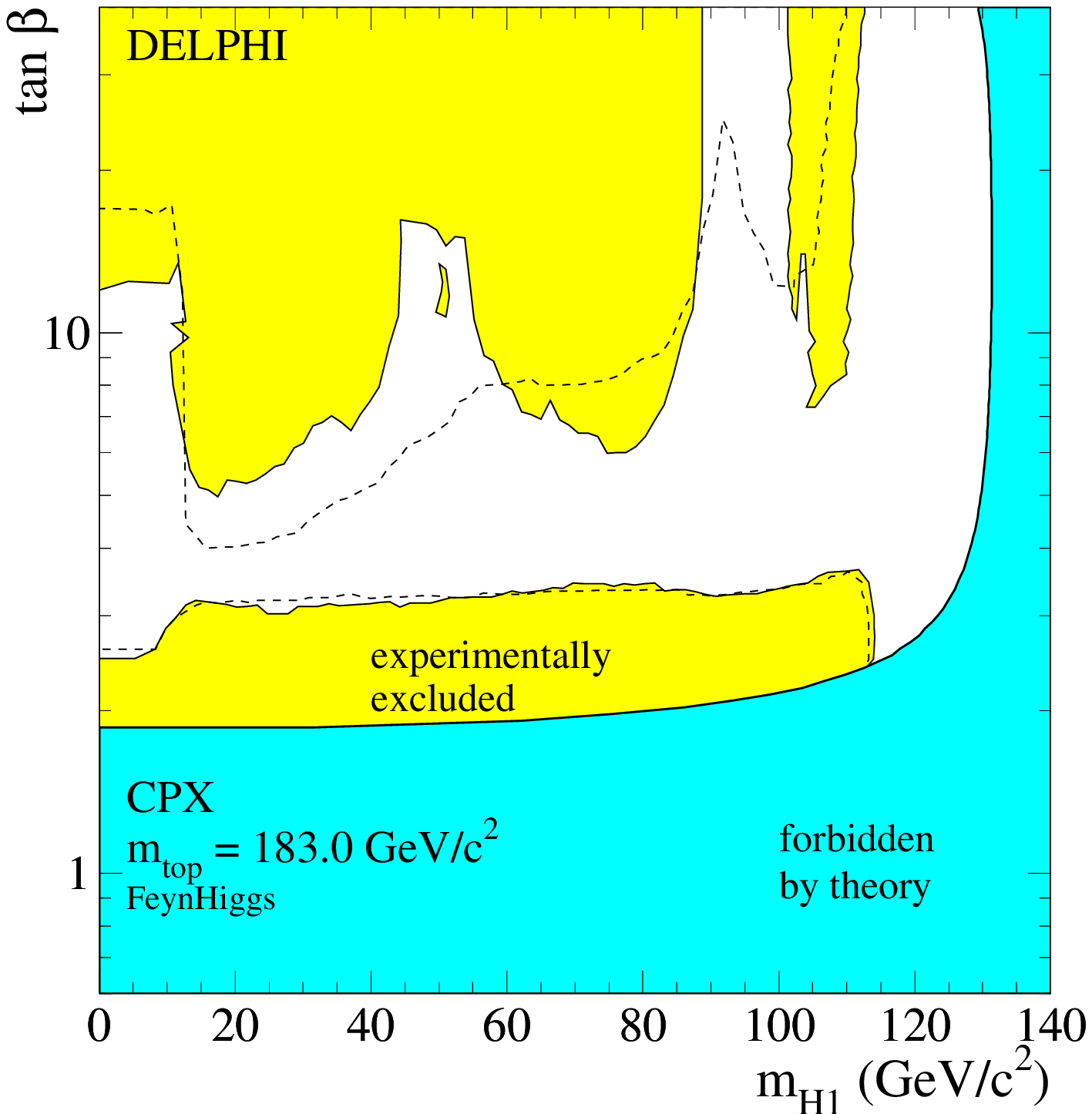,height=9cm} \\
\hspace{-1.cm}
\end{tabular}
\caption[]{
    {\sc CP}-violating {\sc MSSM} scenarios  with corrections 
    as in Ref.~\cite{ref:FDradco_new} 
    for different values of \mtop: regions excluded at 
    95\% {\sc CL} by combining the results of the neutral Higgs 
    boson searches in the whole {\sc DELPHI} data sample  
    (light-grey). 
   The dashed curves show the median expected limits.
   The medium-grey areas are the regions not allowed by theory.
   }
\label{fig:limit_mtop_fey}
\end{center}
\end{figure}

The excluded regions in the (\mHu, \tbeta) plane for the {\sc CPX}
scenario with different \mtop\ values are
presented in Fig.~\ref{fig:limit_mtop_sub} for the renormalization
group approach and in Fig.~\ref{fig:limit_mtop_fey} for the 
Feynman-diagrammatic calculations. 

The results show a strong dependence on the value of \mtop,
as expected since the dominant {\sc CP}-violating effects scale
with m$^4_{\rm{top}}$. In the two theoretical approaches, the exclusion in
the intermediate \tbeta\ range is gradually reduced as \mtop\ increases
and eventually vanishes for \tbeta\ between about 3 and 5
and a top mass of 183~\GeVcc.
This can be traced to the suppression of the \HdZ\ and \HuHd\
cross-sections with increasing values of \mtop, leaving no 
significant rate in any of the three possible production channels. 
At large \tbeta, as \mtop\ increases, the \HuHd\ cross-section is
reduced and the exclusion gets weaker 
in the renormalization group approach while it is almost unchanged
in the Feynman-diagrammatic framework. As already mentioned, this is 
a consequence of the higher \HuZ\ residual cross-sections predicted by
the latter calculations at large \tbeta.

\subsection{Summary}
Scans of the {\sc CPX} scenario and its variants revealed that {\sc CP}
violation in the Higgs sector can have a significant impact on the 
experimental sensitivity of {\sc LEP}. The strong suppression of the
neutral Higgs boson couplings to the Z boson translates into a 
loss of redundancy in the different search channels, and hence leads to
a reduced coverage of the parameter space. The most significant reduction is
observed in the intermediate \tbeta\ region, typically between 3 and 10,
down to the lowest H$_1$ masses. It occurs for phases between 
90$^{\circ}$ and 135$^{\circ}$, 
top mass values equal to 174.3~\GeVcc\ or larger,
and values of the ratio  $|\mu A| / M_{susy}^2 $ equal to 
8 or larger. As a consequence, no absolute mass limits can be derived in
these scenarios. On the other hand, the low \tbeta\ region appears 
still disfavoured, as in the {\sc CP}-conserving models. 
Scans were performed using
two different theoretical approaches for the 
radiative correction calculations. Although the two sets of results show
large differences, they both lead to the same qualitative conclusions.

\section{Conclusions}

Searches for Higgs bosons in 
the whole data sample of the {\sc DELPHI} experiment have been
combined to derive constraints on {\sc MSSM} 
benchmark scenarios, including models with {\sc CP}-violation
in the Higgs sector. 
Experimental results encompass searches
for neutral Higgs bosons in dominant final states expected
in most {\sc MSSM} models, as well as 
searches for charged Higgs bosons and for neutral Higgs bosons 
decaying into hadrons of any flavour,
which bring a gain in sensitivity in restricted regions 
of the parameter space.
An additional improvement is obtained by applying the experimental
results to more production processes than the two expected main channels,
namely the associated production of the lightest Higgs boson with a
Z boson and the pair-production of the two lightest Higgs bosons. 
In the {\sc CP}-conserving {\sc MSSM}, 
the experimental sensitivity at {\sc LEP} relies on the \hZ, \hA\ and \HZ\ 
channels, the last leading to a significant gain in sensitivity in
scenarios where the third neutral Higgs boson, H, is kinematically 
accessible.
In the {\sc CP}-violating {\sc MSSM}, the total signal at {\sc LEP}
is spread mainly over the \HuZ, \HdZ\ and \HuHd\ channels.
Accounting for the simultaneous
production of all possible signals is essential in this type of scenario
where {\sc CP}-violating effects can lead to strong suppression of one
channel or another.

In all  {\sc CP}-conserving scenarios, 
the experimental results allow 
a large fraction of the parameter space to be excluded, 
even in scenarios designed
to test potentially difficult cases (e.g. vanishing 
production cross-sections or decay branching fractions) 
either at {\sc LEP} or at hadron colliders. Limits on masses of
the h and A bosons were deduced as well as upper and lower exclusion
bounds in \tbeta. The dependence of these limits on \mtop\ was
studied in a range between 169.2 to 183.0~\GeVcc.
To quote but one result, the following limits at 95\% of {\sc CL} 
have been established in the framework of the
\mbox{$ m_{\mathrm h}^{\rm max}$} scenario with \mtop~=~174.3~\GeVcc:

\begin{table}[h]
\begin{center}
\begin{tabular}{lcll}  
   \mh $>$ 89.7~\GeVcc  & and & \MA $>$ 90.4~\GeVcc & 
    for any  $\tan\beta$ between 0.4 and 50, \\
   $\tan\beta < 0.72$ & or & $\tan\beta > 1.96 $ &
    for any \mA\ between 0.02 and 1000~\GeVcc. \\
\end{tabular}
\end{center}
\end{table}

\noindent
These mass limits are insensitive to variations of the top quark mass.
The excluded range in \tbeta\ decreases with increasing \mtop\ and
would vanish if \mtop\ was as large as 183.0~\GeVcc.
This scenario provides the most conservative bounds on \tbeta\
among the eight {\sc CP}-conserving scenarios tested.

In the {\sc CP}-violating scenarios, large domains of the kinematically
accessible parameter space remain unexcluded due to strong suppressions
of the couplings between the Z and the Higgs bosons induced by
{\sc CP}-violation. Hence no absolute limits can be set on the Higgs
boson masses in these scenarios.
The unexcluded areas arise in the intermediate
\tbeta\ range, typically between 3 and 10. Their contours vary considerably
with the value of \mtop\ and the {\sc MSSM} parameters which govern
the {\sc CP}-violating effects, $|\mu A|$, $M_{\rm susy}$ and the
phase $arg(A)$. These scenarios have been studied in two
different theoretical frameworks for the radiative correction calculations. 
The impact of {\sc CP}-violation is
observed to be qualitatively the same in the two approaches.



\subsection*{Acknowledgements} \vskip 3 mm

We are greatly indebted to our technical collaborators, to the
members of the CERN-SL Division for the excellent performance of
the LEP collider, and to the funding agencies for their
support in building and operating the DELPHI detector.\\
We acknowledge in particular the support of \\
Austrian Federal Ministry of Education, Science and Culture,
GZ 616.364/2-III/2a/98, \\
FNRS--FWO, Flanders Institute to encourage scientific and technological
research in the industry (IWT) and Belgian Federal Office for Scientific,
Technical and Cultural affairs (OSTC), Belgium, \\
FINEP, CNPq, CAPES, FUJB and FAPERJ, Brazil, \\
Czech Ministry of Industry and Trade, GA CR 202/99/1362,\\
Commission of the European Communities (DG XII), \\
Direction des Sciences de la Mati$\grave{\mbox{\rm e}}$re, CEA, France, \\
Bundesministerium f$\ddot{\mbox{\rm u}}$r Bildung, Wissenschaft,
Forschung und Technologie, Germany,\\
General Secretariat for Research and Technology, Greece, \\
National Science Foundation (NWO) and Foundation for Research on
Matter (FOM), The Netherlands, \\
Norwegian Research Council,  \\
State Committee for Scientific Research, Poland, SPUB-M/CERN/PO3/DZ296/2000,
SPUB-M/CERN/PO3/DZ297/2000, 2P03B 104 19 and 2P03B 69 23(2002-2004)\\
FCT - Funda\c{c}\~ao para a Ci\^encia e Tecnologia, Portugal, \\
Vedecka grantova agentura MS SR, Slovakia, Nr. 95/5195/134, \\
Ministry of Science and Technology of the Republic of Slovenia, \\
CICYT, Spain, AEN99-0950 and AEN99-0761,  \\
The Swedish Research Council,      \\
Particle Physics and Astronomy Research Council, UK, \\
Department of Energy, USA, DE-FG02-01ER41155, \\
EEC RTN contract HPRN-CT-00292-2002.\\

\newpage
\section*{Appendix 1}
We give hereafter efficiencies of the Yukawa \tautaubb\  analysis published 
in Ref.~\cite{ref:2hdm} and applied here to the \hAtt\ signal.

\begin{table} [htbp]
\begin{center}
\begin{tabular}{cc|cc||cc|c} \hline
\multicolumn{2}{c|}{mass (\GeVcc)} & efficiency & &  
\multicolumn{2}{c|}{mass (\GeVcc)} & efficiency \\ 
$m_A$ & $m_h$ & $(\%)$ & & $m_A$ & $m_h$ & $(\%)$ \\ \hline
   4 & 12 &       0.        && 12 & 4 &    0.        \\
   4 & 20 &  2.1$\pm$ 0.3 && 20 & 4 & 1.9$\pm$0.3\\
   4 & 30 &  2.3$\pm$ 0.3 && 30 & 4 & 2.0$\pm$0.3\\
   4 & 40 &  2.6$\pm$ 0.4 && 40 & 4 & 2.0$\pm$0.3\\
   4 & 50 &  1.4$\pm$ 0.3 && 50 & 4 & 1.6$\pm$0.3\\
   4 & 60 &  1.8$\pm$ 0.3 && 60 & 4 & 2.2$\pm$0.3\\
   4 & 70 &  1.3$\pm$ 0.3 && 70 & 4 & 0.9$\pm$0.2\\
   6 & 12 &       0.        && 12 & 6 &    0         \\
   6 & 20 &  2.2$\pm$ 0.3 && 20 & 6 & 2.0$\pm$0.3\\
   6 & 30 &  3.1$\pm$ 0.4 && 30 & 6 & 3.0$\pm$0.4\\
   6 & 40 &  2.6$\pm$ 0.4 && 40 & 6 & 2.8$\pm$0.4\\
   6 & 50 &  2.5$\pm$ 0.4 && 50 & 6 & 2.2$\pm$0.3\\
   6 & 60 &  3.1$\pm$ 0.4 && 60 & 6 & 2.5$\pm$0.4\\
   6 & 70 &  1.6$\pm$ 0.3 && 70 & 6 & 1.0$\pm$0.2\\
   9 & 12 &       0.        && 12 & 9 &    0.        \\
   9 & 20 &  2.9$\pm$ 0.4 && 20 & 9 & 2.7$\pm$0.4\\
   9 & 30 &  3.0$\pm$ 0.4 && 30 & 9 & 3.4$\pm$0.4\\
   9 & 40 &  3.4$\pm$ 0.4 && 40 & 9 & 3.0$\pm$0.4\\
   9 & 50 &  2.3$\pm$ 0.4 && 50 & 9 & 2.9$\pm$0.4\\
   9 & 60 &  2.4$\pm$ 0.4 && 60 & 9 & 1.8$\pm$0.3\\
   9 & 70 &  1.1$\pm$ 0.3 && 70 & 9 & 0.8$\pm$0.2\\
  12 & 12 &       0.&& 12 &12 &    0                 \\
  12 & 20 &  2.9$\pm$ 0.4 && 20 &12 & 2.6$\pm$0.4\\
  12 & 30 &  2.3$\pm$ 0.3 && 30 &12 & 2.4$\pm$0.4\\
  12 & 40 &  2.6$\pm$ 0.4 && 40 &12 & 2.0$\pm$0.3\\
  12 & 50 &  2.4$\pm$ 0.3 && 50 &12 & 2.4$\pm$0.4\\
  12 & 60 &  2.0$\pm$ 0.3 && 60 &12 & 1.9$\pm$0.3\\
  12 & 70 &  0.4$\pm$ 0.2 && 70 &12 & 0.5$\pm$0.2\\
 \end{tabular}
 \end{center}
\caption{\hAtt\ channel :
  efficiencies of the selection (in~\%) at {\sc LEP1} as a function 
  of the masses of the A and h bosons. The analysis, 
  described in Ref.~\cite{ref:2hdm}, was designed
  to search for Yukawa production in the \tautaubb\ final state.    
  The quoted errors are statistical only.}
 \label{bbtt_hA}
\end{table}

\newpage
\section*{Appendix 2}
We give hereafter efficiencies of the \hqq\ analyses published 
in Ref.~\cite{ref:pap99,ref:pap00} and applied to  
(\hAA \rgr \ccbar \ccbar )(\Zqq ) signals 
with low A masses.

\begin{table} [htbp]
\begin{center}
\begin{tabular}{cc|cccc}  \hline 
 \MA & \mh & Efficiency (\%)& 
 \multicolumn{2}{c}{Efficiency (\%) at 206.5~GeV } \\
(\GeVcc)   & (\GeVcc)  & at 199.6~GeV & first period & second period \\ \hline
  4.0 & 10.0   &  0.6 $\pm$ 0.1 &  0.4 $\pm$ 0.1 &  0.4 $\pm$ 0.1\\
  4.0 & 20.0   &  1.2 $\pm$ 0.1 &  1.6 $\pm$ 0.1 &  1.4 $\pm$ 0.1\\
  4.0 & 30.0   &  4.8 $\pm$ 0.2 &  4.9 $\pm$ 0.2 &  4.6 $\pm$ 0.2\\
  4.0 & 50.0   & 14.4 $\pm$ 0.4 & 15.2 $\pm$ 0.4 & 14.5 $\pm$ 0.4\\
  4.0 & 70.0   & 13.0 $\pm$ 0.4 & 13.9 $\pm$ 0.4 & 13.5 $\pm$ 0.4\\
  4.0 & 90.0   & 20.3 $\pm$ 0.4 & 19.3 $\pm$ 0.4 & 18.2 $\pm$ 0.4\\
  4.0 & 105.0  & 33.1 $\pm$ 0.5 & 27.7 $\pm$ 0.5 & 26.9 $\pm$ 0.4\\
  8.0 & 20.0   &  1.9 $\pm$ 0.1 &  2.6 $\pm$ 0.2 &  2.3 $\pm$ 0.2\\
  8.0 & 30.0   &  7.6 $\pm$ 0.3 &  8.3 $\pm$ 0.3 &  7.8 $\pm$ 0.3\\
  8.0 & 50.0   & 20.9 $\pm$ 0.5 & 21.0 $\pm$ 0.4 & 19.7 $\pm$ 0.4\\
  8.0 & 70.0   & 20.8 $\pm$ 0.4 & 20.8 $\pm$ 0.4 & 19.8 $\pm$ 0.4\\
  8.0 & 90.0   & 36.0 $\pm$ 0.5 & 32.8 $\pm$ 0.5 & 31.4 $\pm$ 0.5\\
  8.0 & 105.0  & 51.6 $\pm$ 0.5 & 44.6 $\pm$ 0.5 & 42.4 $\pm$ 0.5\\
\hline
\end{tabular}
\caption[]{(\hAA )(\Zqq ) channel with \Acc:
  efficiencies of the selection (in~\%)  at \rs~=~199.6 and 206.5~\GeV\ 
  as a function of the masses of the A and h bosons, for \mA\ between the
  \ccbar\ and \bbbar\ thresholds. Efficiencies at higher masses can be
  found in Ref.~\cite{ref:pap99,ref:pap00}. 
  We refer the reader to Ref.~\cite{ref:pap00}
  for the definition of the two operational periods of the 2000 data taking 
  campaign. The quoted errors are statistical only.}
\label{ta:aaqqeff}
\end{center}
\end{table}



\clearpage

\newpage

\clearpage

\end{document}